\documentclass[rppweb]{pdg}
\def\isweb{1}

\reviewtitle{Quantum Chromodynamics}
\reviewauthor{J. Huston (Michigan State U.), K. Rabbertz (KIT) and G. Zanderighi (MPI Munich)}
\reviewlabel{qcd}
\ischaptertrue
\apptocmd{\sloppy}{\hbadness 10000\relax}{}{}
\usepackage[utf8]{inputenc}

\usepackage[normalem]{ulem} % provides strike out \sout

\providecommand{\ARIADNE}{\textsc{ARIADNE}\xspace}
\providecommand{\BLACKHAT}{\textsc{BlackHat}\xspace}
\providecommand{\COMIX}{\textsc{Comix}\xspace}
\providecommand{\DYNNLO}{\textsc{Dynnlo}\xspace}
\providecommand{\DYTURBO}{\textsc{Dyturbo}\xspace}
\providecommand{\FEHIP}{\textsc{Fehip}\xspace}
\providecommand{\EERAD}{\textsc{Eerad3}\xspace}
\providecommand{\FEWZ}{\textsc{Fewz}\xspace}
\providecommand{\GOSAM}{\textsc{GoSam}\xspace}
\providecommand{\HELACNLO}{\textsc{Helac-NLO}\xspace}
\providecommand{\HELACPHEGAS}{\textsc{Helac/Phegas}\xspace}
\providecommand{\HNNLO}{\textsc{}\xspace}
\providecommand{\HTURBO}{\textsc{Hturbo}\xspace}
\providecommand{\MATRIX}{\textsc{Matrix}\xspace}
\providecommand{\MCFM}{\textsc{MCFM}\xspace}
\providecommand{\NLOJETPP}{\textsc{NLOJet++}\xspace}
\providecommand{\NJET}{\textsc{Njet}\xspace}
\providecommand{\OPENLOOPS}{\textsc{OpenLoops}\xspace}
\providecommand{\PHOX}{\textsc{Phox}\xspace}
\providecommand{\RECOLA}{\textsc{Recola}\xspace}
\providecommand{\SUSHI}{\textsc{Sushi}\xspace}
\providecommand{\VBFNLO}{\textsc{VBFNLO}\xspace}

\providecommand{\alps}{\ensuremath{\alpha_s}\xspace}
\providecommand{\asmz}{\ensuremath{\alpha_s(\mZ^2)}\xspace}
\providecommand{\MSbar}{\ensuremath{\overline{\textrm{MS}}}\xspace}

\providecommand{\mtau}{\ensuremath{m_\tau}\xspace}
\providecommand{\mtop}{\ensuremath{m_{t}}\xspace}

\usepackage{makeidx}
\makeindex

\ifdefined\isbooklet
\externalcitedocument{qcd-full}
\fi

\IfFileExists{crossref.aux}{\externaldocument{crossref}}{}

\begin{document}

% Set starting page number for book and booklet
\ifdefined\isbook
\setcounter{page}{149}
\fi
\ifdefined\isbooklet
\setcounter{page}{206}
\fi

% Enable bleeder tabs in book mode
\ifdefined\isbook
\rppthumb
\fi

% Include -main or -booklet file, depending on what we're making
\ifdefined\isbooklet
\fontsize{8pt}{9pt}\selectfont
\input{qcd-booklet}
\else
\begin{bibunit}
% $Id: BASENAME-main.tex 23686 2019-02-20 20:00:48Z beringer $
% Main file for PDG review qcd.
%
% This file is included by the top-level file qcd.tex and is where the
% text of your review should be included. If desired, you may split your review into multiple
% files that are included from this file using \input.
%
% Do NOT modify the top-level file qcd.tex - it is generated from
% the PDG database and all manual changes WILL BE LOST!
% ===========================================================================

% Review title
% ------------
% Please use \pdgtitle (rather than e.g. \chapter) to put the title of your review.
%
% To put the review title extracted from the PDG database, use \pdgtitle (no arguments).
% To override the default review title, you can use \pdgtitle[Some different title].
% In the latter case, please ask your overseer to update the review title in the database
\pdgtitle

% Author information for this review
% ----------------------------------
% Please use one of the following forms:
%   \written{month year}
%   \revised{month year}
%   \customauthor{...}
% The first two, \written and \revised take the month and year when the review
% was written or revised as their only argument. The author list is generated
% from the PDG database. This is the preferred way of including author information.
% Only if really needed, you may use the third from, \customauthor, where you can
% specify the full text of the paragraph giving the review author information.
\revised{August 2023}

% Table of contents
% -----------------
% If you want to include a table of contents at the start of your review,
% uncomment the line below. This should only be done for relatively long
% reviews.
\tableofcontents

% Sectioning
% ----------
% This review is a regular review, please use \section for your top-level sectioning.
% For example:
%\section{Your first section title}

% Text of your review
% -------------------
% Please see the instructions at https://pdgdoc.lbl.gov/Pdg/ReviewTool on how to
% include figures, tables, and references. By default we include file examples.tex,
% which provides instructions and examples of how to include figures, tables, how
% to align equations, and more. It also produces an appendix showing all the standard
% PDG symbols and what they produce.
%
% To remove the examples from your review, comment out the following line when you
% start writing your review.
%\input{examples.tex}

\newcount\BleederPointer
\newcount\WhichSection%
%\renewcommand{\thefootnote}{\fnsymbol{footnote}}
%\setfootnotestyle{\fnsymbol}
\BleederPointer=7
\ifnum\WhichSection=7
\magnification=\magstep1
\fi
\overfullrule=5pt

%===============================================================================
% Start of review
%===============================================================================

\section{Basics}
\label{qcd:sec:basics-GS}

Quantum Chromodynamics (QCD), the gauge field theory that describes the strong interactions of colored quarks and
gluons, is the SU(3) component of the SU(3)$\times$SU(2)$\times$U(1) Standard Model of Particle Physics.  The Lagrangian
of QCD\index{Lagrangian, QCD}\index{QCD!Lagrangian} is given by
\begin{equation}\label{qcd:eq:lagrangian}
  {\cal L} =
  \sum_q
  {\bar\psi}_{q,a} (i \gamma^\mu\partial_\mu \delta_{ab} -
  g_s \gamma^\mu t^C_{ab} {\cal A}^C_\mu - m_q \delta_{ab})\psi_{q,b}
  -\frac14 F_{\mu \nu}^A F^{A\,\mu \nu}\, ,
\end{equation}
where repeated indices are summed over.  The $\gamma^\mu$ are the Dirac $\gamma$-matrices.  The $\psi_{q,a}$ are
quark-field spinors for a quark of flavor $q$ and mass $m_q$, with a color-index $a$ that runs from $a=1$ to $N_c=3$,
\ie quarks come in three ``colors.'' Quarks are said to be in the fundamental representation of the SU(3) color group.

The ${\cal A}^{C}_\mu$ correspond to the gluon fields, with $C$ running from $1$ to $N_c^2 - 1 = 8$, \ie there are eight
kinds of gluon.  Gluons transform under the adjoint representation of the SU(3) color group.  The $t^{C}_{ab}$
correspond to eight $3\times3$ matrices and are the generators of the SU(3) group (\cf the section on ``SU(3) isoscalar
factors and representation matrices'' in this {\it Review}, with $t^C_{ab} \equiv \lambda^C_{ab} / 2$).  They encode
the fact that a gluon's interaction with a quark rotates the quark's color in SU(3) space.  The quantity $g_s$ (or
$\alps = \frac{g_s^2}{4\pi}$) is the QCD coupling constant.\index{Coupling constant in QCD}\index{QCD!coupling constant}
Besides quark masses, which have electroweak origin, it is the only fundamental parameter of QCD.  Finally, the field
tensor $F^A_{\mu\nu}$ is given by
\begin{align}
  F^A_{\mu\nu} &= \partial_\mu {\cal A}_{\nu}^A - \partial_\nu {\cal A}_{\mu}^A
                 - g_s\, f_{ABC} {\cal A}^B_\mu {\cal A}^C_\nu, \cr
                 [t^A,t^B] &=  i f_{ABC}t^C,
\end{align}
where the $f_{ABC}$ are the structure constants of the SU(3) group.

Neither quarks nor gluons are observed as free particles. Hadrons are color-singlet (\ie color-neutral) combinations of
quarks, anti-quarks, and gluons.

Ab-initio predictive methods for QCD include lattice gauge theory and perturbative expansions in the coupling.  The
Feynman rules of QCD involve a quark-antiquark-gluon ($q\bar qg$) vertex, a 3-gluon vertex (both proportional to $g_s$),
and a 4-gluon vertex (proportional to $g_s^2$).  A full set of Feynman rules is to be found for example in
Refs.~\cite{Ellis:1991qj,qcd:CHKbook}.

Adopting a standard notation where repeated indices are summed over, useful color-algebra relations include:
$t^A_{ab} t^A_{bc} = C_F \delta_{ac}$, where $C_F \equiv (N_c^2-1)/(2N_c) = 4 / 3$ is the color factor\index{color
  factor}\index{QCD!color factor} (``Casimir'') associated with gluon emission from a quark;
$f_{ACD} f_{BCD} = C_A \delta_{AB}$, where $C_A \equiv N_c = 3$ is the color factor associated with gluon emission from
a gluon; $t^A_{ab} t^B_{ab} = T_R \delta_{AB}$, where $T_R = 1 / 2$ is the color factor for a gluon to split to a
$q\bar q$ pair.

There is freedom for an additional $CP$-violating term to be present in the QCD Lagrangian,
$\theta \frac{\alps}{8\pi} F^{A\,\mu\nu} {\tilde F}^A_{\,\mu\nu}$, where $\theta$ is an additional free parameter, and
${\tilde F}^A_{\mu\nu}$ is the dual of the gluon field tensor,
$\frac12 \epsilon_{\mu \nu \sigma \rho} F^{A\,\sigma\rho}$, with $\epsilon_{\mu \nu \sigma \rho}$ being the fully
antisymmetric Levi-Civita symbol.  Experimental limits on the electric dipole moment (EDM) of ultracold
neutrons\cite{Baker:2006ts,Afach:2015sja} and atomic mercury\cite{Graner:2016ses} constrain the QCD vacuum angle to
satisfy $|\theta| \lesssim 10^{-10}$.  For an overview of current and future experiments measuring EDMs see~\cite{EDMs}.
Further discussion is to be found in Ref.~\cite{Kim:2008hd} and in the Axions section in the Listings of this {\it
  Review}.

This section will concentrate mainly on perturbative aspects of QCD as they relate to collider physics. Related
textbooks and lecture notes include
Refs.~\cite{Ellis:1991qj,Dissertori:2003pj,Brock:1993sz,qcd:CHKbook,Melnikov:2018qpx,Gross:2022hyw}.
Aspects specific to Monte Carlo event generators are reviewed in the dedicated section~\crossref{mcgen}.  Lattice QCD is
also reviewed in a section of its own, Sec.~\crossref{lqcd}, with further discussion of perturbative and
non-perturbative aspects to be found in the sections on ``Quark Masses'', ``The CKM quark-mixing matrix'', ``Structure
functions'', ``Fragmentation Functions'', ``High Energy Soft QCD and Diffraction'',
``Passage of Particles Through Matter'' % GZ ADDED
and ``Heavy-Quark and Soft-Collinear Effective Theory'' in this {\it Review}.

%----------------------------------------------------------------------
\subsection{Running coupling}\index{running coupling in QCD}\index{QCD!running coupling}

In the framework of perturbative QCD (pQCD), predictions for observables are
expressed in terms of the renormalized coupling $\alps(\mu_R^2)$, a
function of an (unphysical) renormalization scale $\mu_R$.
When one takes $\mu_R$ close to the scale of the momentum transfer $Q$ in
a given process, then $\alps(\mu_R^2\simeq Q^2)$ is indicative of the
effective strength of the strong interaction in that process.

The coupling satisfies the following
renormalization group equation (RGE):\index{Renormalization group equation (RGE) of QCD}\index{QCD!renormalization group equation (RGE)}
\begin{equation}\label{qcd:eq:rgeqcd}
  \mu_R^2 \frac{d\alps}{d\mu_R^2} = \beta(\alps) = - (b_0 \alps^2 + b_1 \alps^3 + b_2 \alps^4 + \cdots)\,,
\end{equation}
where $b_0 = (11C_A - 4 n_f T_R)/(12\pi) = (33-2n_f)/(12\pi)$ is referred to as the 1-loop $\beta$-function
coefficient,\index{beta@$\beta$ function of QCD}\index{QCD!beta@$\beta$ function} the 2-loop coefficient is
$b_1 = (17C_A^2 - n_f T_R (10C_A + 6C_F))/(24\pi^2) = (153-19n_f)/(24\pi^2)$, and the 3-loop coefficient is
$b_2 = (2857 - \frac{5033}{9} n_f + \frac{325}{27} n_f^2)/(128 \pi^3)$ %in the $\overline{MS}$-scheme
for the SU(3) values of $C_A$ and $C_F$. Here $n_f$ is the number of quark flavors.  The 4-loop coefficient, $b_3$, is
to be found in Refs.~\cite{vanRitbergen:1997va,Czakon:2004bu}, while the 5-loop coefficient, $b_4$, is in
Refs.~\cite{Baikov:2016tgj,Luthe:2016ima,Herzog:2017ohr,Luthe:2017ttg,Chetyrkin:2017bjc}.  The coefficients $b_2$ and
$b_3$ (and beyond) are renormalization-scheme-dependent and given here in the modified minimal subtraction scheme
(\MSbar)\cite{Bardeen:1978yd},\index{Modified minimal subtraction scheme (\MSbar)}\index{QCD!\MSbar renormalization
  scheme} by far the most widely used scheme in QCD and the one adopted in the following.

The minus sign in \Eq{qcd:eq:rgeqcd} is the origin of ``asymptotic
freedom''\cite{Gross:1973id,Politzer:1973fx},\index{asymptotic freedom}
\ie the fact that the strong coupling becomes weak for
processes involving large momentum transfers (``hard processes'').
For momentum transfers in the $0.1$--$1$\TeV range,
$\alps \sim 0.1$, while the theory is strongly interacting for
scales around and below $1$\GeV.

The $\beta$-function coefficients, the $b_i$, are given for the coupling of an {\it effective theory} in which $n_f$ of
the quark flavors are considered light ($m_q \ll \mu_R$), and in which the remaining heavier quark flavors decouple from
the theory.  One may relate the coupling for the theory with $n_f+1$ light flavors to that with $n_f$ flavors through an
equation of the form
\begin{equation}\label{qcd:eq:vfns}
  \begin{split}
    \alps^{(n_f+1)}(\mu_R^2) =
    \bookalign\alps^{(n_f)}(\mu_R^2) \bookcr{\times}{&\times}\left(1 +
      \sum_{n=1}^\infty \sum_{\ell=0}^{n} c_{n\ell}\, [\alps^{(n_f)}(\mu_R^2)]^n
      \ln^\ell  \frac{\mu_R^2}{m_h^2} \right)\,,
  \end{split}
\end{equation}
where $m_h$ is the mass of the $(n_f\!+\!1)^{\rm{th}}$ flavor, and the first few $c_{n\ell}$ coefficients are
$c_{11}= {\frac{1}{6\pi}}$, $c_{10}=0$, $c_{22} = c_{11}^2$, $c_{21} = {\frac{11}{24\pi^2}}$, and
$c_{20} = -{\frac{11}{72\pi^2}}$ when $m_h$ is the \MSbar mass at scale $m_h$, while $c_{20} = {\frac{7}{24\pi^2}}$ when
$m_h$ is the pole mass (mass definitions are discussed below in Sec.~(\ref{qcd:sec:qcd-quarkmasses}) and in the review
on ``Quark Masses'').  Terms up to $c_{4\ell}$ with $0 \leq\ell\leq 4$ are to be found in
Refs.~\cite{Schroder:2005hy,Chetyrkin:2005ia}.  Numerically, when one chooses $\mu_R=m_h$, the matching is a modest
effect, owing to the zero value for the $c_{10}$ coefficient.  Relations between $n_f$ and $(n_f\!+\!2)$ flavors where
the two heavy flavors are close in mass are given to three loops in Ref.~\cite{Grozin:2011nk}.

Working in an energy range where the number of flavors is taken constant, a simple exact analytic solution exists for
\Eq{qcd:eq:rgeqcd} only if one neglects all but the $b_0$ term, giving
$\alps(\mu_R^2) = (b_0 \ln (\mu_R^2 / {\rm \Lambda}^2))^{-1}$. Here ${\rm \Lambda}$ is a constant of integration, which
corresponds to the scale where the perturbatively-defined coupling would diverge.\index{Lambda@${\rm \Lambda}$ QCD
  parameter}\index{QCD!Lambda@${\rm \Lambda}$ parameter} Its value is indicative of the energy range where
non-perturbative dynamics dominates.  A convenient approximate analytic solution to the RGE that includes terms up to
$b_4$ is given by solving iteratively \Eq{qcd:eq:rgeqcd}
\begin{align}
  \label{qcd:eq:nlo-solution}
  \alps(\mu_R^2)& \simeq
                  \frac{1}{b_0 t}
                  \left(
                  1 - \frac{b_1}{b_0^2}\frac{\ell}{t}
                  + \frac{b_1^2(\ell^2 - \ell - 1) + b_0b_2 }{b_0^4 t^2}\, +
                  \right.\cr
  & + \frac{b_1^3 \left(-2 \ell^3+5 \ell^2+4 \ell
    -1\right)-6 b_0 b_2 b_1 \ell+b_0^2 b_3}{2 b_0^6 t^3} +\cr
    +
  & \frac{18 b_0 b_2 b_1^2 \left(2 \ell^2\!-\!\ell \!-\!1\right)\!+\!b_1^4 \left(6 \ell ^4\!-\!26 \ell ^3\!-\!9 \ell^2\!+\!24 \ell \!+\!7\right)}{6 b_0^8 t^4}\cr
  &+\left.
    \frac{-\!b_0^2 b_3 b_1 (12 \ell \!+\!1)\!+\!2 b_0^2 \left(5 b_2^2\!+\!b_0 b_4\right)}{6 b_0^8 t^4} \right)\!,
\end{align}
with $t \equiv \ln \frac{\mu_R^2}{\rm \Lambda^2}$ and $\ell = \ln t$, again parameterized in terms of a constant
${\rm \Lambda}$.  Note that \Eq{qcd:eq:nlo-solution} is one of several possible approximate 4-loop solutions for
$\alps(\mu_R^2)$, and that a value for ${\rm \Lambda}$ only defines $\alps(\mu_R^2)$ once one knows which particular
approximation is being used.  An alternative to the use of formulas such as \Eq{qcd:eq:nlo-solution} is to solve the RGE
exactly, numerically (including the discontinuities, \Eq{qcd:eq:vfns}, at flavor thresholds).  In such cases the
quantity ${\rm \Lambda}$ does not directly arise (though it can be defined, \cf~Eqs.~(1--3) of
Ref.~\cite{Brida:2016flw}).  For these reasons, in determinations of the coupling, it has become standard practice to
quote the value of \alps at a given scale (typically the mass of the $Z$ boson, \mZ) rather than to quote a value for
${\rm \Lambda}$.

A discussion of determinations of the coupling and a graph illustrating its scale dependence (``running'') are to be
found in~\Section{qcd:sec:coupling-measurement}.

%----------------------------------------------------------------------
\subsection{Quark masses}
\label{qcd:sec:qcd-quarkmasses}

Free quarks have never been observed, which is understood as a result of a long-distance, confining property of the
strong QCD force: up, down, strange, charm, and bottom quarks all {\it hadronize}, \ie become part of a meson or baryon,
on a timescale $\sim 1/{\rm \Lambda}$; the top quark instead decays before it has time to hadronize.  This means that
the question of what one means by the quark mass is a complex one, which requires one to adopt a specific prescription.
A perturbatively defined prescription is the pole mass,\index{pole mass} $m_q$, which corresponds to the position of the
divergence of the quark propagator.  This is close to one's physical picture of mass. However, when relating it to
observable quantities, it suffers from a badly behaved perturbative series which makes it ambiguous to an amount related
to $\rm \Lambda$ (see \eg Ref.~\cite{Beneke:1998ui,Beneke:2016cbu,Hoang:2017btd}).  An alternative is the \MSbar mass,
\index{MSbar@\MSbar mass} $\overline m_q(\mu_R^2)$, which depends on the renormalization scale $\mu_R$.

Results for the masses of heavier quarks are often quoted either as the pole mass or as the \MSbar mass evaluated at a
scale equal to the mass, $\overline m_q(\overline m_q^2)$; light quark masses are often quoted in the \MSbar scheme at a
scale $\mu_R \sim 2$\GeV.  The pole and \MSbar masses are related by a series that starts as
$m_q = \overline m_q(\overline m_q^2)(1 + \frac{4\alps(\overline m_q^2)}{3\pi} + {\cal O}(\alps^2))$, while the
scale-dependence of \MSbar masses is given at lowest order by

\begin{equation}\label{qcd:eq:quark-mass-running}
  \mu_R^2 \frac{d \overline m_q(\mu_R^2)}{d\mu_R^2} =
  \Biggl[ -\frac{\alps(\mu_R^2)}{\pi} + {\cal O}(\alps^2)\Biggr] \overline m_q(\mu_R^2)\,.
\end{equation}
A more detailed discussion is to be found in a dedicated section of the {\it Review}, ``Quark Masses'', with detailed
formulas also in Ref.~\cite{Marquard:2015qpa} and references therein.

In perturbative QCD calculations of scattering processes, it is customary to employ an approximation where the masses of
all quarks, whose magnitudes are much smaller than the momentum transfer involved in the process, are neglected or set
to zero.

% ======================================================================
\section{Structure of QCD predictions}

%======================================================================
\subsection{Fully inclusive cross sections}
\label{qcd:sec:inclcrosssec}

The simplest observables in perturbative QCD are those that do not depend on the initial-state hadrons and are fully
inclusive, disregarding specific details of the final state. An example of such an observable is the total cross section
for the process $e^+e^- \to \rm{hadrons}$ at a center-of-mass energy $Q$. It can be expressed as follows:
\begin{equation}\label{qcd:eq:xsct-ee}
  \frac{\sigma(e^+e^- \to {\rm hadrons},Q)}{\sigma(e^+e^- \to
    \mu^+\mu^-,Q)}
  \equiv R(Q) =
  R_{\rm EW}(Q) (1 + \delta_{\rm QCD}(Q))\,,
\end{equation}
where $R_{\rm EW}(Q)$ is the purely electroweak prediction for the ratio and $\delta_{\rm QCD}(Q)$ is the correction due
to QCD effects.

For the sake of simplicity, we can focus on energies where $Q \ll \mZ$, where the primary contribution to the process
comes from photon exchange (while disregarding electroweak and finite-quark-mass corrections) denoted as
$R_{\rm EW} = N_c \sum_q e_q^2$, with $e_q$ representing the electric charges of the quarks. The QCD correction reads
\begin{equation}\label{qcd:eq:rhadrons-series}
  \delta_{\rm QCD}(Q) = \sum_{n=1}^\infty c_n\cdot
  \left(\frac{\alps(Q^2)}{\pi}\right)^n + {\cal O}\left(\frac{{\rm \Lambda}^4}{Q^4}\right)\;.
\end{equation}
The first four terms in the \alps series expansion are~\cite{Baikov:2012zn}
\begin{subequations}\label{qcd:eq:ascoefs}
  \begin{align}
    c_1 &= 1\,,
          \quad\qquad c_2 = 1.9857 - 0.1152 n_f\,,\\
    c_3 &= -6.63694- 1.20013n_f -
          0.00518n^2_f  - 1.240 \eta\, ,\\
    c_4 &=-156.61 + 18.775 n_f - 0.7974 n^2_f + 0.0215 n_f^3  %\cr
          - (17.828 - 0.575n_f) \eta\, ,
  \end{align}
\end{subequations}
with $\eta = (\sum e_q)^2/(N_c\sum e_q^2)$ and $N_c=3$.  For corresponding expressions including also $Z$ exchange and
finite-quark-mass effects, see Refs.~\cite{Chetyrkin:1994js,Kiyo:2009gb,Baikov:2012er}.

A related series holds also for the QCD corrections to the hadronic decay width of the $\tau$ lepton, which essentially
involves an integral of $R(Q)$ over the allowed range of invariant masses of the hadronic part of the $\tau$ decay (see
\eg Ref.~\cite{Baikov:2008jh}).  The series expansions for QCD corrections to Higgs-boson hadronic (partial) decay
widths in the limit of heavy top quark and massless light flavors at N$^4$LO are given in Ref.~\cite{Herzog:2017dtz}.

One characteristic feature of Eqs.~(\ref{qcd:eq:rhadrons-series}) and (\ref{qcd:eq:ascoefs}) is that the coefficients of
$\alps^n$ increase order by order: calculations in perturbative QCD tend to converge more slowly than would be expected
based just on the size of \alps. The situation is significantly worse near thresholds or in the presence of tight
kinematic cuts.  Another feature is the existence of an extra ``power-correction''\index{power correction}
term ${\cal O}({\rm \Lambda}^4 / Q^4)$ in \Eq{qcd:eq:rhadrons-series}, which accounts for contributions that are
fundamentally non-perturbative.  All high-energy QCD predictions involve power corrections $({\rm \Lambda}/Q)^p$,
although typically the suppression of these corrections with $Q$ is smaller than given in \Eq{qcd:eq:rhadrons-series},
where $p=4$.
The exact power $p$ depends on the observable and, for many processes and observables, it is possible to introduce an
operator product expansion and associate power-suppressed terms with specific higher-dimension (non-perturbative)
operators\cite{Novikov:1980uj}.

\noindent{\bf Scale dependence.}
\index{scale dependence}
In \Eq{qcd:eq:rhadrons-series} the renormalization scale for \alps has been chosen equal to $Q$. The result can also be
expressed in terms of the coupling at an arbitrary renormalization scale $\mu_R$,

\begin{equation}\label{qcd:eq:rhadrons-series-any-mu}
  \delta_{\rm QCD}(Q) = \sum_{n=1}^\infty \overline c_n\!\!\left(\frac{\mu_R^2}{Q^2}\right)\cdot
  \left(\frac{\alps(\mu_R^2)}{\pi}\right)^n + {\cal O}\left(\frac{{\rm \Lambda}^4}{Q^4}\right) \,,
\end{equation}
where $\overline c_1(\mu_R^2/Q^2) \equiv c_1$, $\overline c_2(\mu_R^2/Q^2) = c_2 + \pi b_0 c_1 \ln(\mu_R^2/Q^2)$,
$\overline c_3(\mu_R^2/Q^2) = c_3 + (2b_0 c_2 \pi + b_1 c_1 \pi^2)$
$\times\ln(\mu_R^2/Q^2) + b_0^2 c_1 \pi^2 \ln^2(\mu_R^2/Q^2)$, \etc
Given an infinite number of terms in the \alps expansion, the $\mu_R$ dependence of the $\overline c_n(\mu_R^2/Q^2)$
coefficients will exactly cancel that of $\alps(\mu_R^2)$, and the final result will be independent of the choice of
$\mu_R$: physical observables do not depend on unphysical scales.%
\footnote{ With respect to pQCD there is an important caveat to this statement: at sufficiently high orders,
  perturbative series generally suffer from ``renormalon''\index{Renormalon} divergences $\alps^n n!$ (reviewed in
  Ref.~\cite{Beneke:1998ui}).  This phenomenon is not usually visible with the limited number of perturbative terms
  available today.  However, it is closely connected with non-perturbative contributions and sets a limit on the
  possible precision of perturbative predictions.  The cancellation of scale dependence will also ultimately be affected
  by this renormalon-induced breakdown of perturbation theory.} %

With just terms up to some finite $n=N$, a residual $\mu_R$ dependence will remain, which implies an uncertainty on the
prediction of $R(Q)$ due to the arbitrariness of the scale choice.  This uncertainty will be ${\cal O}(\alps^{N+1})$,
\ie of the same order as the neglected higher-order terms.  For this reason it is customary to use the scale dependence
of QCD predictions as an estimate of the uncertainties due to neglected terms.  One usually takes a central value for
$\mu_R \sim Q$, in order to avoid the poor convergence of the perturbative series that results from the large
$\ln^{n-1}(\mu_R^2/Q^2)$ terms in the $\overline c_n$ coefficients when $\mu_R \ll Q$ or $\mu_R \gg Q$.  Uncertainties
are then commonly determined by varying $\mu_R$ by a factor of two up and down around the central scale choice. This is
not a rigorous prescription for determining the uncertainty, but is motivated by the requirement that there should not
be large logarithms introduced into a calculation by large ratios of scales.  A more detailed discussion on the accuracy
of theoretical predictions and on ways to estimate the theoretical uncertainties can be found
in \Section{qcd:sec:qcd-accuracy}.

% ----------------------------------------------------------------------
\subsection{Processes with initial-state hadrons}

% ......................................................................
\noindent{\bf Deep-Inelastic Scattering.}
\index{deep-inelastic scattering (DIS)}
To illustrate the key features of QCD cross sections in processes with initial-state hadrons, let us consider
deep-inelastic scattering (DIS), $e+p \to e + X$, where an electron $e$ with four-momentum $k$ emits a highly off-shell
photon (momentum $q$) that interacts with the proton (momentum $p$). For photon virtualities $Q^2 \equiv -q^2$ far above
the squared proton mass (but far below the squared $Z$ mass), the differential cross section in terms of the kinematic
variables $Q^2$, $x=Q^2/(2p\cdot q)$ and $y=(q\cdot p)/(k\cdot p)$ is
\begin{equation}\label{qcd:eq:DIS-Xsct}
  \frac{d^2\sigma}{dx dQ^2} = \frac{4 \pi \alpha^2}{2xQ^4} \left[
    (1 + (1-y)^2) F_2(x,Q^2)  - y^2 F_L(x,Q^2)
  \right]\,,
\end{equation}
where $\alpha$ is the electromagnetic coupling and $F_2(x,Q^2)$ and $F_L(x,Q^2)$ are proton structure functions,
\index{proton structure functions} which encode the interaction between the photon and the proton.  In the presence of
parity-violating interactions (\eg $\nu p$ scattering) an additional $F_3$ structure function is present.  For an
extended review, including equations for the full electroweak and polarized cases, see Sec.~\crossref{strucfun} of this
{\it Review}.

Structure functions are not calculable in perturbative QCD, nor is any other cross section that involves QCD
interactions and initial-state hadrons.
To zeroth order in \alps, the structure functions are given directly in terms of non-perturbative parton (quark or
gluon) distribution functions (PDFs),\index{parton distribution function (PDF)}
\begin{equation}
  \label{qcd:eq:F2-quark-parton-mode}
  F_2(x,Q^2) = x \sum_q e_q^2 f_{q/p}(x)\,,\qquad F_L(x,Q^2) = 0\,,
\end{equation}
where $f_{q/p}(x)$ is the non-perturbative PDF for quarks of type $q$ inside the proton, \ie the number density of
quarks of type $q$ inside a fast-moving proton that carry a fraction $x$ of its longitudinal momentum.

PDFs are primarily determined from data in global fits. PDF sets are
available at different orders in perturbation theory (LO, NLO and
NNLO) based on the order at which cross sections used in the fits are
calculated. The evolution equations for the PDFs from one scale to
another matches the perturbative accuracy of the cross sections.
In modern global PDF fits, data are included from DIS, Drell-Yan (DY), jets and
\ttbar processes, and more LHC collider data, with the global PDF fits using 3000-4000 data points. There is a large
change in the PDFs from LO to NLO, with a much smaller change from NLO to NNLO. LO PDFs can be unreliable for collider
predictions, especially at low and high $x$.  The uncertainties for the resulting PDFs are determined primarily from the
experimental uncertainties of the data that serve as input to the global PDF fits. There is also a component due to the
limitations of the parameterizations used, as discussed \eg in Ref.~\cite{Hou:2019efy}, although this may be reduced with the use of more flexible forms. The PDF
uncertainties can either be determined through a Hessian approach or through the use of Monte Carlo replicas. It is now
relatively straightforward to convert results from one approach to the other. There has been a recent combination
(PDF4LHC21) of the CT18, MSHT20 and NNPDF3.1 PDFs, each produced from a global PDF fit, that serves to provide a more
comprehensive view of both the central PDFs and of their uncertainties~\cite{PDF4LHCWorkingGroup:2022cjn}.  Recently,
theoretical uncertainties related to missing higher orders have been included in global PDF determinations but so far
only at NLO \cite{NNPDF:2019vjt,NNPDF:2019ubu,Ball:2021icz,Kassabov:2022orn}.  By tying renormalization scales for
related processes, and factorization scales for all processes, together in the global fit, an estimate of the higher
order uncertainties can be added to the PDF fit. An alternative approach~\cite{Harland-Lang:2018bxd} is to attempt to deal with cross sections rather than PDFs
themselves.

Another approach is to use as much available information as possible from N$^3$LO to examine the impact of such
information on the resultant PDFs ~\cite{McGowan:2022nag}. The result, of course, is sensitive to the higher order
information that is still missing, for example from an incomplete knowledge of the splitting functions and the N$^3$LO
matrix elements.

Since the original work of Ref.~\cite{Ji:2013dva}, one of the intriguing questions is whether lattice calculations can
provide valuable insights for extracting PDFs. Despite initial debates surrounding the underlying
methodologies\cite{Rossi:2017muf,Rossi:2018zkn}, it is now understood that lattice data in Euclidean space can be
connected to light-cone PDFs through factorization theorems, once lattice observables have been renormalized and
extrapolated to the continuum limit~\cite{DelDebbio:2020cbz}. Initial determinations based on lattice data have begun to
emerge\cite{Lin:2014zya,*Alexandrou:2015rja,Lin:2017snn,Alexandrou:2018pbm,Chen:2018xof,Cichy:2019ebf,Cichy:2018mum,Ebert:2019okf,DelDebbio:2020cbz,Constantinou:2020pek,Constantinou:2020hdm,DelDebbio:2020rgv,Fan:2020cpa,HadStruc:2021wmh,Fan:2022kcb},
and the comparisons with determinations based on conventional data show great promise. However, the uncertainties in
lattice-based determinations are still quite large, therefore, for practical applications, PDFs are currently derived
from global fits of experimental data (\cf Sec.~\ref{strucfun} of this {\it Review} and also
Refs.~\cite{Gao:2017yyd,Kovarik:2019xvh}). It is likely that lattice simulations for light-cone PDFs will have an impact
in the short-medium term if they focus on flavor components (strange and charm) and on the kinematic regions that are poorly covered by the
experimental data~\cite{Hou:2022onq}.  Similarly, lattice determinations of transverse momentum dependent PDF (TMDs) and
generalized parton distributions (GPDs) will be more relevant for phenomenology since the experimental data for these
quantities are much less abundant and precise.

The result in Eq.\eqref{qcd:eq:F2-quark-parton-mode}, with PDFs $f_{q/p}(x)$ that are independent of the scale $Q$,
corresponds to the ``quark-parton model''\index{quark-parton model} picture, in which the photon interacts with
point-like free quarks, or equivalently, one has incoherent elastic scattering between the electron and individual
constituents of the proton.  As a consequence, in this picture also $F_2$ and $F_L$ are independent of
$Q$\cite{Bjorken:1969ja}.  When including higher orders in pQCD, one has
\begin{equation}\label{qcd:eq:F2-QCD}
  \begin{split}
    \bookalign F_2(x,Q^2) = x \sum_{n=0}^\infty \frac{\alps^n(\mu_R^2)}{(2\pi)^n} \bookcr{\times}{&\times }\sum_{i=q,g}
    \int_x^1 \frac{dz}{z} C_{2,i}^{(n)}(z, Q^2, \mu_R^2, \mu_F^2)f_{i/p}\Bigl(\frac{x}{z}, \mu_F^2\Bigr) + {\cal
      O}\Bigl(\frac{{\rm \Lambda}^2}{Q^2}\Bigr)\,.%\cr
  \end{split}
\end{equation}

Just as in \Eq{qcd:eq:rhadrons-series-any-mu}, we have a series in powers of $\alps(\mu_R^2)$, each term involving a
coefficient $C_{2,i}^{(n)}$ that can be calculated using Feynman diagrams. At variance with the parton model, the PDFs in
pQCD depend on an additional scale, the factorization scale $\mu_F$,\index{factorization scale} whose significance will
be discussed in the following.  Another important difference is the additional integral over $z$.  The parton that comes
from the proton can undergo a splitting before it enters the hard scattering.  As a result, the $C_{2,i}^{(n)}$
coefficients are functions that depend on the ratio, $z$, of the parton's momentum before and after radiation, and one
must integrate over that ratio.  For the electromagnetic component of DIS with light quarks and gluons, the zeroth order
coefficient functions are $C_{2,q}^{(0)}= e_q^2 \delta(1-z)$ and $C_{2,g}^{(0)} = 0$.  Corrections are known up to
${\cal O}(\alps^3)$ (next-to-next-to-next-to-leading order, N$^3$LO) for both electromagnetic\cite{Vermaseren:2005qc}
and weak currents\cite{Moch:2008fj,Davies:2016ruz}.  For heavy-quark production they are known to
${\cal O}(\alps^2)$\cite{Laenen:1992zk,*Riemersma:1994hv,Blumlein:2019qze}\ (next-to-leading order, NLO, insofar as the
series starts at ${\cal O}(\alps)$). The current status of the theoretical description of unpolarized and polarized DIS
is summarized in Ref.~\cite{Blumlein:2023aso}.  For precise comparisons of LHC cross sections with theoretical
predictions, the photon PDF\index{photon PDF} of the proton is also needed. It has been computed precisely in
Ref.~\cite{Manohar:2016nzj,Manohar:2017eqh} and has now been implemented in most global PDF fits. More recently, the PDF
for leptons\index{lepton PDF} has also been computed~\cite{Buonocore:2020nai}. The photon PDF has an elastic and an inelastic component,
where the photon is generated from the intact proton or from the splitting
of quarks inside the proton, respectively, while the lepton PDF is
generated by photons splitting to collinear leptons.

The majority of the emissions that modify a parton's momentum are collinear (parallel) to that parton, and do not depend on whether
the parton will interact with a photon.  It is natural to view these emissions as modifying
the proton's structure rather than being part of the coefficient function for the parton's interaction with the photon.
Technically, one uses a procedure known as {\it collinear factorization}\index{collinear factorization} to give a
well-defined meaning to this distinction, most commonly through the \MSbar factorization scheme,\index{MSbar@\MSbar
  factorization scheme} defined in the context of dimensional regularization.  The \MSbar factorization scheme involves
an arbitrary choice of {\it factorization scale},\index{factorization scale} $\mu_F$, whose meaning can be understood
roughly as follows: emissions with transverse momenta above $\mu_F$ are included in the
$C_{2,q}^{(n)}(z, Q^2, \mu_R^2,\mu_F^2)$; emissions with transverse momenta below $\mu_F$ are accounted for within the
PDFs, $f_{i/p}(x,\mu_F^2)$.  While collinear factorization is generally believed to be valid for suitable (sufficiently
inclusive) observables in processes with hard scales, Ref.~\cite{Collins:1989gx}, which reviews the factorization proofs
in detail, is cautious in the statements it makes about their exhaustivity, notably for the hadron-collider processes
which we shall discuss below.  Interesting considerations on the current status of our understanding of factorization
can be found in Refs.~\cite{Collins:2007nk,Sterman:2022gyf}.  While the transverse degrees of freedom have been
integrated over for collinear PDFs, it is also possible to produce transverse-momentum dependent (or unintegrated) PDFs
where these degrees of freedom remain.  For a recent comprehensive review of transverse-momentum-dependent parton
distribution functions and fragmentation functions see Ref.~\cite{Boussarie:2023izj}.

For collinear PDFs, the resulting dependence on $\mu_F$ is described by the Dokshitzer-Gribov-Lipatov-Altarelli-Parisi
(DGLAP) equations\cite{Gribov:1972ri,*Lipatov:1974qm,*Altarelli:1977zs,*Dokshitzer:1977sg}{},
\index{Dokshitzer-Gribov-Lipatov-Altarelli-Parisi \\ \mbox{(DGLAP) equations}} which to leading order (LO) read%
\footnote{LO is generally taken to mean the lowest order at which a quantity is non-zero.} %
\begin{equation}\label{qcd:eq:DGLAP}
  \mu_F^2
  \frac{\partial f_{i/p}(x,\mu_F^2)}{\partial \mu_F^2} = \sum_j
  \frac{\alps(\mu_F^2)}{2\pi} \int_x^1 \frac{dz}{z} P_{i\leftarrow j}^{(1)}(z)
  f_{j/p}\left(\frac{x}{z}, \mu_F^2\right) ,
\end{equation}
with, for example, $P^{(1)}_{q \leftarrow g}(z) = T_R (z^2+(1-z)^2)$. The other LO splitting functions\index{splitting
  functions} are listed in Sec.~\ref{strucfun} of this {\it Review}, while results up to NLO, $\alps^2$, and NNLO,
$\alps^3$, are given in Refs.~\cite{Curci:1980uw,*Furmanski:1980cm} and~\cite{Vogt:2004mw,*Moch:2004pa} respectively.
At N$^3$LO accuracy, only partial results are currently available in
Refs.~\cite{Davies:2016jie,Moch:2017uml,Vogt:2018ytw,Vogt:2018miu,Falcioni:2023luc}.  Splitting functions for PDFs in
the helicity dependent case are given in Ref.~\cite{Moch:2014sna}.

Beyond LO, the coefficient functions are also $\mu_F$ dependent, for example
$C_{2,i}^{(1)}(x,Q^2,\mu_R^2,\mu_F^2) = C_{2,i}^{(1)}(x,Q^2,\mu_R^2,Q^2) - \ln\big({\frac{\mu_F^2}{Q^2}}\big) \sum_j
\int_x^1 {\frac{dz}{z}}$ $\times C_{2,j}^{(0)}({\frac{x}{z}}) P_{j\leftarrow i}^{(1)}(z) $.  Progress in higher-order
QED and mixed QED-QCD corrections to the splitting functions can be found in
refs.\cite{deFlorian:2015ujt,*deFlorian:2016gvk,deFlorian:2023zkc}.

As with the renormalization scale, the choice of factorization scale is arbitrary, but if one has an infinite number of
terms in the perturbative series, the $\mu_F$-dependencies of the coefficient functions and PDFs will compensate each
other fully.  Given only $N$ terms of the series, a residual ${\cal O}(\alps^{N+1})$ uncertainty is associated with the
ambiguity in the choice of $\mu_F$. As with $\mu_R$, varying $\mu_F$ provides an input in estimating uncertainties on
predictions.  In inclusive DIS predictions, the default choice for the scales is usually $\mu_R=\mu_F = Q$.

As is the case for the running coupling, in DGLAP evolution one can introduce flavor thresholds near the heavy quark
masses: below a given heavy quark's mass, that quark is not considered to be part of the proton's structure, while above
it is considered to be part of the proton's structure and evolves with massless DGLAP splitting kernels.  With
appropriate parton distribution matching terms at threshold, such a variable flavor number scheme (VFNS),\index{flavor
  number scheme (FNS)} when used with massless coefficient functions, gives the full heavy-quark contributions at high
$Q^2$ scales.  For scales near the threshold, it is instead necessary to appropriately adapt the standard massive
coefficient functions to account for the heavy-quark contribution already included in the
PDFs\cite{Thorne:2006qt,Forte:2010ta,Guzzi:2011ew}.

At sufficiently small $x$ and $Q^2$ in inclusive DIS, resummation of small $x$ logarithms may be
necessary\cite{Fadin:1975cb, Balitsky:1978ic}.  This may in fact have been observed in Refs.~\cite{Ball:2017otu} based
on HERA data\cite{Abramowicz:2015mha}, in a kinematic region where useful information for PDFs for collider predictions
is present.

In situations, in which the center-of-mass energy $\sqrt{s}$ is much larger than all other momentum-transfer scales
in the problem (\eg $Q$ in DIS, $m_b$ for \bbbar production in \pp collisions, \etc), each power of \alps beyond LO can
be accompanied by a power of $\ln(s/Q^2)$ (or $\ln(s/m_b^2)$, \etc).  This is variously referred to as the high-energy,
small-$x$ or Balitsky-Fadin-Kuraev-Lipatov (BFKL) limit\cite{Lipatov:1976zz,Kuraev:1977fs,Balitsky:1978ic}.
\index{Balitsky-Fadin-Kuraev-Lipatov (BFKL) limit} Currently it is possible to account for the dominant and first
sub-dominant~\cite{Fadin:1998py,Ciafaloni:1998gs} power of such logarithms at each order of \alps, and also to estimate
further sub-dominant contributions that are numerically large (see
Refs.~\cite{Altarelli:2008aj,Ciafaloni:2007gf,White:2006yh,Iancu:2015vea} and references therein).  Progress towards
NNLO is discussed in Ref.~\cite{Gromov:2015vua,*Velizhanin:2015xsa,*Caron-Huot:2016tzz,*Fadin:2020lam}.

Physically, the summation of all orders in \alps can be understood as leading to a growth with~$s$ of the gluon density
in the proton.  At sufficiently high energies this implies non-linear effects (commonly referred to as parton
saturation), whose treatment has been the subject of intense study (see for example
Refs.~\cite{Balitsky:1995ub,Kovchegov:1999yj,Davier:2008sk} and references thereto).

\noindent{\bf Hadron-hadron collisions.}

The extension to processes with two initial-state hadrons can be illustrated with the example of the total (inclusive)
cross section for $W$ boson production in collisions of hadrons $h_1$ and $h_2$, which can be written as
\begin{equation}\label{qcd:eq:pp-W-xsct}
  \begin{split}
    \sigma(h_1 h_2 \to W + X) & = \sum_{n=0}^\infty \alps^{n}(\mu_R^2) \bookcr{}{&\!\!\!\!\!\times} \sum_{i,j} \int dx_1
    dx_2\, f_{i/h_1}\Big(x_1,\mu_F^2\Big) f_{j/h_2}\Big(x_2,\mu_F^2\Big) \\&\!\!\!\!\times \hat\sigma_{ij\to
      W+X}^{(n)}\Big(x_1x_2 s, \mu_R^2, \mu_F^2\Big) + {\cal O}\left(\frac{{\rm \Lambda}^2}{\mW^4}\right) \,,
  \end{split}
\end{equation}
\noindent
where $s$ is the squared center-of-mass energy of the collision.  At LO, $n=0$, the hard (partonic) cross section
$\hat\sigma_{ij\to W+X}^{(0)}(x_1x_2 s, \mu_R^2, \mu_F^2)$ is simply proportional to $\delta(x_1 x_2 s - \mW^2)$, in the
narrow $W$-boson width approximation (see Sec.~\crossref{crosssec} of this {\it Review} for detailed expressions for
this and other hard scattering cross sections).  It is non-zero only for choices of $i,j$ that can directly give a $W$,
such as $i=u$, $j=\bar d$.  At higher orders, $n\ge1$, new partonic channels contribute, such as $gq$, and
$x_1 x_2 s \ge \mW^2$ in the narrow $W$-boson width approximation.

Equation~(\ref{qcd:eq:pp-W-xsct}) involves a collinear factorization between the hard cross section and the PDFs, just
like \Eq{qcd:eq:F2-QCD}. As long as the same factorization scheme is used in DIS and \pp or \ppbar (usually the \MSbar
scheme), then PDFs extracted in DIS can be directly used in \pp and \ppbar
predictions\cite{Collins:1985ue,Collins:1989gx}{} (with the anti-quark distributions in an anti-proton being the same as
the quark distributions in a proton).

A number of fully inclusive cross-sections have been computed up to N$^3$LO, \ie including corrections up to relative
order $\alps^3$, in recent years.  These include inclusive Higgs production through gluon fusion in the large $m_t$
limit, calculated at N$^3$LO in Refs.\cite{Anastasiou:2015ema,*Anastasiou:2016cez,Mistlberger:2018etf}.  Higgs boson
pair production via gluon fusion in the same approximation was computed at N$^3$LO in Ref.~\cite{Chen:2019lzz}.
Calculations at this order, differential in the Higgs rapidity~\cite{Dulat:2018bfe,Cieri:2018oms} and fully
differential~\cite{Chen:2021isd} have been presented recently.  Bottom-induced Higgs boson production has also been
computed at N$^3$LO~\cite{Baglio:2022wzu}.  Vector-boson fusion single-~\cite{Dreyer:2016oyx} and
double-pair~\cite{Dreyer:2018qbw} production is also known to N$^3$LO~\cite{Dreyer:2016oyx} in the factorized
approximation.  Neutral~\cite{Duhr:2020seh,Chen:2021vtu} and charged~\cite{Duhr:2020sdp} Drell-Yan processes, as well as
associated Higgs production~\cite{Baglio:2022wzu} have also been computed at N$^3$LO.  Differential rapidity
distributions are available, but not yet with fiducial cuts.  A number of public codes are now available which allow for
phenomenological studies at
N$^3$LO~\cite{Bonvini:2016frm,Dulat:2018rbf,Harlander:2016hcx,Dreyer:2016oyx,Baglio:2022wzu}.
The uncertainty band derived by varying the renormalization and
factorization scales in the perturbative predictions is not always
contained within the corresponding uncertainty band of the previous
order, as might be expected from a well-behaved process.
This has been ascribed to cancellations between channels at NNLO that lead to
underestimates of the uncertainty at that order. This effect does not seem to be present in other processes, such as
Higgs boson production through gluon-gluon fusion. Note that there is currently a mis-match in N$^3$LO calculations in
that PDFs at the same order are not yet available.  Preliminary approximated N$^3$LO PDFs~\cite{McGowan:2022nag}
suggest that N$^3$LO effects can be large, especially as far as heavy-quark distributions and low scales are
concerned, as well as the gluon distribution in a region important for Higgs boson production through gluon-gluon
fusion.  The fully inclusive hard cross sections for several other processes are known to NNLO,%
\footnote{Processes with jets or photons in the final state have divergent cross sections unless one places a cut on the
  jet or (dressed) photon momentum. Accordingly, they are discussed below in \Section{qcd:sec:QCD-FO-predictions}.} %
for instance, Higgs-boson production in association with a vector boson\cite{Brein:2003wg}, Higgs-pair production in the
large $m_t$ approximation\cite{deFlorian:2013jea}, and with full $m_t$ dependence\cite{Borowka:2016ehy},
top-antitop production\cite{Czakon:2013goa,Catani:2019iny},
bottom-anti-bottom production\cite{Catani:2020kkl}
and vector-boson pair
production\cite{Gehrmann:2014fva,Cascioli:2014yka,Grazzini:2017mhc}.  For a comprehensive overview on other recent NNLO
$2 \to 2$ calculations see Ref.~\cite{Heinrich:2020ybq,Huss:2022ful}.  Other NNLO calculations with fiducial cuts will
be discussed in Sec.~\ref{qcd:sec:non-incl-x-sct}.

\noindent{\bf Photoproduction.}\index{photoproduction}

$\gamma p$ (and $\gamma \gamma$) collisions are similar to \pp collisions, with the subtlety that the photon can behave
in two ways: there is ``direct'' photoproduction, in which the photon behaves as a point-like particle and takes part
directly in the hard collision, with hard subprocesses such as $\gamma g\to q\bar q$; there is also resolved
photoproduction, in which the photon behaves like a hadron, with non-perturbative partonic substructure and a
corresponding PDF for its quark and gluon content, $f_{i/\gamma}(x,Q^2)$.  While useful to understand the general
structure of $\gamma p$ collisions, the distinction between direct and resolved photoproduction is not well defined
beyond leading order, as discussed for example in Ref.~\cite{Greco:1993st}.

% ----------------------------------------------------------------------
\vglue -0.05in
\subsection{Cross sections with phase-space restrictions}
\label{qcd:sec:non-incl-x-sct}

QCD final states always consist of hadrons, while perturbative QCD calculations deal with partons.  Physically, an
energetic parton fragments (``showers'') into many further partons, which then, on later timescales, undergo a
transition to hadrons (``hadronization'').  Fixed-order perturbation theory captures only a small part of these
dynamics.  This does not matter for the fully inclusive cross sections discussed above: the showering and hadronization
stages are approximately unitary, \ie they do not substantially change the overall probability of hard scattering,
because they occur long after it has taken place (they introduce at most a correction proportional to a power of the
ratio of timescales involved, \ie a power of ${\rm \Lambda}/Q$, where $Q$ is the hard scattering scale).

Less inclusive measurements, in contrast, may be affected by the extra dynamics.  For those sensitive just to the main
directions of energy flow (jet rates, event shapes, \cf \Sec{qcd:sec:qcd-exp-observables}) fixed-order perturbation
theory is often still adequate, because showering and hadronization do not substantially change the overall energy flow.
This means that one can make a prediction using just a small number of partons, which should correspond well to a
measurement of the same observable carried out on hadrons.  For observables that instead depend on distributions of
individual hadrons (which, \eg, are the inputs to detector simulations), it is mandatory to account for showering and
hadronization.  The range of predictive techniques available for QCD final states reflects this diversity of needs of
different measurements.

While illustrating the different methods, we shall for simplicity mainly use expressions that hold for $e^+e^-$
scattering.  The extension to cases with initial-state partons will be mostly straightforward (space constraints
unfortunately prevent us from addressing diffraction and exclusive hadron-production processes; extensive discussion is
to be found in Sec.~\crossref{softqcd} of this {\it Review} and in Refs.~\cite{Hebecker:1999ej,Belitsky:2005qn}).

% ......................................................................
\vglue -0.1in
\subsubsection{Soft and collinear limits}

Before examining specific predictive methods, it is useful to be aware of a general property of QCD matrix elements in
the soft and collinear limits.
Consider a squared tree-level matrix element $|M_{n}^2(p_1,\ldots,p_n)|$ for the process $e^+e^- \to n$ partons with
momenta $p_1, \ldots, p_n$, and a corresponding phase-space integration measure $d\Phi_n$.
If particle $n$ is a gluon, which becomes collinear (parallel) to another particle $i$ and additionally its momentum
tends to zero (is ``soft''), the matrix element simplifies as follows,
\begin{align}\label{qcd:eq:coll}
  &\lim_{\theta_{in} \to 0,\, E_n \to 0}
    d\Phi_{n} |M_{n}^2(p_1,\ldots, p_{n})|\cr
    \!\!\!\!\!\!&=
                  \  d\Phi_{n-1} |M_{n-1}^2(p_1,\ldots, p_{n-1})|
                  \frac{\alps C_i}{\pi} \frac{d\theta_{in}^2}{\theta_{in}^2} \frac{dE_n}{E_n}\,,
                  \quad
\end{align}
\noindent
where $C_i=C_F$ ($C_A$) if $i$ is a quark (gluon).
This formula has non-integrable divergences both for the inter-parton angle $\theta_{in} \to 0$ and for the gluon energy
$E_n \to 0$, which are mirrored also in the structure of divergences in loop diagrams.
These divergences are important for at least two reasons: firstly, they govern the typical structure of events (inducing
many emissions either with low energy or at small angle with respect to hard partons);
secondly, they will determine which observables can be calculated within fixed-order perturbative QCD.

% ......................................................................
\subsubsection{Fixed-order predictions}
\label{qcd:sec:QCD-FO-predictions}

Let us consider an observable $\cal{O}$ that is a function ${\cal O}_n(p_1,\ldots,p_n)$ of the four-momenta of the $n$
final-state particles in an event (either partons or hadrons).  In what follows, we shall consider the cross section for
events weighted with the value of the observable, $\sigma_{\cal O}$.  As examples, if ${\cal O}_n \equiv 1$ for all $n$,
then $\sigma_{\cal O}$ is just the total cross section; if ${\cal O}_n \equiv \hat\tau(p_1, \ldots, p_n)$ where
$\hat \tau$ is the value of the thrust for that event (see Eq.~\eqref{qcd:eq:thrust-GD} in
Sec.~\ref{qcd:sec:evshapesGD}), then the average value of the thrust is
$\langle \tau \rangle = \sigma_{\cal O}/ \sigma_{\rm tot} $; if
${\cal O}_n \equiv \delta(\tau - \hat\tau(p_1, \ldots, p_n))$ then one gets the differential cross section as a function
of the thrust, $\sigma_{\cal O}\equiv d\sigma / d\tau$.

In the expressions below, we shall omit to write the non-perturbative power correction term, which for most common
observables is proportional to a single power of ${\rm \Lambda}/Q$.

\noindent {\bf Leading Order.}
If the observable ${\cal O}$ is non-zero only for events with at least $n$ final-state particles, then the LO QCD
prediction for the weighted cross section in \ee annihilation is
\begin{equation}\label{qcd:eq:LO-struct}
  \sigma_{\cal O}^{\rm LO} = \alps^{n-2}(\mu_R^2) \int d\Phi_n
  |M_n^2 (p_1,\ldots,p_n)| \, {\cal O}_n (p_1,\ldots, p_n)\,,
\end{equation}
\noindent
where the squared tree-level matrix element, $|M_{n}^2(p_1,\ldots,p_n)|$, including relevant symmetry factors, has been
summed over all subprocesses (\eg $\ee \to q\bar q q\bar q$, $\ee \to q\bar q gg$) and has had all factors of \alps
extracted in front.  In processes other than \ee collisions, the center-of-mass energy of the LO process is generally
not fixed, and so the powers of the coupling are often brought inside the integrals, with the scale $\mu_R$ chosen event
by event, as a function of the event kinematics.

Other than in the simplest cases (see the review on Cross Sections in this {\it Review}), the matrix elements in
\Eq{qcd:eq:LO-struct} are usually calculated automatically with programs such as
\Comphep\cite{Boos:2004kh}, \MADGRAPH\cite{Alwall:2014hca,*qcd:Alwall:2014hca-url},
\ALPGEN\cite{Mangano:2002ea,*qcd:Mangano:2002ea-url}, \COMIX/\SHERPA\cite{Gleisberg:2008fv,*qcd:Gleisberg:2008fv-url},
and \HELACPHEGAS\cite{Cafarella:2007pc,*qcd:Cafarella:2007pc-url}.  Some of these (\Comphep, \MADGRAPH) use formulas
obtained from direct evaluations of Feynman diagrams. Others (\ALPGEN, \HELACPHEGAS and \COMIX/\SHERPA) use methods
designed to be particularly efficient at high multiplicities, such as Berends-Giele recursion\cite{Berends:1987me}{},
which builds up amplitudes for complex processes from simpler ones (see also
Refs.~\cite{Dixon:1996wi,Britto:2004ap,Cachazo:2004kj}{} for other tree-level calculational methods).

The phase-space integration is usually carried out by Monte Carlo sampling, because of the high dimensionality of the
integration and in order to deal with the possibly involved kinematic cuts that are used in the corresponding
experimental measurements.  Perturbatively calculable observables should be insensitive to the emission of soft and
collinear radiation.  Because of the divergences in the matrix element, \Eq{qcd:eq:coll}, at lowest order the integral
converges only if the observable vanishes for kinematic configurations in which one of the $n$ particles is arbitrarily
soft or it is collinear to another particle.  As an example, the cross section for producing any configuration of $n$
partons will lead to an infinite integral, whereas a finite result will be obtained for the cross section for producing
$n$ deposits of energy (or jets, see \Sec{qcd:sec:jetalgoGD}), each above some energy threshold and well separated from
each other in angle.

At a practical level, LO calculations can be carried out for $2\to n$ processes with $n \lesssim 6-10$.  The exact upper
limit depends on the process, the method used to evaluate the matrix elements (recursive methods are more efficient),
and the extent to which the phase-space integration can be optimized to work around the large variations in the values
of the matrix elements.  \vskip 10pt

\noindent {\bf NLO.}
Given an observable that is non-zero starting from $n$ final-state particles, its prediction at NLO involves
supplementing the LO result, Eq.~(\ref{qcd:eq:LO-struct}), with the $2\to(n+1)$-particle squared tree-level matrix
element ($|M_{n+1}^2|$), and the interference of a $2\to n$ tree-level and $2\to n$ 1-loop amplitude
($2{\rm{Re}}(M_{n} M^*_{n,\rm{1-loop}}$)),
\begin{equation}\label{qcd:eq:NLO-struct}
  \begin{split}
    \sigma_{\cal O}^{\rm NLO}&= \sigma_{\cal O}^{\rm LO} + \alps^{n-1}(\mu_R^2) \int \!\!d\Phi_{n+1}
    |M_{n+1}^2(p_1,\ldots, p_{n+1})| \bookcr{}{&\times} {\cal O}_{n+1}(p_1,\ldots, p_{n+1})\, \webcr{}{&} +\
    \alps^{n-1}(\mu_R^2) \bookcr{}{&\times} \int \!\!d\Phi_n \, 2{\rm{Re}}\left[ \right.  M_{n}(p_1,\ldots, p_n) \,
    M^*_{n,{\rm{1-loop}}}(p_1,\ldots, p_n) \left.\right] \, \bookcr{}{&\times} {\cal O}_n(p_1,\ldots, p_n)\ .
  \end{split}
\end{equation}
Relative to LO calculations, two important issues appear in the NLO calculations.  Firstly, the extra complexity of
loop-calculations relative to tree-level calculations means that automated calculations started to appear only about
fifteen years ago (see below).  Secondly, loop amplitudes are infinite in four dimensions, while tree-level amplitudes
are finite, but their {\it integrals over the phase space} are infinite, due to the divergences of \Eq{qcd:eq:coll}.
These two sources of infinities have the same soft and collinear origins and cancel after the integration only if the
observable $\cal O$ satisfies the property of infrared and collinear safety,\index{infrared and collinear
  safety}\index{collinear and infrared safety} which means that the observable is non-sensitive to soft emissions or to
collinear splittings, \ie
\begin{equation}\label{qcd:eq:IRS-def-GD-reused}
  \begin{split}
    {\cal O}_{n+1}(p_1,\ldots,p_s,\ldots,p_n)\ \ \ \ \ & \rightarrow {\cal
      O}_n(p_1,\ldots,p_{s-1},p_{s+1},\ldots,p_n)\quad
    \bookcr{}{&} {\rm if}\; p_s\rightarrow 0  \\
    {\cal O}_{n+1}(p_1,\ldots,p_a,p_b,\ldots,p_{n})\ & \rightarrow {\cal O}_n(p_1,\ldots,p_a+p_b,\ldots,p_n) \quad
    % \cr &
    \bookcr{}{&} {\rm if}\; p_a\, ||\, p_b \, .
  \end{split}
\end{equation}
\noindent
Examples of infrared-safe quantities include event-shape distributions and jet cross sections (with appropriate jet
algorithms, see below).  Unsafe quantities include the distribution of the momentum of the hardest QCD particle (which
is not conserved under collinear splitting), observables that require the complete absence of radiation in some region
of phase space (\eg rapidity gaps or 100\% isolation cuts, which are affected by soft emissions), or the particle
multiplicity (affected by both soft and collinear emissions).  The non-cancellation of divergences at NLO due to
infrared or collinear unsafety compromises the usefulness not only of the NLO calculation, but also that of a LO
calculation, since LO is only an acceptable approximation if one can prove that higher-order terms are smaller.
Infrared and collinear unsafety usually also implies large non-perturbative effects.

As with LO calculations, the phase-space integrals in \Eq{qcd:eq:NLO-struct} are usually carried out by Monte Carlo
integration, so as to facilitate the study of arbitrary observables.  Various methods exist to obtain numerically
efficient cancellation among the different infinities.  These include notably dipole\cite{Catani:1996vz}{},
FKS\cite{Frixione:1995ms} and antenna\cite{Kosower:1997zr,*Campbell:1998nn,*Kosower:2003bh} subtraction.

Thanks to new ideas like the OPP method\cite{Ossola:2006us}, generalized\cite{Britto:2004nc} and
$D$-dimensional\cite{Ellis:2008ir} unitarity, on-shell methods\cite{Berger:2009zb}, and on-the-fly reduction
algorithms\cite{Cascioli:2011va}, recent years have seen a breakthrough in the calculation of one-loop matrix elements
(for reviews on unitarity based method see Ref.~\cite{Bern:2007dw,Ellis:2011cr}).  Thanks to these innovative methods,
automated tools for NLO calculations have been developed and a number of programs are available publicly:
\MGMCatNLO\cite{Alwall:2014hca} and \HELACNLO\cite{Bevilacqua:2011xh,*qcd:Bevilacqua:2011xh-sr1} provide full frameworks
for NLO calculations; \GOSAM\cite{Cullen:2014yla,*qcd:Cullen:2014url}{},
\NJET\cite{Badger:2012pg,*qcd:Badger:2012pg-url}{}, \OPENLOOPS\cite{Buccioni:2017yxi,*qcd:Buccioni:2017yxi-url} and
\RECOLA\cite{Actis:2016mpe} calculate just the 1-loop part and are typically interfaced with an external tool for a
combination with the appropriate tree-level amplitudes.  Other tools such as
\NLOJETPP\cite{Nagy:2003tz,*qcd:Nagy:2003tz-url}, \MCFM\cite{Campbell:2000bg},
\VBFNLO\cite{Arnold:2011wj,**qcd:Arnold:2011wj-sr1}, the \PHOX family\cite{Binoth:1999qq,**qcd:Binoth:1999qq-sr1} or
\BLACKHAT\cite{Bern:2012my} implement analytic calculations and provide full frameworks to compute NLO cross sections
for selected classes of processes.  Recently, a lot of attention has also been paid to the calculation of NLO
electroweak corrections.\index{electroweak corrections} Electroweak corrections are especially important for
transverse momenta significantly above the $W$ and $Z$ masses, because they are enhanced by two powers of $\ln p_t/\mW$
for each power of the electroweak coupling, and close to Sudakov peaks\footnote{For the definition of Sudakov form factors or Sudakov peaks see e.g. Refs.~\cite{Ellis:1991qj,qcd:CHKbook,Gross:2022hyw}.}, where most of the data lie and the best
experimental precision can be achieved.  In some cases the above programs can be used to calculate also NLO electroweak
or beyond-standard-model
corrections\cite{Cullen:2012eh,Kallweit:2014xda,Denner:2014bna,Chiesa:2015mya,Frixione:2015zaa,Biedermann:2017yoi,Frederix:2018nkq}.

Given the progress in QCD and EW fixed-order computations, the largest unknown from fixed-order corrections is often
given by the mixed QCD-electroweak corrections of ${\cal O}(\alps \alpha)$. These mixed two-loop corrections are often
available in an approximate
form\cite{Anastasiou:2008tj,Dittmaier:2014qza,Dittmaier:2015rxo,deFlorian:2018wcj,Bonetti:2017ovy,Bonetti:2018ukf,Anastasiou:2018adr,Hirschi:2019fkz,Becchetti:2020wof,Bonetti:2020hqh,Behring:2020cqi,Dittmaier:2020vra,Buccioni:2020cfi,Buonocore:2021rxx}.
The first complete computation of the mixed QCD–EW corrections to the neutral-current Drell–Yan process appeared
recently~\cite{Bonciani:2021zzf,Armadillo:2022bgm,Buccioni:2022kgy}.  For a review on EW corrections to collider
processes see Ref.~\cite{Denner:2019vbn}.

\noindent {\bf NNLO.}
NNLO is considerably more complicated than NLO as it involves a further order in \alps, consisting of: the squared
$(n+2)$-parton tree-level amplitude, the interference of the $(n+1)$-parton tree-level and 1-loop amplitudes, the
interference of the $n$-parton tree-level and 2-loop amplitudes, and the squared $n$-parton 1-loop amplitude.

Each of these elements involves large numbers of soft and collinear divergences, satisfying relations analogous to
\Eq{qcd:eq:coll} which now involve multiple collinear or soft particles and higher loop orders (see \eg
Refs.~\cite{Bern:1994zx,Campbell:1997hg,Catani:1998nv}{}).  Arranging for the cancellation of the divergences after
numerical Monte Carlo integration has been one of the significant challenges of NNLO calculations, as has been the
determination of the relevant 2-loop amplitudes.  For the cancellations of divergences a wide range of methods has been
developed.  Some of
them\cite{Binoth:2000ps,Anastasiou:2003gr,GehrmannDeRidder:2005cm,Somogyi:2006da,Czakon:2010td,Caola:2017dug,Magnea:2018hab,Prisco:2020kyb,Bertolotti:2022aih,Anastasiou:2022eym}
retain the approach, inherent in NLO methods, of directly combining the separate loop and tree-level amplitudes.  Others
combine a suitably chosen, partially inclusive $2\to n$ NNLO calculation with a fully differential $2\to n+1$ NLO
calculation\cite{Catani:2007vq,*qcd:Catani:2007vq-url,Bonciani:2015sha,Boughezal:2015dva,Boughezal:2015eha,Gaunt:2015pea,Cacciari:2015jma,Capatti:2022tit}.
The $q_T$-subtraction method was extended to deal with mixed QCD–QED corrections at NNLO~\cite{Cieri:2020ikq}.  For an
overview of NNLO subtraction methods see Ref~\cite{TorresBobadilla:2020ekr}.

Quite a number of processes have been calculated differentially at NNLO so far.  The state of the art for $e^+e^-$
collisions is
$e^+e^- \to
3\,$jets\cite{Gehrmann-DeRidder:2007nzq,*GehrmannDeRidder:2007hr,*GehrmannDeRidder:2008ug,Gehrmann-DeRidder:2014hxk,*qcd:Gehrmann-DeRidder:2014hxk-url,Weinzierl:2008iv,*Weinzierl:2009ms,DelDuca:2016ily,*DelDuca:2016csb}{}.

For DIS, dijet production is known at NNLO\cite{Currie:2016ytq,Currie:2017tpe}{} and the description jet production has
been recently pushed even to N$^3$LO using the Projection-to-Born method\cite{Currie:2018fgr,Gehrmann:2018odt}.  For
hadron colliders, all $2\to1$ processes are known, specifically vector
boson\cite{Melnikov:2006kv,*qcd:Melnikov:2016kv-url,Catani:2009sm,*qcd:Catani:2009sm-url}{} and Higgs boson production
in the large $m_t$ limit~\cite{Anastasiou:2005qj,*qcd:Anastasiou:2005qj-url,Catani:2007vq}{}.  The finite top-mass
corrections at this order have also been computed~\cite{Czakon:2021yub}. This calculation eliminates one important
source of theoretical uncertainty to inclusive Higgs production at the LHC.  Substantial progress has been made in the
years for hadron-collider $2\to2$ processes, with calculations having been performed for nearly all relevant processes:
$ZZ$\cite{Cascioli:2014yka}{} $WW$\cite{Gehrmann:2014fva} and $WZ$\cite{Grazzini:2017ckn}{},
$\gamma\gamma$\cite{Catani:2011qz,Campbell:2016yrh}{}, $Z\gamma$\cite{Grazzini:2013bna} and
$W\gamma$\cite{Grazzini:2015nwa}{} (many of these color singlet processes are available also in
\MCFM\cite{Campbell:1999ah,Boughezal:2016wmq}{} or MATRIX\cite{Grazzini:2017mhc}), inclusive
photon\cite{Campbell:2016lzl,Chen:2019zmr}, $\gamma+\,$jet\cite{Campbell:2017dqk,Chen:2019zmr},
$W+\,$jet\cite{Boughezal:2015dva,Gehrmann-DeRidder:2017mvr},
$Z+\,$jet\cite{Ridder:2015dxa,Boughezal:2015ded,Campbell:2017dqk}
$H+\,$jet\cite{Boughezal:2015dra,Boughezal:2015aha,Caola:2015wna,Chen:2016zka}{}, $WH$\cite{Ferrera:2011bk} and
$ZH$\cite{Ferrera:2014lca}{}, $s$-channel~\cite{Liu:2018gxa} and $t$-channel
single-top\cite{Brucherseifer:2014ama,Berger:2016oht,Berger:2017zof,Campbell:2020fhf}{}, \ttbar
production\cite{Czakon:2014xsa,Czakon:2016ckf}, dijet production\cite{Currie:2016bfm,Chen:2022tpk}{}, $W$ production in
association with a $c$-jet~\cite{Czakon:2020coa,Czakon:2022khx} and $HH$\cite{deFlorian:2013jea}{} (in large-top-mass
approximation, see also the exact (two-loop) NLO result\cite{Borowka:2016ehy}).  Recently, also the NNLO corrections to
identified $B$-hadron hadro-production have been computed~\cite{Czakon:2021ohs}.  The frontier of NNLO calculation has
now reached the complexity of $2\to 3$ processes. The first $2\to3$ process known at NNLO has been Higgs production
through vector-boson fusion, using an approximation in which the two underlying DIS-like $q\to qV$ scatterings are
factorized, the so-called structure function approximation\cite{Cacciari:2015jma,Cruz-Martinez:2018rod}.  Corrections
beyond the structure function approximation are expected to be small, on the order of a percent or
less\cite{Liu:2019tuy}.  More recently, first genuine $2\to 3$ LHC processes have been described at NNLO accuracy,
including three photon~\cite{Chawdhry:2019bji,Kallweit:2020gcp}, two photons and one jet~\cite{Chawdhry:2021hkp}, two
jets and one photon~\cite{Badger:2023mgf}, three-jets~\cite{Czakon:2021mjy} and $Wb\overline{b}$
production~\cite{Hartanto:2022qhh}. The calculation of three-jet production might be relevant for future extractions of
\asmz from three-jet observables at the LHC\@~\cite{Alvarez:2023fhi}.

Cross sections at the LHC are most often measured with fiducial cuts, for example on the transverse momenta and
rapidities of the measured objects, restricting the measurement to regions where the objects have a good efficiency to
be detected and are well-reconstructed. Ideally, the theoretical predictions should also be constructed at the same
fiducial level; the other possibility is to extrapolate the experimental fiducial results to the full phase space. Such
an extrapolation, however, requires the extrapolation (typically using a parton shower Monte Carlo) to be accurate over
the full phase space.

Comparisons of fixed order predictions to fiducial measurements can sometimes result in the presence of large logarithms
which degrade the accuracy of the prediction. Such is the case, for example, in the calculation of the Higgs rapidity
distribution, in the diphoton final state. The imposition of transverse momenta and rapidity cuts on the two photons
leads to an uncertainty notably greater at N$^3$LO then at NNLO, due to the presence of these logs. One possible
solution is to change the form of the kinematic cut on the photons, to a product of the two photon transverse
momenta~\cite{Salam:2021tbm}, reducing the impact of the logs; another solution is to perform a resummation of the logs,
restoring the expected precision~\cite{Ebert:2020dfc}.

The Les Houches precision wishlist compiles predictions needed to fully exploit the data that will be taken at the High
Luminosity LHC\cite{Amoroso:2020lgh}. Most of the needed calculations require accuracy of at least NNLO QCD and NLO EW,
and many require the prediction of $2\to3$ processes, such as $W/Z+\ge 2$ jets, $H+\ge 2$ jets, and $ttH$ to NNLO. The
latter process has been computed at NNLO in Ref.~\cite{Catani:2022mfv} in the approximation where the Higgs boson in the
double-virtual contribution is soft. This calculation considerably reduces the theory uncertainty on this process.

As discussed in this section, calculations at NLO can now be relatively easily generated by non-experts using the
programs mentioned above.  By now there are also a number of publicly available tools to compute a range of processes at
NNLO accuracy in QCD, such as \EERAD{}\cite{Gehrmann-DeRidder:2014hxk} (for $e^+e^-$ collisions),
\HNNLO{}\cite{Catani:2007vq}, \DYNNLO{}\cite{Catani:2009sm}, \DYTURBO{}\cite{Camarda:2019zyx},
\FEHIP{}\cite{Anastasiou:2005qj}, \FEWZ{}\cite{Gavin:2010az}, \HTURBO{}\cite{Camarda:2022wti},
\MATRIX{}\cite{Grazzini:2017mhc}, \MCFM{}\cite{Boughezal:2016wmq} or \SUSHI{}\cite{Harlander:2012pb,Harlander:2016hcx}.
However, in some cases NNLO calculations can be too complex and CPU-intensive to allow such an approach. In these cases,
the relevant matrix element information for a specific observable can be stored by means of interpolation grids
developed originally for cross sections at NLO~\cite{Carli:2010rw,Kluge:2006xs,Britzger:2012bs} and recently extended to
include electroweak corrections~\cite{Carrazza:2020gss}. The application to NNLO has been demonstrated for \ttbar and
for DIS jet cross sections in Refs.~\cite{Czakon:2017dip,Britzger:2019kkb}. Each such interpolation grid corresponds to
one fixed observable with specific selection criteria and binning, but is flexible with respect to the renormalization
or factorization scale, the PDF, or the \alps evolution chosen for the cross-section computation.  An even more flexible
method, at the cost of requiring large amounts of storage space, saves huge numbers of partonic events in the form of
ROOT n-tuples, which allow predictions to be generated on-the-fly for many observables of a particular
process~\cite{Maitre:2020blv,Maitre:2016sov}. Recent developments of both techniques for \pp collisions at NNLO are
described in Ref.~\cite{Amoroso:2020lgh}.

% ......................................................................
\subsubsection{Resummation}
\label{qcd:sec:resummation}
\index{resummation}

Many experimental measurements place tight constraints on emissions in the final state. For example, in \ee events, that
(one minus) the thrust should be less than some value $\tau \ll 1$, or, in $pp \to Z$, events that the $Z$-boson
transverse momentum or the transverse momentum of the accompanying jet should be much smaller than the $Z$-boson mass.
A further example is the production of heavy particles or jets near threshold (so that little energy is left over for
real emissions) in DIS and \pp collisions.

In such cases, the constraint vetoes a significant part of the integral over the soft and collinear divergence of
\Eq{qcd:eq:coll}.  As a result, there is only a partial cancellation between real emission terms (subject to the
constraint) and loop (virtual) contributions (not subject to the constraint), causing each order of \alps to be
accompanied by a large coefficient $\sim L^2$, where \eg $L =\ln \tau$ or $L = \ln(\mZ/p_{t}^{Z})$.  One ends up with a
perturbative series, whose terms go as $\sim (\alps L^2)^n$.  It is not uncommon that $\alps L^2 \gg 1$, so that the
perturbative series converges very poorly if at all.%
\footnote{To be precise one should be aware of two causes of the divergence of perturbative series. That which
  interests us here is associated with the presence of a new large parameter (\eg ratio of scales).  It is distinct from
  the ``renormalon'' induced factorial divergences of perturbation theory discussed above.} %
In such cases one may carry out a ``resummation'', which accounts for the dominant logarithmically enhanced terms to all
orders in \alps, by making use of known properties of matrix elements for multiple soft and collinear emissions, and of
the all-orders properties of the divergent parts of virtual corrections, following original works such as
Refs.~\cite{Dokshitzer:1978hw,Parisi:1979se,Curci:1979bg,Bassetto:1979nt,Collins:1981uk,Collins:1981va,Kodaira:1981nh,Kodaira:1982az,Collins:1984kg,Catani:1992ua}
and also through soft-collinear effective theory\cite{Bauer:2000yr,Bauer:2001yt}\ (\cf also the section on ``Heavy-Quark
and Soft-Collinear Effective Theory'' in this {\it Review}, as well as Ref.~\cite{Becher:2014oda}).

For cases with double logarithmic enhancements (two powers of logarithm per power of \alps), there are two
classification schemes for resummation accuracy. Writing the cross section including the constraint as $\sigma(L)$ and
the unconstrained (total) cross section as $\sigma_{\rm tot}$, the series expansion takes the form
\begin{equation}\label{qcd:eq:logs-logs-logs}
  \sigma(L) \simeq \sigma_{\rm tot} \sum_{n=0}^\infty \sum_{k=0}^{2n} R_{nk}
  \alps^n(\mu_R^2) L^k, \qquad L \gg 1\,,
\end{equation}
\noindent
and leading log (LL) resummation means that one accounts for all terms with $k=2n$, next-to-leading-log (NLL) includes
additionally all terms with $k=2n-1$, \etc Often $\sigma(L)$ (or its Fourier or Mellin transform) {\it exponentiates},%
\footnote{Whether or not this happens depends on the quantity being resummed. A classic example involves two-jet rate
  in $e^+ e^-$ collisions as a function of a jet-resolution parameter $y_{\rm{cut}}$. The logarithms of $1/y_{\rm{cut}}$
  exponentiate for the $k_t$ (Durham) jet algorithm\cite{Catani:1991hj}, but not\cite{Brown:1990nm} for the JADE\
  algorithm\cite{Bartel:1986ua}{} (both are discussed below in \Sec{qcd:sec:jetalgoGD}).} %
\begin{equation}\label{qcd:eq:exp-logs-logs}
  \sigma(L) \simeq \sigma_{\rm tot} \exp\left[\sum_{n=1}^\infty \sum_{k=0}^{n+1} G_{nk}
    \alps^n(\mu_R^2) L^k\right], \qquad L \gg 1\,,
\end{equation}
\noindent
where one notes the different upper limit on $k$ ($\le n+1$) compared to \Eq{qcd:eq:logs-logs-logs}.  This is a more
powerful form of resummation: the $G_{12}$ term alone reproduces the full LL series in \Eq{qcd:eq:logs-logs-logs}.  With
the form \Eq{qcd:eq:exp-logs-logs} one still uses the nomenclature LL, but this now means that all terms with $k=n+1$
are included, and NLL implies all terms with $k=n$, \etc

For a large number of observables, NLL resummations are available in the sense of \Eq{qcd:eq:exp-logs-logs} (see
Refs.~\cite{Kidonakis:1998nf,Bonciani:2003nt,Banfi:2004yd} and references therein).  NNLL has been achieved for the DY
and Higgs-boson $p_t$
distributions\cite{deFlorian:2000pr,Bozzi:2005wk,*qcd:Bozzi:2005wk-url,Bozzi:2010xn,Becher:2010tm}{} (also available in
the CuTe\cite{qcd:CuTe}{}, HRes\cite{deFlorian:2012mx,*qcd:deFlorian:2012mx-url}{} and ResBos\cite{Balazs:1997xd}
families of programs and also differentially in vector-boson decay products\cite{Catani:2015vma}{}) and related
variables\cite{Banfi:2012du}{}, for the $p_t$ of vector-boson pairs\cite{Grazzini:2015wpa,Campbell:2022uzw}{}, for the
back-to-back energy-energy correlation in \ee\cite{deFlorian:2004mp}{}, the jet broadening in $e^+e^-$
collisions\cite{Becher:2012qc}{}, the jet-veto survival probability in Higgs and $Z$ boson production in \pp
collisions\cite{Banfi:2012jm,*Becher:2013xia,Stewart:2013faa,Campbell:2023cha,**qcd:xxx},%
\footnote{A veto on the jet phase space can be severe, for example by requiring exactly zero jets above a given
  transverse momentum cut accompanying a Higgs boson, or relatively mild, for example by placing a transverse momentum
  cut of 30\GeV on the measurement of the production of a Higgs boson with one or more jets. In general, inclusive cross
  sections are preferable, as uncertainties on both the theoretical and experimental sides are smaller.} %
an event-shape type observable known as the beam thrust\cite{Stewart:2010pd}{}, hadron-collider jet masses in specific
limits\cite{Chien:2012ur,*Jouttenus:2013hs}{} (see also Ref.~\cite{Dasgupta:2012hg}{}), the production of top anti-top
pairs near threshold\cite{Ahrens:2010zv,Aliev:2010zk,Kidonakis:2010dk}{} (and references therein), and high-$p_t$ $W$
and $Z$ production\cite{Becher:2011fc}.  Automation of NNLL jet-veto resummations for different processes has been
achieved in Ref.~\cite{Becher:2014aya} (\cf also the NLL automation in Ref.~\cite{Gerwick:2014gya}), while automation
for a certain class of $e^+e^-$ observables has been achieved in Ref.~\cite{Banfi:2014sua}.  N$^3$LL resummations are
available for the thrust variable, C-parameter and heavy-jet mass in \ee
annihilations\cite{Becher:2008cf,Hoang:2014wka,Chien:2010kc}, for $p_t$ distribution of the Higgs
boson\cite{Bizon:2017rah,Chen:2018pzu} and weak gauge bosons\cite{Bizon:2019zgf,Neumann:2022lft,Re:2021con} and for
Higgs- and vector-boson production near threshold\cite{Catani:2014uta}.  In order to make better contact with
experimental measurements, recent years have seen an increasing interest in resummations in exclusive phase-space
regions and joint
resummations\cite{Banfi:2015pju,Larkoski:2015uaa,Lustermans:2016nvk,Muselli:2017bad,Bizon:2018foh,Bonvini:2018ixe,Procura:2018zpn,Lustermans:2019plv,Monni:2019yyr,Becher:2020ugp}.
Finally, there has also been considerable progress in resummed calculations for jet substructure, whose observables
involve more complicated definitions than is the case for standard resummations, see \eg
Refs.\cite{Dasgupta:2013ihk,Larkoski:2014wba,Larkoski:2017jix,Marzani:2019hun,Kogler:2018hem} and references therein (see
also Sec.~\ref{sec:QCDjetsub}).

% ......................................................................
\subsubsection{Fragmentation functions}
\index{fragmentation function}

Since the parton-hadron transition is non-perturbative, it is not possible to perturbatively calculate quantities such
as the energy-spectra of specific hadrons in high-energy collisions.  However, one can factorize perturbative and
non-perturbative contributions via the concept of fragmentation functions. These are the final-state analogue of the
parton distribution functions which are used for initial-state hadrons.  Like parton distribution functions, they depend
on a (fragmentation) factorization scale and satisfy a DGLAP evolution equation.

It should be added that if one ignores the non-perturbative difficulties and just calculates the energy and angular
spectrum of partons in perturbative QCD with some low cutoff scale $\sim {\rm \Lambda}$ (using resummation to sum large
logarithms of $\sqrt{s}/{\rm \Lambda}$), then this reproduces many features of the corresponding hadron
spectra\cite{qcd:Basics}.  This is often taken to suggest that hadronization is ``local'', in this sense it mainly
involves partons that are close both in position and in momentum.

Section~\crossref{frag} of this {\it Review} provides further information and references on these topics, including also
the question of heavy-quark fragmentation.

% ......................................................................
\subsubsection{Parton-shower Monte Carlo event generators}

Parton-shower Monte Carlo (MC) event generators like
\PYTHIA\cite{Sjostrand:2000wi,Sjostrand:2006za,*qcd:Sjostrand:2006za-url,Sjostrand:2014zea,*qcd:Sjostrand:2014zea-url},
\HERWIG\cite{Webber:1983if,Corcella:2000bw,*qcd:Corcella:2000bw-url,Bahr:2008pv,*qcd:Bahr:2008pv-url}, and
\SHERPA\cite{Gleisberg:2008ta}%
\footnote{The program \ARIADNE{\protect\cite{Lonnblad:1992tz}} has also been widely used for simulating $e^+e^-$ and DIS
  collisions.} %
provide fully exclusive simulations of QCD events at the level of measurable particles, the so-called ``particle level''
or ``hadron level''.\index{particle level} \index{hadron level} Here, ``measurable'' refers to color-neutral particles
with mean lifetimes $\tau$ long enough to be associated with tracks or decay vertices in particle detectors. Usually,
this requires mean decay lengths $c\tau$ of around $10\,$mm. As such MC event generators are a crucial tool for all
applications that involve simulating the response of detectors to QCD events.  Here we give only a brief outline of how
they work and refer the reader to Sec.~\crossref{mcgen} and Ref.~\cite{Buckley:2011ms} for a full overview.

In general, we expect parton-shower matched predictions to differ from the underlying fixed-order results in regions
where (1) there is a large sensitivity to jet shapes (for instance small R jets), (2) there is a restriction in phase
space such that soft gluon resummation effects become important, (3) the observable contains multiple disparate scales,
(4) there are perturbative instabilities at fixed order, \eg related to kinematical cuts, and (5) the observable is
sensitive to higher multiplicity states than those described by the fixed-order calculation (for an explicit study of
some of these effects see \eg \cite{Bellm:2019yyh}).

The MC generation of an event involves several stages. It starts with the random generation of the kinematics and
partonic channels of whatever {\it hard scattering process} the user has requested at some high scale $Q_0$ (for complex
processes, this may be carried out by an external program).
This is followed by a {\it parton shower},\index{parton shower}
usually based on the successive random generation of gluon emissions (and $g \to q\bar q$ splittings).
Leading contributions to the shower have emissions that are ordered according to some ordering variable. Common choices
of scale for the ordering of emissions are virtuality, transverse momentum or angle.
Each emission is generated at a scale lower than the previous emission, following a (soft and collinear resummed)
perturbative QCD distribution, which depends on the momenta of all previous emissions.
Parton showering stops at a scale of order 1\GeV, at which point a {\it hadronization model}\index{hadronization
  model}
is used to convert the resulting partons into hadrons.
One widely-used model involves stretching a color ``string'' across quarks and gluons, and breaking it up into
hadrons\cite{Andersson:1983ia,Sjostrand:1984ic}.
Another breaks each gluon into a \qqbar pair and then groups quarks and anti-quarks into colorless ``clusters'', which
then give the hadrons\cite{Webber:1983if}.
As both models are tuned primarily to LEP data, the cluster and string models provide similar results for most
observables sensitive to hadronization\cite{Bellm:2019yyh}.
For \pp and $\gamma p$ processes, modeling is also needed to treat the collision between the two hadron remnants, which
generates an {\it underlying event} (UE),\index{underlying event (UE)}
usually implemented via additional $2\to2$ scatterings (``multiple parton interactions'', MPI)\index{Multiple
  parton-parton interactions, MPI}
at a scale of a few \GeV, following Ref.~\cite{Sjostrand:1987su}.
The parameter values for the MPI models must be determined from fits to minimum-bias and/or underlying-event observables of LHC collision
data. As the different MC event generators are adapted to essentially the same measurements, ideally the respective MPI
implementations should lead to similar predictions for each program.  One complication, however, is some non-universality
of the underlying event among different physics processes.

A deficiency of the soft and collinear approximations that underlie parton showers is that they may fail to reproduce
the full pattern of hard wide-angle emissions, important, for example, in many new physics searches.
It is therefore common to use LO multi-parton matrix elements to generate hard high-multiplicity partonic configurations
as additional starting points for the showering, supplemented with some prescription (CKKW\cite{Catani:2001cc},
MLM\cite{Alwall:2007fs}{}) for consistently merging samples with different initial multiplicities.
Monte Carlo generators, as described above, compute cross sections for the requested hard process that are correct at
LO\@.

A wide variety of processes are available in MC implementations that are correct to NLO, using the
\MCatNLO~\cite{Frixione:2002ik} or \POWHEG\cite{Nason:2004rx} prescriptions, through the
\MGMCatNLO~\cite{Alwall:2014hca}, \POWHEGBOX~\cite{Alioli:2010xd,*qcd:Alioli:2010xd-url} and
\SHERPA~\cite{Gleisberg:2008fv} programs.
Techniques have also been developed to combine NLO plus parton shower accuracy for different multiplicities of final-state
jets\cite{Platzer:2012bs,*Frederix:2012ps,*Hamilton:2012rf}.
While NLO+PS accurate predictions can now be implemented in the above tools rather easily and are available for a range
of processes, given the advances in NNLO calculations it is natural to seek for NNLO+PS accurate Monte Carlos as
well. The two main approaches today are the MiNNLO~\cite{Monni:2019whf} and Geneva~\cite{Alioli:2013hqa} ones.  NNLO
plus shower accuracy was achieved first for Drell-Yan and Higgs
production~\cite{Hamilton:2013fea,Karlberg:2014qua,*Hoeche:2014aia,*Hoche:2014dla,*Alioli:2015toa,*Alioli:2021qbf,*Alioli:2023har}
and is now available for several $2\to 2$ color singlet
processes~\cite{Astill:2016hpa,Astill:2018ivh,Re:2018vac,Alioli:2019qzz,Lombardi:2020wju,Lombardi:2021wug,Alioli:2020qrd,Lombardi:2021rvg,Alioli:2021egp,Alioli:2022dkj,Lindert:2022qdd,Gavardi:2022ixt,Buonocore:2021fnj,Haisch:2022nwz,Zanoli:2021iyp},
as well as for heavy-quark pair-production~\cite{Mazzitelli:2020jio,Mazzitelli:2021mmm,Mazzitelli:2023znt}.

It is important to understand/verify the accuracy of the parton shower predictions in the Monte Carlo programs.
A general framework for assessing the logarithmic accuracy of parton-shower algorithms has been formulated, based on
 their ability to reproduce the singularity structure of multi-parton matrix elements, and their ability to reproduce
logarithmic resummation results~\cite{Dasgupta:2018nvj}.  The dominant contributions relevant for the extension of
parton showers to higher logarithmic accuracy have been
computed~\cite{Hartgring:2013jma,Li:2016yez,Hoche:2017iem,Dulat:2018vuy,Banfi:2018mcq} and included in some algorithms.

There exist ways to improve on current parton-shower algorithms~\cite{Bewick:2019rbu,Dasgupta:2020fwr} and to
demonstrate the parton shower accuracy through a comparison to analytic next-to-leading logarithmic calculations for a range
of observables~\cite{Dasgupta:2020fwr}.
Considerable progress has been achieved in recent years~\cite{Bewick:2019rbu,Hamilton:2020rcu,Forshaw:2020wrq,
  Nagy:2020rmk, Nagy:2020dvz, Karlberg:2021kwr, Hamilton:2021dyz,Herren:2022jej,
  Dasgupta:2020fwr,Nagy:2022bph,vanBeekveld:2022zhl,vanBeekveld:2022ukn,vanBeekveld:2023lfu}.
References~\cite{Dasgupta:2020fwr, Herren:2022jej,Nagy:2020rmk,Nagy:2020dvz,Nagy:2022bph} introduced the first shower
implementations towards NLL, but these implementations did not include spin-correlation effects and lacked full color
dependence at LL level. Ref.\cite{Hamilton:2020rcu} incorporated full color effects at the NLL level, while spin
correlations at NLL were included in Refs.\cite{Karlberg:2021kwr, Hamilton:2021dyz}, resulting in an NLL accurate parton
shower for proton-proton collisions~\cite{vanBeekveld:2022zhl, vanBeekveld:2022ukn}. The modifications necessary for
processes such as Deep Inelastic Scattering (DIS) and Vector Boson Fusion (VBF) are addressed
in~\cite{vanBeekveld:2023lfu}. The next steps involve incorporating non-perturbative effects, accounting for quark
masses, and tuning NLL parton shower on experimental data.

% ......................................................................
\subsection{Accuracy of predictions}
\label{qcd:sec:qcd-accuracy}

Estimating the accuracy of perturbative QCD predictions is not an exact science.  It is often said that LO calculations
are accurate to within a factor of two. This is based on experience with NLO corrections in the cases where these are
available.  In processes involving new partonic scattering channels at NLO and/or large ratios of scales (such as jet
observables in processes with vector bosons, or the production of high-$p_t$ jets containing $B$-hadrons), the ratio of
the NLO to LO predictions, commonly called the ``$K$ factor'',\index{K@$K$ factor} can be substantially larger than
two.  NLO corrections tend to be large for processes for which there is a great deal of color annihilation in the
interaction.  In addition, NLO corrections tend to decrease as more final state legs are added.

For calculations beyond LO, one approach to estimate the perturbative uncertainty is to base it on the last known
perturbative correction; this may lead to misleading results if new sub-processes are present at the next-higher order.
A more widely used method is to estimate it from the change in the prediction when varying the renormalization and
factorization scales around a central value $Q$ that is taken close to the physical scale of the process.  A
conventional range of variation is $Q/2 < \mu_R,\mu_F < 2Q$, varying the two scales independently with the restriction
${\frac{1}{2}} \mu_R < \mu_F < 2 \mu_R$\cite{Cacciari:2003fi}.
This constraint limits the risk of misleadingly small uncertainties due to fortuitous cancellations between the $\mu_F$
and $\mu_R$ dependence when both are varied together, while avoiding the appearance of large logarithms of
$\mu_R^2 / \mu_F^2$ when both are varied completely independently.  Where possible, it can be instructive to examine the
two-dimensional scale distributions ($\mu_R$ \vs $\mu_F$) to obtain a better understanding of the interplay between
$\mu_R$ and $\mu_F$.  This procedure should not be assumed to always estimate the full uncertainty from missing higher
orders, but it does indicate the size of one important known source of higher-order ambiguity.%
\footnote{%
  Various studies have been carried out on how to estimate uncertainties from missing higher orders that go beyond scale
  variations\cite{Cacciari:2011ze,David:2013gaa,Bagnaschi:2014wea,Duhr:2021mfd,Ghosh:2022lrf}.} %

Most $2\rightarrow2$ processes at the LHC are now known to NNLO. This typically results in a large reduction in the
uncertainty from that obtained at NLO. However, care must be taken in the estimate of the uncertainty in processes
containing jets, as accidental cancellations of the scale uncertainties may result an artificial reduction of the scale
uncertainty; for some jet $R$ values, the uncertainty may even be zero. There are several possibilities for providing a
more realistic estimate of the scale uncertainty for such processes ~\cite{Bellm:2019yyh}, but none have been widely
adopted at the LHC.

Calculations that involve resummations usually have an additional source of uncertainty associated with the choice of
argument of the logarithms being resummed, \eg $\ln (2 \frac{p_{t}^{Z}}{ \mZ})$ as opposed to
$\ln (\frac{1}{2} \frac{p_{t}^{Z}}{\mZ})$, as well as a prescription to switch off resummation effects when the
logarithm is not large.  In addition to varying renormalization and factorization scales, it is common practice to vary
the argument of the logarithm by a suitable factor in either direction with respect to the default argument.

The accuracy of QCD predictions is limited also by non-perturbative (or hadronization) corrections,
\index{non-perturbative corrections}\index{hadronization corrections} which typically scale as a power of
${\rm \Lambda}/Q$.%
\footnote{In some circumstances, the scale in the denominator could be a smaller kinematic or physical scale that
  depends on the observable.} %
For measurements that are directly sensitive to the structure of the hadronic final state, the corrections are usually
linear in ${\rm \Lambda}/Q$.  The non-perturbative corrections are further enhanced in processes with a significant
underlying event (\ie in \pp and \ppbar collisions) and the impact of
non-perturbative corrections is larger in cases where the perturbative cross sections fall steeply as a
function of $p_t$ or some other kinematic variable, for example in inclusive jet $p_t$ or dijet mass spectra.
Under high-luminosity
running conditions, such as 13\TeV at the LHC, there can be on the order of 50 minimum-bias interactions occurring at
each beam-beam crossing. This additional energy needs to be subtracted, and is typically removed by means of a
rapidity-dependent transverse energy density determined on an event-by-event basis~\cite{Soyez:2012hv}. This
subtraction, of necessity, also removes the underlying event, which must be added back in by means of MC event generator
modelling if one wants to restore the measured event to the particle level.

Non-perturbative corrections are commonly estimated from the difference between Monte Carlo events at the ``parton
level''\index{parton level} and at particle level.  Parton level refers to the stage of the parton shower, where the
evolution is stopped at an energy scale of typical hadron masses of a few \GeV.  An issue to be aware of is that
``parton level'' is not a uniquely defined concept.  For example, in a MC event generator such a procedure depends on an
arbitrary and tunable internal cutoff scale that separates the parton showering from the hadronization.  In contrast, no
such cutoff scale exists in an NLO or NNLO partonic calculation.  The uncertainties in these corrections are often
estimated by comparing different tunes of the various MC event generators.  It should be noted that such estimates are
not guaranteed to fully cover the true uncertainties.

Alternative methods exist for estimating hadronization corrections, that attempt to analytically deduce non-perturbative
effects in one observable based on measurements of other observables (see the
reviews\cite{Beneke:1998ui,Dasgupta:2003iq}).  While they directly address the problem of different possible definitions
of parton level, it should also be said that they are far less flexible than Monte Carlo programs and not always able to
provide equally good descriptions of the data.

One of the main issues is whether the fixed-order partonic final state of a NLO or NNLO prediction can match the parton
shower in its ability to describe the experimental jet shape (minus any underlying event). Calculations at NNLO provide
a better match to the parton shower predictions than do NLO ones, as might be expected from the additional gluon
available to describe the jet shape.  The hadronization predictions appear to work for both orders, but at an unknown
accuracy.

% ======================================================================
\section{Experimental studies of QCD}
\label{qcd:sec:exptest-GD}

Since we are not able to directly measure partons (quarks or gluons), but only hadrons and their decay products, a
central issue for every experimental study of perturbative QCD is establishing a correspondence between observables
obtained at the parton and the hadron level. The only theoretically sound correspondence is achieved by means of {\it
  infrared and collinear safe}\index{infrared and collinear safety}\index{collinear and infrared safety} quantities,
which allow one to obtain finite predictions at any order of perturbative QCD.

As stated above, the simplest case of infrared- and collinear-safe observables are inclusive cross sections. More generally,
when measuring fully inclusive observables, the final state is not analyzed at all regarding its (topological,
kinematical) structure or its composition. Basically the relevant information consists in the rate of a process ending
up in a partonic or hadronic final state. In \ee annihilation, widely used examples are the ratios of partial widths or
branching ratios for the electroweak decay of particles into hadrons or leptons, such as $Z$ or $\tau$ decays, (\cf
Sec.~\ref{qcd:sec:inclcrosssec}).  Such ratios are often favored over absolute cross sections or partial widths because
of large cancellations of experimental and theoretical systematic uncertainties. The strong suppression of
non-perturbative effects, ${\cal O}({\rm \Lambda}^4/Q^4)$, is one of the attractive features of such observables,
however, at the same time, the sensitivity to radiative QCD corrections is small, which for example leads to a larger
statistical uncertainty when using them for the determination of the strong coupling constant. In the case of $\tau$
decays not only the hadronic branching ratio is of interest, but also moments of the spectral functions of hadronic $\tau$
decays, which sample different parts of the decay spectrum and thus provide additional information.  Other examples of
fully inclusive observables are structure functions (and related sum rules) in DIS.  These are extensively discussed in
Sec.~\ref{strucfun} of this {\it Review}.

On the other hand, often the structure or composition of the final state are analyzed, and cross sections differential in
one or more variables characterizing this structure are of interest.  Examples are jet rates, jet substructure, event
shapes or transverse momentum distributions of jets or vector bosons in hadron collisions.  The case of fragmentation
functions, \ie the measurement of hadron production as a function of the hadron momentum relative to some hard
scattering scale, is discussed in Sec.~\ref{frag} of this {\it Review}.

It is worth mentioning that, besides the correspondence between the parton and hadron level, also a correspondence
between the hadron level and the actually measured quantities in the detector has to be established. The simplest
examples are corrections for finite experimental acceptance and efficiencies. Whereas acceptance corrections essentially
are of theoretical nature, since they involve extrapolations from the measurable (partial) to the full phase space,
other corrections such as for efficiency, resolution and response are of experimental nature.  For example, measurements
of differential cross sections such as jet rates require corrections in order to relate, \eg, the energy deposits in a
calorimeter to the jets at the hadron level.  Typically detector simulations and/or data-driven methods are used in
order to obtain these corrections. Care should be taken here in order to have a clear separation between the
parton-to-hadron level and hadron-to-detector level corrections. Finally, for the sake of an easy comparison to the
results of other experiments and/or theoretical calculations, it is suggested to provide, whenever possible,
measurements corrected for detector effects and/or all necessary information related to the detector response (\eg, the
detector response matrix). Any fiducial phase space for measurements should be defined as close as possible to the
detector-level selection in order to minimize model-dependent extrapolations. A versatile repository for storing such
information is the use of Rivet routines~\cite{Bierlich:2019rhm}.

% ======================================================================
\subsection{Hadronic final-state observables}
\label{qcd:sec:qcd-exp-observables}

% ----------------------------------------------------------------------
\subsubsection{Jets}
\label{qcd:sec:jetalgoGD}

\hyphenation{bun-ches} In hard interactions, final-state partons and hadrons appear predominantly in collimated bunches,
which are generically called {\it jets}.\index{QCD!jets}
To a first approximation, a jet can be thought of as a hard parton that has undergone soft and collinear showering and
then hadronization.  Jets are used both for testing our understanding and predictions of high-energy QCD processes, and
also for identifying the hard partonic structure of decays of massive particles such as top quarks and W, Z and Higgs
bosons.

In order to map observed hadrons onto a set of jets, one uses a {\it jet definition}.\index{jet definition}
The mapping involves explicit choices: for example when a gluon is radiated from a quark, for what range of kinematics
should the gluon be part of the quark jet, or instead form a separate jet?  Good jet definitions are infrared and
collinear safe, simple to use in theoretical and experimental contexts, applicable to any type of inputs (parton or
hadron momenta, charged particle tracks, and/or energy deposits in the detectors) and lead to jets that are not too
sensitive to non-perturbative effects.

An extensive treatment of the topic of jet definitions is given in Ref.~\cite{Moretti:1998qx}{} (for $e^+e^-$
collisions) and Refs.~\cite{Salam:2009jx,Ellis:2007ib,Cacciari:2015jwa}.  Here we briefly review the two main classes:
cone algorithms, extensively used at hadron colliders before the LHC,\index{jet algorithm!cone} and sequential recombination
algorithms,\index{jet algorithm!sequential recombination}
more widespread in \ee and $ep$ colliders and at the LHC.

Very generically, most (iterative) cone algorithms start with some seed particle $i$, sum the momenta of all particles
$j$ within a cone of opening-angle $R$, typically defined in terms of differences in rapidity and azimuthal angle. They then take the
direction of this sum as a new seed and repeat until the direction of the cone is stable, and call the contents of the
resulting stable cone a jet if its transverse momentum is above some threshold $p_{t,{\rm min}}$.  The parameters $R$
and $p_{t,{\rm min}}$ should be chosen according to the needs of a given analysis.

There are many variants of the cone algorithm, and they differ in the set of seeds they use and the manner in which they
ensure a one-to-one mapping of particles to jets, given that two stable cones may share particles (``overlap'').  The
use of seed particles is a problem w.r.t.\ infrared and collinear safety. Seeded algorithms are generally not compatible
with higher-order (or sometimes even leading-order) QCD calculations, especially in multi-jet contexts, as well as
potentially subject to large non-perturbative corrections and instabilities.  Seeded algorithms (JetCLU, MidPoint, and
various other experiment-specific iterative cone algorithms) are therefore to be deprecated.  Such algorithms are not
used at the LHC, but were at the Fermilab Tevatron, where data still provide useful information, for example for global
PDF fits.\footnote{In the data, the difference between the use of the Midpoint algorithm and the use of the SISCone algorithm is
  small~\cite{CDF:2005vhb}, allowing the use of the SISCone algorithm for any theory comparisons at higher order in QCD.} %
A modern alternative is to use a seedless variant, SISCone\cite{Salam:2007xv}.

Sequential recombination algorithms at hadron colliders (and in DIS) are characterized by a distance
$d_{ij} = \min(k^{2p}_{t,i}, k^{2p}_{t,j}) \Delta^2_{ij} / R^2$ between all pairs of particles $i,j$, where
$\Delta_{ij}$ is their separation in the rapidity-azimuthal plane, $k_{t,i}$ is the transverse momentum w.r.t.\ the
incoming beams, and $R$ is a free parameter. At the LHC, $R$ is typically in the range from 0.4 to 0.8, although
analyses can also use jet sizes up to 1.0-1.2.  They also involve a ``beam'' distance $d_{iB} = k_{t,i}^{2p}$.  One
identifies the smallest of all the $d_{ij}$ and $d_{iB}$, and if it is a $d_{ij}$, then $i$ and $j$ are merged into a
new pseudo-particle (with some prescription, a recombination scheme,\index{recombination scheme}
for the definition of the merged four-momentum).  If the smallest distance is a $d_{iB}$, then $i$ is removed from the
list of particles and called a jet.  As with cone algorithms, one usually considers only jets above some
transverse-momentum threshold $p_{t,{\rm min}}$.  The parameter $p$ determines the kind of algorithm: $p=1$ corresponds
to the ({\it inclusive-})$k_t$ algorithm\cite{Catani:1991hj,Catani:1993hr,Ellis:1993tq}, $p=0$ defines the {\it
  Cambridge-Aachen} algorithm\cite{Dokshitzer:1997in,Wobisch:1998wt}, while for the {\it anti-}$k_t$ algorithm
$p=-1$\cite{Cacciari:2008gp}. All these variants are infrared and collinear safe.  Whereas the former two lead to
irregularly shaped jet boundaries, the latter results in cone-like boundaries, except in situations where there are
nearby jets.  The {\it anti-}$k_t$ algorithm has become the de-facto standard for the LHC experiments.

In \ee annihilation the $k_t$ algorithm\cite{Catani:1991hj} uses
$y_{ij} = 2\,\min(E_{i}^2, E_{j}^2) (1-\cos \theta_{ij})/Q^2$ as distance measure between two particles/partons $i$ and
$j$ and repeatedly merges the pair with smallest $y_{ij}$, until all $y_{ij}$ distances are above some threshold
$y_{\rm cut}$, the jet resolution parameter. $Q$ is a measure of the overall hardness of the event.  The
(pseudo)-particles that remain at this point are called the jets.  Here it is $y_{\rm cut}$ (rather than $R$ and
$p_{t,\min}$) that should be chosen according to the needs of the analysis.  The two-jet rate in the $k_t$ algorithm has
the property that logarithms $\ln(1/y_{\rm{cut}})$ exponentiate.  This is one reason why it is preferred over the
earlier JADE\ algorithm\cite{Bartel:1986ua}, which uses the distance measure
$y_{ij} = 2\,E_{i}\,E_{j}\, (1-\cos \theta_{ij})/Q^2$.  Note that other variants of sequential recombination algorithms
for \ee annihilations, using different definitions of the resolution measure $y_{ij}$, exhibit much larger sensitivities
to fragmentation and hadronization effects than the $k_t$ and JADE\ algorithms\cite{Bethke:1991wk}.  Efficient
implementations of the above algorithms are available through the {\it FastJet}
package\cite{Cacciari:2005hq,*Cacciari:2011ma}.

While building infrared (IR) safe jets is generally considered a solved problem, this is not quite the case when one
also wishes to assign a flavor to a jet. In fact, the general experimental definition considers a flavored-jet to be a
jet that contains at least one flavor tag (such as a $B$ or $D$ meson) above a given transverse momentum
threshold. Because of collinear or soft wide-angle $g\to q\bar q$ splittings, it is easy to see that such a definition
is neither collinear nor infrared safe.  This problem was addressed in Ref.~\cite{Banfi:2006hf,Banfi:2007gu} in the
context of heavy-flavor production at the Tevatron. However, the jet-algorithm proposed in that work was impractical to
implement experimentally because it was based on the $k_t$-algorithm. Furthermore, it required tagging two nearby
flavored hadrons and involved a rather complex beam-distance measure.  Since most experimental studies instead rely on
the anti-$k_t$ algorithm, recent investigations have focused on developing algorithms that preserve the anti-$k_t$
kinematics of the jets while assigning jet flavors in an infrared safe
way~\cite{Czakon:2022wam,Gauld:2022lem,Caletti:2022glq}. It turns out that the problem is more involved than
anticipated, and a formulation of infrared-safe anti-$k_t$-like jets could only be achieved by introducing an
interleaved flavor neutralization procedure~\cite{Caola:2023wpj}. An unfolding procedure will be necessary to convert
experimental measurements of flavor-$k_t$ jets to a form that can be directly compared to theoretical predictions.

% ======================================================================
\subsubsection{Event Shapes}
\label{qcd:sec:evshapesGD}
\index{event shape}

Event-shape variables are functions of the four momenta of the particles in the final state and characterize the
topology of an event's energy flow.  They are sensitive to QCD radiation (and correspondingly to the strong coupling)
insofar as gluon emission changes the shape of the energy flow.

The classic example of an event shape is the {\it thrust}\cite{Brandt:1964sa,Farhi:1977sg}\index{thrust} in \ee
annihilations, defined as
\begin{equation}\label{qcd:eq:thrust-GD}
  \hat\tau = \max_{\vec{n}_\tau} \frac{\sum_i |\vec{p}_i \cdot \vec{n}_\tau|}{\sum_i |\vec{p}_i|}\; ,
\end{equation}
\noindent
where $\vec{p}_i$ are the momenta of the particles or the jets in the final-state and the maximum is obtained for the
thrust axis $\vec{n}_\tau$.  In the Born limit of the production of a perfect back-to-back \qqbar pair, the limit
$\hat\tau \rightarrow 1$ is obtained, whereas a perfectly spherical many-particle configuration leads to
$\hat\tau \rightarrow 1/2$. Further event shapes of similar nature have been extensively measured at LEP and at HERA,
and for their definitions and reviews we refer to
Refs.~\cite{Ellis:1991qj,Dissertori:2003pj,Dasgupta:2003iq,Biebel:2001dm,Kluth:2006bw}.  The energy-energy correlation
function\cite{Basham:1978bw},\index{energy-energy correlation} namely the energy-weighted angular distribution of
produced hadron pairs, and its associated asymmetry are further shape variables which have been studied in detail at \ee
colliders. For hadron colliders the appropriate modification consists in only taking the transverse momentum
component\cite{Ali:1984yp}.  The event shape variable {\it N-jettiness} has been proposed\cite{Stewart:2010tn}, that
measures the degree to which the hadrons in the final state are aligned along $N$ jet axes or the beam direction.  It
vanishes in the limit of exactly $N$ infinitely narrow jets.

Phenomenological discussions of event shapes at hadron colliders can be found in
Refs.~\cite{Stewart:2010tn,Banfi:2004nk,Banfi:2010xy,Becher:2015lmy,Gao:2019ojf}.  Measurements of hadronic event-shape
distributions have been published by CDF\cite{Aaltonen:2011et},
ATLAS\cite{Aad:2012np,Aad:2012fza,ATLAS:2015yaa,Aad:2016ria,Aaboud:2017fml,Aaboud:2018hie,ATLAS:2023tgo} and
CMS\cite{Khachatryan:2011dx,Chatrchyan:2013tna,Khachatryan:2014ika,Sirunyan:2018adt}.

Event shapes are used for many purposes. These include measuring the strong coupling (see \eg
Ref.~\cite{ATLAS:2023tgo}), tuning the parameters of Monte Carlo programs, investigating analytical models of
hadronization and distinguishing QCD events from events that might involve decays of new particles (giving event-shape
values closer to the spherical limit).

% ======================================================================
\subsubsection{Jet substructure, quark \vs gluon jets}
\label{sec:QCDjetsub}
\index{jet substructure}\index{gluon induced jets}\index{quark induced jets}

Jet substructure, which can be resolved by finding subjets or by measuring jet shapes, is sensitive to the details of
QCD radiation in the shower development inside a jet and has been extensively used to study differences in the
properties of quark and gluon induced jets, strongly related to their different color charges. There is
clear experimental evidence that gluon jets have a softer particle spectrum and are ``broader'' than (light-) quark jets
(as expected from perturbative QCD) when looking at observables such as the jet shape\index{jet shape}
$\Psi(r/R)$ (see \eg Ref.~\cite{Ali:2010tw}). This is the fraction of transverse momentum contained within a sub-cone of
cone-size $r$ for jets of cone-size $R$. It is sensitive to the relative fractions of quark and gluon jets in an
inclusive jet sample and receives contributions from soft-gluon initial-state radiation and the underlying
event. Therefore, it has been widely employed for validation and tuning of Monte Carlo parton-shower
models. Furthermore, this quantity turns out to be sensitive to the modification of the gluon radiation pattern in heavy
ion collisions (see \eg Ref.~\cite{Chatrchyan:2013kwa}).

Jet shape measurements using proton-proton collision data have been presented for inclusive jet
samples\cite{Aad:2011kq,Chatrchyan:2012mec,ALICE:2014dla} and for top-quark production\cite{Aad:2013fba}.  Further
discussions, references and summaries can be found in Refs.~\cite{Kluth:2006bw,Glasman:2008yf,Carli:2015qta,Gras:2017jty} and Sec.~4
of Ref.~\cite{Abdesselam:2010pt}.

The use of jet substructure has also been investigated in order to distinguish QCD jets from jets that originate from
hadronic decays of boosted massive particles (high-$p_t$ electroweak bosons, top quarks and hypothesized new particles).
A considerable number of experimental studies have been carried out with Tevatron and LHC data, in order to investigate
the performance of the proposed algorithms for resolving jet substructure and to apply them to searches for new physics,
as well as to the reconstruction of boosted top quarks, vector bosons and the Higgs boson.  For reviews of this rapidly
growing field, see Sec.~5.3 of Ref.~\cite{Salam:2009jx},
Refs.~\cite{Krohn:2009th,Abdesselam:2010pt,Altheimer:2012mn,Altheimer:2013yza,Stichel:2014oka,Adams:2015hiv,deOliveira:2015xxd,Kogler:2018hem,Marzani:2019hun}
and references thereto.
One convenient representation
for visualizing jet substructure is through the Lund plane~\cite{Dreyer:2018nbf}. From a theoretical perspective, the Lund plane would be
constructed using quarks and gluons, but similar observables can be constructed through the use of jets, as described in detail in
Ref.~\cite{ATLAS:2020bbn}.

Neural network techniques and deep learning methods have also been applied to jet and top physics and jet substructure,
see \eg Refs.~\cite{Louppe:2017ipp,Guest:2018yhq,Kasieczka:2019dbj,Komiske:2018cqr}.  Perhaps no other sub-field has
benefited as much from machine learning techniques as the study of jet substructure. As a jet can have O(100)
constituents each with kinematic and other information, jet substructure analysis is naturally a highly multivariate
problem.  Deep learning techniques can use all of the available information to study jets in their natural high
dimensionality. Such techniques have not only improved discrimination between different final states/types of jets, but
have also improved our understanding of perturbative QCD. See for example the review in Ref.~\cite{Larkoski:2017jix}.

% ======================================================================
\subsection{QCD measurements at colliders}

There exists a wealth of data on QCD-related measurements in \ee, $ep$, \pp, and \ppbar collisions, to which a short
overview like this would not be able to do any justice. Reviews of the subject have been published in
Refs.~\cite{Biebel:2001dm,Kluth:2006bw} for \ee colliders and in Ref.~\cite{SchornerSadenius:2012de} for $ep$
scattering, whereas for hadron colliders overviews are given in, \eg, Refs.~\cite{Ellis:2007ib,Carli:2015qta} and
Refs.~\cite{Campbell:2006wx,Mangano:2010zza,Butterworth:2012fj,qcd:CHKbook,Gross:2022hyw}.

Below we concentrate our discussion on measurements that are most sensitive to hard QCD processes with focus on jet
production.

% ----------------------------------------------------------------------
\subsubsection{\ee colliders}

Analyses of jet production in \ee collisions are mostly based on data from the JADE experiment at center-of-mass
energies between $14$ and $44$\GeV, as well as on LEP collider data at the $Z$ resonance and up to $209$\GeV.  The
analyses cover the measurements of (differential or exclusive) jet rates (with multiplicities typically up to 4, 5 or 6
jets), the study of three-jet events and particle production between the jets, as well as four-jet production and
angular correlations in four-jet events.

Event-shape distributions from \ee data have been an important input to the tuning of parton shower MC models, typically
matched to matrix elements for three-jet production.  In general these models provide good descriptions of the
available, highly precise data.  Especially for the large LEP data sample at the $Z$ peak, the statistical uncertainties
are mostly negligible and the experimental systematic uncertainties are at the percent level or even below. These are
usually dominated by the uncertainties related to the MC model dependence of the efficiency and acceptance corrections
(often referred to as ``detector corrections'').

Observables measured in \ee collisions have been used for determinations of the strong coupling constant\index{Strong
  coupling constant} (\cf \Section{qcd:sec:coupling-measurement} below) and for putting constraints on the QCD color
factors\index{color factor}\index{QCD!color factor} (\cf Sec.~\ref{qcd:sec:basics-GS} for their definitions), thus
probing the non-Abelian nature of QCD.
Angular correlations in four-jet events are sensitive at leading order.
Some
sensitivity to these color factors, although only at NLO, is also obtained from event-shape distributions. Scaling
violations of fragmentation functions and the different subjet structure in quark and gluon induced jets also give
access to these color factors.  A compilation of results\cite{Kluth:2006bw} quotes world average values of
$C_A = 2.89 \pm 0.03\,(\rm{stat}) \pm 0.21\,(\rm{syst})$ and $C_F = 1.30 \pm 0.01\,(\rm{stat}) \pm 0.09\,(\rm{syst})$, with
a correlation coefficient of 82\%. These results are in perfect agreement with the expectations from SU(3) of $C_A = 3$
and $C_F = 4/3$.

% ----------------------------------------------------------------------
\subsubsection{DIS and photoproduction}

Jet measurements in $ep$ collisions, both in the DIS and photoproduction regimes, allow for tests of QCD factorization
(as they involve only one initial state proton and thus one PDF function), and provide sensitivity to both the gluon PDF
and to the strong coupling constant. Calculations are available at NNLO in both
regimes~\cite{Currie:2016ytq,Currie:2017tpe}.  An N$^3$LO calculation using the Projection-to-Born method was also presented
in Ref.~\cite{Currie:2018fgr,Gehrmann:2018odt}.  Experimental uncertainties of the order of 5--10\% have been achieved,
whereas statistical uncertainties are negligible to a large
extent. For comparison to theoretical predictions, at large jet $p_t$ the PDF uncertainty dominates the theoretical
uncertainty (typically of order 5--10\%, in some regions of phase space up to 20\%), therefore jet observables become
useful inputs for PDF fits.

In general, the data are well described by the NLO and NNLO matrix-element calculations, combined with DGLAP evolution
equations, in particular at large $Q^2$ and central values of jet pseudo-rapidity.  At low values of $Q^2$ and $x$, in
particular for large jet pseudo-rapidities, certain features of the data have been interpreted as requiring BFKL-type
evolution, though the predictions for such schemes are still limited.  It is worth noting that there is no indication
that the BKFL approximation is needed within the currently probed phase space in the $x,Q^2$ plane, and an alternative
approach\cite{Carli:2010cg}, which implements the merging of LO matrix-element based event generation with a parton
shower (using the \SHERPA framework), successfully describes the data in all kinematical regions, including the low
$Q^2$, low $x$ domain.  At moderately small $x$ values, it should perhaps not be surprising that the BFKL approach and
fixed-order matrix-element merging with parton showers may both provide adequate descriptions of the data, because some
part of the multi-parton phase space that they model is common to both approaches.

In the case of photoproduction, a wealth of measurements with low $p_t$ jets were performed in order to constrain the
photon content of the proton (which is by now determined with percent accuracy thanks to the LUX
approach~\cite{Manohar:2016nzj,Manohar:2017eqh}).  A few examples of measurements can be found in
Refs.~\cite{Chekanov:2007ab,Chekanov:2007ac,Aktas:2006qe,ZEUS:2011aa,Abramowicz:2012jz} for photoproduction and in
Refs.~\cite{Aaron:2009vs,Aaron:2007xx,Chekanov:2007pa,Chekanov:2008af,Abramowicz:2010cka,Abramowicz:2010ke,Chekanov:2008ab,
  Aaron:2010ac,H1:2014cbm,Andreev:2017vxu} for DIS\@.

% ----------------------------------------------------------------------
\subsubsection{Hadron-hadron colliders}

The spectrum of observables and the number of measurements performed at hadron colliders is enormous, probing many
regions of phase space and covering a huge range of cross sections, as exemplified in
\Fig{qcd:fig:LHC-xsec-summary-QCDreview} for a wide class of processes by the ATLAS experiment, and specialised to
top-quark related processes by the CMS experiment at the LHC\@. In general, the theory agreement with data is excellent
for a wide variety of processes, indicating the success of perturbative QCD with the PDF and the strong coupling as
inputs.  For the sake of brevity, in the following only certain classes of those measurements will be discussed, which
permit various aspects of the QCD studies to be addressed.  Most of our discussion will focus on LHC results, which are
available for center-of-mass energies of 2.76, 5, 7, 8 and 13\TeV with integrated luminosities of up to
$140\, \rm{fb}^{-1}$.  As of writing of this update, new results at 13.6\TeV are starting to appear.  Generally
speaking, besides representing precision tests of the Standard Model and QCD in particular, these measurements serve
several purposes, such as: (i) probing pQCD and its various approximations and implementations in MC models to quantify
the order of magnitude of not yet calculated contributions and to gauge their precision when used as background
predictions to searches for new physics, or (ii) extracting/constraining model parameters such as the strong coupling
constant or PDFs.

The final states measured at the LHC include single, double and triple gauge boson production, top production (single
top, top pair and four top production), Higgs boson production, alone and in conjunction with a W or Z boson, and with a
top quark pair.  Many/most of these events are accompanied by additional jets.  So far only limits have been placed on
double Higgs production. The volume of LHC results prohibits a comprehensive description in this {\it Review}; hence,
only a few highlights will be presented.

\begin{pdgxfigure}[place=p,wide=true,width=0.95\textwidth]
  \includegraphics{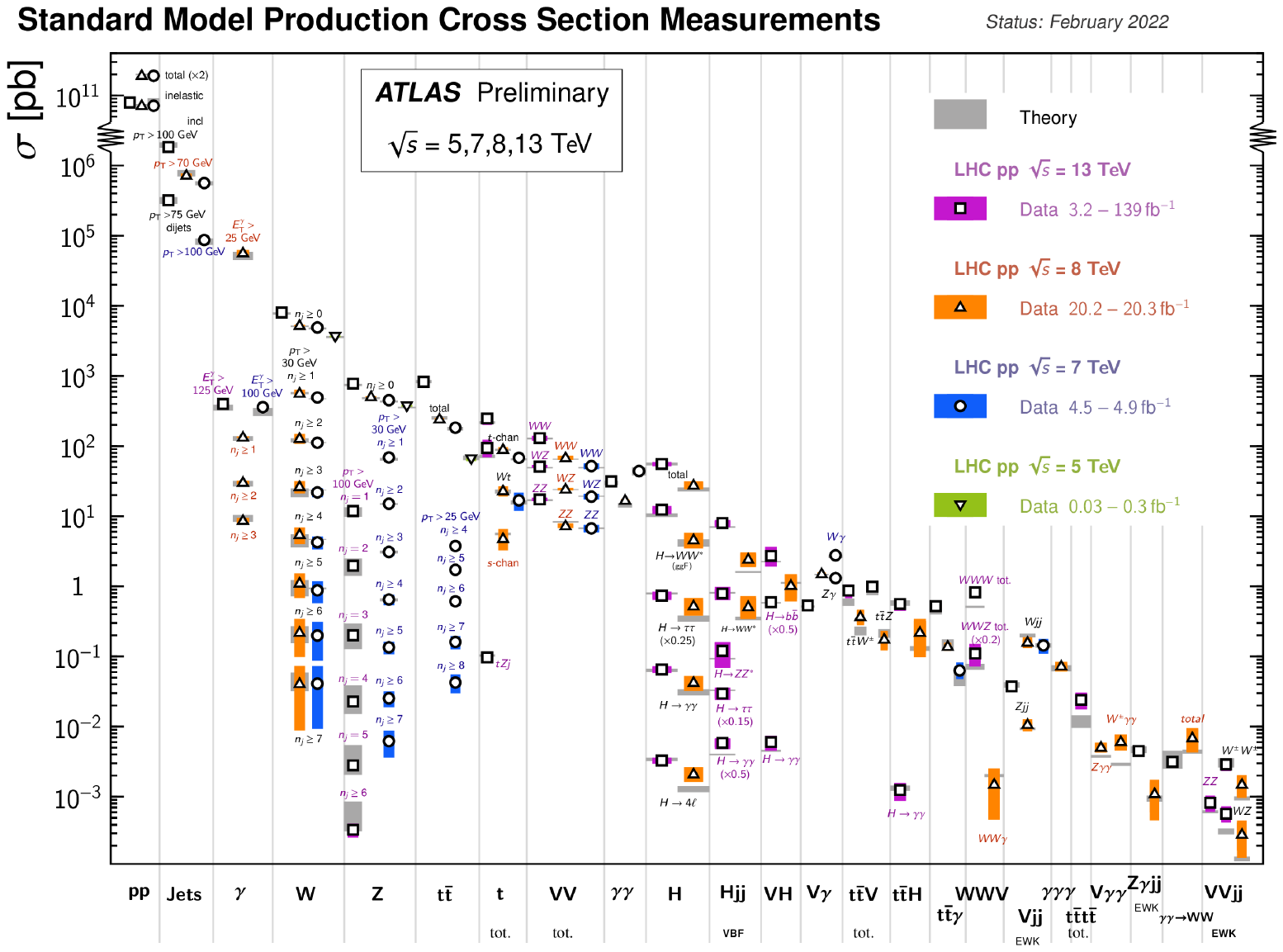}
  \includegraphics{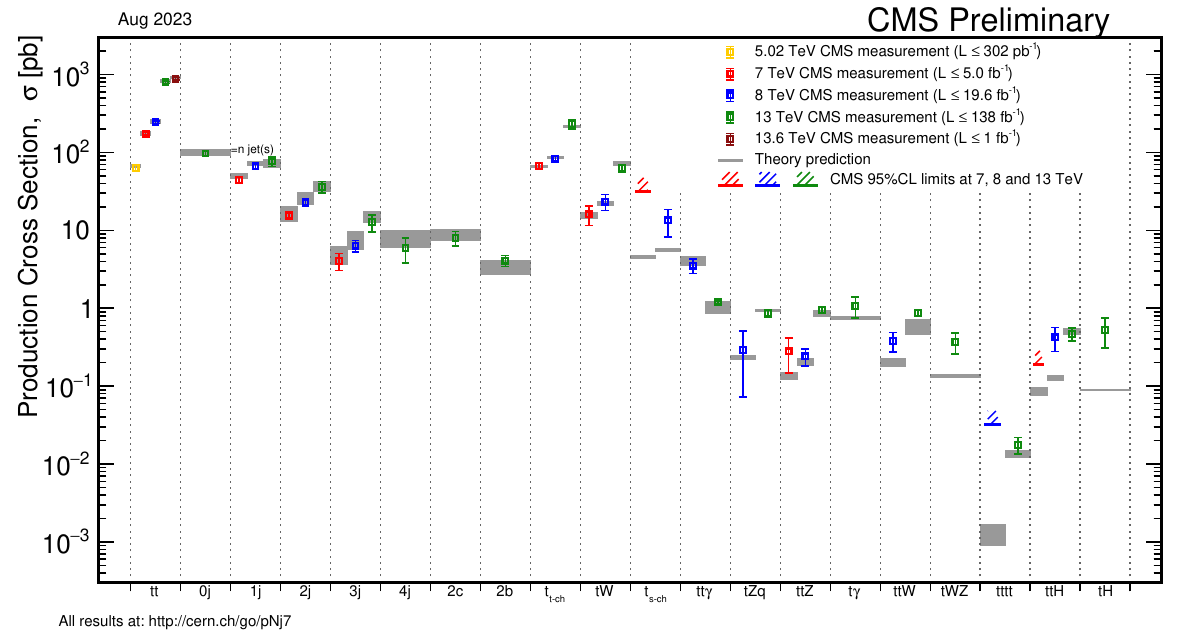}
  \caption{Overview of cross section measurements for a wide class of processes by the
    ATLAS{\protect\cite{qcd:ATLAS-xsec-overview}} experiment, and specialised to top-quark related processes by the
    CMS{\protect\cite{qcd:CMS-xsec-overview}} experiment at the LHC, for center-of-mass energies of 5, 7, 8, 13, and
    13.6\TeV. Also shown are the theoretical predictions and their uncertainties.}
  \label{qcd:fig:LHC-xsec-summary-QCDreview}
\end{pdgxfigure}

Among the most important cross sections measured, and the one with the largest dynamic range, is the inclusive jet
spectrum as a function of the jet transverse momentum ($p_t$), for several rapidity regions and for $p_t$ up to 700\GeV
at the Tevatron and $\sim 3.5$\TeV at the LHC.  It is worth noting that this latter upper limit in $p_t$ corresponds to a
distance scale of $\sim 10^{-19}$ m: no other experiment so far is able to directly probe smaller distance scales of
nature than this measurement.  The Tevatron inclusive jet measurements in Run 2
(Refs.~\cite{Abulencia:2007ez,Aaltonen:2008eq,Abazov:2008ae,Abazov:2011vi}) were carried out with the MidPoint jet
clustering algorithm (or its equivalent) and, in a few cases, with the $k_t$ jet clustering algorithm.  Most of the LHC
measurements use the {\it anti-}$k_t$ algorithm, with a variety of jet radii. The use of multiple jet radii in the same
analysis allows a better understanding of the underlying QCD dynamics.  Measurements by ALICE, ATLAS and CMS have been
published in
Refs.~\cite{Abelev:2013fn,Aad:2013lpa,Aad:2014vwa,Aaboud:2017dvo,Khachatryan:2015luy,Chatrchyan:2012bja,Khachatryan:2016mlc,Khachatryan:2016wdh,ATLAS:2017ble,CMS:2020caw,CMS:2021yzl}.

In general, we observe a good description of the data by the NLO and NNLO QCD predictions over about 11 orders of
magnitude in cross section, as long as care is taken for the functional form of the central scale choice\cite{Currie:2018xkj}.  The
experimental systematic uncertainties are dominated by the jet energy scale uncertainty, quoted to be in the range of a
few percent (see for instance the review in Ref.~\cite{Schwartzman:2015ada}), leading to uncertainties of $\sim5-30\%$
on the cross section, increasing with $p_t$ and rapidity.  The PDF uncertainties dominate the theoretical uncertainty at
large $p_t$ and rapidity.  In fact, inclusive jet data are one of the most important inputs to global PDF fits, in
particular for constraining the high-$x$ gluon PDF\cite{Hou:2019jgw,AbdulKhalek:2020jut,Jing:2023isu}.  Constraints on
the PDFs can also be obtained from ratios of inclusive cross sections at different center-of-mass
energies\cite{Aad:2013lpa,Khachatryan:2016mlc}.  Ratios of jet cross sections are a means to (at least
partially) cancel the jet energy scale uncertainties and thus provide jet observables with significantly improved
precision.

Dijet events are typically analyzed in terms of their invariant mass or average dijet $p_t$ and angular distributions,
which allows for tests of NLO and NNLO QCD predictions (see \eg
Refs.~\cite{Chatrchyan:2012bja,Aad:2013tea,Sirunyan:2017skj} for recent LHC results), and for setting stringent limits
on deviations from the Standard Model, such as quark compositeness or contact interactions (some examples can be found
in
Refs.~\cite{Aaltonen:2008dn,Abazov:2009ac,Chatrchyan:2012bf,Khachatryan:2014cja,Sirunyan:2017ygf,ATLAS:2012pu,Aaboud:2017yvp,ATLAS:2017ble}).
Furthermore, dijet azimuthal correlations between the two leading jets, normalized to the total dijet cross section, are
an extremely valuable tool for studying the spectrum of gluon radiation in the event.  The azimuthal separation of the
two leading jets is sensitive to multi-jet production, avoiding at the same time large systematic uncertainties from the
jet energy calibration.  For example, results from the Tevatron\cite{Abazov:2004hm,Abazov:2012jhu} and the
LHC\cite{daCosta:2011ni,Khachatryan:2011zj,Khachatryan:2016hkr,Sirunyan:2017jnl,Aaboud:2018hie,Sirunyan:2019rpc} show
that the LO (non-trivial) prediction for this observable, with at most three partons in the final state, is not able to
describe the data for an azimuthal separation below $2\pi/3$, where NLO contributions (with 4 partons) restore the
agreement with data. In addition, this observable can be employed to tune Monte Carlo predictions of soft gluon
radiation.  Further examples of dijet observables that probe special corners of phase space are those that involve
forward (large rapidity) jets and where a large rapidity separation, possibly also a rapidity gap, is required between
the two jets. Reviews of such measurements can be found in Ref.~\cite{Carli:2015qta}, showing that no single prediction
is capable of describing the data in all phase-space regions. In particular, no conclusive evidence for BFKL effects in
these observables has been established so far.

Beyond dijet final states, measurements of the production of three or more jets, including cross section ratios, have
been performed (see Refs.~\cite{Carli:2015qta,Kokkas:2015gfa} for recent reviews), as a means of testing perturbative
QCD predictions, determining the strong coupling constant, and probing/tuning MC models, in particular those combining
multi-parton matrix elements with parton showers~\cite{ATLAS:2020vup,ATLAS:2023tgo}.  The calculation of three-jet
production to NNLO~\cite{Czakon:2021mjy} will allow more precise predictions of the three-jet to two-jet ratio, and thus
the extraction of \asmz from this observable. This calculation also allows for the use of transverse energy-energy
correlations (TEEC) for the determination of \asmz~\cite{ATLAS:2023tgo}.

$W$ and $Z$ production serve as benchmark cross sections at the LHC. The large boson mass stabilizes perturbative
predictions, which results in better theoretical accuracy. In terms of experimental precision, measurements of inclusive
vector boson ($W,Z$) production provide the most precisely determined observables at hadron colliders so far. This is
because the experimental signatures are based on charged leptons which are measured much more accurately than jets or
photons. At the
LHC\cite{Aaboud:2016btc,Aad:2016naf,Aaij:2015gna,CMS:2011aa,Chatrchyan:2014mua,Aaij:2012vn,Aaij:2015zlq,Aaij:2016mgv},
the dominant uncertainty stems from the luminosity determination ($\le$2--4\%), while other uncertainties (\eg
statistical or from lepton efficiencies) are controlled at the $\sim$0.5--3\% level.  The uncertainty from the
acceptance correction of about $\sim$1--2\% can be reduced by measuring so-called fiducial cross sections, \ie by
applying kinematic cuts also to the particle level of the theoretical predictions.  Measurements combining
the electron and muon final states were able to achieve a precision a few per mille level in the dilepton final state for the normalized cross section for $p_T^{ll}$ less than 30\GeV~\cite{ATLAS:2019zci,CMS:2019raw}.
This level of precision can also be achieved by
measuring
cross section ratios ($W / Z$ or $W^+/W^-$).
On the theory side, as discussed earlier in this {\it Review}, the production of these color-singlet states
has been calculated to N$^3$LO~\cite{Duhr:2020seh,Chen:2021vtu,Duhr:2020sdp}.  Since the dominant theoretical
uncertainty is related to the choice of PDFs, these high-precision data provide useful handles for PDF determinations.

Further insights are obtained from measurements of differential vector boson production, as a function of the invariant
dilepton mass, the boson's rapidity or its transverse momentum. For example, the dilepton invariant mass distribution
has been measured\cite{Chatrchyan:2011cm,CMS:2014jea,Aad:2013iua,Aad:2014qja,Aad:2016zzw,Sirunyan:2018owv} for masses
between 15 and 3000\GeV, covering more than 8 orders of magnitude in cross section. NNLO QCD predictions, together with
modern PDF sets and including higher-order electroweak and QED final-state radiation corrections, describe the data to
within 5--10\% over this large range, whereas NLO predictions show larger deviations, unless matched to a parton shower.

Similar conclusions can be drawn from the observed rapidity distribution of the dilepton system (see \eg
Refs.~\cite{Aaboud:2016btc,CMS:2014jea,Aaboud:2017ffb}) or, in the case of $W$ production, from the observed charged
lepton rapidity distribution and its charge asymmetry. The latter is particularly sensitive to differences among PDF
sets\cite{Aaboud:2016btc,Chatrchyan:2013mza,Khachatryan:2016pev,Aad:2019rou}, also thanks to the high precision achieved
by the ATLAS and CMS experiments for central rapidity ranges. These measurements are extended to the very forward
region, up to 4.5 in lepton rapidity, by the LHCb experiment\cite{Aaij:2015zlq,Aaij:2016mgv,LHCb:2016zpq}.

An overview of these kinds of measurement can be found in Ref.~\cite{Carli:2015qta}.  There one can also find a
discussion of and references to LHC results from studies of the vector boson's transverse momentum distribution, $p^V_t$
(see also Refs.~\cite{Aad:2015auj,Khachatryan:2016nbe,Sirunyan:2017igm}).  This observable covers a wide kinematic range
and probes different aspects of higher-order QCD effects.  It is sensitive to jet production in association with the
vector boson, without suffering from the large jet energy scale uncertainties.  In the $p^V_t$ region of several tens of
\GeV to over 1\TeV, the NNLO predictions for V+jet can be used to predict the high-$p_t$ boson production cross
section.%
\footnote{For these calculations, there is a requirement of the presence of a jet, but the $p_t$ cut is typically small
  (30\GeV) compared to the high $p_t$ region being discussed here.} %
The NNLO predictions agree with the data to within about 10\%, and agree somewhat better at high transverse momentum
than do the NLO predictions\cite{Aaboud:2017soa}.  At transverse momenta below $\sim$20\GeV, the fixed-order predictions
fail and soft-gluon resummation is needed to restore the agreement with data.  The soft gluon resummation can either be
performed analytically, or effectively using parton showering implemented in Monte Carlo programs.  While analytic
approaches reach a higher perturbative precision, they typically refer to inclusive measurements without fiducial cuts.

The addition of further jets to the final state extends the kinematic range as well as increasing the complexity of the
calculation/measurements.  The number of results obtained both at the Tevatron and at the LHC is extensive, summaries
can be found in Refs.~\cite{Carli:2015qta,Blumenschein:2015iqa}, and more recent results can be found in
Refs.~\cite{Aaboud:2017soa,Aad:2019hga,Khachatryan:2016fue,Sirunyan:2017wgx,Sirunyan:2018cpw}.  The measurements cover a
very large phase space, \eg with jet transverse momenta between $30$\GeV and $\sim1.5$\TeV and jet rapidities up to
$|y|<4.4$\cite{Aaboud:2017soa,ATLAS:2022nrp}. Jet multiplicities as high as seven jets accompanying the vector boson
have already been probed at the LHC, together with a substantial number of other kinematical observables, such as
angular correlations among the various jets or among the jets and the vector boson, or the sum of jet transverse
momenta, $H_T$. Whereas the jet $p_t$ and $H_T$ distributions are dominated by jet energy scale uncertainties at levels
similar to those discussed above for inclusive jet production, angular correlations and jet multiplicity ratios have
been measured with a precision of $\sim10\%$, see \eg Refs.~\cite{Aad:2013ysa,Chatrchyan:2013tna}.

NLO calculations for up to five jets\cite{Bern:2013gka} in addition to the vector boson are in good agreement with the
data over that phase space, where the calculations are applicable.  Predictions for V+jet at NNLO improve the
description of the data for distributions involving the vector boson or the leading jet.  MC models that implement
parton shower matching to matrix elements (either at LO or NLO) have mixed results.

The challenges get even more severe in the case of vector boson plus heavy quark ($b$,~$c$) production: on the theory
side because an additional scale is introduced by the heavy quark mass, and different schemes exist for the handling of
heavy quarks and their mass effects in the initial and/or final state; and on the experimental side because additional
uncertainties related to the heavy-flavor tagging must be considered~\cite{ATLAS:2022uav} (see also the discussion
regarding flavor-jet definition in Sec.~\ref{qcd:sec:jetalgoGD}).  A review of heavy quark production at the LHC is
presented in Ref.~\cite{Voutilainen:2015lqa}, where the di-$b$-jet $p_t$ and mass spectra are found to be well modeled
within uncertainties by most generators for $b$-jet production with or without associated $W$ and $Z$ bosons.  However,
sizable differences between data and predictions are seen in the modeling of events with single $b$ jets, particularly
at large $b$-jet $p_t$, where gluon splitting processes become dominant, as also confirmed by studies of $b$-hadron and
$b$-jet angular correlations.

The precision reached in photon measurements is in between that for lepton and jet measurements. The photon energy and
angles can be measured at about the same precision as the lepton energy and angles in Drell-Yan production, but there
are greater challenges encountered in photon reconstruction (for example isolation) and in purity determination.  Note,
though, that the photon purity approaches unity as the photon $p_t$ increases.  At high $p_t$, it becomes increasingly
difficult for a jet to fragment into an isolated neutral electromagnetic cluster which mimics the photon signature.  The
inclusive photon cross section can be
measured\cite{Aad:2013tea,Aad:2016xcr,ATLAS:2019buk,Chatrchyan:2011ue,Chatrchyan:2013mwa}, as well as the production of
a photon accompanied by one or more
jets\cite{Aad:2013zba,Aaboud:2017kff,Aaboud:2016sdm,Aaboud:2017kff,Chatrchyan:2013mwa,Sirunyan:2018gro,Sirunyan:2019uya,ATLAS:2019iaa}. The
kinematic range for photon production is less than that for jet production because of the presence of the
electromagnetic coupling, but still reaches about 2\TeV.  Better agreement is obtained with NNLO predictions for photon
production than for NLO predictions, except when the latter are matched to matrix element plus parton shower
predictions.  Photon production in association with a heavy-flavor jet is a useful input for the determination of the
$b$ and $c$ quark PDFs\cite{Aaboud:2017skj}.

Electroweak corrections\index{electroweak corrections}
are expected to become more and more relevant now that the \TeV energy range starts to be explored, and EW corrections
can now also be computed automatically~\cite{Chiesa:2015mya}. For a comprehensive review on electroweak corrections see
Ref.~\cite{Denner:2019vbn}. For example, such corrections were found\cite{Kuhn:2005gv,Lindert:2017olm} to be sizable
(tens of percent) when studying the ratio $(d\sigma^\gamma/dp_t) / (d\sigma^Z/dp_t) $ in $\gamma\,(Z) +$jet production,
$p_t$ being the boson's transverse momentum, and might account for (some of) the differences observed in a CMS
measurement\cite{Khachatryan:2015ira} of this quantity.

A number of interesting developments, in terms of probing higher-order QCD effects, have occurred in the sector of
diboson production, in particular for the $WW$ and $\gamma\gamma$ cases.  Regarding the former, an early disagreement of
about $10\%$ between the LHC measurements and the NLO predictions had led to a number of speculations of possible new
physics effects in this channel.  However, more recent ATLAS and CMS
measurements\cite{Aad:2016wpd,Aaboud:2017qkn,Aaboud:2019nkz,Khachatryan:2015sga} are in agreement with the NNLO
prediction\cite{Gehrmann:2014fva}.  The statistical reach of the LHC has resulted in the discovery of triple massive
gauge boson production\cite{ATLAS:2019dny,ATLAS:2022xnu,CMS:2020hjs}.

In the case of diphoton production, ATLAS\cite{Aad:2012tba,Aaboud:2017vol,ATLAS:2021mbt} and
CMS\cite{Chatrchyan:2014fsa} have provided accurate measurements, in particular for phase-space regions that are
sensitive to radiative QCD corrections (multi-jet production), such as small azimuthal photon separation. While there
are large deviations between data and NLO predictions in this region, a calculation\cite{Catani:2011qz} at NNLO accuracy
manages to mostly fill this gap. This is an interesting example where scale variations can not provide a reliable
estimate of missing contributions beyond NLO, since at NNLO new channels appear in the initial state (gluon fusion in
this case). These missing channels can be included in a matrix element plus parton shower calculation in which two
additional jets are included at NLO. The result exhibits a similar level of agreement as that obtained at NNLO.
Three-photon production has also been measured\cite{Aaboud:2017lxm} and is in good agreement with NNLO theory
predictions~\cite{Chawdhry:2019bji,Kallweit:2020gcp}.

In terms of heaviest particle involved, top-quark production at the LHC has become an important tool for probing
higher-order QCD calculations, thanks to very impressive achievements both on the experimental and theoretical side, as
extensively summarized in Ref.~\cite{Kroninger:2015oma}.  Regarding \ttbar production, the most precise inclusive cross
section measurements are achieved using the dilepton ($e\,\mu$) final state, with a total uncertainty of
4\%\cite{Aaboud:2016iot,Aaboud:2016syx,Sirunyan:2018goh,CMS:2019esx,ATLAS:2023gsl}. This is of about the same size as
the uncertainty on the most advanced theoretical
predictions\cite{Czakon:2013goa,Czakon:2015owf,Catani:2019iny,Catani:2019hip}, obtained at NNLO with additional
soft-gluon resummation at NNLL accuracy\cite{Czakon:2018nun}.  There is excellent agreement between data and the QCD
predictions.

The \ttbar final state allows multiple observables to be measured.  A large number of differential cross section
measurements have been performed at 7, 8 and 13\TeV center-of-mass energy, studying distributions such as the top-quark
$p_t$ and rapidity, the transverse momentum and invariant mass of the \ttbar system (probing the \TeV range), or the
number of additional jets. These measurements have been compared to a wide range of predictions, at fixed order up to
NNLO as well as using LO or NLO matrix elements matched to parton showers.  Each of the observables provides information
on the high $x$ gluon.  However, there are tensions among the multiple observables that can lead to difficulties in PDF
fits.  Four top production has been measured in multi-lepton final states~\cite{ATLAS:2023ajo} by the ATLAS and CMS
collaboration, with results consistent with the standard model prediction~\cite{Bevilacqua:2012em}.

Thanks to both the precise measurements of, and predictions for, the inclusive top-pair cross section, which is
sensitive to the strong coupling constant and the top-quark mass, this observable has been used to measure the strong
coupling constant at NNLO accuracy from hadron collider
data\cite{Chatrchyan:2013haa,Klijnsma:2017eqp}~(\cf \Section{qcd:sec:coupling-measurement} below), as well as to obtain
a measurement of the top-quark's pole mass without employing direct reconstruction
methods\cite{Chatrchyan:2013haa,ATLAS:2014nxi,Sirunyan:2017uhy}.

The Higgs boson provides a tool for QCD studies, especially as the dominant production mechanism is $gg$ fusion, which
is subject to very large QCD corrections. Higgs boson production has been measured in the $ZZ$, $\gamma\gamma$,
$b\bar{b}$, $WW$ and $\tau \tau$ decay channels. A measurement of the cross section in the $ZZ$ and $\gamma\gamma$
channels is one of the first measurements at 13.6\TeV to be published~\cite{ATLAS:2023tnc}.  The experimental cross
section is now known with a precision approaching 10\%\cite{Aaboud:2018ezd,Sirunyan:2018sgc}, similar to the size of the
theoretical uncertainty\cite{deFlorian:2016spz}, of which the PDF+\alps uncertainty is the largest component. Part of
this systematic is the mis-match in orders between the PDF determination (NNLO) and the cross section evaluation
(N$^3$LO), as discussed earlier. The experimental precision has allowed detailed fiducial and differential cross section
measurements. For example, with the diphoton final state, the transverse momentum of the Higgs boson can be measured out
to the order of 650\GeV\cite{Aaboud:2018xdt,Sirunyan:2018kta,ATLAS:2022fnp,CMS:2022wpo}, where top quark mass effects
become important. The production of a Higgs boson with up to 4 jets has been
measured\cite{Aaboud:2018xdt,Sirunyan:2018sgc}. The experimental cross sections have been compared to NNLO predictions
(for $H+\ge1$ jet), NLO for 2 and 3 jets, and NNLO+NNLL for the transverse momentum distribution. In addition, finite
top quark mass effects have been taken into account at NLO. The use of the boosted $H\rightarrow b\bar{b}$ topology
allows probes of Higgs boson transverse momenta on the order of 600\GeV and greater \cite{Sirunyan:2018sgc}.  So far the
agreement with the perturbative QCD corrections is good.

% ======================================================================
\section{Determinations of the strong coupling constant}
\label{qcd:sec:coupling-measurement}

Beside the quark masses, the only free parameter in the QCD Lagrangian is the strong coupling constant \alps.
\index{Strong coupling constant} %
The coupling constant in itself is not a physical observable, but rather a quantity defined in the context of
perturbation theory, which enters predictions for experimentally measurable observables, such as $R$
in~\Eq{qcd:eq:xsct-ee}.  The value of the strong coupling constant must be inferred from such measurements and is
subject to experimental and theoretical uncertainties.  The incomplete knowledge of \alps propagates into uncertainties
in numerous precision tests of the Standard Model.  Here, we present an update of the 2022 PDG average value of \asmz
and its uncertainty\cite{ParticleDataGroup:2022pth}.

Many experimental observables are used to determine \alps.  A number of recent determinations are collected in
Refs.~\cite{dEnterria:2019its,dEnterria:2022hzv}. Further discussions and considerations on determinations of \alps can
also be found in Refs.~\cite{Salam:2017qdl,Pich:2018lmu,Pich:2020gzz}. %
Such considerations include:
\begin{description}
\item[$\bullet$] %
  The observable's sensitivity to \alps as compared to the experimental precision. For example, for the \ee cross
  section to hadrons (\cf $R$ in Sec.~\ref{qcd:sec:inclcrosssec}), QCD effects are only a small correction, since the
  perturbative series starts at order $\alps^{0}$; three-jet production or event shapes in \ee annihilations are directly
  sensitive to \alps since they start at order \alps; the hadronic decay width of heavy quarkonia,
  $\Gamma(\Upsilon\rightarrow\rm{hadrons})$, is very sensitive to \alps since its leading order term is
  $\propto \alps^3$.

\item[$\bullet$] %
  The accuracy of the perturbative prediction, or equivalently of the relation between \alps and the value of the
  observable.
  Several observables have been
  known to NNLO for quite some time.  These include, for instance, inclusive observables, as well as three-jet rates and
  event shapes in \ee collisions, inclusive jet and dijet production in DIS, and inclusive jet, dijet, \ttbar,
  $W$/$Z$+jet and three-jet production cross sections in \pp or \ppbar collisions.  The \ee hadronic cross section and
  $\tau$, $W$ and $Z$ branching fractions to hadrons are even known to N$^3$LO, if one denotes the LO as the first non-trivial term.
  In certain cases, fixed-order predictions are supplemented with resummation.  The precise magnitude of the associated
  theory uncertainties usually is estimated as discussed in Sec.~\ref{qcd:sec:qcd-accuracy}.

\item[$\bullet$] %
  The size of non-perturbative effects. Sufficiently inclusive quantities, like the \ee cross section to hadrons, have
  small non-perturbative contributions $\sim {\rm \Lambda}^4/Q^4$. Others, such as event-shape distributions, have
  typically contributions $\sim {\rm \Lambda}/Q$.

\item[$\bullet$] %
  The scale at which the measurement is performed. An uncertainty $\delta$ on a measurement of $\alps(Q^2)$, at a scale
  $Q$, translates to an uncertainty $\delta' = (\alps^2(\mZ^2)/\alps^2(Q^2)) \cdot\delta$ on \asmz. For example, this
  enhances the already important impact of precise low-$Q$ measurements, such as from $\tau$ decays, in combinations
  performed at the \mZ scale.
\end{description}

\noindent{}The selection of results from which to determine the world average value of \asmz is restricted to those that
are %
\begin{description}
\item[-] published in a peer-reviewed journal at the time of writing this report,
\item[-] based on the most complete perturbative QCD predictions of at least NNLO accuracy,
\item[-] accompanied by reliable estimates of all experimental and theoretical uncertainties.
\end{description}

Numerous measurements from jet production in DIS and at hadron colliders are still excluded from the average presented
here, because the determination of \asmz from those data sets has not yet been upgraded to NNLO\@. A few new results did
appear comparing to theory at NNLO and are included in the corresponding section.  NLO analyses will still be discussed
in this {\it Review}, as they are important ingredients for the experimental evidence of the energy dependence of \alps,
\ie for asymptotic freedom, one of the key features of QCD\@.

In order to calculate the world average value of \asmz, we apply, as in earlier editions, an intermediate step of
pre-averaging results within the sub-fields now labeled %
``Hadronic $\tau$ decays and low $Q^2$ continuum'' ($\tau$ decays and low $Q^2$), %
``Heavy quarkonia decays'' ($Q\bar{Q}$ bound states), %
``PDF fits'' (PDF fits), %
``Hadronic final states of \ee annihilations'' (\ee jets \& shapes), %
``Observables from hadron-induced collisions'' (hadron colliders), and %
``Electroweak precision fit'' (electroweak) %
as explained in the following sections. %
For each sub-field, the {\it unweighted average} of all selected results is taken as the pre-average value of \asmz, and
the {\it unweighted average} of the quoted total uncertainties is assigned to be the respective overall error of this
pre-average.  Asymmetric total uncertainties are symmetrised beforehand by adopting the larger of the two values as the
($\pm$) uncertainty.  For the ``Lattice QCD'' (lattice) sub-field we do not perform a pre-averaging; instead, we adopt
for this sub-field the average value and uncertainty derived by the Flavour Lattice Averaging Group (FLAG)
\index{Flavour Lattice Averaging Group (FLAG)} in Ref.~\cite{FlavourLatticeAveragingGroupFLAG:2021npn}.

Assuming that the six sub-fields (excluding lattice) are largely independent of each other, we determine a non-lattice
world average value using a `{\it $\chi^2$ averaging}' method.  In a last step we perform an unweighted average of the
values and uncertainties of \asmz from our non-lattice result and the lattice result presented in the FLAG~2021
report~\cite{FlavourLatticeAveragingGroupFLAG:2021npn}.

%-----------------------------------------------------------------------
\subsection{Hadronic \texorpdfstring{\boldmath$\tau$\unboldmath}{tau} decays and low
  \texorpdfstring{\boldmath$Q^2$\unboldmath}{Q2} continuum:}

\noindent{}Based on complete N$^3$LO predictions\cite{Baikov:2008jh}, analyses of the $\tau$ hadronic decay width and
spectral functions have been performed, \eg in
Refs.~\cite{Baikov:2008jh,Beneke:2008ad,Maltman:2008nf,Narison:2009vy,Caprini:2009vf,Pich:2013lsa,Boito:2014sta}, and
lead to precise determinations of \alps at the energy scale of $\mtau^2$. They are based on different approaches to
treat perturbative and non-perturbative contributions, the impacts of which have been a matter of intense discussions
for a long time, see \eg Refs.~\cite{Pich:2013lsa,Altarelli:2013bpa,Boito:2014sta,Pich:2016bdg}.  In particular, in
$\tau$ decays there is a significant difference between results obtained using fixed-order (FOPT) or contour-improved
perturbation theory (CIPT),\index{Fixed-order perturbation theory (FOPT)}\index{contour-improved perturbation theory
  (CIPT)} such that analyses based on CIPT generally arrive at larger values of $\alps(\mtau^2)$ than those based on
FOPT\@.

In addition, some results show discrepancies in $\alps(\mtau^2)$ among groups of authors using the same data sets and
perturbative calculations, most likely due to different treatments of the non-perturbative contributions, \cf
Ref.~\cite{Boito:2014sta} with Refs.~\cite{Pich:2013lsa,Davier:2013sfa}.  References~\cite{Boito:2016oam,Boito:2019iwh}
question the validity of using a truncated operator product expansion (OPE) at $Q^2=\mtau^2$ based on the disagreement
found between experimental values of the spectral moments and the theory representations based on the truncated OPE fits
at $Q^2>\mtau^2$.

Recent developments now have shed light on the disagreement between FOPT- and CIPT-based
analyses. References~\cite{Hoang:2020mkw,Hoang:2021nlz,Golterman:2023oml} argue that CIPT-based calculations require a
dedicated estimation of non-perturbative effects instead of ``standard'' ones used so far.  Otherwise an asymptotic
separation between results using the two perturbative approaches would remain. Potential ways forward are suggested in
Refs.~\cite{Benitez-Rathgeb:2022hfj,Benitez-Rathgeb:2022yqb}. As a consequence we remove for the time being
determinations of the strong coupling constant based on CIPT from the derivation of the central value and the
uncertainty of \asmz.

We determine the pre-average value of \asmz for this sub-field only from studies that employ FOPT expansions and remove
any eventual averaging with CIPT central values or increased uncertainties due to the differences in CIPT vs.\ FOPT\@.
As the results from Refs.~\cite{Baikov:2008jh,Davier:2013sfa,Pich:2016bdg} are not totally independent, we pre-averaged
as a first step these three results in the previous edition of this {\it Review}.  Lacking, however, an estimate of the
theory uncertainty for the FOPT method, Ref.~\cite{Davier:2013sfa} had to be left out, such that the first entry to this
category of \alps determinations is pre-averaged from $\asmz = 0.1183\pm 0.0026$\cite{Baikov:2008jh} and
$\asmz = 0.1181\pm 0.0015$\cite{Pich:2016bdg} to $\asmz = 0.1182\pm 0.0021$ (summarized as BP 2008-16 FO in
\Fig{qcd:fig:pre-averages}).

Subsequently, this is combined with $\asmz = 0.1158\pm 0.0022$\cite{Boito:2018yvl} and
$\asmz = 0.1171\pm 0.0010$\cite{Boito:2020xli}, which replaces the previous result from Ref.~\cite{Boito:2014sta}.  We
also include the result from $\tau$ decay and lifetime measurements, obtained in Sec.~{\it Electroweak Model and
  constraints on New Physics} of the 2022 edition of this {\it Review}~\cite{ParticleDataGroup:2022pth},
$\asmz = 0.1171 \pm {0.0018}$.  The latter result, being a global fit of $\tau$ data, involves some correlations with
the other extractions of this category. However, since we perform an unweighted average of the central value and
uncertainty, the effects of the potential correlations are reduced.  Finally, a new determination reported in
Ref.~\cite{Ayala:2022cxo} is added to this category: $\asmz = 0.1183^{+0.0009}_{-0.0012}$.  A recent publication evaluating
data on $R_{\rm had}$ in the continuum below the charm threshold exhibits huge experimental uncertainties and therefore
has not been considered~\cite{Shen:2023qgz}.

All these results are summarized in \Fig{qcd:fig:pre-averages}. Determining the unweighted average of the central values
and their overall uncertainties, we arrive at $\asmz = 0.1173 \pm 0.0017$, which we will use as the first input for
determining the world average value of \asmz. This corresponds to $\alps(\mtau^2) = 0.314 \pm 0.014$.

\begin{pdgxfigure}[width=0.60\linewidth,bookwidth=0.75\linewidth]
  \includegraphics{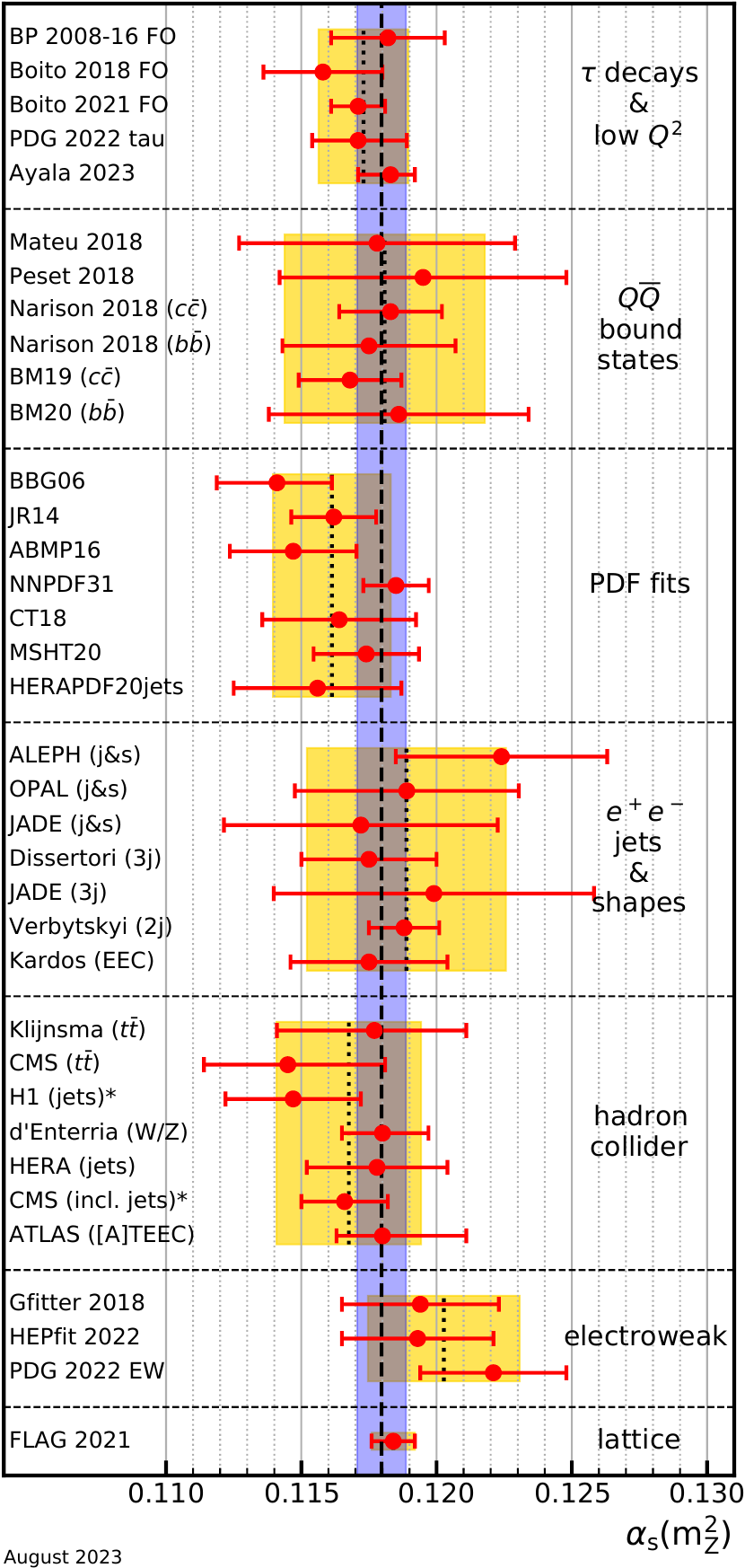}
  \caption{Summary of determinations of \asmz with uncertainty in the seven sub-fields as discussed in the text. The
    yellow (light shaded) bands and dotted lines indicate the pre-average values of each sub-field. The dashed line and
    blue (dark shaded) band represent the final world average value of \asmz. The ``*'' symbol within the ``hadron
    colliders'' sub-field indicates a determination including a simultaneous fit of PDFs.}
  \label{qcd:fig:pre-averages}\index{world average of \asmz}
\end{pdgxfigure}

%-----------------------------------------------------------------------
\subsection{Heavy quarkonia decays:}

\noindent{}Two determinations of \asmz have been performed~\cite{Mateu:2017hlz,Peset:2018ria} that are based on N$^3$LO
accurate predictions.  Reference~\cite{Mateu:2017hlz} performs a simultaneous fit of the strong coupling and the bottom
mass $\overline{m_b}$, including states with principal quantum number up to $n \le 2$ in order to break the degeneracy
between \alps and $\overline{m_b}$, finding $\asmz = 0.1178 \pm 0.0051$.  Reference~\cite{Peset:2018ria} instead uses as
input of the fit the renormalon-free combination of masses of the meson $B_c$, the bottomonium $\eta_b$ and the
charmonium $\eta_c$, $M_{B_{c}}- M_{\eta_{b}}/2 - M_{\eta_{c}}/2$, which is weakly dependent on the heavy quark masses,
but shows a good dependence on \alps. Using this observable, they obtain $\asmz = 0.1195 \pm 0.0053$.  Two further
values are derived at NNLO in Ref.~\cite{Narison:2018dcr,Narison:2018xbj} from mass splittings and sum rules giving
$\asmz = 0.1183 \pm 0.0019$ and $\asmz = 0.1175 \pm 0.0032$ when evolved from the relevant charmonium respectively
bottomonium mass scales to $\mZ^2$.  Finally, by means of quarkonium sum rules, Refs.~\cite{Boito:2019pqp,Boito:2020lyp}
quote $\asmz = 0.1168 \pm 0.0019$ and $\asmz = 0.1186 \pm 0.0048$ for charmonium and bottomonium respectively.  These
six determinations satisfy our criteria to be included in the heavy-quarkonia category of the world average.  Their
unweighted combination leads to the pre-average for this category of $\asmz = 0.1181 \pm 0.0037$.

%-----------------------------------------------------------------------
\subsection{PDF fits:}

\noindent{}Another class of studies, analyzing structure functions at NNLO QCD (and partly beyond), provide results that
serve as relevant inputs for the world average of \alps.  Some of these studies do $not$, however, explicitly include
estimates of theoretical uncertainties when quoting fit results of \alps. In such cases we add, in quadrature, half of
the difference between the results obtained in NNLO and NLO to the quoted errors.

A combined analysis of non-singlet structure functions from DIS\cite{Blumlein:2006be}, based on QCD predictions up to
N$^3$LO in some of its parts, results in $\asmz = 0.1141 \pm 0.0022$ (BBG).  Studies of singlet and non-singlet
structure functions, based on NNLO predictions, result in $\asmz = 0.1162 \pm 0.0017$\cite{Jimenez-Delgado:2014twa}
(JR14).  The AMBP group\cite{Alekhin:2017kpj,Alekhin:2018pai} determined a set of parton distribution functions using
data from HERA, NOMAD, CHORUS, from the Tevatron and the LHC using the Drell-Yan process and the hadro-production of
single-top and top-quark pairs, and determined $\asmz = 0.1147 \pm 0.0024$\cite{Alekhin:2017kpj}.

The MSHT group~\cite{Cridge:2021qfd}, also including hadron collider data, determined a set of parton density
functions (MSHT20) together with $\asmz = 0.1174 \pm 0.0013$.  Similarly, the CT group~\cite{Hou:2019efy} determined the
CT18 parton density set together with $\asmz = 0.1164 \pm 0.0026$. The NNPDF group\cite{Ball:2018iqk} presented NNPDF3.1
parton distribution functions together with $\asmz = 0.1185 \pm 0.0012$. New in this category is the result of
$\asmz = 0.1156 \pm 0.0031$ reported as HERAPDF2.0Jets in Ref.~\cite{H1:2021xxi}. In addition to DIS data from HERA also
jet cross sections are considered in the fit to theory at NNLO\@.

We note that criticism has been expressed on some of the above extractions.  Among the issues raised, we mention the
neglect of singlet contributions at $x \ge 0.3$ in pure non-singlet fits\cite{Thorne:2011kq}, the impact and detailed
treatment of particular classes of data in the fits\cite{Thorne:2011kq,Alekhin:2011ey}, possible biases due to
insufficiently flexible parameterizations of the PDFs\cite{NNPDF:2011aa} and the use of a fixed-flavor number
scheme\cite{Ball:2013gsa,Thorne:2013hpa}. For recent more extensive discussions see
\eg Ref.~\cite{PDF4LHCWorkingGroup:2022cjn}.

Summarizing the results from world data on structure functions, taking the {\it unweighted average} of the central
values and errors of all selected results, leads to a pre-average value of $\asmz = 0.1161 \pm 0.0022$, see also
\Fig{qcd:fig:pre-averages}.

%-----------------------------------------------------------------------
\subsection{Hadronic final states of \texorpdfstring{\boldmath$e^+e^-$\unboldmath}{e+e-} annihilations:}

\noindent{}Re-analyses of jets and event shapes in $e^+e^-$ annihilation (j\&s), measured around the $Z$ peak and at
LEP2 center-of-mass energies up to 209\GeV, using NNLO predictions matched to NLL resummation and Monte Carlo models to
correct for hadronization effects, resulted in $\asmz = 0.1224\pm0.0039$ (ALEPH)\cite{Dissertori:2009ik}, and in
$\asmz = 0.1189\pm0.0043$ (OPAL)\cite{OPAL:2011aa}.  Similarly, an analysis of JADE data\cite{Bethke:2008hf} at
center-of-mass energies between 14 and 46\GeV gives $\asmz = 0.1172\pm0.0051$, with contributions from the hadronization
model and from perturbative QCD uncertainties of 0.0035 and 0.0030, respectively.  Precise determinations of \alps
from three-jet production alone (3j), at NNLO, resulted in $\asmz = 0.1175\pm0.0025$\cite{Dissertori:2009qa} from ALEPH data
and in $\asmz = 0.1199\pm0.0059$\cite{Schieck:2012mp} from JADE.
A recent determination is based on an NNLO+NNLL accurate calculation that allows to fit the region of lower three-jet rate
(2j) using data collected at LEP and PETRA at different energies. This fit gives
$\asmz = 0.1188\pm0.0013$\cite{Verbytskyi:2019zhh}, where the dominant uncertainty is the hadronization uncertainty,
which is estimated from Monte Carlo simulations.
A fit of energy-energy-correlation (EEC), also based on an NNLO+NNLL calculation, together with a Monte Carlo based
modeling of hadronization corrections gives $\asmz = 0.1175 \pm 0.0029$\cite{Kardos:2018kqj}.  These results are
summarized in the \ee sector of \Fig{qcd:fig:pre-averages}.

Another class of \alps determinations is based on analytic modeling of non-perturbative and hadronization effects,
rather than on Monte Carlo models\cite{Davison:2008vx,Abbate:2010xh,Gehrmann:2012sc,Hoang:2015hka}, using methods like
power corrections, factorization of soft-collinear effective field theory, dispersive models and low scale QCD effective
couplings.  In these studies, the world data on thrust distributions (T), or the C-parameter distributions (C), are
analyzed and fitted to perturbative QCD predictions at NNLO matched with resummation of leading logs up to N$^3$LL
accuracy, see Sec.~\ref{qcd:sec:resummation}.  The results are $\asmz = 0.1135 \pm 0.0011$\cite{Abbate:2010xh} and
$\asmz = 0.1134^{+0.0031}_{-0.0025}$\cite{Gehrmann:2012sc} from thrust, and
$\asmz = 0.1123 \pm 0.0015$\cite{Hoang:2015hka} from C-parameter.

A long-standing question was why this latter class of determinations led systematically to rather small values of \asmz
as compared to the former one based on Monte Carlo models to correct for hadronization effects. New insights have now
been gained in this regard.
In a recent calculation of the leading non-perturbative contribution
to the C-parameter in the three-jet symmetric limit based on an
effective coupling approach it was found that it differs by a factor
of two from the one in the two-jet limit~\cite{Luisoni:2020efy}.
Subsequently, using the same effective coupling approach, the leading $1/Q$ power correction
was computed for thrust and the C-parameter in the full three-jet region under the assumption of a large (negative)
$n_f$ limit~\cite{Caola:2021kzt,*Caola:2022vea}. Expanding on this approach, Ref.\cite{Nason:2023asn} extended the
calculation of non-perturbative corrections to include the heavy-jet mass, the difference of jet masses, wide
broadening, and, with some caveat, the three-jet resolution variable in the Durham algorithm. The key finding in
Ref.~\cite{Nason:2023asn} is that non-perturbative corrections computed in the three-jet region significantly deviate
from those computed in the two-jet limit and hence the aforementioned fits based on power corrections in the two-jet
limit result in smaller values of \asmz.  Another important observation is that the inclusion of resummation effects
introduces a relatively substantial ambiguity outside the two-jet limit. Additionally, other factors such as the
choice of mass-scheme used to extend the definition of event shapes to massive hadrons can have significant effects.

These findings are inconsistent with the very small experimental, hadronization, and theoretical uncertainties of only
2, 5, and 9 per-mille, respectively, as reported in Refs.~\cite{Abbate:2010xh,Hoang:2015hka}. For these reasons, we
exclude the results of Refs.~\cite{Abbate:2010xh,Gehrmann:2012sc,Hoang:2015hka} from the average.  Determinations based
on corrections for non-perturbative hadronization effects using QCD-inspired Monte Carlo generators have also faced
criticism due to the differing nature of parton-level simulations compared to fixed-order calculations. However, these
determinations typically exhibit a more conservative theoretical uncertainty.

Not included in the computation of the world average but worth mentioning are a computation of the NLO corrections to
5-jet production and comparison to the measured 5-jet rates at LEP\cite{Frederix:2010ne}, giving
$\asmz = 0.1156^{+0.0041}_{-0.0034}$, and a computation of non-perturbative and perturbative QCD contributions to the
scale evolution of quark and gluon jet multiplicities, including resummation, resulting in
$\asmz = 0.1199 \pm 0.0026$\cite{Bolzoni:2013rsa}.

The unweighted average of the considered determinations as shown in the \ee sector of Fig.~\ref{qcd:fig:pre-averages}
yields $\asmz = 0.1189 \pm 0.0037$.

%-----------------------------------------------------------------------
\subsection{Observables from hadron-induced collisions:}

\noindent{}Until recently, determinations of \alps using hadron collider data, mostly from jet or \ttbar production
processes, could be performed at NLO only. NNLO calculations have now become available for
\ttbar~\cite{Czakon:2013goa,Czakon:2015owf,Catani:2019hip} and for inclusive jet, dijet, and three-jet
production~\cite{Currie:2016bfm,Currie:2017eqf,Czakon:2019tmo,Czakon:2021mjy,Alvarez:2023fhi}.  For \ttbar production,
in addition, logarithms to NNLL have been resummed~\cite{Czakon:2018nun}.  Both should be supplemented by electroweak
corrections~\cite{Dittmaier:2012kx,Frederix:2016ost,Czakon:2017wor,Reyer:2019obz}, which become important for high-\pt
collisions at the LHC. $Z$+jet production, studied with respect to an \alps determination at NLO from multi-jet events
in Ref.~\cite{Johnson:2017ttl}, is also known at NNLO for the 1-jet case~\cite{Boughezal:2015ded,Ridder:2016nkl}.

Determinations of \alps from production cross sections at hadron colliders also require a knowledge of the relevant PDFs
for those \alps values. Two strategies are pursued for the extraction of \alps, one using pre-determined PDFs as input
and a second strategy fitting the proton PDFs together with the strong coupling constant.  Each global PDF group
produces PDF sets not only for a value of $0.118$ that is close to the world average, but also for a wide range above
and below in increments of 0.001, which can be used in such determinations. The latter technique of simultaneously
fitting \alps and the PDFs is technically more accurate, given that the new data used in the determination of \alps may
modify the PDFs in a manner not taken into account by the \alps-variation PDFs provided by the fitting
collaborations. The former technique may result in a bias of unknown magnitude~\cite{Forte:2020pyp}. As the LHC
experiments have the ability to combine a PDF fit with the \alps determination, for example with tools like described in
Ref.~\cite{Alekhin:2014irh}, we expect more experimental joint determinations of PDFs and of \alps for future iterations
of this review.

The first determination of \alps at NNLO accuracy in QCD has been reported by CMS from the \ttbar production cross
section at $\sqrt{s} = 7$\TeV~\cite{Chatrchyan:2013haa}. In former {\it Reviews} this opened up a new sub-field on its
own. In the meantime, multiple datasets on \ttbar production from Tevatron at $\sqrt{s} = 1.96$\TeV and from LHC at
$\sqrt{s} = 7$, 8, and $13$\TeV have been analyzed simultaneously to determine \alps~\cite{Klijnsma:2017eqp} to
$\asmz = 0.1177^{+0.0034}_{-0.0036}$, where the largest uncertainties are associated with missing higher orders and with
PDFs, and where an additional complication arises from the top-mass dependence.  Since this combined analysis contains
among other things an updated measurement as compared to the dataset used by CMS, the latter is replaced in the
averaging by this combined result. A second entry into this sub-field is given by an analysis of \ttbar production data
at $\sqrt{s} = 13$\TeV from the CMS Collaboration\cite{Sirunyan:2018goh}. From the four values derived for different PDF
sets, the unweighted average is taken: $\asmz = 0.1145^{+0.0036}_{-0.0031}$.  A second analysis by CMS using
differential distributions of \ttbar production~\cite{CMS:2019esx} has been performed at NLO only and is not further
considered.

From collisions at HERA the \alps determination at NNLO using inclusive jet and dijet measurements in addition to DIS
data of the H1 Collaboration~\cite{Andreev:2017vxu} has been corrected for an issue with the theory prediction reported
in Ref.~\cite{Currie:2017tpe}. We choose the result $\asmz = 0.1147\pm{0.0025}$ that is derived from a simultaneous fit
of the strong coupling constant together with proton PDFs. We note that results of this section derived from such a
simultaneous fit will be marked with a ``*'' in the corresponding figures.  A second
determination~\cite{Britzger:2019kkb} combines multiple datasets on inclusive jet production of the H1 and ZEUS
collaborations into one fit using interpolation grids at NNLO\@.  The result of $\asmz = 0.1178{\pm{0.0026}}$ has been
updated for the same issue reported above~\cite{Currie:2017tpe} and is included in the unweighted average, although some
of the inclusive jet data have already been used in the previous analysis. We note that the ZEUS Collaboration has
presented preliminary results from a new inclusive jet measurement~\cite{Lorkowski:2023lnc}.

A first new determination of \asmz at NNLO from inclusive jet production at the LHC has been presented by the CMS
Collaboration~\cite{CMS:2021yzl}. A simultaneous fit of the strong coupling constant and the proton PDFs to the HERA DIS
and the new LHC jet data gives: $\asmz = 0.1166 \pm {0.0016}$. The ATLAS Collaboration has published an extraction of
\asmz at NNLO from the transverse energy-energy correlation (TEEC) and its asymmetry (ATEEC)~\cite{ATLAS:2023tgo}. The
results of this first derivation from an event shape observable requiring three-jet predictions at NNLO are
$\asmz = 0.1175^{+0.0035}_{-0.0018}$ for the TEEC and $\asmz = 0.1185^{+0.0027}_{-0.0015}$ for the ATEEC,
respectively. We include the unweighted average of the two numbers $\asmz = 0.1180^{+0.0031}_{-0.0017}$ as a new result
in this category. Very recently, two preliminary determinations of the strong coupling constant, albeit at NLO, have
been reported by CMS from energy correlators inside jets and from azimuthal correlations among
jets~\cite{CMS-PAS-SMP-22-015,CMS-PAS-SMP-22-005}.

Finally, Ref.~\cite{dEnterria:2019aat} extracts the strong coupling constant from a fit at NNLO to measurements of
inclusive $Z$ and $W$ boson production by experiments at the LHC and Tevatron colliders. From the four values quoted for
different PDF sets we include the unweighted average for the CT14 and MMHT14 PDFs $\asmz = 0.1180^{+0.0017}_{-0.0015}.$
The other results suffer from either a bad fit quality or \asmz values outside the range of validity for the PDF
set.
Furthermore, two determinations using measurements of the
$Z$ boson's recoil at low \pt have been submitted using data of the CDF respectively
ATLAS experiments~\cite{Camarda:2022qdg,ATLAS:2023lhg}. The very
small size of the estimated uncertainties has stimulated some
discussion.

As unweighted pre-average for this sub-field we now obtain $\asmz = 0.1168 \pm 0.0027$. If the stricter requirement of
simultaneous fits with PDFs is imposed, then only the H1 and CMS results are left giving $\asmz = 0.1157 \pm 0.0021$ for
this sub-field. In sect.~\ref{qcd:sec:world} below, we will also report the outcome for this choice.

Many further \alps determinations from jet measurements have not yet been advanced to NNLO accuracy.  A selection of
results from inclusive jet~\cite{Affolder:2001hn,Chekanov:2006yc,Abazov:2009nc,Malaescu:2012ts,H1:2014cbm,
  Khachatryan:2014waa,Khachatryan:2016mlc,Britzger:2017maj}, dijet~\cite{Sirunyan:2017skj}, and multi-jet
measurements~\cite{Chekanov:2005ve,Chatrchyan:2013txa,H1:2014cbm,Abazov:2012lua,CMS:2014mna,Andreev:2016tgi,
  ATLAS:2015yaa, Aaboud:2017fml,Aaboud:2018hie} is presented in \Fig{qcd:fig:nlo-averages}, where the uncertainty in
most cases is dominated by the impact of missing higher orders estimated through scale variations.  From the CMS
Collaboration we quote for the inclusive jet production at $\sqrt{s} = 7$ and $8\TeV$, and for dijet production at
$8\TeV$ the values that have been derived in a simultaneous fit with the PDFs and marked with ``*'' in the figure.  The
last point of the inclusive jet sub-field from Ref.~\cite{Britzger:2017maj} is derived from a simultaneous fit to six
datasets from different experiments and partially includes data used already for the other data points, \eg the CMS
result at $7$\TeV.

The multi-jet \alps determinations are based on three-jet cross sections (m3j), three- to two-jet cross-section ratios (R32),
dijet angular decorrelations (RdR, RdPhi), and the aforementioned transverse energy-energy-correlations, but at
$\sqrt{s} = 7$ and $8\TeV$ only. The H1 result is extracted from a fit to inclusive one-, two-, and three-jet cross sections
(nj) simultaneously.

All NLO results are within their large uncertainties in agreement with the world average and the associated analyses
provide valuable new values for the scale dependence of \alps at energy scales now extending beyond $2.0$\TeV as shown
in Fig.~\ref{qcd:fig:runningas}.

\begin{pdgxfigure}[wide=false, bookwidth=0.70\columnwidth]
  \includegraphics{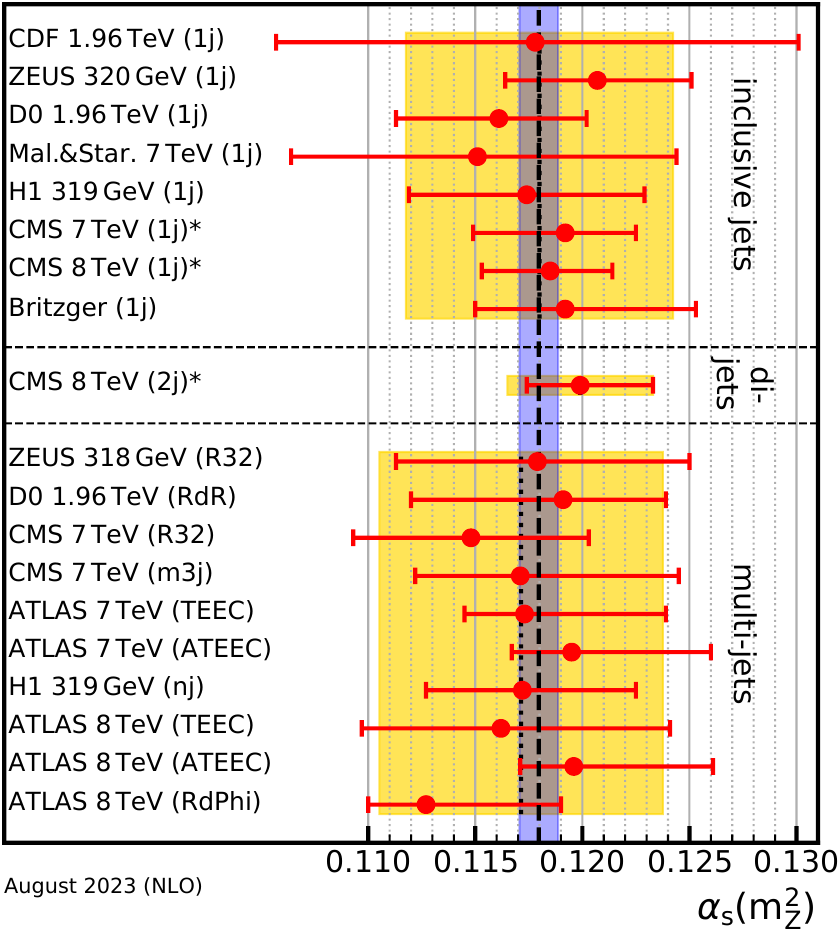}
  \caption{Summary of determinations of \asmz at NLO from inclusive jet, dijet, and multi-jet measurements at hadron
    colliders.  The uncertainty is dominated by estimates of the impact of missing higher orders.  The yellow (light
    shaded) bands and dotted lines indicate average values for each sub-field. The dashed line and blue (dark shaded)
    band represent the final world average value of \asmz. The ``*'' symbol indicates determinations including a
    simultaneous fit of PDFs.}
  \label{qcd:fig:nlo-averages}
\end{pdgxfigure}

%-----------------------------------------------------------------------
\subsection{Electroweak precision fit:}

\noindent{}For this category, we take the global electroweak fit of Ref.~\cite{Haller:2018nnx}, which includes kinematic
top quark and $W$ boson mass measurements from the LHC, determinations of the effective leptonic electroweak mixing
angles from the Tevatron, a Higgs mass measurement from ATLAS and CMS, and an evaluation of the hadronic contribution to
the running of the electromagnetic coupling at the Z-boson mass.  We also consider the fit of $\asmz$ from the global
fit to electroweak data presented in the supplemental material of Ref.~\cite{deBlas:2022hdk}.  We choose to include the
result of their ``conservative scenario''%
\footnote{We note that this result agrees with the previous fit in the ``conservative scenario'' of
  Ref.~\cite{deBlas:2021wap}, which precedes the latest CDF $W$ mass determination.} %
in accounting for the uncertainties of \mtop and \mW, which avoids a potential bias from inconsistencies between the
world average of the $W$ boson mass and the new measurement reported by the CDF Collaboration in
Ref.~\cite{CDF:2022hxs}.  In addition, we use the newer results of the electroweak fit at the Z mass pole from LEP and
SLC data presented in Sec.~{\it Electroweak Model and constraints on New Physics} of the 2022 edition of this {\it
  Review}.  All three determinations, $\asmz = 0.1194 \pm 0.0029$\cite{Haller:2018nnx},
$\asmz = 0.1193 \pm 0.0028$\cite{deBlas:2022hdk}, and $\asmz = 0.1221 \pm 0.0027$\cite{ParticleDataGroup:2022pth}, are
also in perfect agreement with the original result obtained from LEP and SLD data\cite{ALEPH:2005ab}.  Our pre-averaging
gives $\asmz = 0.1203 \pm 0.0028$.

We note, however, that results from electroweak precision data strongly depend on the strict validity of Standard Model
predictions and the existence of the minimal Higgs mechanism to implement electroweak symmetry breaking.  Any - even
small - deviation of nature from this model could strongly influence this extraction of \alps.

%-----------------------------------------------------------------------
\subsection{Lattice QCD:}

\noindent{}Several methods exist to extract the strong coupling constant from lattice QCD, as reviewed also in Sec.~{\it
  Lattice QCD} of this {\it Review} and in Ref.~\cite{DelDebbio:2021ryq}.  The Flavour Lattice Averaging Group has
considered the most up-to-date determinations and combined them to produce an update of their average
\alps~\cite{FlavourLatticeAveragingGroupFLAG:2021npn}. Their final result is obtained by considering a multitude of
possible input calculations and by retaining in their final average only those that fulfill their predefined quality
criteria, detailed in the Sec.~9.2.1 of Ref.~\cite{FlavourLatticeAveragingGroupFLAG:2021npn}.  In summary, a
determination of \alps needs to satisfy the following requirements:
\begin{itemize}
\item The determination of \alps should be based on a comparison of a short-distance quantity ${\cal Q}$ at scale $\mu$
  with a well-defined continuum limit, which does not involve UV or IR divergences, when the quantity is expressed by
  using an expansion in terms of a short distance definition of the strong coupling (\eg in the $\overline{MS}$ scheme);
\item The scale $\mu$, at which the determination is carried out, has to be sufficiently large such that the error
  associated with the perturbative truncation remains small enough;
\item If ${\cal Q}$ is defined by physical quantities in infinite volume, one needs to satisfy the constraints
  $L/a \gg \mu/\Lambda_{\rm QCD}$, where $L$ is the lattice size and $a$ the lattice spacing;
\item Only results for $n_f \geq 3$ are considered.
\item As is the case for the PDG average, a calculation must be published in a peer-reviewed journal to be eligible for
  inclusion into the FLAG average.
\end{itemize}
We note that the FLAG criteria applied now are unchanged compared to the FLAG~2019
Review~\cite{FlavourLatticeAveragingGroup:2019iem} and are considered to be relatively loose. More stringent criteria
have already been formulated by FLAG, and it is likely that in future their averages include only those results that
satisfy these stricter criteria.

Altogether, from a large number of old and new results six from the 2019 Review also enter the final FLAG~2021
average~\cite{Bruno:2017gxd,Aoki:2009tf,McNeile:2010ji,Maltman:2008bx,Chakraborty:2014aca}%
\footnote{Reference~\cite{McNeile:2010ji} contains two results.} %
and three new ones qualify for inclusion~\cite{Ayala:2020odx,Bazavov:2019qoo,Cali:2020hrj}.  These determinations,
together with their uncertainties, are displayed in Fig.~\ref{qcd:fig:lattice-averages}. The yellow (light shaded) band
and dotted line indicate the FLAG~2021 average, while the dashed line and blue (dark shaded) band represent the world
average (see later). The level of agreement of individual results to the world average, or to the non-lattice world
average is very similar.

Similarly to what is done here, the FLAG Collaboration built pre-averages of results within the various classes. The
five categories that contribute to the average are: %
step-scaling methods ($\asmz = 0.11848 \pm {0.00081}$), %
the potential at short distances ($\asmz = 0.11782 \pm {0.00165}$), %
Wilson loops ($\asmz = 0.11871 \pm {0.00128}$), %
heavy-quark current two-point functions ($\asmz = 0.11826 \pm {0.00200}$) %
and, for the first time, light quark vacuum polarization ($\asmz = 0.11863 \pm {0.00360}$). %
Other categories such as the calculation of QCD vertices, or of the eigenvalue spectrum of the Dirac operator have not yet
published results that fulfill all requirements to be included in the average.

We note that, in addition to presenting new results and improvements on previous works, the ALPHA
Collaboration~\cite{DallaBrida:2019mqg} has introduced a novel approach to non-perturbative renormalization called
decoupling. This strategy shifts the perspective on results involving unphysical flavor numbers, particularly
$n_f=0$. By performing a non-perturbative matching calculation, these results can be non-perturbatively related to
results with $n_f > 0$. Consequently, obtaining precise and controlled $n_f = 0$ results becomes of great importance,
with significant implications for future FLAG reports.

The final value is obtained by performing a {\it weighted average} of the pre-averages. The final uncertainty however is
not the combined uncertainty of the pre-averages (which is $0.0006$), since the errors on almost all determinations are
dominated by the perturbative truncation error.  Instead, the error on the pre-range for \alps from the step-scaling
method is taken, since perturbative truncation errors are sub-dominant in this method.  The final FLAG~2021 average
(rounded to four digits) is
\begin{equation}\label{qcd:eq:asaveragelat}
  \asmz = 0.1184 \pm 0.0008\qquad\textrm{(FLAG 2021 average)}\,,
\end{equation}
which is fully compatible with the FLAG~2019 result of $\asmz = 0.1182 \pm 0.0008$.

We believe that this result expresses to a large extent the consensus of the lattice community and that the imposed
criteria and the rigorous assessment of systematic uncertainties qualify for a direct inclusion of this FLAG average
here. As in the previous review, we therefore adopt the FLAG average with its uncertainty as our value of \alps for the
lattice category. Moreover, this lattice result will not be directly combined with any other sub-field average, but with
our non-lattice average to give our final world average value for \alps.

\begin{pdgxfigure}[wide=false, bookwidth=0.70\columnwidth]
  \includegraphics{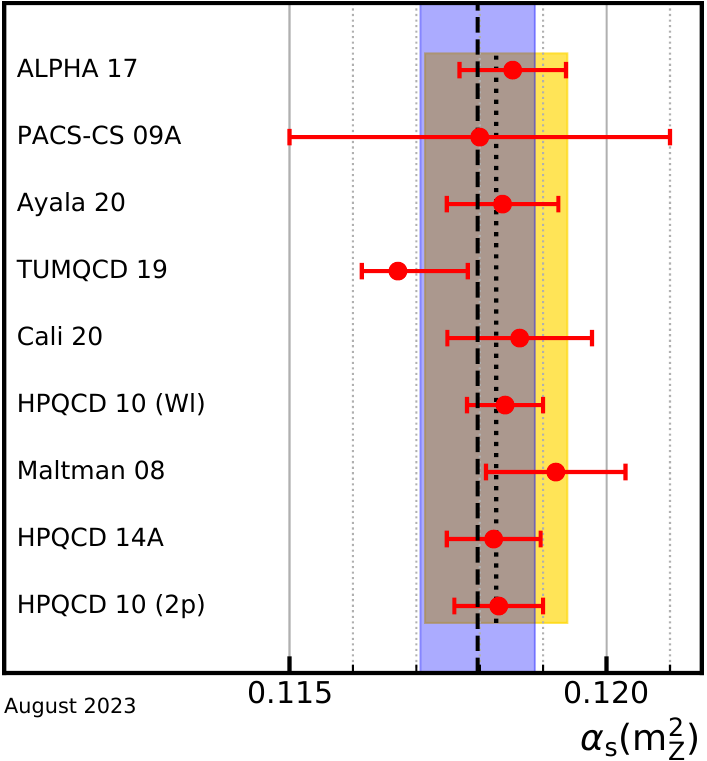}
  \caption{Lattice determinations that enter the FLAG~2021 average. The yellow (light shaded) band and dotted line
    indicate the {\it unweighted average} value for this sub-field. The dashed line and blue (dark shaded) band
    represent our final world average value of \asmz.}
  \label{qcd:fig:lattice-averages}
\end{pdgxfigure}

%-----------------------------------------------------------------------
\subsection{Determination of the world average value of \texorpdfstring{\boldmath\asmz\unboldmath}{alpha-s(MZ)}:}
\label{qcd:sec:world}

\noindent{}Obtaining a world average value for \asmz is a non-trivial exercise.  A certain arbitrariness and subjective
component is inevitable because of the choice of measurements to be included in the average, the treatment of
(non-Gaussian) systematic uncertainties of mostly theoretical nature, as well as the treatment of correlations among the
various inputs, of theoretical as well as experimental origin.

We have chosen to determine pre-averages for sub-fields of measurements that are considered to exhibit a maximum degree
of independence among each other, considering experimental as well as theoretical issues. The seven pre-averages,
illustrated also in \Fig{qcd:fig:pre-averages}, are listed in column two of Table~\ref{qcd:tab:preaverages}. We recall
that these are exclusively obtained from extractions that are based on (at least) NNLO QCD predictions, and are
published in peer-reviewed journals at the time of completing this {\it Review}.  To obtain our final world average, we
first combine six pre-averages, excluding the lattice result, using a {\it $\chi^2$ averaging method}. This gives
\begin{equation}
  \asmz = 0.1175 \pm 0.0010\qquad\textrm{(PDG 2023 without lattice)}\,.
  \label{qcd:eq:asglobalaveragenolat}
\end{equation}
This result is fully compatible with the lattice pre-average \Eq{qcd:eq:asaveragelat} and has a comparable error. To
avoid a possible over-reduction, we combine these two numbers using an unweighted average and take as an uncertainty the
average between these two uncertainties. This gives our final world average value\index{world average of \asmz}
\begin{equation}\label{qcd:eq:asglobalaverage}
  \asmz = 0.1180 \pm 0.0009\qquad\textrm{(PDG 2023 average)}\,.
\end{equation}

If for the sub-field of hadron colliders we are more restrictive and instead only accept results from a simultaneous fit
of PDFs, we arrive at $0.1157 \pm 0.0021$ for this sub-field leading to $0.1172 \pm 0.0010$ (without lattice) and
$\asmz = 0.1178 \pm 0.0009$ for the final average. Both the new world average value and the restricted result are
compatible with each other and changed only marginally as compared to the values reported in the last edition of this
{\it Review}.

It also stands to question whether the sub-fields of PDF fits and hadron colliders still are as independent as
originally assumed. To test the potential impact, the fit has been repeated while grouping all \asmz determinations of
both sub-fields into a common one. We obtain $0.1164 \pm 0.0024$ for the new larger sub-field and $0.1178 \pm 0.0011$
for the combination with all other sub-fields except lattice, which is well within the estimated uncertainties.
Moreover, we present in column four of Table~\ref{qcd:tab:preaverages} the combined result for \asmz when the respective
sub-field is omitted from the combination. The variation in values obtained for \asmz is $\pm 0.0004$, which is less
than half our estimated uncertainty.

Since two long-standing issues causing offsets among determinations of \asmz in the sub-fields of $\tau$ decays and low
$Q^2$ and \ee jets \& shapes have been resolved, it may be argued that a weighted fit among the non-lattice sub-fields
may be warranted.  We compare the outcome of weighted fits with our standard procedure in columns two and three of
Table~\ref{qcd:tab:preaverages}. We observe that the weighted averages are rather close to the unweighted ones.
However, the uncertainties become significantly smaller. This approach may be too aggressive as it ignores the
correlations among the data, methods, and theory ingredients of the various determinations. We feel that the uncertainty
of $\pm 0.0005$ is an underestimation of the true error.  We also note that in the unweighted combination the estimated
uncertainty for each sub-field is larger than the spread of the results as given by the standard deviation. In the
weighted fit this crosscheck fails in four out of six cases.

The last several years have seen clarification of some persistent concerns and a wealth of new results at NNLO,
providing not only a rather precise and reasonably stable world average value of \asmz, but also a clear signature and
proof of the energy dependence of \alps in full agreement with the QCD prediction of asymptotic freedom.\index{running
  coupling in QCD}\index{QCD!running coupling} This is demonstrated in \Fig{qcd:fig:runningas}, where results of
$\alps(Q^2)$ obtained at discrete energy scales $Q$ are summarized, which now mostly include those based on NNLO QCD\@.%
\footnote{The uncertainties of the HERA jets points were, by mistake, shown at only half their size. The uncertainties
  of the ALEPH points of the \ee jets/shapes sub-field now correspond to the total uncertainty.}  Thanks to the results
from the LHC, the energy scales, at which \alps is determined, now extend even beyond 2\TeV.%
\footnote{We note, however, that the relevant energy scale of a measurement is not uniquely defined. In addition to
  being multiplied by factors of \eg $1/2$ or $2$, for instance in studies of the ratio of three- to two-jet cross
  sections at the LHC, the relevant scale can be taken to be the average of the transverse momenta of the two leading
  jets~\cite{Chatrchyan:2013txa}, or alternatively might be chosen to be the transverse momentum of the $3^{rd}$ jet.}%
The points in this plot are extracted from Refs.~\cite{ParticleDataGroup:2022pth,Boito:2018yvl,Narison:2018xbj,
  Britzger:2019kkb,Dissertori:2009ik,Bethke:2008hf,Schieck:2012mp,OPAL:2011aa,Abazov:2009nc,Abazov:2012lua,Aaboud:2018hie,
  CMS:2014mna,Khachatryan:2016mlc,Klijnsma:2017eqp,ATLAS:2023tgo}.

\newpage
\begin{pdgxtable}[place=h!,wide=true]
  \caption{Unweighted and weighted pre-averages of \asmz for each sub-field in columns two and three. The bottom line
    corresponds to the combined result (without lattice gauge theory) using the $\chi^2$ averaging method. The same
    $\chi^2$ averaging is used for column four combining all unweighted averages except for the sub-field of column one.
    See text for more details.}
  \label{qcd:tab:preaverages}
  \centering
  \begin{pdgxtabular}{lccc}
    \pdgtableheader{averages per sub-field & unweighted & weighted & unweighted without subfield}
    $\tau$ decays \& low $Q^2$ & $0.1173 \pm 0.0017$ & $0.1174 \pm 0.0009$ & $0.1177 \pm 0.0013$ \\
    $Q\bar{Q}$ bound states    & $0.1181 \pm 0.0037$ & $0.1177 \pm 0.0011$ & $0.1175 \pm 0.0011$ \\
    PDF fits                   & $0.1161 \pm 0.0022$ & $0.1168 \pm 0.0014$ & $0.1179 \pm 0.0011$ \\
    \ee jets \& shapes         & $0.1189 \pm 0.0037$ & $0.1187 \pm 0.0017$ & $0.1174 \pm 0.0011$ \\
    hadron colliders           & $0.1168 \pm 0.0027$ & $0.1169 \pm 0.0014$ & $0.1177 \pm 0.0011$ \\
    electroweak                & $0.1203 \pm 0.0028$ & $0.1203 \pm 0.0016$ & $0.1171 \pm 0.0011$ \\\hline
    PDG 2023 (without lattice) & $0.1175 \pm 0.0010$ & $0.1178 \pm 0.0005$ & n/a \\
  \end{pdgxtabular}
\end{pdgxtable}

\begin{pdgxfigure}[place=h!,wide=true,bookwidth=0.9\linewidth]
  \includegraphics{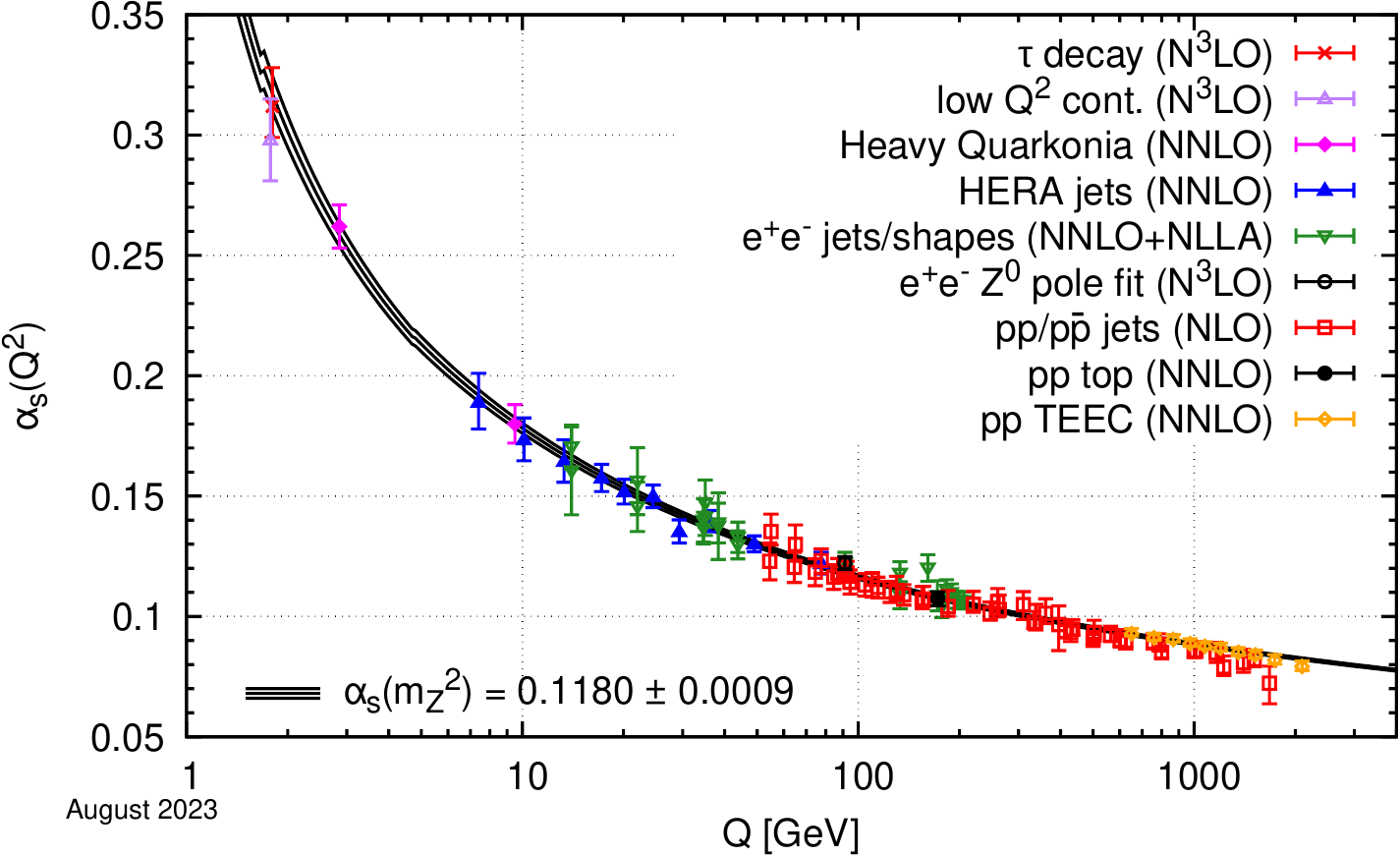}
  \caption{Summary of determinations of \alps as a function of the energy scale $Q$ compared to the running of the
    coupling computed at five loops taking as an input the current PDG average, $\asmz = 0.1180\pm 0.0009$.  Compared to
    the previous edition, numerous points have been updated or added.}
  \label{qcd:fig:runningas}\index{running coupling in QCD}\index{QCD!running coupling}
\end{pdgxfigure}

In this combination, as in past combinations, we have considered lattice QCD calculations of \alps independently of
experimental/phenomenological determinations. In the future, when the lattice continuum extrapolations are under better
control, it may be useful to group lattice QCD determinations of \alps with experimental determinations of \alps that
have systematics of similar origin, in a similar manner as we currently group, for example, hadron collider results
together~\cite{DelDebbio:2021ryq}.

%-----------------------------------------------------------------------

% References
% ----------
% The following line includes your bibliography using BibTeX. In case you do not
% yet use BibTex, you can put your bibliography below (using a series of \bibitem entries).
% Please  note that using BibTeX will become mandatory in the future.
%\IfFileExists{qcd.bib}{\putbib[qcd]}{}
%\FloatBarrier
\clearpage
\bibliography{qcd}

\ifx\mcitethebibliography\mciteundefinedmacro
\PackageError{unsrtM.bst}{mciteplus.sty has not been loaded}
{This bibstyle requires the use of the mciteplus package.}\fi
\def\enquote#1{``#1''}
\expandafter\ifx\csname url\endcsname\relax
  \def\url#1{{\tt #1}}\fi
\expandafter\ifx\csname urlprefix\endcsname\relax\def\urlprefix{URL }\fi
\expandafter\ifx\csname eprint\endcsname\relax\def\eprint#1{\url{#1}}\fi
\begin{mcitethebibliography}{100}

\bibitem{Ellis:1991qj}
R.~K. Ellis, W.~J. Stirling and B.~R. Webber, Camb. Monogr. Part. Phys. Nucl.
  Phys. Cosmol. {\bf 8}, 1 (1996)\relax
\mciteBstWouldAddEndPuncttrue
\mciteSetBstMidEndSepPunct{\mcitedefaultmidpunct}
{\mcitedefaultendpunct}{\mcitedefaultseppunct}\relax
\EndOfBibitem
\bibitem{qcd:CHKbook}
J.~Campbell, J.~Huston, F.~Krauss {\it ``The Black Book of Quantum
  Chromodynamics, a Primer for the QCD Era,''}\rm Oxford University Press, UK
  (2017)\relax
\mciteBstWouldAddEndPuncttrue
\mciteSetBstMidEndSepPunct{\mcitedefaultmidpunct}
{\mcitedefaultendpunct}{\mcitedefaultseppunct}\relax
\EndOfBibitem
\bibitem{Baker:2006ts}
C.~A. Baker {\em et~al.\/},
  \href{http://doi.org/10.1103/PhysRevLett.97.131801}{Phys. Rev. Lett.}
  \href{http://doi.org/10.1103/PhysRevLett.97.131801}{{\bf 97}, 131801} (2006),
  \href{https://arxiv.org/abs/hep-ex/0602020}{[hep-ex/0602020]}\relax
\mciteBstWouldAddEndPuncttrue
\mciteSetBstMidEndSepPunct{\mcitedefaultmidpunct}
{\mcitedefaultendpunct}{\mcitedefaultseppunct}\relax
\EndOfBibitem
\bibitem{Afach:2015sja}
J.~M. Pendlebury {\em et~al.\/},
  \href{http://doi.org/10.1103/PhysRevD.92.092003}{Phys. Rev.}
  \href{http://doi.org/10.1103/PhysRevD.92.092003}{{\bf D92}, 9, 092003}
  (2015), \href{https://arxiv.org/abs/1509.04411}{[arXiv:1509.04411]}\relax
\mciteBstWouldAddEndPuncttrue
\mciteSetBstMidEndSepPunct{\mcitedefaultmidpunct}
{\mcitedefaultendpunct}{\mcitedefaultseppunct}\relax
\EndOfBibitem
\bibitem{Graner:2016ses}
B.~Graner {\em et~al.\/},
  \href{http://doi.org/10.1103/PhysRevLett.119.119901}{Phys. Rev. Lett.}
  \href{http://doi.org/10.1103/PhysRevLett.119.119901}{{\bf 116}, 16, 161601}
  (2016), [Erratum: Phys. Rev. Lett.119,no.11,119901(2017)],
  \href{https://arxiv.org/abs/1601.04339}{[arXiv:1601.04339]}\relax
\mciteBstWouldAddEndPuncttrue
\mciteSetBstMidEndSepPunct{\mcitedefaultmidpunct}
{\mcitedefaultendpunct}{\mcitedefaultseppunct}\relax
\EndOfBibitem
\bibitem{EDMs}
\url{https://www.psi.ch/en/nedm/edms-world-wide}\relax
\mciteBstWouldAddEndPuncttrue
\mciteSetBstMidEndSepPunct{\mcitedefaultmidpunct}
{\mcitedefaultendpunct}{\mcitedefaultseppunct}\relax
\EndOfBibitem
\bibitem{Kim:2008hd}
J.~E. Kim and G.~Carosi, \href{http://doi.org/10.1103/RevModPhys.82.557}{Rev.
  Mod. Phys.} \href{http://doi.org/10.1103/RevModPhys.82.557}{{\bf 82}, 557}
  (2010), \href{https://arxiv.org/abs/0807.3125}{[arXiv:0807.3125]}\relax
\mciteBstWouldAddEndPuncttrue
\mciteSetBstMidEndSepPunct{\mcitedefaultmidpunct}
{\mcitedefaultendpunct}{\mcitedefaultseppunct}\relax
\EndOfBibitem
\bibitem{Dissertori:2003pj}
G.~Dissertori, I.~G. Knowles and M.~Schmelling, {\em {High energy experiments
  and theory, {\rm Oxford, UK: Clarendon}}\/} (2003)\relax
\mciteBstWouldAddEndPuncttrue
\mciteSetBstMidEndSepPunct{\mcitedefaultmidpunct}
{\mcitedefaultendpunct}{\mcitedefaultseppunct}\relax
\EndOfBibitem
\bibitem{Brock:1993sz}
R.~Brock {\em et~al.\/} (CTEQ),
  \href{http://doi.org/10.1103/RevModPhys.67.157}{Rev. Mod. Phys.}
  \href{http://doi.org/10.1103/RevModPhys.67.157}{{\bf 67}, 157} (1995)\relax
\mciteBstWouldAddEndPuncttrue
\mciteSetBstMidEndSepPunct{\mcitedefaultmidpunct}
{\mcitedefaultendpunct}{\mcitedefaultseppunct}\relax
\EndOfBibitem
\bibitem{Melnikov:2018qpx}
K.~Melnikov, \href{http://doi.org/10.23730/CYRSP-2018-003.37}{CERN Yellow Rep.
  School Proc.} \href{http://doi.org/10.23730/CYRSP-2018-003.37}{{\bf 3}, 37}
  (2018)\relax
\mciteBstWouldAddEndPuncttrue
\mciteSetBstMidEndSepPunct{\mcitedefaultmidpunct}
{\mcitedefaultendpunct}{\mcitedefaultseppunct}\relax
\EndOfBibitem
\bibitem{Gross:2022hyw}
F.~Gross {\em et~al.\/}  (2022),
  \href{https://arxiv.org/abs/2212.11107}{[arXiv:2212.11107]}\relax
\mciteBstWouldAddEndPuncttrue
\mciteSetBstMidEndSepPunct{\mcitedefaultmidpunct}
{\mcitedefaultendpunct}{\mcitedefaultseppunct}\relax
\EndOfBibitem
\bibitem{vanRitbergen:1997va}
T.~van Ritbergen, J.~A.~M. Vermaseren and S.~A. Larin,
  \href{http://doi.org/10.1016/S0370-2693(97)00370-5}{Phys. Lett.}
  \href{http://doi.org/10.1016/S0370-2693(97)00370-5}{{\bf B400}, 379} (1997),
  \href{https://arxiv.org/abs/hep-ph/9701390}{[hep-ph/9701390]}\relax
\mciteBstWouldAddEndPuncttrue
\mciteSetBstMidEndSepPunct{\mcitedefaultmidpunct}
{\mcitedefaultendpunct}{\mcitedefaultseppunct}\relax
\EndOfBibitem
\bibitem{Czakon:2004bu}
M.~Czakon, \href{http://doi.org/10.1016/j.nuclphysb.2005.01.012}{Nucl. Phys.}
  \href{http://doi.org/10.1016/j.nuclphysb.2005.01.012}{{\bf B710}, 485}
  (2005), \href{https://arxiv.org/abs/hep-ph/0411261}{[hep-ph/0411261]}\relax
\mciteBstWouldAddEndPuncttrue
\mciteSetBstMidEndSepPunct{\mcitedefaultmidpunct}
{\mcitedefaultendpunct}{\mcitedefaultseppunct}\relax
\EndOfBibitem
\bibitem{Baikov:2016tgj}
P.~A. Baikov, K.~G. Chetyrkin and J.~H. K\"uhn,
  \href{http://doi.org/10.1103/PhysRevLett.118.082002}{Phys. Rev. Lett.}
  \href{http://doi.org/10.1103/PhysRevLett.118.082002}{{\bf 118}, 8, 082002}
  (2017), \href{https://arxiv.org/abs/1606.08659}{[arXiv:1606.08659]}\relax
\mciteBstWouldAddEndPuncttrue
\mciteSetBstMidEndSepPunct{\mcitedefaultmidpunct}
{\mcitedefaultendpunct}{\mcitedefaultseppunct}\relax
\EndOfBibitem
\bibitem{Luthe:2016ima}
T.~Luthe {\em et~al.\/}, \href{http://doi.org/10.1007/JHEP07(2016)127}{JHEP}
  \href{http://doi.org/10.1007/JHEP07(2016)127}{{\bf 07}, 127} (2016),
  \href{https://arxiv.org/abs/1606.08662}{[arXiv:1606.08662]}\relax
\mciteBstWouldAddEndPuncttrue
\mciteSetBstMidEndSepPunct{\mcitedefaultmidpunct}
{\mcitedefaultendpunct}{\mcitedefaultseppunct}\relax
\EndOfBibitem
\bibitem{Herzog:2017ohr}
F.~Herzog {\em et~al.\/}, \href{http://doi.org/10.1007/JHEP02(2017)090}{JHEP}
  \href{http://doi.org/10.1007/JHEP02(2017)090}{{\bf 02}, 090} (2017),
  \href{https://arxiv.org/abs/1701.01404}{[arXiv:1701.01404]}\relax
\mciteBstWouldAddEndPuncttrue
\mciteSetBstMidEndSepPunct{\mcitedefaultmidpunct}
{\mcitedefaultendpunct}{\mcitedefaultseppunct}\relax
\EndOfBibitem
\bibitem{Luthe:2017ttg}
T.~Luthe {\em et~al.\/}, \href{http://doi.org/10.1007/JHEP10(2017)166}{JHEP}
  \href{http://doi.org/10.1007/JHEP10(2017)166}{{\bf 10}, 166} (2017),
  \href{https://arxiv.org/abs/1709.07718}{[arXiv:1709.07718]}\relax
\mciteBstWouldAddEndPuncttrue
\mciteSetBstMidEndSepPunct{\mcitedefaultmidpunct}
{\mcitedefaultendpunct}{\mcitedefaultseppunct}\relax
\EndOfBibitem
\bibitem{Chetyrkin:2017bjc}
K.~G. Chetyrkin {\em et~al.\/},
  \href{http://doi.org/10.1007/JHEP12(2017)006}{JHEP}
  \href{http://doi.org/10.1007/JHEP12(2017)006}{{\bf 10}, 179} (2017),
  [Addendum: JHEP12,006(2017)],
  \href{https://arxiv.org/abs/1709.08541}{[arXiv:1709.08541]}\relax
\mciteBstWouldAddEndPuncttrue
\mciteSetBstMidEndSepPunct{\mcitedefaultmidpunct}
{\mcitedefaultendpunct}{\mcitedefaultseppunct}\relax
\EndOfBibitem
\bibitem{Bardeen:1978yd}
W.~A. Bardeen {\em et~al.\/},
  \href{http://doi.org/10.1103/PhysRevD.18.3998}{Phys. Rev.}
  \href{http://doi.org/10.1103/PhysRevD.18.3998}{{\bf D18}, 3998} (1978)\relax
\mciteBstWouldAddEndPuncttrue
\mciteSetBstMidEndSepPunct{\mcitedefaultmidpunct}
{\mcitedefaultendpunct}{\mcitedefaultseppunct}\relax
\EndOfBibitem
\bibitem{Gross:1973id}
D.~J. Gross and F.~Wilczek,
  \href{http://doi.org/10.1103/PhysRevLett.30.1343}{Phys. Rev. Lett.}
  \href{http://doi.org/10.1103/PhysRevLett.30.1343}{{\bf 30}, 1343} (1973),
  [,271(1973)]\relax
\mciteBstWouldAddEndPuncttrue
\mciteSetBstMidEndSepPunct{\mcitedefaultmidpunct}
{\mcitedefaultendpunct}{\mcitedefaultseppunct}\relax
\EndOfBibitem
\bibitem{Politzer:1973fx}
H.~D. Politzer, \href{http://doi.org/10.1103/PhysRevLett.30.1346}{Phys. Rev.
  Lett.} \href{http://doi.org/10.1103/PhysRevLett.30.1346}{{\bf 30}, 1346}
  (1973), [,274(1973)]\relax
\mciteBstWouldAddEndPuncttrue
\mciteSetBstMidEndSepPunct{\mcitedefaultmidpunct}
{\mcitedefaultendpunct}{\mcitedefaultseppunct}\relax
\EndOfBibitem
\bibitem{Schroder:2005hy}
Y.~Schroder and M.~Steinhauser,
  \href{http://doi.org/10.1088/1126-6708/2006/01/051}{JHEP}
  \href{http://doi.org/10.1088/1126-6708/2006/01/051}{{\bf 01}, 051} (2006),
  \href{https://arxiv.org/abs/hep-ph/0512058}{[hep-ph/0512058]}\relax
\mciteBstWouldAddEndPuncttrue
\mciteSetBstMidEndSepPunct{\mcitedefaultmidpunct}
{\mcitedefaultendpunct}{\mcitedefaultseppunct}\relax
\EndOfBibitem
\bibitem{Chetyrkin:2005ia}
K.~G. Chetyrkin, J.~H. Kuhn and C.~Sturm,
  \href{http://doi.org/10.1016/j.nuclphysb.2006.03.020}{Nucl. Phys.}
  \href{http://doi.org/10.1016/j.nuclphysb.2006.03.020}{{\bf B744}, 121}
  (2006), \href{https://arxiv.org/abs/hep-ph/0512060}{[hep-ph/0512060]}\relax
\mciteBstWouldAddEndPuncttrue
\mciteSetBstMidEndSepPunct{\mcitedefaultmidpunct}
{\mcitedefaultendpunct}{\mcitedefaultseppunct}\relax
\EndOfBibitem
\bibitem{Grozin:2011nk}
A.~G. Grozin {\em et~al.\/},
  \href{http://doi.org/10.1007/JHEP09(2011)066}{JHEP}
  \href{http://doi.org/10.1007/JHEP09(2011)066}{{\bf 09}, 066} (2011),
  \href{https://arxiv.org/abs/1107.5970}{[arXiv:1107.5970]}\relax
\mciteBstWouldAddEndPuncttrue
\mciteSetBstMidEndSepPunct{\mcitedefaultmidpunct}
{\mcitedefaultendpunct}{\mcitedefaultseppunct}\relax
\EndOfBibitem
\bibitem{Brida:2016flw}
M.~Dalla~Brida {\em et~al.\/} (ALPHA),
  \href{http://doi.org/10.1103/PhysRevLett.117.182001}{Phys. Rev. Lett.}
  \href{http://doi.org/10.1103/PhysRevLett.117.182001}{{\bf 117}, 18, 182001}
  (2016), \href{https://arxiv.org/abs/1604.06193}{[arXiv:1604.06193]}\relax
\mciteBstWouldAddEndPuncttrue
\mciteSetBstMidEndSepPunct{\mcitedefaultmidpunct}
{\mcitedefaultendpunct}{\mcitedefaultseppunct}\relax
\EndOfBibitem
\bibitem{Beneke:1998ui}
M.~Beneke, \href{http://doi.org/10.1016/S0370-1573(98)00130-6}{Phys. Rept.}
  \href{http://doi.org/10.1016/S0370-1573(98)00130-6}{{\bf 317}, 1} (1999),
  \href{https://arxiv.org/abs/hep-ph/9807443}{[hep-ph/9807443]}\relax
\mciteBstWouldAddEndPuncttrue
\mciteSetBstMidEndSepPunct{\mcitedefaultmidpunct}
{\mcitedefaultendpunct}{\mcitedefaultseppunct}\relax
\EndOfBibitem
\bibitem{Beneke:2016cbu}
M.~Beneke {\em et~al.\/},
  \href{http://doi.org/10.1016/j.physletb.2017.10.054}{Phys. Lett.}
  \href{http://doi.org/10.1016/j.physletb.2017.10.054}{{\bf B775}, 63} (2017),
  \href{https://arxiv.org/abs/1605.03609}{[arXiv:1605.03609]}\relax
\mciteBstWouldAddEndPuncttrue
\mciteSetBstMidEndSepPunct{\mcitedefaultmidpunct}
{\mcitedefaultendpunct}{\mcitedefaultseppunct}\relax
\EndOfBibitem
\bibitem{Hoang:2017btd}
A.~H. Hoang, C.~Lepenik and M.~Preisser,
  \href{http://doi.org/10.1007/JHEP09(2017)099}{JHEP}
  \href{http://doi.org/10.1007/JHEP09(2017)099}{{\bf 09}, 099} (2017),
  \href{https://arxiv.org/abs/1706.08526}{[arXiv:1706.08526]}\relax
\mciteBstWouldAddEndPuncttrue
\mciteSetBstMidEndSepPunct{\mcitedefaultmidpunct}
{\mcitedefaultendpunct}{\mcitedefaultseppunct}\relax
\EndOfBibitem
\bibitem{Marquard:2015qpa}
P.~Marquard {\em et~al.\/},
  \href{http://doi.org/10.1103/PhysRevLett.114.142002}{Phys. Rev. Lett.}
  \href{http://doi.org/10.1103/PhysRevLett.114.142002}{{\bf 114}, 14, 142002}
  (2015), \href{https://arxiv.org/abs/1502.01030}{[arXiv:1502.01030]}\relax
\mciteBstWouldAddEndPuncttrue
\mciteSetBstMidEndSepPunct{\mcitedefaultmidpunct}
{\mcitedefaultendpunct}{\mcitedefaultseppunct}\relax
\EndOfBibitem
\bibitem{Baikov:2012zn}
P.~A. Baikov {\em et~al.\/},
  \href{http://doi.org/10.1016/j.physletb.2012.06.052}{Phys. Lett.}
  \href{http://doi.org/10.1016/j.physletb.2012.06.052}{{\bf B714}, 62} (2012),
  \href{https://arxiv.org/abs/1206.1288}{[arXiv:1206.1288]}\relax
\mciteBstWouldAddEndPuncttrue
\mciteSetBstMidEndSepPunct{\mcitedefaultmidpunct}
{\mcitedefaultendpunct}{\mcitedefaultseppunct}\relax
\EndOfBibitem
\bibitem{Chetyrkin:1994js}
K.~G. Chetyrkin, J.~H. Kuhn and A.~Kwiatkowski  (1996), [Phys.
  Rept.277,189(1996)],
  \href{https://arxiv.org/abs/hep-ph/9503396}{[hep-ph/9503396]}\relax
\mciteBstWouldAddEndPuncttrue
\mciteSetBstMidEndSepPunct{\mcitedefaultmidpunct}
{\mcitedefaultendpunct}{\mcitedefaultseppunct}\relax
\EndOfBibitem
\bibitem{Kiyo:2009gb}
Y.~Kiyo {\em et~al.\/},
  \href{http://doi.org/10.1016/j.nuclphysb.2009.08.010}{Nucl. Phys.}
  \href{http://doi.org/10.1016/j.nuclphysb.2009.08.010}{{\bf B823}, 269}
  (2009), \href{https://arxiv.org/abs/0907.2120}{[arXiv:0907.2120]}\relax
\mciteBstWouldAddEndPuncttrue
\mciteSetBstMidEndSepPunct{\mcitedefaultmidpunct}
{\mcitedefaultendpunct}{\mcitedefaultseppunct}\relax
\EndOfBibitem
\bibitem{Baikov:2012er}
P.~A. Baikov {\em et~al.\/},
  \href{http://doi.org/10.1103/PhysRevLett.108.222003}{Phys. Rev. Lett.}
  \href{http://doi.org/10.1103/PhysRevLett.108.222003}{{\bf 108}, 222003}
  (2012), \href{https://arxiv.org/abs/1201.5804}{[arXiv:1201.5804]}\relax
\mciteBstWouldAddEndPuncttrue
\mciteSetBstMidEndSepPunct{\mcitedefaultmidpunct}
{\mcitedefaultendpunct}{\mcitedefaultseppunct}\relax
\EndOfBibitem
\bibitem{Baikov:2008jh}
P.~A. Baikov, K.~G. Chetyrkin and J.~H. Kuhn,
  \href{http://doi.org/10.1103/PhysRevLett.101.012002}{Phys. Rev. Lett.}
  \href{http://doi.org/10.1103/PhysRevLett.101.012002}{{\bf 101}, 012002}
  (2008), \href{https://arxiv.org/abs/0801.1821}{[arXiv:0801.1821]}\relax
\mciteBstWouldAddEndPuncttrue
\mciteSetBstMidEndSepPunct{\mcitedefaultmidpunct}
{\mcitedefaultendpunct}{\mcitedefaultseppunct}\relax
\EndOfBibitem
\bibitem{Herzog:2017dtz}
F.~Herzog {\em et~al.\/}, \href{http://doi.org/10.1007/JHEP08(2017)113}{JHEP}
  \href{http://doi.org/10.1007/JHEP08(2017)113}{{\bf 08}, 113} (2017),
  \href{https://arxiv.org/abs/1707.01044}{[arXiv:1707.01044]}\relax
\mciteBstWouldAddEndPuncttrue
\mciteSetBstMidEndSepPunct{\mcitedefaultmidpunct}
{\mcitedefaultendpunct}{\mcitedefaultseppunct}\relax
\EndOfBibitem
\bibitem{Novikov:1980uj}
V.~A. Novikov {\em et~al.\/},
  \href{http://doi.org/10.1016/0550-3213(80)90290-4}{Nucl. Phys.}
  \href{http://doi.org/10.1016/0550-3213(80)90290-4}{{\bf B174}, 378}
  (1980)\relax
\mciteBstWouldAddEndPuncttrue
\mciteSetBstMidEndSepPunct{\mcitedefaultmidpunct}
{\mcitedefaultendpunct}{\mcitedefaultseppunct}\relax
\EndOfBibitem
\bibitem{Hou:2019efy}
T.-J. Hou {\em et~al.\/},
  \href{http://doi.org/10.1103/PhysRevD.103.014013}{Phys. Rev. D}
  \href{http://doi.org/10.1103/PhysRevD.103.014013}{{\bf 103}, 1, 014013}
  (2021), \href{https://arxiv.org/abs/1912.10053}{[arXiv:1912.10053]}\relax
\mciteBstWouldAddEndPuncttrue
\mciteSetBstMidEndSepPunct{\mcitedefaultmidpunct}
{\mcitedefaultendpunct}{\mcitedefaultseppunct}\relax
\EndOfBibitem
\bibitem{PDF4LHCWorkingGroup:2022cjn}
R.~D. Ball {\em et~al.\/} (PDF4LHC Working Group),
  \href{http://doi.org/10.1088/1361-6471/ac7216}{J. Phys. G}
  \href{http://doi.org/10.1088/1361-6471/ac7216}{{\bf 49}, 8, 080501} (2022),
  \href{https://arxiv.org/abs/2203.05506}{[arXiv:2203.05506]}\relax
\mciteBstWouldAddEndPuncttrue
\mciteSetBstMidEndSepPunct{\mcitedefaultmidpunct}
{\mcitedefaultendpunct}{\mcitedefaultseppunct}\relax
\EndOfBibitem
\bibitem{NNPDF:2019vjt}
R.~Abdul~Khalek {\em et~al.\/} (NNPDF),
  \href{http://doi.org/10.1140/epjc/s10052-019-7364-5}{Eur. Phys. J.}
  \href{http://doi.org/10.1140/epjc/s10052-019-7364-5}{{\bf C}, 79:838} (2019),
  \href{https://arxiv.org/abs/1905.04311}{[arXiv:1905.04311]}\relax
\mciteBstWouldAddEndPuncttrue
\mciteSetBstMidEndSepPunct{\mcitedefaultmidpunct}
{\mcitedefaultendpunct}{\mcitedefaultseppunct}\relax
\EndOfBibitem
\bibitem{NNPDF:2019ubu}
R.~Abdul~Khalek {\em et~al.\/} (NNPDF),
  \href{http://doi.org/10.1140/epjc/s10052-019-7401-4}{Eur. Phys. J. C}
  \href{http://doi.org/10.1140/epjc/s10052-019-7401-4}{{\bf 79}, 11, 931}
  (2019), \href{https://arxiv.org/abs/1906.10698}{[arXiv:1906.10698]}\relax
\mciteBstWouldAddEndPuncttrue
\mciteSetBstMidEndSepPunct{\mcitedefaultmidpunct}
{\mcitedefaultendpunct}{\mcitedefaultseppunct}\relax
\EndOfBibitem
\bibitem{Ball:2021icz}
R.~D. Ball and R.~L. Pearson,
  \href{http://doi.org/10.1140/epjc/s10052-021-09602-x}{Eur. Phys. J. C}
  \href{http://doi.org/10.1140/epjc/s10052-021-09602-x}{{\bf 81}, 9, 830}
  (2021), \href{https://arxiv.org/abs/2105.05114}{[arXiv:2105.05114]}\relax
\mciteBstWouldAddEndPuncttrue
\mciteSetBstMidEndSepPunct{\mcitedefaultmidpunct}
{\mcitedefaultendpunct}{\mcitedefaultseppunct}\relax
\EndOfBibitem
\bibitem{Kassabov:2022orn}
Z.~Kassabov, M.~Ubiali and C.~Voisey,
  \href{http://doi.org/10.1007/JHEP03(2023)148}{JHEP}
  \href{http://doi.org/10.1007/JHEP03(2023)148}{{\bf 03}, 148} (2023),
  \href{https://arxiv.org/abs/2207.07616}{[arXiv:2207.07616]}\relax
\mciteBstWouldAddEndPuncttrue
\mciteSetBstMidEndSepPunct{\mcitedefaultmidpunct}
{\mcitedefaultendpunct}{\mcitedefaultseppunct}\relax
\EndOfBibitem
\bibitem{Harland-Lang:2018bxd}
L.~A. Harland-Lang and R.~S. Thorne,
  \href{http://doi.org/10.1140/epjc/s10052-019-6731-6}{Eur. Phys. J. C}
  \href{http://doi.org/10.1140/epjc/s10052-019-6731-6}{{\bf 79}, 3, 225}
  (2019), \href{https://arxiv.org/abs/1811.08434}{[arXiv:1811.08434]}\relax
\mciteBstWouldAddEndPuncttrue
\mciteSetBstMidEndSepPunct{\mcitedefaultmidpunct}
{\mcitedefaultendpunct}{\mcitedefaultseppunct}\relax
\EndOfBibitem
\bibitem{McGowan:2022nag}
J.~McGowan {\em et~al.\/},
  \href{http://doi.org/10.1140/epjc/s10052-023-11236-0}{Eur. Phys. J. C}
  \href{http://doi.org/10.1140/epjc/s10052-023-11236-0}{{\bf 83}, 3, 185}
  (2023), [Erratum: Eur.Phys.J.C 83, 302 (2023)],
  \href{https://arxiv.org/abs/2207.04739}{[arXiv:2207.04739]}\relax
\mciteBstWouldAddEndPuncttrue
\mciteSetBstMidEndSepPunct{\mcitedefaultmidpunct}
{\mcitedefaultendpunct}{\mcitedefaultseppunct}\relax
\EndOfBibitem
\bibitem{Ji:2013dva}
X.~Ji, \href{http://doi.org/10.1103/PhysRevLett.110.262002}{Phys. Rev. Lett.}
  \href{http://doi.org/10.1103/PhysRevLett.110.262002}{{\bf 110}, 262002}
  (2013), \href{https://arxiv.org/abs/1305.1539}{[arXiv:1305.1539]}\relax
\mciteBstWouldAddEndPuncttrue
\mciteSetBstMidEndSepPunct{\mcitedefaultmidpunct}
{\mcitedefaultendpunct}{\mcitedefaultseppunct}\relax
\EndOfBibitem
\bibitem{Rossi:2017muf}
G.~C. Rossi and M.~Testa,
  \href{http://doi.org/10.1103/PhysRevD.96.014507}{Phys. Rev.}
  \href{http://doi.org/10.1103/PhysRevD.96.014507}{{\bf D96}, 1, 014507}
  (2017), \href{https://arxiv.org/abs/1706.04428}{[arXiv:1706.04428]}\relax
\mciteBstWouldAddEndPuncttrue
\mciteSetBstMidEndSepPunct{\mcitedefaultmidpunct}
{\mcitedefaultendpunct}{\mcitedefaultseppunct}\relax
\EndOfBibitem
\bibitem{Rossi:2018zkn}
G.~Rossi and M.~Testa, \href{http://doi.org/10.1103/PhysRevD.98.054028}{Phys.
  Rev. D} \href{http://doi.org/10.1103/PhysRevD.98.054028}{{\bf 98}, 5, 054028}
  (2018), \href{https://arxiv.org/abs/1806.00808}{[arXiv:1806.00808]}\relax
\mciteBstWouldAddEndPuncttrue
\mciteSetBstMidEndSepPunct{\mcitedefaultmidpunct}
{\mcitedefaultendpunct}{\mcitedefaultseppunct}\relax
\EndOfBibitem
\bibitem{DelDebbio:2020cbz}
L.~Del~Debbio, T.~Giani and C.~J. Monahan,
  \href{http://doi.org/10.1007/JHEP09(2020)021}{JHEP}
  \href{http://doi.org/10.1007/JHEP09(2020)021}{{\bf 09}, 021} (2020),
  \href{https://arxiv.org/abs/2007.02131}{[arXiv:2007.02131]}\relax
\mciteBstWouldAddEndPuncttrue
\mciteSetBstMidEndSepPunct{\mcitedefaultmidpunct}
{\mcitedefaultendpunct}{\mcitedefaultseppunct}\relax
\EndOfBibitem
\bibitem{Lin:2014zya}
H.-W. Lin {\em et~al.\/},
  \href{http://doi.org/10.1103/PhysRevD.91.054510}{Phys. Rev.}
  \href{http://doi.org/10.1103/PhysRevD.91.054510}{{\bf D91}, 054510} (2015),
  \href{https://arxiv.org/abs/1402.1462}{[arXiv:1402.1462]}\relax
\mciteBstWouldAddEndPuncttrue
\mciteSetBstMidEndSepPunct{\mcitedefaultmidpunct}
{\mcitedefaultendpunct}{\mcitedefaultseppunct}\relax
\EndOfBibitem
\bibitem{Alexandrou:2015rja}
C.~Alexandrou {\em et~al.\/},
  \href{http://doi.org/10.1103/PhysRevD.92.014502}{Phys. Rev.}
  \href{http://doi.org/10.1103/PhysRevD.92.014502}{{\bf D92}, 014502} (2015),
  \href{https://arxiv.org/abs/1504.07455}{[arXiv:1504.07455]}\relax
\mciteBstWouldAddEndPuncttrue
\mciteSetBstMidEndSepPunct{\mcitedefaultmidpunct}
{\mcitedefaultendpunct}{\mcitedefaultseppunct}\relax
\EndOfBibitem
\bibitem{Lin:2017snn}
H.-W. Lin {\em et~al.\/},
  \href{http://doi.org/10.1016/j.ppnp.2018.01.007}{Prog. Part. Nucl. Phys.}
  \href{http://doi.org/10.1016/j.ppnp.2018.01.007}{{\bf 100}, 107} (2018),
  \href{https://arxiv.org/abs/1711.07916}{[arXiv:1711.07916]}\relax
\mciteBstWouldAddEndPuncttrue
\mciteSetBstMidEndSepPunct{\mcitedefaultmidpunct}
{\mcitedefaultendpunct}{\mcitedefaultseppunct}\relax
\EndOfBibitem
\bibitem{Alexandrou:2018pbm}
C.~Alexandrou {\em et~al.\/},
  \href{http://doi.org/10.1103/PhysRevLett.121.112001}{Phys. Rev. Lett.}
  \href{http://doi.org/10.1103/PhysRevLett.121.112001}{{\bf 121}, 11, 112001}
  (2018), \href{https://arxiv.org/abs/1803.02685}{[arXiv:1803.02685]}\relax
\mciteBstWouldAddEndPuncttrue
\mciteSetBstMidEndSepPunct{\mcitedefaultmidpunct}
{\mcitedefaultendpunct}{\mcitedefaultseppunct}\relax
\EndOfBibitem
\bibitem{Chen:2018xof}
J.-W. Chen {\em et~al.\/}  (2018),
  \href{https://arxiv.org/abs/1803.04393}{[arXiv:1803.04393]}\relax
\mciteBstWouldAddEndPuncttrue
\mciteSetBstMidEndSepPunct{\mcitedefaultmidpunct}
{\mcitedefaultendpunct}{\mcitedefaultseppunct}\relax
\EndOfBibitem
\bibitem{Cichy:2019ebf}
K.~Cichy, L.~Del~Debbio and T.~Giani,
  \href{http://doi.org/10.1007/JHEP10(2019)137}{JHEP}
  \href{http://doi.org/10.1007/JHEP10(2019)137}{{\bf 10}, 137} (2019),
  \href{https://arxiv.org/abs/1907.06037}{[arXiv:1907.06037]}\relax
\mciteBstWouldAddEndPuncttrue
\mciteSetBstMidEndSepPunct{\mcitedefaultmidpunct}
{\mcitedefaultendpunct}{\mcitedefaultseppunct}\relax
\EndOfBibitem
\bibitem{Cichy:2018mum}
K.~Cichy and M.~Constantinou, \href{http://doi.org/10.1155/2019/3036904}{Adv.
  High Energy Phys.} \href{http://doi.org/10.1155/2019/3036904}{{\bf 2019},
  3036904} (2019),
  \href{https://arxiv.org/abs/1811.07248}{[arXiv:1811.07248]}\relax
\mciteBstWouldAddEndPuncttrue
\mciteSetBstMidEndSepPunct{\mcitedefaultmidpunct}
{\mcitedefaultendpunct}{\mcitedefaultseppunct}\relax
\EndOfBibitem
\bibitem{Ebert:2019okf}
M.~A. Ebert, I.~W. Stewart and Y.~Zhao,
  \href{http://doi.org/10.1007/JHEP09(2019)037}{JHEP}
  \href{http://doi.org/10.1007/JHEP09(2019)037}{{\bf 09}, 037} (2019),
  \href{https://arxiv.org/abs/1901.03685}{[arXiv:1901.03685]}\relax
\mciteBstWouldAddEndPuncttrue
\mciteSetBstMidEndSepPunct{\mcitedefaultmidpunct}
{\mcitedefaultendpunct}{\mcitedefaultseppunct}\relax
\EndOfBibitem
\bibitem{Constantinou:2020pek}
M.~Constantinou, \href{http://doi.org/10.1140/epja/s10050-021-00353-7}{Eur.
  Phys. J. A} \href{http://doi.org/10.1140/epja/s10050-021-00353-7}{{\bf 57},
  2, 77} (2021),
  \href{https://arxiv.org/abs/2010.02445}{[arXiv:2010.02445]}\relax
\mciteBstWouldAddEndPuncttrue
\mciteSetBstMidEndSepPunct{\mcitedefaultmidpunct}
{\mcitedefaultendpunct}{\mcitedefaultseppunct}\relax
\EndOfBibitem
\bibitem{Constantinou:2020hdm}
M.~Constantinou {\em et~al.\/},
  \href{http://doi.org/10.1016/j.ppnp.2021.103908}{Prog. Part. Nucl. Phys.}
  \href{http://doi.org/10.1016/j.ppnp.2021.103908}{{\bf 121}, 103908} (2021),
  \href{https://arxiv.org/abs/2006.08636}{[arXiv:2006.08636]}\relax
\mciteBstWouldAddEndPuncttrue
\mciteSetBstMidEndSepPunct{\mcitedefaultmidpunct}
{\mcitedefaultendpunct}{\mcitedefaultseppunct}\relax
\EndOfBibitem
\bibitem{DelDebbio:2020rgv}
L.~Del~Debbio {\em et~al.\/},
  \href{http://doi.org/10.1007/JHEP02(2021)138}{JHEP}
  \href{http://doi.org/10.1007/JHEP02(2021)138}{{\bf 02}, 138} (2021),
  \href{https://arxiv.org/abs/2010.03996}{[arXiv:2010.03996]}\relax
\mciteBstWouldAddEndPuncttrue
\mciteSetBstMidEndSepPunct{\mcitedefaultmidpunct}
{\mcitedefaultendpunct}{\mcitedefaultseppunct}\relax
\EndOfBibitem
\bibitem{Fan:2020cpa}
Z.~Fan, R.~Zhang and H.-W. Lin,
  \href{http://doi.org/10.1142/S0217751X21500809}{Int. J. Mod. Phys. A}
  \href{http://doi.org/10.1142/S0217751X21500809}{{\bf 36}, 13, 2150080}
  (2021), \href{https://arxiv.org/abs/2007.16113}{[arXiv:2007.16113]}\relax
\mciteBstWouldAddEndPuncttrue
\mciteSetBstMidEndSepPunct{\mcitedefaultmidpunct}
{\mcitedefaultendpunct}{\mcitedefaultseppunct}\relax
\EndOfBibitem
\bibitem{HadStruc:2021wmh}
T.~Khan {\em et~al.\/} (HadStruc),
  \href{http://doi.org/10.1103/PhysRevD.104.094516}{Phys. Rev. D}
  \href{http://doi.org/10.1103/PhysRevD.104.094516}{{\bf 104}, 9, 094516}
  (2021), \href{https://arxiv.org/abs/2107.08960}{[arXiv:2107.08960]}\relax
\mciteBstWouldAddEndPuncttrue
\mciteSetBstMidEndSepPunct{\mcitedefaultmidpunct}
{\mcitedefaultendpunct}{\mcitedefaultseppunct}\relax
\EndOfBibitem
\bibitem{Fan:2022kcb}
Z.~Fan, W.~Good and H.-W. Lin  (2022),
  \href{https://arxiv.org/abs/2210.09985}{[arXiv:2210.09985]}\relax
\mciteBstWouldAddEndPuncttrue
\mciteSetBstMidEndSepPunct{\mcitedefaultmidpunct}
{\mcitedefaultendpunct}{\mcitedefaultseppunct}\relax
\EndOfBibitem
\bibitem{Gao:2017yyd}
J.~Gao, L.~Harland-Lang and J.~Rojo,
  \href{http://doi.org/10.1016/j.physrep.2018.03.002}{Phys. Rept.}
  \href{http://doi.org/10.1016/j.physrep.2018.03.002}{{\bf 742}, 1} (2018),
  \href{https://arxiv.org/abs/1709.04922}{[arXiv:1709.04922]}\relax
\mciteBstWouldAddEndPuncttrue
\mciteSetBstMidEndSepPunct{\mcitedefaultmidpunct}
{\mcitedefaultendpunct}{\mcitedefaultseppunct}\relax
\EndOfBibitem
\bibitem{Kovarik:2019xvh}
K.~Kova\v{r}\'\i{}k, P.~M. Nadolsky and D.~E. Soper,
  \href{http://doi.org/10.1103/RevModPhys.92.045003}{Rev. Mod. Phys.}
  \href{http://doi.org/10.1103/RevModPhys.92.045003}{{\bf 92}, 4, 045003}
  (2020), \href{https://arxiv.org/abs/1905.06957}{[arXiv:1905.06957]}\relax
\mciteBstWouldAddEndPuncttrue
\mciteSetBstMidEndSepPunct{\mcitedefaultmidpunct}
{\mcitedefaultendpunct}{\mcitedefaultseppunct}\relax
\EndOfBibitem
\bibitem{Hou:2022onq}
T.-J. Hou {\em et~al.\/},
  \href{http://doi.org/10.1103/PhysRevD.107.076018}{Phys. Rev. D}
  \href{http://doi.org/10.1103/PhysRevD.107.076018}{{\bf 107}, 7, 076018}
  (2023), \href{https://arxiv.org/abs/2211.11064}{[arXiv:2211.11064]}\relax
\mciteBstWouldAddEndPuncttrue
\mciteSetBstMidEndSepPunct{\mcitedefaultmidpunct}
{\mcitedefaultendpunct}{\mcitedefaultseppunct}\relax
\EndOfBibitem
\bibitem{Bjorken:1969ja}
J.~D. Bjorken and E.~A. Paschos,
  \href{http://doi.org/10.1103/PhysRev.185.1975}{Phys. Rev.}
  \href{http://doi.org/10.1103/PhysRev.185.1975}{{\bf 185}, 1975} (1969)\relax
\mciteBstWouldAddEndPuncttrue
\mciteSetBstMidEndSepPunct{\mcitedefaultmidpunct}
{\mcitedefaultendpunct}{\mcitedefaultseppunct}\relax
\EndOfBibitem
\bibitem{Vermaseren:2005qc}
J.~A.~M. Vermaseren, A.~Vogt and S.~Moch,
  \href{http://doi.org/10.1016/j.nuclphysb.2005.06.020}{Nucl. Phys.}
  \href{http://doi.org/10.1016/j.nuclphysb.2005.06.020}{{\bf B724}, 3} (2005),
  \href{https://arxiv.org/abs/hep-ph/0504242}{[hep-ph/0504242]}\relax
\mciteBstWouldAddEndPuncttrue
\mciteSetBstMidEndSepPunct{\mcitedefaultmidpunct}
{\mcitedefaultendpunct}{\mcitedefaultseppunct}\relax
\EndOfBibitem
\bibitem{Moch:2008fj}
S.~Moch, J.~A.~M. Vermaseren and A.~Vogt,
  \href{http://doi.org/10.1016/j.nuclphysb.2009.01.001}{Nucl. Phys.}
  \href{http://doi.org/10.1016/j.nuclphysb.2009.01.001}{{\bf B813}, 220}
  (2009), \href{https://arxiv.org/abs/0812.4168}{[arXiv:0812.4168]}\relax
\mciteBstWouldAddEndPuncttrue
\mciteSetBstMidEndSepPunct{\mcitedefaultmidpunct}
{\mcitedefaultendpunct}{\mcitedefaultseppunct}\relax
\EndOfBibitem
\bibitem{Davies:2016ruz}
J.~Davies {\em et~al.\/}, \href{http://doi.org/10.22323/1.265.0059}{PoS}
  \href{http://doi.org/10.22323/1.265.0059}{{\bf DIS2016}, 059} (2016),
  \href{https://arxiv.org/abs/1606.08907}{[arXiv:1606.08907]}\relax
\mciteBstWouldAddEndPuncttrue
\mciteSetBstMidEndSepPunct{\mcitedefaultmidpunct}
{\mcitedefaultendpunct}{\mcitedefaultseppunct}\relax
\EndOfBibitem
\bibitem{Laenen:1992zk}
E.~Laenen {\em et~al.\/},
  \href{http://doi.org/10.1016/0550-3213(93)90201-Y}{Nucl. Phys.}
  \href{http://doi.org/10.1016/0550-3213(93)90201-Y}{{\bf B392}, 162}
  (1993)\relax
\mciteBstWouldAddEndPuncttrue
\mciteSetBstMidEndSepPunct{\mcitedefaultmidpunct}
{\mcitedefaultendpunct}{\mcitedefaultseppunct}\relax
\EndOfBibitem
\bibitem{Riemersma:1994hv}
S.~Riemersma, J.~Smith and W.~L. van Neerven,
  \href{http://doi.org/10.1016/0370-2693(95)00036-K}{Phys. Lett.}
  \href{http://doi.org/10.1016/0370-2693(95)00036-K}{{\bf B347}, 143} (1995),
  \href{https://arxiv.org/abs/hep-ph/9411431}{[hep-ph/9411431]}\relax
\mciteBstWouldAddEndPuncttrue
\mciteSetBstMidEndSepPunct{\mcitedefaultmidpunct}
{\mcitedefaultendpunct}{\mcitedefaultseppunct}\relax
\EndOfBibitem
\bibitem{Blumlein:2019qze}
J.~Blümlein {\em et~al.\/}  (2019),
  \href{https://arxiv.org/abs/1903.06155}{[arXiv:1903.06155]}\relax
\mciteBstWouldAddEndPuncttrue
\mciteSetBstMidEndSepPunct{\mcitedefaultmidpunct}
{\mcitedefaultendpunct}{\mcitedefaultseppunct}\relax
\EndOfBibitem
\bibitem{Blumlein:2023aso}
J.~Bl\"umlein  (2023),
  \href{https://arxiv.org/abs/2306.01362}{[arXiv:2306.01362]}\relax
\mciteBstWouldAddEndPuncttrue
\mciteSetBstMidEndSepPunct{\mcitedefaultmidpunct}
{\mcitedefaultendpunct}{\mcitedefaultseppunct}\relax
\EndOfBibitem
\bibitem{Manohar:2016nzj}
A.~Manohar {\em et~al.\/},
  \href{http://doi.org/10.1103/PhysRevLett.117.242002}{Phys. Rev. Lett.}
  \href{http://doi.org/10.1103/PhysRevLett.117.242002}{{\bf 117}, 24, 242002}
  (2016), \href{https://arxiv.org/abs/1607.04266}{[arXiv:1607.04266]}\relax
\mciteBstWouldAddEndPuncttrue
\mciteSetBstMidEndSepPunct{\mcitedefaultmidpunct}
{\mcitedefaultendpunct}{\mcitedefaultseppunct}\relax
\EndOfBibitem
\bibitem{Manohar:2017eqh}
A.~V. Manohar {\em et~al.\/},
  \href{http://doi.org/10.1007/JHEP12(2017)046}{JHEP}
  \href{http://doi.org/10.1007/JHEP12(2017)046}{{\bf 12}, 046} (2017),
  \href{https://arxiv.org/abs/1708.01256}{[arXiv:1708.01256]}\relax
\mciteBstWouldAddEndPuncttrue
\mciteSetBstMidEndSepPunct{\mcitedefaultmidpunct}
{\mcitedefaultendpunct}{\mcitedefaultseppunct}\relax
\EndOfBibitem
\bibitem{Buonocore:2020nai}
L.~Buonocore {\em et~al.\/},
  \href{http://doi.org/10.1007/JHEP08(2020)019}{JHEP}
  \href{http://doi.org/10.1007/JHEP08(2020)019}{{\bf 08}, 08, 019} (2020),
  \href{https://arxiv.org/abs/2005.06477}{[arXiv:2005.06477]}\relax
\mciteBstWouldAddEndPuncttrue
\mciteSetBstMidEndSepPunct{\mcitedefaultmidpunct}
{\mcitedefaultendpunct}{\mcitedefaultseppunct}\relax
\EndOfBibitem
\bibitem{Collins:1989gx}
J.~C. Collins, D.~E. Soper and G.~F. Sterman,
  \href{http://doi.org/10.1142/9789814503266_0001}{Adv. Ser. Direct. High
  Energy Phys.} \href{http://doi.org/10.1142/9789814503266_0001}{{\bf 5}, 1}
  (1989), \href{https://arxiv.org/abs/hep-ph/0409313}{[hep-ph/0409313]}\relax
\mciteBstWouldAddEndPuncttrue
\mciteSetBstMidEndSepPunct{\mcitedefaultmidpunct}
{\mcitedefaultendpunct}{\mcitedefaultseppunct}\relax
\EndOfBibitem
\bibitem{Collins:2007nk}
J.~Collins and J.-W. Qiu,
  \href{http://doi.org/10.1103/PhysRevD.75.114014}{Phys. Rev. D}
  \href{http://doi.org/10.1103/PhysRevD.75.114014}{{\bf 75}, 114014} (2007),
  \href{https://arxiv.org/abs/0705.2141}{[arXiv:0705.2141]}\relax
\mciteBstWouldAddEndPuncttrue
\mciteSetBstMidEndSepPunct{\mcitedefaultmidpunct}
{\mcitedefaultendpunct}{\mcitedefaultseppunct}\relax
\EndOfBibitem
\bibitem{Sterman:2022gyf}
G.~Sterman, in \enquote{{Snowmass 2021},}  (2022),
  \href{https://arxiv.org/abs/2207.06507}{[arXiv:2207.06507]}\relax
\mciteBstWouldAddEndPuncttrue
\mciteSetBstMidEndSepPunct{\mcitedefaultmidpunct}
{\mcitedefaultendpunct}{\mcitedefaultseppunct}\relax
\EndOfBibitem
\bibitem{Boussarie:2023izj}
R.~Boussarie {\em et~al.\/}  (2023),
  \href{https://arxiv.org/abs/2304.03302}{[arXiv:2304.03302]}\relax
\mciteBstWouldAddEndPuncttrue
\mciteSetBstMidEndSepPunct{\mcitedefaultmidpunct}
{\mcitedefaultendpunct}{\mcitedefaultseppunct}\relax
\EndOfBibitem
\bibitem{Gribov:1972ri}
V.~N. Gribov and L.~N. Lipatov, Sov. J. Nucl. Phys. {\bf 15}, 438 (1972), [Yad.
  Fiz.15,781(1972)]\relax
\mciteBstWouldAddEndPuncttrue
\mciteSetBstMidEndSepPunct{\mcitedefaultmidpunct}
{\mcitedefaultendpunct}{\mcitedefaultseppunct}\relax
\EndOfBibitem
\bibitem{Lipatov:1974qm}
L.~N. Lipatov, Sov. J. Nucl. Phys. {\bf 20}, 94 (1975), [Yad.
  Fiz.20,181(1974)]\relax
\mciteBstWouldAddEndPuncttrue
\mciteSetBstMidEndSepPunct{\mcitedefaultmidpunct}
{\mcitedefaultendpunct}{\mcitedefaultseppunct}\relax
\EndOfBibitem
\bibitem{Altarelli:1977zs}
G.~Altarelli and G.~Parisi,
  \href{http://doi.org/10.1016/0550-3213(77)90384-4}{Nucl. Phys.}
  \href{http://doi.org/10.1016/0550-3213(77)90384-4}{{\bf B126}, 298}
  (1977)\relax
\mciteBstWouldAddEndPuncttrue
\mciteSetBstMidEndSepPunct{\mcitedefaultmidpunct}
{\mcitedefaultendpunct}{\mcitedefaultseppunct}\relax
\EndOfBibitem
\bibitem{Dokshitzer:1977sg}
Y.~L. Dokshitzer, Sov. Phys. JETP {\bf 46}, 641 (1977), [Zh. Eksp. Teor.
  Fiz.73,1216(1977)]\relax
\mciteBstWouldAddEndPuncttrue
\mciteSetBstMidEndSepPunct{\mcitedefaultmidpunct}
{\mcitedefaultendpunct}{\mcitedefaultseppunct}\relax
\EndOfBibitem
\bibitem{Curci:1980uw}
G.~Curci, W.~Furmanski and R.~Petronzio,
  \href{http://doi.org/10.1016/0550-3213(80)90003-6}{Nucl. Phys.}
  \href{http://doi.org/10.1016/0550-3213(80)90003-6}{{\bf B175}, 27}
  (1980)\relax
\mciteBstWouldAddEndPuncttrue
\mciteSetBstMidEndSepPunct{\mcitedefaultmidpunct}
{\mcitedefaultendpunct}{\mcitedefaultseppunct}\relax
\EndOfBibitem
\bibitem{Furmanski:1980cm}
W.~Furmanski and R.~Petronzio,
  \href{http://doi.org/10.1016/0370-2693(80)90636-X}{Phys. Lett.}
  \href{http://doi.org/10.1016/0370-2693(80)90636-X}{{\bf 97B}, 437}
  (1980)\relax
\mciteBstWouldAddEndPuncttrue
\mciteSetBstMidEndSepPunct{\mcitedefaultmidpunct}
{\mcitedefaultendpunct}{\mcitedefaultseppunct}\relax
\EndOfBibitem
\bibitem{Vogt:2004mw}
A.~Vogt, S.~Moch and J.~A.~M. Vermaseren,
  \href{http://doi.org/10.1016/j.nuclphysb.2004.04.024}{Nucl. Phys.}
  \href{http://doi.org/10.1016/j.nuclphysb.2004.04.024}{{\bf B691}, 129}
  (2004), \href{https://arxiv.org/abs/hep-ph/0404111}{[hep-ph/0404111]}\relax
\mciteBstWouldAddEndPuncttrue
\mciteSetBstMidEndSepPunct{\mcitedefaultmidpunct}
{\mcitedefaultendpunct}{\mcitedefaultseppunct}\relax
\EndOfBibitem
\bibitem{Moch:2004pa}
S.~Moch, J.~A.~M. Vermaseren and A.~Vogt,
  \href{http://doi.org/10.1016/j.nuclphysb.2004.03.030}{Nucl. Phys.}
  \href{http://doi.org/10.1016/j.nuclphysb.2004.03.030}{{\bf B688}, 101}
  (2004), \href{https://arxiv.org/abs/hep-ph/0403192}{[hep-ph/0403192]}\relax
\mciteBstWouldAddEndPuncttrue
\mciteSetBstMidEndSepPunct{\mcitedefaultmidpunct}
{\mcitedefaultendpunct}{\mcitedefaultseppunct}\relax
\EndOfBibitem
\bibitem{Davies:2016jie}
J.~Davies {\em et~al.\/},
  \href{http://doi.org/10.1016/j.nuclphysb.2016.12.012}{Nucl. Phys. B}
  \href{http://doi.org/10.1016/j.nuclphysb.2016.12.012}{{\bf 915}, 335} (2017),
  \href{https://arxiv.org/abs/1610.07477}{[arXiv:1610.07477]}\relax
\mciteBstWouldAddEndPuncttrue
\mciteSetBstMidEndSepPunct{\mcitedefaultmidpunct}
{\mcitedefaultendpunct}{\mcitedefaultseppunct}\relax
\EndOfBibitem
\bibitem{Moch:2017uml}
S.~Moch {\em et~al.\/}, \href{http://doi.org/10.1007/JHEP10(2017)041}{JHEP}
  \href{http://doi.org/10.1007/JHEP10(2017)041}{{\bf 10}, 041} (2017),
  \href{https://arxiv.org/abs/1707.08315}{[arXiv:1707.08315]}\relax
\mciteBstWouldAddEndPuncttrue
\mciteSetBstMidEndSepPunct{\mcitedefaultmidpunct}
{\mcitedefaultendpunct}{\mcitedefaultseppunct}\relax
\EndOfBibitem
\bibitem{Vogt:2018ytw}
A.~Vogt {\em et~al.\/}, \href{http://doi.org/10.22323/1.290.0046}{PoS}
  \href{http://doi.org/10.22323/1.290.0046}{{\bf RADCOR2017}, 046} (2018),
  \href{https://arxiv.org/abs/1801.06085}{[arXiv:1801.06085]}\relax
\mciteBstWouldAddEndPuncttrue
\mciteSetBstMidEndSepPunct{\mcitedefaultmidpunct}
{\mcitedefaultendpunct}{\mcitedefaultseppunct}\relax
\EndOfBibitem
\bibitem{Vogt:2018miu}
A.~Vogt {\em et~al.\/}, \href{http://doi.org/10.22323/1.303.0050}{PoS}
  \href{http://doi.org/10.22323/1.303.0050}{{\bf LL2018}, 050} (2018),
  \href{https://arxiv.org/abs/1808.08981}{[arXiv:1808.08981]}\relax
\mciteBstWouldAddEndPuncttrue
\mciteSetBstMidEndSepPunct{\mcitedefaultmidpunct}
{\mcitedefaultendpunct}{\mcitedefaultseppunct}\relax
\EndOfBibitem
\bibitem{Falcioni:2023luc}
G.~Falcioni {\em et~al.\/},
  \href{http://doi.org/10.1016/j.physletb.2023.137944}{Phys. Lett. B}
  \href{http://doi.org/10.1016/j.physletb.2023.137944}{{\bf 842}, 137944}
  (2023), \href{https://arxiv.org/abs/2302.07593}{[arXiv:2302.07593]}\relax
\mciteBstWouldAddEndPuncttrue
\mciteSetBstMidEndSepPunct{\mcitedefaultmidpunct}
{\mcitedefaultendpunct}{\mcitedefaultseppunct}\relax
\EndOfBibitem
\bibitem{Moch:2014sna}
S.~Moch, J.~A.~M. Vermaseren and A.~Vogt,
  \href{http://doi.org/10.1016/j.nuclphysb.2014.10.016}{Nucl. Phys.}
  \href{http://doi.org/10.1016/j.nuclphysb.2014.10.016}{{\bf B889}, 351}
  (2014), \href{https://arxiv.org/abs/1409.5131}{[arXiv:1409.5131]}\relax
\mciteBstWouldAddEndPuncttrue
\mciteSetBstMidEndSepPunct{\mcitedefaultmidpunct}
{\mcitedefaultendpunct}{\mcitedefaultseppunct}\relax
\EndOfBibitem
\bibitem{deFlorian:2015ujt}
D.~de~Florian, G.~F.~R. Sborlini and G.~Rodrigo,
  \href{http://doi.org/10.1140/epjc/s10052-016-4131-8}{Eur. Phys. J.}
  \href{http://doi.org/10.1140/epjc/s10052-016-4131-8}{{\bf C76}, 5, 282}
  (2016), \href{https://arxiv.org/abs/1512.00612}{[arXiv:1512.00612]}\relax
\mciteBstWouldAddEndPuncttrue
\mciteSetBstMidEndSepPunct{\mcitedefaultmidpunct}
{\mcitedefaultendpunct}{\mcitedefaultseppunct}\relax
\EndOfBibitem
\bibitem{deFlorian:2016gvk}
D.~de~Florian, G.~F.~R. Sborlini and G.~Rodrigo,
  \href{http://doi.org/10.1007/JHEP10(2016)056}{JHEP}
  \href{http://doi.org/10.1007/JHEP10(2016)056}{{\bf 10}, 056} (2016),
  \href{https://arxiv.org/abs/1606.02887}{[arXiv:1606.02887]}\relax
\mciteBstWouldAddEndPuncttrue
\mciteSetBstMidEndSepPunct{\mcitedefaultmidpunct}
{\mcitedefaultendpunct}{\mcitedefaultseppunct}\relax
\EndOfBibitem
\bibitem{deFlorian:2023zkc}
D.~de~Florian and L.~P. Conte  (2023),
  \href{https://arxiv.org/abs/2305.14144}{[arXiv:2305.14144]}\relax
\mciteBstWouldAddEndPuncttrue
\mciteSetBstMidEndSepPunct{\mcitedefaultmidpunct}
{\mcitedefaultendpunct}{\mcitedefaultseppunct}\relax
\EndOfBibitem
\bibitem{Thorne:2006qt}
R.~S. Thorne, \href{http://doi.org/10.1103/PhysRevD.73.054019}{Phys. Rev.}
  \href{http://doi.org/10.1103/PhysRevD.73.054019}{{\bf D73}, 054019} (2006),
  \href{https://arxiv.org/abs/hep-ph/0601245}{[hep-ph/0601245]}\relax
\mciteBstWouldAddEndPuncttrue
\mciteSetBstMidEndSepPunct{\mcitedefaultmidpunct}
{\mcitedefaultendpunct}{\mcitedefaultseppunct}\relax
\EndOfBibitem
\bibitem{Forte:2010ta}
S.~Forte {\em et~al.\/},
  \href{http://doi.org/10.1016/j.nuclphysb.2010.03.014}{Nucl. Phys.}
  \href{http://doi.org/10.1016/j.nuclphysb.2010.03.014}{{\bf B834}, 116}
  (2010), \href{https://arxiv.org/abs/1001.2312}{[arXiv:1001.2312]}\relax
\mciteBstWouldAddEndPuncttrue
\mciteSetBstMidEndSepPunct{\mcitedefaultmidpunct}
{\mcitedefaultendpunct}{\mcitedefaultseppunct}\relax
\EndOfBibitem
\bibitem{Guzzi:2011ew}
M.~Guzzi {\em et~al.\/}, \href{http://doi.org/10.1103/PhysRevD.86.053005}{Phys.
  Rev.} \href{http://doi.org/10.1103/PhysRevD.86.053005}{{\bf D86}, 053005}
  (2012), \href{https://arxiv.org/abs/1108.5112}{[arXiv:1108.5112]}\relax
\mciteBstWouldAddEndPuncttrue
\mciteSetBstMidEndSepPunct{\mcitedefaultmidpunct}
{\mcitedefaultendpunct}{\mcitedefaultseppunct}\relax
\EndOfBibitem
\bibitem{Fadin:1975cb}
V.~S. Fadin, E.~A. Kuraev and L.~N. Lipatov,
  \href{http://doi.org/10.1016/0370-2693(75)90524-9}{Phys. Lett.}
  \href{http://doi.org/10.1016/0370-2693(75)90524-9}{{\bf 60B}, 50}
  (1975)\relax
\mciteBstWouldAddEndPuncttrue
\mciteSetBstMidEndSepPunct{\mcitedefaultmidpunct}
{\mcitedefaultendpunct}{\mcitedefaultseppunct}\relax
\EndOfBibitem
\bibitem{Balitsky:1978ic}
I.~I. Balitsky and L.~N. Lipatov, Sov. J. Nucl. Phys. {\bf 28}, 822 (1978),
  [Yad. Fiz.28,1597(1978)]\relax
\mciteBstWouldAddEndPuncttrue
\mciteSetBstMidEndSepPunct{\mcitedefaultmidpunct}
{\mcitedefaultendpunct}{\mcitedefaultseppunct}\relax
\EndOfBibitem
\bibitem{Ball:2017otu}
R.~D. Ball {\em et~al.\/},
  \href{http://doi.org/10.1140/epjc/s10052-018-5774-4}{Eur. Phys. J.}
  \href{http://doi.org/10.1140/epjc/s10052-018-5774-4}{{\bf C78}, 4, 321}
  (2018), \href{https://arxiv.org/abs/1710.05935}{[arXiv:1710.05935]}\relax
\mciteBstWouldAddEndPuncttrue
\mciteSetBstMidEndSepPunct{\mcitedefaultmidpunct}
{\mcitedefaultendpunct}{\mcitedefaultseppunct}\relax
\EndOfBibitem
\bibitem{Abramowicz:2015mha}
H.~Abramowicz {\em et~al.\/} (H1, ZEUS),
  \href{http://doi.org/10.1140/epjc/s10052-015-3710-4}{Eur. Phys. J.}
  \href{http://doi.org/10.1140/epjc/s10052-015-3710-4}{{\bf C75}, 12, 580}
  (2015), \href{https://arxiv.org/abs/1506.06042}{[arXiv:1506.06042]}\relax
\mciteBstWouldAddEndPuncttrue
\mciteSetBstMidEndSepPunct{\mcitedefaultmidpunct}
{\mcitedefaultendpunct}{\mcitedefaultseppunct}\relax
\EndOfBibitem
\bibitem{Lipatov:1976zz}
L.~N. Lipatov, Sov. J. Nucl. Phys. {\bf 23}, 338 (1976), [Yad.
  Fiz.23,642(1976)]\relax
\mciteBstWouldAddEndPuncttrue
\mciteSetBstMidEndSepPunct{\mcitedefaultmidpunct}
{\mcitedefaultendpunct}{\mcitedefaultseppunct}\relax
\EndOfBibitem
\bibitem{Kuraev:1977fs}
E.~A. Kuraev, L.~N. Lipatov and V.~S. Fadin, Sov. Phys. JETP {\bf 45}, 199
  (1977), [Zh. Eksp. Teor. Fiz.72,377(1977)]\relax
\mciteBstWouldAddEndPuncttrue
\mciteSetBstMidEndSepPunct{\mcitedefaultmidpunct}
{\mcitedefaultendpunct}{\mcitedefaultseppunct}\relax
\EndOfBibitem
\bibitem{Fadin:1998py}
V.~S. Fadin and L.~N. Lipatov,
  \href{http://doi.org/10.1016/S0370-2693(98)00473-0}{Phys. Lett.}
  \href{http://doi.org/10.1016/S0370-2693(98)00473-0}{{\bf B429}, 127} (1998),
  \href{https://arxiv.org/abs/hep-ph/9802290}{[hep-ph/9802290]}\relax
\mciteBstWouldAddEndPuncttrue
\mciteSetBstMidEndSepPunct{\mcitedefaultmidpunct}
{\mcitedefaultendpunct}{\mcitedefaultseppunct}\relax
\EndOfBibitem
\bibitem{Ciafaloni:1998gs}
M.~Ciafaloni and G.~Camici,
  \href{http://doi.org/10.1016/S0370-2693(98)00551-6}{Phys. Lett.}
  \href{http://doi.org/10.1016/S0370-2693(98)00551-6}{{\bf B430}, 349} (1998),
  \href{https://arxiv.org/abs/hep-ph/9803389}{[hep-ph/9803389]}\relax
\mciteBstWouldAddEndPuncttrue
\mciteSetBstMidEndSepPunct{\mcitedefaultmidpunct}
{\mcitedefaultendpunct}{\mcitedefaultseppunct}\relax
\EndOfBibitem
\bibitem{Altarelli:2008aj}
G.~Altarelli, R.~D. Ball and S.~Forte,
  \href{http://doi.org/10.1016/j.nuclphysb.2008.03.003}{Nucl. Phys.}
  \href{http://doi.org/10.1016/j.nuclphysb.2008.03.003}{{\bf B799}, 199}
  (2008), \href{https://arxiv.org/abs/0802.0032}{[arXiv:0802.0032]}\relax
\mciteBstWouldAddEndPuncttrue
\mciteSetBstMidEndSepPunct{\mcitedefaultmidpunct}
{\mcitedefaultendpunct}{\mcitedefaultseppunct}\relax
\EndOfBibitem
\bibitem{Ciafaloni:2007gf}
M.~Ciafaloni {\em et~al.\/},
  \href{http://doi.org/10.1088/1126-6708/2007/08/046}{JHEP}
  \href{http://doi.org/10.1088/1126-6708/2007/08/046}{{\bf 08}, 046} (2007),
  \href{https://arxiv.org/abs/0707.1453}{[arXiv:0707.1453]}\relax
\mciteBstWouldAddEndPuncttrue
\mciteSetBstMidEndSepPunct{\mcitedefaultmidpunct}
{\mcitedefaultendpunct}{\mcitedefaultseppunct}\relax
\EndOfBibitem
\bibitem{White:2006yh}
C.~D. White and R.~S. Thorne,
  \href{http://doi.org/10.1103/PhysRevD.75.034005}{Phys. Rev.}
  \href{http://doi.org/10.1103/PhysRevD.75.034005}{{\bf D75}, 034005} (2007),
  \href{https://arxiv.org/abs/hep-ph/0611204}{[hep-ph/0611204]}\relax
\mciteBstWouldAddEndPuncttrue
\mciteSetBstMidEndSepPunct{\mcitedefaultmidpunct}
{\mcitedefaultendpunct}{\mcitedefaultseppunct}\relax
\EndOfBibitem
\bibitem{Iancu:2015vea}
E.~Iancu {\em et~al.\/},
  \href{http://doi.org/10.1016/j.physletb.2015.03.068}{Phys. Lett.}
  \href{http://doi.org/10.1016/j.physletb.2015.03.068}{{\bf B744}, 293} (2015),
  \href{https://arxiv.org/abs/1502.05642}{[arXiv:1502.05642]}\relax
\mciteBstWouldAddEndPuncttrue
\mciteSetBstMidEndSepPunct{\mcitedefaultmidpunct}
{\mcitedefaultendpunct}{\mcitedefaultseppunct}\relax
\EndOfBibitem
\bibitem{Gromov:2015vua}
N.~Gromov, F.~Levkovich-Maslyuk and G.~Sizov,
  \href{http://doi.org/10.1103/PhysRevLett.115.251601}{Phys. Rev. Lett.}
  \href{http://doi.org/10.1103/PhysRevLett.115.251601}{{\bf 115}, 25, 251601}
  (2015), \href{https://arxiv.org/abs/1507.04010}{[arXiv:1507.04010]}\relax
\mciteBstWouldAddEndPuncttrue
\mciteSetBstMidEndSepPunct{\mcitedefaultmidpunct}
{\mcitedefaultendpunct}{\mcitedefaultseppunct}\relax
\EndOfBibitem
\bibitem{Velizhanin:2015xsa}
V.~N. Velizhanin  (2015),
  \href{https://arxiv.org/abs/1508.02857}{[arXiv:1508.02857]}\relax
\mciteBstWouldAddEndPuncttrue
\mciteSetBstMidEndSepPunct{\mcitedefaultmidpunct}
{\mcitedefaultendpunct}{\mcitedefaultseppunct}\relax
\EndOfBibitem
\bibitem{Caron-Huot:2016tzz}
S.~Caron-Huot and M.~Herranen,
  \href{http://doi.org/10.1007/JHEP02(2018)058}{JHEP}
  \href{http://doi.org/10.1007/JHEP02(2018)058}{{\bf 02}, 058} (2018),
  \href{https://arxiv.org/abs/1604.07417}{[arXiv:1604.07417]}\relax
\mciteBstWouldAddEndPuncttrue
\mciteSetBstMidEndSepPunct{\mcitedefaultmidpunct}
{\mcitedefaultendpunct}{\mcitedefaultseppunct}\relax
\EndOfBibitem
\bibitem{Fadin:2020lam}
V.~Fadin, {\em {Chapter 4: BFKL \textemdash{} Past and Future}\/}, 63--90
  (2021), \href{https://arxiv.org/abs/2012.11931}{[arXiv:2012.11931]}\relax
\mciteBstWouldAddEndPuncttrue
\mciteSetBstMidEndSepPunct{\mcitedefaultmidpunct}
{\mcitedefaultendpunct}{\mcitedefaultseppunct}\relax
\EndOfBibitem
\bibitem{Balitsky:1995ub}
I.~Balitsky, \href{http://doi.org/10.1016/0550-3213(95)00638-9}{Nucl. Phys.}
  \href{http://doi.org/10.1016/0550-3213(95)00638-9}{{\bf B463}, 99} (1996),
  \href{https://arxiv.org/abs/hep-ph/9509348}{[hep-ph/9509348]}\relax
\mciteBstWouldAddEndPuncttrue
\mciteSetBstMidEndSepPunct{\mcitedefaultmidpunct}
{\mcitedefaultendpunct}{\mcitedefaultseppunct}\relax
\EndOfBibitem
\bibitem{Kovchegov:1999yj}
Y.~V. Kovchegov, \href{http://doi.org/10.1103/PhysRevD.60.034008}{Phys. Rev.}
  \href{http://doi.org/10.1103/PhysRevD.60.034008}{{\bf D60}, 034008} (1999),
  \href{https://arxiv.org/abs/hep-ph/9901281}{[hep-ph/9901281]}\relax
\mciteBstWouldAddEndPuncttrue
\mciteSetBstMidEndSepPunct{\mcitedefaultmidpunct}
{\mcitedefaultendpunct}{\mcitedefaultseppunct}\relax
\EndOfBibitem
\bibitem{Davier:2008sk}
M.~Davier {\em et~al.\/},
  \href{http://doi.org/10.1140/epjc/s10052-008-0666-7}{Eur. Phys. J. C}
  \href{http://doi.org/10.1140/epjc/s10052-008-0666-7}{{\bf 56}, 305} (2008),
  \href{https://arxiv.org/abs/0803.0979}{[arXiv:0803.0979]}\relax
\mciteBstWouldAddEndPuncttrue
\mciteSetBstMidEndSepPunct{\mcitedefaultmidpunct}
{\mcitedefaultendpunct}{\mcitedefaultseppunct}\relax
\EndOfBibitem
\bibitem{Collins:1985ue}
J.~C. Collins, D.~E. Soper and G.~F. Sterman,
  \href{http://doi.org/10.1016/0550-3213(85)90565-6}{Nucl. Phys.}
  \href{http://doi.org/10.1016/0550-3213(85)90565-6}{{\bf B261}, 104}
  (1985)\relax
\mciteBstWouldAddEndPuncttrue
\mciteSetBstMidEndSepPunct{\mcitedefaultmidpunct}
{\mcitedefaultendpunct}{\mcitedefaultseppunct}\relax
\EndOfBibitem
\bibitem{Anastasiou:2015ema}
C.~Anastasiou {\em et~al.\/},
  \href{http://doi.org/10.1103/PhysRevLett.114.212001}{Phys. Rev. Lett.}
  \href{http://doi.org/10.1103/PhysRevLett.114.212001}{{\bf 114}, 212001}
  (2015), \href{https://arxiv.org/abs/1503.06056}{[arXiv:1503.06056]}\relax
\mciteBstWouldAddEndPuncttrue
\mciteSetBstMidEndSepPunct{\mcitedefaultmidpunct}
{\mcitedefaultendpunct}{\mcitedefaultseppunct}\relax
\EndOfBibitem
\bibitem{Anastasiou:2016cez}
C.~Anastasiou {\em et~al.\/},
  \href{http://doi.org/10.1007/JHEP05(2016)058}{JHEP}
  \href{http://doi.org/10.1007/JHEP05(2016)058}{{\bf 05}, 058} (2016),
  \href{https://arxiv.org/abs/1602.00695}{[arXiv:1602.00695]}\relax
\mciteBstWouldAddEndPuncttrue
\mciteSetBstMidEndSepPunct{\mcitedefaultmidpunct}
{\mcitedefaultendpunct}{\mcitedefaultseppunct}\relax
\EndOfBibitem
\bibitem{Mistlberger:2018etf}
B.~Mistlberger, \href{http://doi.org/10.1007/JHEP05(2018)028}{JHEP}
  \href{http://doi.org/10.1007/JHEP05(2018)028}{{\bf 05}, 028} (2018),
  \href{https://arxiv.org/abs/1802.00833}{[arXiv:1802.00833]}\relax
\mciteBstWouldAddEndPuncttrue
\mciteSetBstMidEndSepPunct{\mcitedefaultmidpunct}
{\mcitedefaultendpunct}{\mcitedefaultseppunct}\relax
\EndOfBibitem
\bibitem{Chen:2019lzz}
L.-B. Chen {\em et~al.\/},
  \href{http://doi.org/10.1016/j.physletb.2020.135292}{Phys. Lett. B}
  \href{http://doi.org/10.1016/j.physletb.2020.135292}{{\bf 803}, 135292}
  (2020), \href{https://arxiv.org/abs/1909.06808}{[arXiv:1909.06808]}\relax
\mciteBstWouldAddEndPuncttrue
\mciteSetBstMidEndSepPunct{\mcitedefaultmidpunct}
{\mcitedefaultendpunct}{\mcitedefaultseppunct}\relax
\EndOfBibitem
\bibitem{Dulat:2018bfe}
F.~Dulat, B.~Mistlberger and A.~Pelloni,
  \href{http://doi.org/10.1103/PhysRevD.99.034004}{Phys. Rev.}
  \href{http://doi.org/10.1103/PhysRevD.99.034004}{{\bf D99}, 3, 034004}
  (2019), \href{https://arxiv.org/abs/1810.09462}{[arXiv:1810.09462]}\relax
\mciteBstWouldAddEndPuncttrue
\mciteSetBstMidEndSepPunct{\mcitedefaultmidpunct}
{\mcitedefaultendpunct}{\mcitedefaultseppunct}\relax
\EndOfBibitem
\bibitem{Cieri:2018oms}
L.~Cieri {\em et~al.\/}, \href{http://doi.org/10.1007/JHEP02(2019)096}{JHEP}
  \href{http://doi.org/10.1007/JHEP02(2019)096}{{\bf 02}, 096} (2019),
  \href{https://arxiv.org/abs/1807.11501}{[arXiv:1807.11501]}\relax
\mciteBstWouldAddEndPuncttrue
\mciteSetBstMidEndSepPunct{\mcitedefaultmidpunct}
{\mcitedefaultendpunct}{\mcitedefaultseppunct}\relax
\EndOfBibitem
\bibitem{Chen:2021isd}
X.~Chen {\em et~al.\/}  (2021),
  \href{https://arxiv.org/abs/2102.07607}{[arXiv:2102.07607]}\relax
\mciteBstWouldAddEndPuncttrue
\mciteSetBstMidEndSepPunct{\mcitedefaultmidpunct}
{\mcitedefaultendpunct}{\mcitedefaultseppunct}\relax
\EndOfBibitem
\bibitem{Baglio:2022wzu}
J.~Baglio {\em et~al.\/}, \href{http://doi.org/10.1007/JHEP12(2022)066}{JHEP}
  \href{http://doi.org/10.1007/JHEP12(2022)066}{{\bf 12}, 066} (2022),
  \href{https://arxiv.org/abs/2209.06138}{[arXiv:2209.06138]}\relax
\mciteBstWouldAddEndPuncttrue
\mciteSetBstMidEndSepPunct{\mcitedefaultmidpunct}
{\mcitedefaultendpunct}{\mcitedefaultseppunct}\relax
\EndOfBibitem
\bibitem{Dreyer:2016oyx}
F.~A. Dreyer and A.~Karlberg,
  \href{http://doi.org/10.1103/PhysRevLett.117.072001}{Phys. Rev. Lett.}
  \href{http://doi.org/10.1103/PhysRevLett.117.072001}{{\bf 117}, 7, 072001}
  (2016), \href{https://arxiv.org/abs/1606.00840}{[arXiv:1606.00840]}\relax
\mciteBstWouldAddEndPuncttrue
\mciteSetBstMidEndSepPunct{\mcitedefaultmidpunct}
{\mcitedefaultendpunct}{\mcitedefaultseppunct}\relax
\EndOfBibitem
\bibitem{Dreyer:2018qbw}
F.~A. Dreyer and A.~Karlberg,
  \href{http://doi.org/10.1103/PhysRevD.98.114016}{Phys. Rev. D}
  \href{http://doi.org/10.1103/PhysRevD.98.114016}{{\bf 98}, 11, 114016}
  (2018), \href{https://arxiv.org/abs/1811.07906}{[arXiv:1811.07906]}\relax
\mciteBstWouldAddEndPuncttrue
\mciteSetBstMidEndSepPunct{\mcitedefaultmidpunct}
{\mcitedefaultendpunct}{\mcitedefaultseppunct}\relax
\EndOfBibitem
\bibitem{Duhr:2020seh}
C.~Duhr, F.~Dulat and B.~Mistlberger,
  \href{http://doi.org/10.1103/PhysRevLett.125.172001}{Phys. Rev. Lett.}
  \href{http://doi.org/10.1103/PhysRevLett.125.172001}{{\bf 125}, 17, 172001}
  (2020), \href{https://arxiv.org/abs/2001.07717}{[arXiv:2001.07717]}\relax
\mciteBstWouldAddEndPuncttrue
\mciteSetBstMidEndSepPunct{\mcitedefaultmidpunct}
{\mcitedefaultendpunct}{\mcitedefaultseppunct}\relax
\EndOfBibitem
\bibitem{Chen:2021vtu}
X.~Chen {\em et~al.\/}  (2021),
  \href{https://arxiv.org/abs/2107.09085}{[arXiv:2107.09085]}\relax
\mciteBstWouldAddEndPuncttrue
\mciteSetBstMidEndSepPunct{\mcitedefaultmidpunct}
{\mcitedefaultendpunct}{\mcitedefaultseppunct}\relax
\EndOfBibitem
\bibitem{Duhr:2020sdp}
C.~Duhr, F.~Dulat and B.~Mistlberger,
  \href{http://doi.org/10.1007/JHEP11(2020)143}{JHEP}
  \href{http://doi.org/10.1007/JHEP11(2020)143}{{\bf 11}, 143} (2020),
  \href{https://arxiv.org/abs/2007.13313}{[arXiv:2007.13313]}\relax
\mciteBstWouldAddEndPuncttrue
\mciteSetBstMidEndSepPunct{\mcitedefaultmidpunct}
{\mcitedefaultendpunct}{\mcitedefaultseppunct}\relax
\EndOfBibitem
\bibitem{Bonvini:2016frm}
M.~Bonvini {\em et~al.\/}, \href{http://doi.org/10.1007/JHEP08(2016)105}{JHEP}
  \href{http://doi.org/10.1007/JHEP08(2016)105}{{\bf 08}, 105} (2016),
  \href{https://arxiv.org/abs/1603.08000}{[arXiv:1603.08000]}\relax
\mciteBstWouldAddEndPuncttrue
\mciteSetBstMidEndSepPunct{\mcitedefaultmidpunct}
{\mcitedefaultendpunct}{\mcitedefaultseppunct}\relax
\EndOfBibitem
\bibitem{Dulat:2018rbf}
F.~Dulat, A.~Lazopoulos and B.~Mistlberger,
  \href{http://doi.org/10.1016/j.cpc.2018.06.025}{Comput. Phys. Commun.}
  \href{http://doi.org/10.1016/j.cpc.2018.06.025}{{\bf 233}, 243} (2018),
  \href{https://arxiv.org/abs/1802.00827}{[arXiv:1802.00827]}\relax
\mciteBstWouldAddEndPuncttrue
\mciteSetBstMidEndSepPunct{\mcitedefaultmidpunct}
{\mcitedefaultendpunct}{\mcitedefaultseppunct}\relax
\EndOfBibitem
\bibitem{Harlander:2016hcx}
R.~V. Harlander, S.~Liebler and H.~Mantler,
  \href{http://doi.org/10.1016/j.cpc.2016.10.015}{Comput. Phys. Commun.}
  \href{http://doi.org/10.1016/j.cpc.2016.10.015}{{\bf 212}, 239} (2017),
  \href{https://arxiv.org/abs/1605.03190}{[arXiv:1605.03190]}\relax
\mciteBstWouldAddEndPuncttrue
\mciteSetBstMidEndSepPunct{\mcitedefaultmidpunct}
{\mcitedefaultendpunct}{\mcitedefaultseppunct}\relax
\EndOfBibitem
\bibitem{Brein:2003wg}
O.~Brein, A.~Djouadi and R.~Harlander,
  \href{http://doi.org/10.1016/j.physletb.2003.10.112}{Phys. Lett.}
  \href{http://doi.org/10.1016/j.physletb.2003.10.112}{{\bf B579}, 149} (2004),
  \href{https://arxiv.org/abs/hep-ph/0307206}{[hep-ph/0307206]}\relax
\mciteBstWouldAddEndPuncttrue
\mciteSetBstMidEndSepPunct{\mcitedefaultmidpunct}
{\mcitedefaultendpunct}{\mcitedefaultseppunct}\relax
\EndOfBibitem
\bibitem{deFlorian:2013jea}
D.~de~Florian and J.~Mazzitelli,
  \href{http://doi.org/10.1103/PhysRevLett.111.201801}{Phys. Rev. Lett.}
  \href{http://doi.org/10.1103/PhysRevLett.111.201801}{{\bf 111}, 201801}
  (2013), \href{https://arxiv.org/abs/1309.6594}{[arXiv:1309.6594]}\relax
\mciteBstWouldAddEndPuncttrue
\mciteSetBstMidEndSepPunct{\mcitedefaultmidpunct}
{\mcitedefaultendpunct}{\mcitedefaultseppunct}\relax
\EndOfBibitem
\bibitem{Borowka:2016ehy}
S.~Borowka {\em et~al.\/},
  \href{http://doi.org/10.1103/PhysRevLett.117.079901}{Phys. Rev. Lett.}
  \href{http://doi.org/10.1103/PhysRevLett.117.079901}{{\bf 117}, 1, 012001}
  (2016), [Erratum: Phys. Rev. Lett.117,no.7,079901(2016)],
  \href{https://arxiv.org/abs/1604.06447}{[arXiv:1604.06447]}\relax
\mciteBstWouldAddEndPuncttrue
\mciteSetBstMidEndSepPunct{\mcitedefaultmidpunct}
{\mcitedefaultendpunct}{\mcitedefaultseppunct}\relax
\EndOfBibitem
\bibitem{Czakon:2013goa}
M.~Czakon, P.~Fiedler and A.~Mitov,
  \href{http://doi.org/10.1103/PhysRevLett.110.252004}{Phys. Rev. Lett.}
  \href{http://doi.org/10.1103/PhysRevLett.110.252004}{{\bf 110}, 252004}
  (2013), \href{https://arxiv.org/abs/1303.6254}{[arXiv:1303.6254]}\relax
\mciteBstWouldAddEndPuncttrue
\mciteSetBstMidEndSepPunct{\mcitedefaultmidpunct}
{\mcitedefaultendpunct}{\mcitedefaultseppunct}\relax
\EndOfBibitem
\bibitem{Catani:2019iny}
S.~Catani {\em et~al.\/},
  \href{http://doi.org/10.1103/PhysRevD.99.051501}{Phys. Rev.}
  \href{http://doi.org/10.1103/PhysRevD.99.051501}{{\bf D99}, 5, 051501}
  (2019), \href{https://arxiv.org/abs/1901.04005}{[arXiv:1901.04005]}\relax
\mciteBstWouldAddEndPuncttrue
\mciteSetBstMidEndSepPunct{\mcitedefaultmidpunct}
{\mcitedefaultendpunct}{\mcitedefaultseppunct}\relax
\EndOfBibitem
\bibitem{Catani:2020kkl}
S.~Catani {\em et~al.\/}, \href{http://doi.org/10.1007/JHEP03(2021)029}{JHEP}
  \href{http://doi.org/10.1007/JHEP03(2021)029}{{\bf 03}, 029} (2021),
  \href{https://arxiv.org/abs/2010.11906}{[arXiv:2010.11906]}\relax
\mciteBstWouldAddEndPuncttrue
\mciteSetBstMidEndSepPunct{\mcitedefaultmidpunct}
{\mcitedefaultendpunct}{\mcitedefaultseppunct}\relax
\EndOfBibitem
\bibitem{Gehrmann:2014fva}
T.~Gehrmann {\em et~al.\/},
  \href{http://doi.org/10.1103/PhysRevLett.113.212001}{Phys. Rev. Lett.}
  \href{http://doi.org/10.1103/PhysRevLett.113.212001}{{\bf 113}, 21, 212001}
  (2014), \href{https://arxiv.org/abs/1408.5243}{[arXiv:1408.5243]}\relax
\mciteBstWouldAddEndPuncttrue
\mciteSetBstMidEndSepPunct{\mcitedefaultmidpunct}
{\mcitedefaultendpunct}{\mcitedefaultseppunct}\relax
\EndOfBibitem
\bibitem{Cascioli:2014yka}
F.~Cascioli {\em et~al.\/},
  \href{http://doi.org/10.1016/j.physletb.2014.06.056}{Phys. Lett.}
  \href{http://doi.org/10.1016/j.physletb.2014.06.056}{{\bf B735}, 311} (2014),
  \href{https://arxiv.org/abs/1405.2219}{[arXiv:1405.2219]}\relax
\mciteBstWouldAddEndPuncttrue
\mciteSetBstMidEndSepPunct{\mcitedefaultmidpunct}
{\mcitedefaultendpunct}{\mcitedefaultseppunct}\relax
\EndOfBibitem
\bibitem{Grazzini:2017mhc}
M.~Grazzini, S.~Kallweit and M.~Wiesemann,
  \href{http://doi.org/10.1140/epjc/s10052-018-5771-7}{Eur. Phys. J.}
  \href{http://doi.org/10.1140/epjc/s10052-018-5771-7}{{\bf C78}, 7, 537}
  (2018), \href{https://arxiv.org/abs/1711.06631}{[arXiv:1711.06631]}\relax
\mciteBstWouldAddEndPuncttrue
\mciteSetBstMidEndSepPunct{\mcitedefaultmidpunct}
{\mcitedefaultendpunct}{\mcitedefaultseppunct}\relax
\EndOfBibitem
\bibitem{Heinrich:2020ybq}
G.~Heinrich, \href{http://doi.org/10.1016/j.physrep.2021.03.006}{Phys. Rept.}
  \href{http://doi.org/10.1016/j.physrep.2021.03.006}{{\bf 922}, 1} (2021),
  \href{https://arxiv.org/abs/2009.00516}{[arXiv:2009.00516]}\relax
\mciteBstWouldAddEndPuncttrue
\mciteSetBstMidEndSepPunct{\mcitedefaultmidpunct}
{\mcitedefaultendpunct}{\mcitedefaultseppunct}\relax
\EndOfBibitem
\bibitem{Huss:2022ful}
A.~Huss {\em et~al.\/}, \href{http://doi.org/10.1088/1361-6471/acbaec}{J. Phys.
  G} \href{http://doi.org/10.1088/1361-6471/acbaec}{{\bf 50}, 4, 043001}
  (2023), \href{https://arxiv.org/abs/2207.02122}{[arXiv:2207.02122]}\relax
\mciteBstWouldAddEndPuncttrue
\mciteSetBstMidEndSepPunct{\mcitedefaultmidpunct}
{\mcitedefaultendpunct}{\mcitedefaultseppunct}\relax
\EndOfBibitem
\bibitem{Greco:1993st}
M.~Greco and A.~Vicini,
  \href{http://doi.org/10.1016/0550-3213(94)90117-1}{Nucl. Phys.}
  \href{http://doi.org/10.1016/0550-3213(94)90117-1}{{\bf B415}, 386}
  (1994)\relax
\mciteBstWouldAddEndPuncttrue
\mciteSetBstMidEndSepPunct{\mcitedefaultmidpunct}
{\mcitedefaultendpunct}{\mcitedefaultseppunct}\relax
\EndOfBibitem
\bibitem{Hebecker:1999ej}
A.~Hebecker, \href{http://doi.org/10.1016/S0370-1573(00)00005-3}{Phys. Rept.}
  \href{http://doi.org/10.1016/S0370-1573(00)00005-3}{{\bf 331}, 1} (2000),
  \href{https://arxiv.org/abs/hep-ph/9905226}{[hep-ph/9905226]}\relax
\mciteBstWouldAddEndPuncttrue
\mciteSetBstMidEndSepPunct{\mcitedefaultmidpunct}
{\mcitedefaultendpunct}{\mcitedefaultseppunct}\relax
\EndOfBibitem
\bibitem{Belitsky:2005qn}
A.~V. Belitsky and A.~V. Radyushkin,
  \href{http://doi.org/10.1016/j.physrep.2005.06.002}{Phys. Rept.}
  \href{http://doi.org/10.1016/j.physrep.2005.06.002}{{\bf 418}, 1} (2005),
  \href{https://arxiv.org/abs/hep-ph/0504030}{[hep-ph/0504030]}\relax
\mciteBstWouldAddEndPuncttrue
\mciteSetBstMidEndSepPunct{\mcitedefaultmidpunct}
{\mcitedefaultendpunct}{\mcitedefaultseppunct}\relax
\EndOfBibitem
\bibitem{Boos:2004kh}
E.~Boos {\em et~al.\/} (CompHEP),
  \href{http://doi.org/10.1016/j.nima.2004.07.096}{Nucl. Instrum. Meth.}
  \href{http://doi.org/10.1016/j.nima.2004.07.096}{{\bf A534}, 250} (2004),
  \href{https://arxiv.org/abs/hep-ph/0403113}{[hep-ph/0403113]}\relax
\mciteBstWouldAddEndPuncttrue
\mciteSetBstMidEndSepPunct{\mcitedefaultmidpunct}
{\mcitedefaultendpunct}{\mcitedefaultseppunct}\relax
\EndOfBibitem
\bibitem{Alwall:2014hca}
J.~Alwall {\em et~al.\/}, \href{http://doi.org/10.1007/JHEP07(2014)079}{JHEP}
  \href{http://doi.org/10.1007/JHEP07(2014)079}{{\bf 07}, 079} (2014),
  \href{https://arxiv.org/abs/1405.0301}{[arXiv:1405.0301]}\relax
\mciteBstWouldAddEndPuncttrue
\mciteSetBstMidEndSepPunct{\mcitedefaultmidpunct}
{\mcitedefaultendpunct}{\mcitedefaultseppunct}\relax
\EndOfBibitem
\bibitem{qcd:Alwall:2014hca-url}
\url{https://launchpad.net/mg5amcnlo}\relax
\mciteBstWouldAddEndPuncttrue
\mciteSetBstMidEndSepPunct{\mcitedefaultmidpunct}
{\mcitedefaultendpunct}{\mcitedefaultseppunct}\relax
\EndOfBibitem
\bibitem{Mangano:2002ea}
M.~L. Mangano {\em et~al.\/},
  \href{http://doi.org/10.1088/1126-6708/2003/07/001}{JHEP}
  \href{http://doi.org/10.1088/1126-6708/2003/07/001}{{\bf 07}, 001} (2003),
  \href{https://arxiv.org/abs/hep-ph/0206293}{[hep-ph/0206293]}\relax
\mciteBstWouldAddEndPuncttrue
\mciteSetBstMidEndSepPunct{\mcitedefaultmidpunct}
{\mcitedefaultendpunct}{\mcitedefaultseppunct}\relax
\EndOfBibitem
\bibitem{qcd:Mangano:2002ea-url}
\url{http://cern.ch/mlm/alpgen/}\relax
\mciteBstWouldAddEndPuncttrue
\mciteSetBstMidEndSepPunct{\mcitedefaultmidpunct}
{\mcitedefaultendpunct}{\mcitedefaultseppunct}\relax
\EndOfBibitem
\bibitem{Gleisberg:2008fv}
T.~Gleisberg and S.~Hoeche,
  \href{http://doi.org/10.1088/1126-6708/2008/12/039}{JHEP}
  \href{http://doi.org/10.1088/1126-6708/2008/12/039}{{\bf 12}, 039} (2008),
  \href{https://arxiv.org/abs/0808.3674}{[arXiv:0808.3674]}\relax
\mciteBstWouldAddEndPuncttrue
\mciteSetBstMidEndSepPunct{\mcitedefaultmidpunct}
{\mcitedefaultendpunct}{\mcitedefaultseppunct}\relax
\EndOfBibitem
\bibitem{qcd:Gleisberg:2008fv-url}
\url{https://sherpa-team.gitlab.io/}\relax
\mciteBstWouldAddEndPuncttrue
\mciteSetBstMidEndSepPunct{\mcitedefaultmidpunct}
{\mcitedefaultendpunct}{\mcitedefaultseppunct}\relax
\EndOfBibitem
\bibitem{Cafarella:2007pc}
A.~Cafarella, C.~G. Papadopoulos and M.~Worek,
  \href{http://doi.org/10.1016/j.cpc.2009.04.023}{Comput. Phys. Commun.}
  \href{http://doi.org/10.1016/j.cpc.2009.04.023}{{\bf 180}, 1941} (2009),
  \href{https://arxiv.org/abs/0710.2427}{[arXiv:0710.2427]}\relax
\mciteBstWouldAddEndPuncttrue
\mciteSetBstMidEndSepPunct{\mcitedefaultmidpunct}
{\mcitedefaultendpunct}{\mcitedefaultseppunct}\relax
\EndOfBibitem
\bibitem{qcd:Cafarella:2007pc-url}
\url{http://cern.ch/helac-phegas/}\relax
\mciteBstWouldAddEndPuncttrue
\mciteSetBstMidEndSepPunct{\mcitedefaultmidpunct}
{\mcitedefaultendpunct}{\mcitedefaultseppunct}\relax
\EndOfBibitem
\bibitem{Berends:1987me}
F.~A. Berends and W.~T. Giele,
  \href{http://doi.org/10.1016/0550-3213(88)90442-7}{Nucl. Phys.}
  \href{http://doi.org/10.1016/0550-3213(88)90442-7}{{\bf B306}, 759}
  (1988)\relax
\mciteBstWouldAddEndPuncttrue
\mciteSetBstMidEndSepPunct{\mcitedefaultmidpunct}
{\mcitedefaultendpunct}{\mcitedefaultseppunct}\relax
\EndOfBibitem
\bibitem{Dixon:1996wi}
L.~J. Dixon, in \enquote{{QCD and beyond. Proceedings, Theoretical Advanced
  Study Institute in Elementary Particle Physics, TASI-95, Boulder, USA, June
  4-30, 1995},} 539--584 (1996),
  \href{https://arxiv.org/abs/hep-ph/9601359}{[hep-ph/9601359]},
  \urlprefix\url{http://www-public.slac.stanford.edu/sciDoc/docMeta.aspx?slacPubNumber=SLAC-PUB-7106}\relax
\mciteBstWouldAddEndPuncttrue
\mciteSetBstMidEndSepPunct{\mcitedefaultmidpunct}
{\mcitedefaultendpunct}{\mcitedefaultseppunct}\relax
\EndOfBibitem
\bibitem{Britto:2004ap}
R.~Britto, F.~Cachazo and B.~Feng,
  \href{http://doi.org/10.1016/j.nuclphysb.2005.02.030}{Nucl. Phys.}
  \href{http://doi.org/10.1016/j.nuclphysb.2005.02.030}{{\bf B715}, 499}
  (2005), \href{https://arxiv.org/abs/hep-th/0412308}{[hep-th/0412308]}\relax
\mciteBstWouldAddEndPuncttrue
\mciteSetBstMidEndSepPunct{\mcitedefaultmidpunct}
{\mcitedefaultendpunct}{\mcitedefaultseppunct}\relax
\EndOfBibitem
\bibitem{Cachazo:2004kj}
F.~Cachazo, P.~Svrcek and E.~Witten,
  \href{http://doi.org/10.1088/1126-6708/2004/09/006}{JHEP}
  \href{http://doi.org/10.1088/1126-6708/2004/09/006}{{\bf 09}, 006} (2004),
  \href{https://arxiv.org/abs/hep-th/0403047}{[hep-th/0403047]}\relax
\mciteBstWouldAddEndPuncttrue
\mciteSetBstMidEndSepPunct{\mcitedefaultmidpunct}
{\mcitedefaultendpunct}{\mcitedefaultseppunct}\relax
\EndOfBibitem
\bibitem{Catani:1996vz}
S.~Catani and M.~H. Seymour,
  \href{http://doi.org/10.1016/S0550-3213(96)00589-5}{Nucl. Phys.}
  \href{http://doi.org/10.1016/S0550-3213(96)00589-5}{{\bf B485}, 291} (1997),
  [Erratum: Nucl. Phys.B510,503(1998)],
  \href{https://arxiv.org/abs/hep-ph/9605323}{[hep-ph/9605323]}\relax
\mciteBstWouldAddEndPuncttrue
\mciteSetBstMidEndSepPunct{\mcitedefaultmidpunct}
{\mcitedefaultendpunct}{\mcitedefaultseppunct}\relax
\EndOfBibitem
\bibitem{Frixione:1995ms}
S.~Frixione, Z.~Kunszt and A.~Signer,
  \href{http://doi.org/10.1016/0550-3213(96)00110-1}{Nucl. Phys.}
  \href{http://doi.org/10.1016/0550-3213(96)00110-1}{{\bf B467}, 399} (1996),
  \href{https://arxiv.org/abs/hep-ph/9512328}{[hep-ph/9512328]}\relax
\mciteBstWouldAddEndPuncttrue
\mciteSetBstMidEndSepPunct{\mcitedefaultmidpunct}
{\mcitedefaultendpunct}{\mcitedefaultseppunct}\relax
\EndOfBibitem
\bibitem{Kosower:1997zr}
D.~A. Kosower, \href{http://doi.org/10.1103/PhysRevD.57.5410}{Phys. Rev.}
  \href{http://doi.org/10.1103/PhysRevD.57.5410}{{\bf D57}, 5410} (1998),
  \href{https://arxiv.org/abs/hep-ph/9710213}{[hep-ph/9710213]}\relax
\mciteBstWouldAddEndPuncttrue
\mciteSetBstMidEndSepPunct{\mcitedefaultmidpunct}
{\mcitedefaultendpunct}{\mcitedefaultseppunct}\relax
\EndOfBibitem
\bibitem{Campbell:1998nn}
J.~M. Campbell, M.~A. Cullen and E.~W.~N. Glover,
  \href{http://doi.org/10.1007/s100520050529}{Eur. Phys. J.}
  \href{http://doi.org/10.1007/s100520050529}{{\bf C9}, 245} (1999),
  \href{https://arxiv.org/abs/hep-ph/9809429}{[hep-ph/9809429]}\relax
\mciteBstWouldAddEndPuncttrue
\mciteSetBstMidEndSepPunct{\mcitedefaultmidpunct}
{\mcitedefaultendpunct}{\mcitedefaultseppunct}\relax
\EndOfBibitem
\bibitem{Kosower:2003bh}
D.~A. Kosower, \href{http://doi.org/10.1103/PhysRevD.71.045016}{Phys. Rev.}
  \href{http://doi.org/10.1103/PhysRevD.71.045016}{{\bf D71}, 045016} (2005),
  \href{https://arxiv.org/abs/hep-ph/0311272}{[hep-ph/0311272]}\relax
\mciteBstWouldAddEndPuncttrue
\mciteSetBstMidEndSepPunct{\mcitedefaultmidpunct}
{\mcitedefaultendpunct}{\mcitedefaultseppunct}\relax
\EndOfBibitem
\bibitem{Ossola:2006us}
G.~Ossola, C.~G. Papadopoulos and R.~Pittau,
  \href{http://doi.org/10.1016/j.nuclphysb.2006.11.012}{Nucl. Phys.}
  \href{http://doi.org/10.1016/j.nuclphysb.2006.11.012}{{\bf B763}, 147}
  (2007), \href{https://arxiv.org/abs/hep-ph/0609007}{[hep-ph/0609007]}\relax
\mciteBstWouldAddEndPuncttrue
\mciteSetBstMidEndSepPunct{\mcitedefaultmidpunct}
{\mcitedefaultendpunct}{\mcitedefaultseppunct}\relax
\EndOfBibitem
\bibitem{Britto:2004nc}
R.~Britto, F.~Cachazo and B.~Feng,
  \href{http://doi.org/10.1016/j.nuclphysb.2005.07.014}{Nucl. Phys.}
  \href{http://doi.org/10.1016/j.nuclphysb.2005.07.014}{{\bf B725}, 275}
  (2005), \href{https://arxiv.org/abs/hep-th/0412103}{[hep-th/0412103]}\relax
\mciteBstWouldAddEndPuncttrue
\mciteSetBstMidEndSepPunct{\mcitedefaultmidpunct}
{\mcitedefaultendpunct}{\mcitedefaultseppunct}\relax
\EndOfBibitem
\bibitem{Ellis:2008ir}
R.~K. Ellis {\em et~al.\/},
  \href{http://doi.org/10.1016/j.nuclphysb.2009.07.023}{Nucl. Phys.}
  \href{http://doi.org/10.1016/j.nuclphysb.2009.07.023}{{\bf B822}, 270}
  (2009), \href{https://arxiv.org/abs/0806.3467}{[arXiv:0806.3467]}\relax
\mciteBstWouldAddEndPuncttrue
\mciteSetBstMidEndSepPunct{\mcitedefaultmidpunct}
{\mcitedefaultendpunct}{\mcitedefaultseppunct}\relax
\EndOfBibitem
\bibitem{Berger:2009zb}
C.~F. Berger and D.~Forde,
  \href{http://doi.org/10.1146/annurev.nucl.012809.104538}{Ann. Rev. Nucl.
  Part. Sci.} \href{http://doi.org/10.1146/annurev.nucl.012809.104538}{{\bf
  60}, 181} (2010),
  \href{https://arxiv.org/abs/0912.3534}{[arXiv:0912.3534]}\relax
\mciteBstWouldAddEndPuncttrue
\mciteSetBstMidEndSepPunct{\mcitedefaultmidpunct}
{\mcitedefaultendpunct}{\mcitedefaultseppunct}\relax
\EndOfBibitem
\bibitem{Cascioli:2011va}
F.~Cascioli, P.~Maierhofer and S.~Pozzorini,
  \href{http://doi.org/10.1103/PhysRevLett.108.111601}{Phys. Rev. Lett.}
  \href{http://doi.org/10.1103/PhysRevLett.108.111601}{{\bf 108}, 111601}
  (2012), \href{https://arxiv.org/abs/1111.5206}{[arXiv:1111.5206]}\relax
\mciteBstWouldAddEndPuncttrue
\mciteSetBstMidEndSepPunct{\mcitedefaultmidpunct}
{\mcitedefaultendpunct}{\mcitedefaultseppunct}\relax
\EndOfBibitem
\bibitem{Bern:2007dw}
Z.~Bern, L.~J. Dixon and D.~A. Kosower,
  \href{http://doi.org/10.1016/j.aop.2007.04.014}{Annals Phys.}
  \href{http://doi.org/10.1016/j.aop.2007.04.014}{{\bf 322}, 1587} (2007),
  \href{https://arxiv.org/abs/0704.2798}{[arXiv:0704.2798]}\relax
\mciteBstWouldAddEndPuncttrue
\mciteSetBstMidEndSepPunct{\mcitedefaultmidpunct}
{\mcitedefaultendpunct}{\mcitedefaultseppunct}\relax
\EndOfBibitem
\bibitem{Ellis:2011cr}
R.~K. Ellis {\em et~al.\/},
  \href{http://doi.org/10.1016/j.physrep.2012.01.008}{Phys. Rept.}
  \href{http://doi.org/10.1016/j.physrep.2012.01.008}{{\bf 518}, 141} (2012),
  \href{https://arxiv.org/abs/1105.4319}{[arXiv:1105.4319]}\relax
\mciteBstWouldAddEndPuncttrue
\mciteSetBstMidEndSepPunct{\mcitedefaultmidpunct}
{\mcitedefaultendpunct}{\mcitedefaultseppunct}\relax
\EndOfBibitem
\bibitem{Bevilacqua:2011xh}
G.~Bevilacqua {\em et~al.\/},
  \href{http://doi.org/10.1016/j.cpc.2012.10.033}{Comput. Phys. Commun.}
  \href{http://doi.org/10.1016/j.cpc.2012.10.033}{{\bf 184}, 986} (2013),
  \href{https://arxiv.org/abs/1110.1499}{[arXiv:1110.1499]}\relax
\mciteBstWouldAddEndPuncttrue
\mciteSetBstMidEndSepPunct{\mcitedefaultmidpunct}
{\mcitedefaultendpunct}{\mcitedefaultseppunct}\relax
\EndOfBibitem
\bibitem{qcd:Bevilacqua:2011xh-sr1}
\url{http://cern.ch/helac-phegas/}\relax
\mciteBstWouldAddEndPuncttrue
\mciteSetBstMidEndSepPunct{\mcitedefaultmidpunct}
{\mcitedefaultendpunct}{\mcitedefaultseppunct}\relax
\EndOfBibitem
\bibitem{Cullen:2014yla}
G.~Cullen {\em et~al.\/},
  \href{http://doi.org/10.1140/epjc/s10052-014-3001-5}{Eur. Phys. J.}
  \href{http://doi.org/10.1140/epjc/s10052-014-3001-5}{{\bf C74}, 8, 3001}
  (2014), \href{https://arxiv.org/abs/1404.7096}{[arXiv:1404.7096]}\relax
\mciteBstWouldAddEndPuncttrue
\mciteSetBstMidEndSepPunct{\mcitedefaultmidpunct}
{\mcitedefaultendpunct}{\mcitedefaultseppunct}\relax
\EndOfBibitem
\bibitem{qcd:Cullen:2014url}
\url{http://gosam.hepforge.org/}\relax
\mciteBstWouldAddEndPuncttrue
\mciteSetBstMidEndSepPunct{\mcitedefaultmidpunct}
{\mcitedefaultendpunct}{\mcitedefaultseppunct}\relax
\EndOfBibitem
\bibitem{Badger:2012pg}
S.~Badger {\em et~al.\/},
  \href{http://doi.org/10.1016/j.cpc.2013.03.018}{Comput. Phys. Commun.}
  \href{http://doi.org/10.1016/j.cpc.2013.03.018}{{\bf 184}, 1981} (2013),
  \href{https://arxiv.org/abs/1209.0100}{[arXiv:1209.0100]}\relax
\mciteBstWouldAddEndPuncttrue
\mciteSetBstMidEndSepPunct{\mcitedefaultmidpunct}
{\mcitedefaultendpunct}{\mcitedefaultseppunct}\relax
\EndOfBibitem
\bibitem{qcd:Badger:2012pg-url}
\url{https://bitbucket.org/njet/wiki/Home/}\relax
\mciteBstWouldAddEndPuncttrue
\mciteSetBstMidEndSepPunct{\mcitedefaultmidpunct}
{\mcitedefaultendpunct}{\mcitedefaultseppunct}\relax
\EndOfBibitem
\bibitem{Buccioni:2017yxi}
F.~Buccioni, S.~Pozzorini and M.~Zoller,
  \href{http://doi.org/10.1140/epjc/s10052-018-5562-1}{Eur. Phys. J.}
  \href{http://doi.org/10.1140/epjc/s10052-018-5562-1}{{\bf C78}, 1, 70}
  (2018), \href{https://arxiv.org/abs/1710.11452}{[arXiv:1710.11452]}\relax
\mciteBstWouldAddEndPuncttrue
\mciteSetBstMidEndSepPunct{\mcitedefaultmidpunct}
{\mcitedefaultendpunct}{\mcitedefaultseppunct}\relax
\EndOfBibitem
\bibitem{qcd:Buccioni:2017yxi-url}
\url{https://openloops.hepforge.org/}\relax
\mciteBstWouldAddEndPuncttrue
\mciteSetBstMidEndSepPunct{\mcitedefaultmidpunct}
{\mcitedefaultendpunct}{\mcitedefaultseppunct}\relax
\EndOfBibitem
\bibitem{Actis:2016mpe}
S.~Actis {\em et~al.\/},
  \href{http://doi.org/10.1016/j.cpc.2017.01.004}{Comput. Phys. Commun.}
  \href{http://doi.org/10.1016/j.cpc.2017.01.004}{{\bf 214}, 140} (2017),
  \href{https://arxiv.org/abs/1605.01090}{[arXiv:1605.01090]}\relax
\mciteBstWouldAddEndPuncttrue
\mciteSetBstMidEndSepPunct{\mcitedefaultmidpunct}
{\mcitedefaultendpunct}{\mcitedefaultseppunct}\relax
\EndOfBibitem
\bibitem{Nagy:2003tz}
Z.~Nagy, \href{http://doi.org/10.1103/PhysRevD.68.094002}{Phys. Rev.}
  \href{http://doi.org/10.1103/PhysRevD.68.094002}{{\bf D68}, 094002} (2003),
  \href{https://arxiv.org/abs/hep-ph/0307268}{[hep-ph/0307268]}\relax
\mciteBstWouldAddEndPuncttrue
\mciteSetBstMidEndSepPunct{\mcitedefaultmidpunct}
{\mcitedefaultendpunct}{\mcitedefaultseppunct}\relax
\EndOfBibitem
\bibitem{qcd:Nagy:2003tz-url}
\url{http://www.desy.de/~znagy/Site/NLOJet++.html}\relax
\mciteBstWouldAddEndPuncttrue
\mciteSetBstMidEndSepPunct{\mcitedefaultmidpunct}
{\mcitedefaultendpunct}{\mcitedefaultseppunct}\relax
\EndOfBibitem
\bibitem{Campbell:2000bg}
J.~M. Campbell and R.~K. Ellis,
  \href{http://doi.org/10.1103/PhysRevD.62.114012}{Phys. Rev.}
  \href{http://doi.org/10.1103/PhysRevD.62.114012}{{\bf D62}, 114012} (2000),
  \href{https://arxiv.org/abs/hep-ph/0006304}{[hep-ph/0006304]}\relax
\mciteBstWouldAddEndPuncttrue
\mciteSetBstMidEndSepPunct{\mcitedefaultmidpunct}
{\mcitedefaultendpunct}{\mcitedefaultseppunct}\relax
\EndOfBibitem
\bibitem{Arnold:2011wj}
J.~Baglio {\em et~al.\/}  (2011),
  \href{https://arxiv.org/abs/1107.4038}{[arXiv:1107.4038]}\relax
\mciteBstWouldAddEndPuncttrue
\mciteSetBstMidEndSepPunct{\mcitedefaultmidpunct}
{\mcitedefaultendpunct}{\mcitedefaultseppunct}\relax
\EndOfBibitem
\bibitem{*qcd:Arnold:2011wj-sr1}
\url{http://www-itp.particle.uni-karlsruhe.de/$\sim$vbfnloweb}\relax
\mciteBstWouldAddEndPuncttrue
\mciteSetBstMidEndSepPunct{\mcitedefaultmidpunct}
{\mcitedefaultendpunct}{\mcitedefaultseppunct}\relax
\EndOfBibitem
\bibitem{Binoth:1999qq}
T.~Binoth {\em et~al.\/}, \href{http://doi.org/10.1007/s100520050024}{Eur.
  Phys. J.} \href{http://doi.org/10.1007/s100520050024}{{\bf C16}, 311} (2000),
  \href{https://arxiv.org/abs/hep-ph/9911340}{[hep-ph/9911340]}\relax
\mciteBstWouldAddEndPuncttrue
\mciteSetBstMidEndSepPunct{\mcitedefaultmidpunct}
{\mcitedefaultendpunct}{\mcitedefaultseppunct}\relax
\EndOfBibitem
\bibitem{*qcd:Binoth:1999qq-sr1}
\url{http://lapth.in2p3.fr/PHOX\_FAMILY/}\relax
\mciteBstWouldAddEndPuncttrue
\mciteSetBstMidEndSepPunct{\mcitedefaultmidpunct}
{\mcitedefaultendpunct}{\mcitedefaultseppunct}\relax
\EndOfBibitem
\bibitem{Bern:2012my}
Z.~Bern {\em et~al.\/}, \href{http://doi.org/10.22323/1.151.0018}{PoS}
  \href{http://doi.org/10.22323/1.151.0018}{{\bf LL2012}, 018} (2012),
  \href{https://arxiv.org/abs/1210.6684}{[arXiv:1210.6684]}\relax
\mciteBstWouldAddEndPuncttrue
\mciteSetBstMidEndSepPunct{\mcitedefaultmidpunct}
{\mcitedefaultendpunct}{\mcitedefaultseppunct}\relax
\EndOfBibitem
\bibitem{Cullen:2012eh}
G.~Cullen, N.~Greiner and G.~Heinrich,
  \href{http://doi.org/10.1140/epjc/s10052-013-2388-8}{Eur. Phys. J.}
  \href{http://doi.org/10.1140/epjc/s10052-013-2388-8}{{\bf C73}, 4, 2388}
  (2013), \href{https://arxiv.org/abs/1212.5154}{[arXiv:1212.5154]}\relax
\mciteBstWouldAddEndPuncttrue
\mciteSetBstMidEndSepPunct{\mcitedefaultmidpunct}
{\mcitedefaultendpunct}{\mcitedefaultseppunct}\relax
\EndOfBibitem
\bibitem{Kallweit:2014xda}
S.~Kallweit {\em et~al.\/}, \href{http://doi.org/10.1007/JHEP04(2015)012}{JHEP}
  \href{http://doi.org/10.1007/JHEP04(2015)012}{{\bf 04}, 012} (2015),
  \href{https://arxiv.org/abs/1412.5157}{[arXiv:1412.5157]}\relax
\mciteBstWouldAddEndPuncttrue
\mciteSetBstMidEndSepPunct{\mcitedefaultmidpunct}
{\mcitedefaultendpunct}{\mcitedefaultseppunct}\relax
\EndOfBibitem
\bibitem{Denner:2014bna}
A.~Denner {\em et~al.\/}, \href{http://doi.org/10.1007/JHEP04(2015)018}{JHEP}
  \href{http://doi.org/10.1007/JHEP04(2015)018}{{\bf 04}, 018} (2015),
  \href{https://arxiv.org/abs/1412.7421}{[arXiv:1412.7421]}\relax
\mciteBstWouldAddEndPuncttrue
\mciteSetBstMidEndSepPunct{\mcitedefaultmidpunct}
{\mcitedefaultendpunct}{\mcitedefaultseppunct}\relax
\EndOfBibitem
\bibitem{Chiesa:2015mya}
M.~Chiesa, N.~Greiner and F.~Tramontano,
  \href{http://doi.org/10.1088/0954-3899/43/1/013002}{J. Phys.}
  \href{http://doi.org/10.1088/0954-3899/43/1/013002}{{\bf G43}, 1, 013002}
  (2016), \href{https://arxiv.org/abs/1507.08579}{[arXiv:1507.08579]}\relax
\mciteBstWouldAddEndPuncttrue
\mciteSetBstMidEndSepPunct{\mcitedefaultmidpunct}
{\mcitedefaultendpunct}{\mcitedefaultseppunct}\relax
\EndOfBibitem
\bibitem{Frixione:2015zaa}
S.~Frixione {\em et~al.\/}, \href{http://doi.org/10.1007/JHEP06(2015)184}{JHEP}
  \href{http://doi.org/10.1007/JHEP06(2015)184}{{\bf 06}, 184} (2015),
  \href{https://arxiv.org/abs/1504.03446}{[arXiv:1504.03446]}\relax
\mciteBstWouldAddEndPuncttrue
\mciteSetBstMidEndSepPunct{\mcitedefaultmidpunct}
{\mcitedefaultendpunct}{\mcitedefaultseppunct}\relax
\EndOfBibitem
\bibitem{Biedermann:2017yoi}
B.~Biedermann {\em et~al.\/},
  \href{http://doi.org/10.1140/epjc/s10052-017-5054-8}{Eur. Phys. J.}
  \href{http://doi.org/10.1140/epjc/s10052-017-5054-8}{{\bf C77}, 492} (2017),
  \href{https://arxiv.org/abs/1704.05783}{[arXiv:1704.05783]}\relax
\mciteBstWouldAddEndPuncttrue
\mciteSetBstMidEndSepPunct{\mcitedefaultmidpunct}
{\mcitedefaultendpunct}{\mcitedefaultseppunct}\relax
\EndOfBibitem
\bibitem{Frederix:2018nkq}
R.~Frederix {\em et~al.\/}, \href{http://doi.org/10.1007/JHEP11(2021)085}{JHEP}
  \href{http://doi.org/10.1007/JHEP11(2021)085}{{\bf 07}, 185} (2018),
  [Erratum: JHEP 11, 085 (2021)],
  \href{https://arxiv.org/abs/1804.10017}{[arXiv:1804.10017]}\relax
\mciteBstWouldAddEndPuncttrue
\mciteSetBstMidEndSepPunct{\mcitedefaultmidpunct}
{\mcitedefaultendpunct}{\mcitedefaultseppunct}\relax
\EndOfBibitem
\bibitem{Anastasiou:2008tj}
C.~Anastasiou, R.~Boughezal and F.~Petriello,
  \href{http://doi.org/10.1088/1126-6708/2009/04/003}{JHEP}
  \href{http://doi.org/10.1088/1126-6708/2009/04/003}{{\bf 04}, 003} (2009),
  \href{https://arxiv.org/abs/0811.3458}{[arXiv:0811.3458]}\relax
\mciteBstWouldAddEndPuncttrue
\mciteSetBstMidEndSepPunct{\mcitedefaultmidpunct}
{\mcitedefaultendpunct}{\mcitedefaultseppunct}\relax
\EndOfBibitem
\bibitem{Dittmaier:2014qza}
S.~Dittmaier, A.~Huss and C.~Schwinn,
  \href{http://doi.org/10.1016/j.nuclphysb.2014.05.027}{Nucl. Phys.}
  \href{http://doi.org/10.1016/j.nuclphysb.2014.05.027}{{\bf B885}, 318}
  (2014), \href{https://arxiv.org/abs/1403.3216}{[arXiv:1403.3216]}\relax
\mciteBstWouldAddEndPuncttrue
\mciteSetBstMidEndSepPunct{\mcitedefaultmidpunct}
{\mcitedefaultendpunct}{\mcitedefaultseppunct}\relax
\EndOfBibitem
\bibitem{Dittmaier:2015rxo}
S.~Dittmaier, A.~Huss and C.~Schwinn,
  \href{http://doi.org/10.1016/j.nuclphysb.2016.01.006}{Nucl. Phys.}
  \href{http://doi.org/10.1016/j.nuclphysb.2016.01.006}{{\bf B904}, 216}
  (2016), \href{https://arxiv.org/abs/1511.08016}{[arXiv:1511.08016]}\relax
\mciteBstWouldAddEndPuncttrue
\mciteSetBstMidEndSepPunct{\mcitedefaultmidpunct}
{\mcitedefaultendpunct}{\mcitedefaultseppunct}\relax
\EndOfBibitem
\bibitem{deFlorian:2018wcj}
D.~de~Florian, M.~Der and I.~Fabre,
  \href{http://doi.org/10.1103/PhysRevD.98.094008}{Phys. Rev.}
  \href{http://doi.org/10.1103/PhysRevD.98.094008}{{\bf D98}, 9, 094008}
  (2018), \href{https://arxiv.org/abs/1805.12214}{[arXiv:1805.12214]}\relax
\mciteBstWouldAddEndPuncttrue
\mciteSetBstMidEndSepPunct{\mcitedefaultmidpunct}
{\mcitedefaultendpunct}{\mcitedefaultseppunct}\relax
\EndOfBibitem
\bibitem{Bonetti:2017ovy}
M.~Bonetti, K.~Melnikov and L.~Tancredi,
  \href{http://doi.org/10.1103/PhysRevD.97.034004}{Phys. Rev.}
  \href{http://doi.org/10.1103/PhysRevD.97.034004}{{\bf D97}, 3, 034004}
  (2018), \href{https://arxiv.org/abs/1711.11113}{[arXiv:1711.11113]}\relax
\mciteBstWouldAddEndPuncttrue
\mciteSetBstMidEndSepPunct{\mcitedefaultmidpunct}
{\mcitedefaultendpunct}{\mcitedefaultseppunct}\relax
\EndOfBibitem
\bibitem{Bonetti:2018ukf}
M.~Bonetti, K.~Melnikov and L.~Tancredi,
  \href{http://doi.org/10.1103/PhysRevD.97.056017}{Phys. Rev.}
  \href{http://doi.org/10.1103/PhysRevD.97.056017}{{\bf D97}, 5, 056017}
  (2018), [Erratum: Phys. Rev.D97,no.9,099906(2018)],
  \href{https://arxiv.org/abs/1801.10403}{[arXiv:1801.10403]}\relax
\mciteBstWouldAddEndPuncttrue
\mciteSetBstMidEndSepPunct{\mcitedefaultmidpunct}
{\mcitedefaultendpunct}{\mcitedefaultseppunct}\relax
\EndOfBibitem
\bibitem{Anastasiou:2018adr}
C.~Anastasiou {\em et~al.\/},
  \href{http://doi.org/10.1007/JHEP03(2019)162}{JHEP}
  \href{http://doi.org/10.1007/JHEP03(2019)162}{{\bf 03}, 162} (2019),
  \href{https://arxiv.org/abs/1811.11211}{[arXiv:1811.11211]}\relax
\mciteBstWouldAddEndPuncttrue
\mciteSetBstMidEndSepPunct{\mcitedefaultmidpunct}
{\mcitedefaultendpunct}{\mcitedefaultseppunct}\relax
\EndOfBibitem
\bibitem{Hirschi:2019fkz}
V.~Hirschi, S.~Lionetti and A.~Schweitzer,
  \href{http://doi.org/10.1007/JHEP05(2019)002}{JHEP}
  \href{http://doi.org/10.1007/JHEP05(2019)002}{{\bf 05}, 002} (2019),
  \href{https://arxiv.org/abs/1902.10167}{[arXiv:1902.10167]}\relax
\mciteBstWouldAddEndPuncttrue
\mciteSetBstMidEndSepPunct{\mcitedefaultmidpunct}
{\mcitedefaultendpunct}{\mcitedefaultseppunct}\relax
\EndOfBibitem
\bibitem{Becchetti:2020wof}
M.~Becchetti {\em et~al.\/},
  \href{http://doi.org/10.1103/PhysRevD.103.054037}{Phys. Rev. D}
  \href{http://doi.org/10.1103/PhysRevD.103.054037}{{\bf 103}, 5, 054037}
  (2021), \href{https://arxiv.org/abs/2010.09451}{[arXiv:2010.09451]}\relax
\mciteBstWouldAddEndPuncttrue
\mciteSetBstMidEndSepPunct{\mcitedefaultmidpunct}
{\mcitedefaultendpunct}{\mcitedefaultseppunct}\relax
\EndOfBibitem
\bibitem{Bonetti:2020hqh}
M.~Bonetti {\em et~al.\/}, \href{http://doi.org/10.1007/JHEP11(2020)045}{JHEP}
  \href{http://doi.org/10.1007/JHEP11(2020)045}{{\bf 11}, 045} (2020),
  \href{https://arxiv.org/abs/2007.09813}{[arXiv:2007.09813]}\relax
\mciteBstWouldAddEndPuncttrue
\mciteSetBstMidEndSepPunct{\mcitedefaultmidpunct}
{\mcitedefaultendpunct}{\mcitedefaultseppunct}\relax
\EndOfBibitem
\bibitem{Behring:2020cqi}
A.~Behring {\em et~al.\/},
  \href{http://doi.org/10.1103/PhysRevD.103.013008}{Phys. Rev. D}
  \href{http://doi.org/10.1103/PhysRevD.103.013008}{{\bf 103}, 1, 013008}
  (2021), \href{https://arxiv.org/abs/2009.10386}{[arXiv:2009.10386]}\relax
\mciteBstWouldAddEndPuncttrue
\mciteSetBstMidEndSepPunct{\mcitedefaultmidpunct}
{\mcitedefaultendpunct}{\mcitedefaultseppunct}\relax
\EndOfBibitem
\bibitem{Dittmaier:2020vra}
S.~Dittmaier, T.~Schmidt and J.~Schwarz,
  \href{http://doi.org/10.1007/JHEP12(2020)201}{JHEP}
  \href{http://doi.org/10.1007/JHEP12(2020)201}{{\bf 12}, 201} (2020),
  \href{https://arxiv.org/abs/2009.02229}{[arXiv:2009.02229]}\relax
\mciteBstWouldAddEndPuncttrue
\mciteSetBstMidEndSepPunct{\mcitedefaultmidpunct}
{\mcitedefaultendpunct}{\mcitedefaultseppunct}\relax
\EndOfBibitem
\bibitem{Buccioni:2020cfi}
F.~Buccioni {\em et~al.\/},
  \href{http://doi.org/10.1016/j.physletb.2020.135969}{Phys. Lett. B}
  \href{http://doi.org/10.1016/j.physletb.2020.135969}{{\bf 811}, 135969}
  (2020), \href{https://arxiv.org/abs/2005.10221}{[arXiv:2005.10221]}\relax
\mciteBstWouldAddEndPuncttrue
\mciteSetBstMidEndSepPunct{\mcitedefaultmidpunct}
{\mcitedefaultendpunct}{\mcitedefaultseppunct}\relax
\EndOfBibitem
\bibitem{Buonocore:2021rxx}
L.~Buonocore {\em et~al.\/},
  \href{http://doi.org/10.1103/PhysRevD.103.114012}{Phys. Rev. D}
  \href{http://doi.org/10.1103/PhysRevD.103.114012}{{\bf 103}, 114012} (2021),
  \href{https://arxiv.org/abs/2102.12539}{[arXiv:2102.12539]}\relax
\mciteBstWouldAddEndPuncttrue
\mciteSetBstMidEndSepPunct{\mcitedefaultmidpunct}
{\mcitedefaultendpunct}{\mcitedefaultseppunct}\relax
\EndOfBibitem
\bibitem{Bonciani:2021zzf}
R.~Bonciani {\em et~al.\/},
  \href{http://doi.org/10.1103/PhysRevLett.128.012002}{Phys. Rev. Lett.}
  \href{http://doi.org/10.1103/PhysRevLett.128.012002}{{\bf 128}, 1, 012002}
  (2022), \href{https://arxiv.org/abs/2106.11953}{[arXiv:2106.11953]}\relax
\mciteBstWouldAddEndPuncttrue
\mciteSetBstMidEndSepPunct{\mcitedefaultmidpunct}
{\mcitedefaultendpunct}{\mcitedefaultseppunct}\relax
\EndOfBibitem
\bibitem{Armadillo:2022bgm}
T.~Armadillo {\em et~al.\/},
  \href{http://doi.org/10.1007/JHEP05(2022)072}{JHEP}
  \href{http://doi.org/10.1007/JHEP05(2022)072}{{\bf 05}, 072} (2022),
  \href{https://arxiv.org/abs/2201.01754}{[arXiv:2201.01754]}\relax
\mciteBstWouldAddEndPuncttrue
\mciteSetBstMidEndSepPunct{\mcitedefaultmidpunct}
{\mcitedefaultendpunct}{\mcitedefaultseppunct}\relax
\EndOfBibitem
\bibitem{Buccioni:2022kgy}
F.~Buccioni {\em et~al.\/}, \href{http://doi.org/10.1007/JHEP06(2022)022}{JHEP}
  \href{http://doi.org/10.1007/JHEP06(2022)022}{{\bf 06}, 022} (2022),
  \href{https://arxiv.org/abs/2203.11237}{[arXiv:2203.11237]}\relax
\mciteBstWouldAddEndPuncttrue
\mciteSetBstMidEndSepPunct{\mcitedefaultmidpunct}
{\mcitedefaultendpunct}{\mcitedefaultseppunct}\relax
\EndOfBibitem
\bibitem{Denner:2019vbn}
A.~Denner and S.~Dittmaier,
  \href{http://doi.org/10.1016/j.physrep.2020.04.001}{Phys. Rept.}
  \href{http://doi.org/10.1016/j.physrep.2020.04.001}{{\bf 864}, 1} (2020),
  \href{https://arxiv.org/abs/1912.06823}{[arXiv:1912.06823]}\relax
\mciteBstWouldAddEndPuncttrue
\mciteSetBstMidEndSepPunct{\mcitedefaultmidpunct}
{\mcitedefaultendpunct}{\mcitedefaultseppunct}\relax
\EndOfBibitem
\bibitem{Bern:1994zx}
Z.~Bern {\em et~al.\/},
  \href{http://doi.org/10.1016/0550-3213(94)90179-1}{Nucl. Phys.}
  \href{http://doi.org/10.1016/0550-3213(94)90179-1}{{\bf B425}, 217} (1994),
  \href{https://arxiv.org/abs/hep-ph/9403226}{[hep-ph/9403226]}\relax
\mciteBstWouldAddEndPuncttrue
\mciteSetBstMidEndSepPunct{\mcitedefaultmidpunct}
{\mcitedefaultendpunct}{\mcitedefaultseppunct}\relax
\EndOfBibitem
\bibitem{Campbell:1997hg}
J.~M. Campbell and E.~W.~N. Glover,
  \href{http://doi.org/10.1016/S0550-3213(98)00295-8}{Nucl. Phys.}
  \href{http://doi.org/10.1016/S0550-3213(98)00295-8}{{\bf B527}, 264} (1998),
  \href{https://arxiv.org/abs/hep-ph/9710255}{[hep-ph/9710255]}\relax
\mciteBstWouldAddEndPuncttrue
\mciteSetBstMidEndSepPunct{\mcitedefaultmidpunct}
{\mcitedefaultendpunct}{\mcitedefaultseppunct}\relax
\EndOfBibitem
\bibitem{Catani:1998nv}
S.~Catani and M.~Grazzini,
  \href{http://doi.org/10.1016/S0370-2693(98)01513-5}{Phys. Lett.}
  \href{http://doi.org/10.1016/S0370-2693(98)01513-5}{{\bf B446}, 143} (1999),
  \href{https://arxiv.org/abs/hep-ph/9810389}{[hep-ph/9810389]}\relax
\mciteBstWouldAddEndPuncttrue
\mciteSetBstMidEndSepPunct{\mcitedefaultmidpunct}
{\mcitedefaultendpunct}{\mcitedefaultseppunct}\relax
\EndOfBibitem
\bibitem{Binoth:2000ps}
T.~Binoth and G.~Heinrich,
  \href{http://doi.org/10.1016/S0550-3213(00)00429-6}{Nucl. Phys.}
  \href{http://doi.org/10.1016/S0550-3213(00)00429-6}{{\bf B585}, 741} (2000),
  \href{https://arxiv.org/abs/hep-ph/0004013}{[hep-ph/0004013]}\relax
\mciteBstWouldAddEndPuncttrue
\mciteSetBstMidEndSepPunct{\mcitedefaultmidpunct}
{\mcitedefaultendpunct}{\mcitedefaultseppunct}\relax
\EndOfBibitem
\bibitem{Anastasiou:2003gr}
C.~Anastasiou, K.~Melnikov and F.~Petriello,
  \href{http://doi.org/10.1103/PhysRevD.69.076010}{Phys. Rev.}
  \href{http://doi.org/10.1103/PhysRevD.69.076010}{{\bf D69}, 076010} (2004),
  \href{https://arxiv.org/abs/hep-ph/0311311}{[hep-ph/0311311]}\relax
\mciteBstWouldAddEndPuncttrue
\mciteSetBstMidEndSepPunct{\mcitedefaultmidpunct}
{\mcitedefaultendpunct}{\mcitedefaultseppunct}\relax
\EndOfBibitem
\bibitem{GehrmannDeRidder:2005cm}
A.~Gehrmann-De~Ridder, T.~Gehrmann and E.~W.~N. Glover,
  \href{http://doi.org/10.1088/1126-6708/2005/09/056}{JHEP}
  \href{http://doi.org/10.1088/1126-6708/2005/09/056}{{\bf 09}, 056} (2005),
  \href{https://arxiv.org/abs/hep-ph/0505111}{[hep-ph/0505111]}\relax
\mciteBstWouldAddEndPuncttrue
\mciteSetBstMidEndSepPunct{\mcitedefaultmidpunct}
{\mcitedefaultendpunct}{\mcitedefaultseppunct}\relax
\EndOfBibitem
\bibitem{Somogyi:2006da}
G.~Somogyi, Z.~Trocsanyi and V.~Del~Duca,
  \href{http://doi.org/10.1088/1126-6708/2007/01/070}{JHEP}
  \href{http://doi.org/10.1088/1126-6708/2007/01/070}{{\bf 01}, 070} (2007),
  \href{https://arxiv.org/abs/hep-ph/0609042}{[hep-ph/0609042]}\relax
\mciteBstWouldAddEndPuncttrue
\mciteSetBstMidEndSepPunct{\mcitedefaultmidpunct}
{\mcitedefaultendpunct}{\mcitedefaultseppunct}\relax
\EndOfBibitem
\bibitem{Czakon:2010td}
M.~Czakon, \href{http://doi.org/10.1016/j.physletb.2010.08.036}{Phys. Lett.}
  \href{http://doi.org/10.1016/j.physletb.2010.08.036}{{\bf B693}, 259} (2010),
  \href{https://arxiv.org/abs/1005.0274}{[arXiv:1005.0274]}\relax
\mciteBstWouldAddEndPuncttrue
\mciteSetBstMidEndSepPunct{\mcitedefaultmidpunct}
{\mcitedefaultendpunct}{\mcitedefaultseppunct}\relax
\EndOfBibitem
\bibitem{Caola:2017dug}
F.~Caola, K.~Melnikov and R.~R\"ontsch,
  \href{http://doi.org/10.1140/epjc/s10052-017-4774-0}{Eur. Phys. J. C}
  \href{http://doi.org/10.1140/epjc/s10052-017-4774-0}{{\bf 77}, 4, 248}
  (2017), \href{https://arxiv.org/abs/1702.01352}{[arXiv:1702.01352]}\relax
\mciteBstWouldAddEndPuncttrue
\mciteSetBstMidEndSepPunct{\mcitedefaultmidpunct}
{\mcitedefaultendpunct}{\mcitedefaultseppunct}\relax
\EndOfBibitem
\bibitem{Magnea:2018hab}
L.~Magnea {\em et~al.\/}, \href{http://doi.org/10.1007/JHEP12(2018)107}{JHEP}
  \href{http://doi.org/10.1007/JHEP12(2018)107}{{\bf 12}, 107} (2018),
  [Erratum: JHEP 06, 013 (2019)],
  \href{https://arxiv.org/abs/1806.09570}{[arXiv:1806.09570]}\relax
\mciteBstWouldAddEndPuncttrue
\mciteSetBstMidEndSepPunct{\mcitedefaultmidpunct}
{\mcitedefaultendpunct}{\mcitedefaultseppunct}\relax
\EndOfBibitem
\bibitem{Prisco:2020kyb}
R.~M. Prisco and F.~Tramontano,
  \href{http://doi.org/10.1007/JHEP06(2021)089}{JHEP}
  \href{http://doi.org/10.1007/JHEP06(2021)089}{{\bf 06}, 089} (2021),
  \href{https://arxiv.org/abs/2012.05012}{[arXiv:2012.05012]}\relax
\mciteBstWouldAddEndPuncttrue
\mciteSetBstMidEndSepPunct{\mcitedefaultmidpunct}
{\mcitedefaultendpunct}{\mcitedefaultseppunct}\relax
\EndOfBibitem
\bibitem{Bertolotti:2022aih}
G.~Bertolotti {\em et~al.\/}  (2022),
  \href{https://arxiv.org/abs/2212.11190}{[arXiv:2212.11190]}\relax
\mciteBstWouldAddEndPuncttrue
\mciteSetBstMidEndSepPunct{\mcitedefaultmidpunct}
{\mcitedefaultendpunct}{\mcitedefaultseppunct}\relax
\EndOfBibitem
\bibitem{Anastasiou:2022eym}
C.~Anastasiou and G.~Sterman,
  \href{http://doi.org/10.1007/JHEP05(2023)242}{JHEP}
  \href{http://doi.org/10.1007/JHEP05(2023)242}{{\bf 05}, 242} (2023),
  \href{https://arxiv.org/abs/2212.12162}{[arXiv:2212.12162]}\relax
\mciteBstWouldAddEndPuncttrue
\mciteSetBstMidEndSepPunct{\mcitedefaultmidpunct}
{\mcitedefaultendpunct}{\mcitedefaultseppunct}\relax
\EndOfBibitem
\bibitem{Catani:2007vq}
S.~Catani and M.~Grazzini,
  \href{http://doi.org/10.1103/PhysRevLett.98.222002}{Phys. Rev. Lett.}
  \href{http://doi.org/10.1103/PhysRevLett.98.222002}{{\bf 98}, 222002} (2007),
  \href{https://arxiv.org/abs/hep-ph/0703012}{[hep-ph/0703012]}\relax
\mciteBstWouldAddEndPuncttrue
\mciteSetBstMidEndSepPunct{\mcitedefaultmidpunct}
{\mcitedefaultendpunct}{\mcitedefaultseppunct}\relax
\EndOfBibitem
\bibitem{qcd:Catani:2007vq-url}
\url{http://theory.fi.infn.it/grazzini/codes.html}\relax
\mciteBstWouldAddEndPuncttrue
\mciteSetBstMidEndSepPunct{\mcitedefaultmidpunct}
{\mcitedefaultendpunct}{\mcitedefaultseppunct}\relax
\EndOfBibitem
\bibitem{Bonciani:2015sha}
R.~Bonciani {\em et~al.\/},
  \href{http://doi.org/10.1140/epjc/s10052-015-3793-y}{Eur. Phys. J. C}
  \href{http://doi.org/10.1140/epjc/s10052-015-3793-y}{{\bf 75}, 12, 581}
  (2015), \href{https://arxiv.org/abs/1508.03585}{[arXiv:1508.03585]}\relax
\mciteBstWouldAddEndPuncttrue
\mciteSetBstMidEndSepPunct{\mcitedefaultmidpunct}
{\mcitedefaultendpunct}{\mcitedefaultseppunct}\relax
\EndOfBibitem
\bibitem{Boughezal:2015dva}
R.~Boughezal {\em et~al.\/},
  \href{http://doi.org/10.1103/PhysRevLett.115.062002}{Phys. Rev. Lett.}
  \href{http://doi.org/10.1103/PhysRevLett.115.062002}{{\bf 115}, 6, 062002}
  (2015), \href{https://arxiv.org/abs/1504.02131}{[arXiv:1504.02131]}\relax
\mciteBstWouldAddEndPuncttrue
\mciteSetBstMidEndSepPunct{\mcitedefaultmidpunct}
{\mcitedefaultendpunct}{\mcitedefaultseppunct}\relax
\EndOfBibitem
\bibitem{Boughezal:2015eha}
R.~Boughezal, X.~Liu and F.~Petriello,
  \href{http://doi.org/10.1103/PhysRevD.91.094035}{Phys. Rev. D}
  \href{http://doi.org/10.1103/PhysRevD.91.094035}{{\bf 91}, 9, 094035} (2015),
  \href{https://arxiv.org/abs/1504.02540}{[arXiv:1504.02540]}\relax
\mciteBstWouldAddEndPuncttrue
\mciteSetBstMidEndSepPunct{\mcitedefaultmidpunct}
{\mcitedefaultendpunct}{\mcitedefaultseppunct}\relax
\EndOfBibitem
\bibitem{Gaunt:2015pea}
J.~Gaunt {\em et~al.\/}, \href{http://doi.org/10.1007/JHEP09(2015)058}{JHEP}
  \href{http://doi.org/10.1007/JHEP09(2015)058}{{\bf 09}, 058} (2015),
  \href{https://arxiv.org/abs/1505.04794}{[arXiv:1505.04794]}\relax
\mciteBstWouldAddEndPuncttrue
\mciteSetBstMidEndSepPunct{\mcitedefaultmidpunct}
{\mcitedefaultendpunct}{\mcitedefaultseppunct}\relax
\EndOfBibitem
\bibitem{Cacciari:2015jma}
M.~Cacciari {\em et~al.\/},
  \href{http://doi.org/10.1103/PhysRevLett.115.082002}{Phys. Rev. Lett.}
  \href{http://doi.org/10.1103/PhysRevLett.115.082002}{{\bf 115}, 8, 082002}
  (2015), [Erratum: Phys. Rev. Lett.120,no.13,139901(2018)],
  \href{https://arxiv.org/abs/1506.02660}{[arXiv:1506.02660]}\relax
\mciteBstWouldAddEndPuncttrue
\mciteSetBstMidEndSepPunct{\mcitedefaultmidpunct}
{\mcitedefaultendpunct}{\mcitedefaultseppunct}\relax
\EndOfBibitem
\bibitem{Capatti:2022tit}
Z.~Capatti, V.~Hirschi and B.~Ruijl,
  \href{http://doi.org/10.1007/JHEP10(2022)120}{JHEP}
  \href{http://doi.org/10.1007/JHEP10(2022)120}{{\bf 10}, 120} (2022),
  \href{https://arxiv.org/abs/2203.11038}{[arXiv:2203.11038]}\relax
\mciteBstWouldAddEndPuncttrue
\mciteSetBstMidEndSepPunct{\mcitedefaultmidpunct}
{\mcitedefaultendpunct}{\mcitedefaultseppunct}\relax
\EndOfBibitem
\bibitem{Cieri:2020ikq}
L.~Cieri {\em et~al.\/}, \href{http://doi.org/10.1007/JHEP09(2020)155}{JHEP}
  \href{http://doi.org/10.1007/JHEP09(2020)155}{{\bf 09}, 155} (2020),
  \href{https://arxiv.org/abs/2005.01315}{[arXiv:2005.01315]}\relax
\mciteBstWouldAddEndPuncttrue
\mciteSetBstMidEndSepPunct{\mcitedefaultmidpunct}
{\mcitedefaultendpunct}{\mcitedefaultseppunct}\relax
\EndOfBibitem
\bibitem{TorresBobadilla:2020ekr}
W.~J. Torres~Bobadilla {\em et~al.\/},
  \href{http://doi.org/10.1140/epjc/s10052-021-08996-y}{Eur. Phys. J. C}
  \href{http://doi.org/10.1140/epjc/s10052-021-08996-y}{{\bf 81}, 3, 250}
  (2021), \href{https://arxiv.org/abs/2012.02567}{[arXiv:2012.02567]}\relax
\mciteBstWouldAddEndPuncttrue
\mciteSetBstMidEndSepPunct{\mcitedefaultmidpunct}
{\mcitedefaultendpunct}{\mcitedefaultseppunct}\relax
\EndOfBibitem
\bibitem{Gehrmann-DeRidder:2007nzq}
A.~Gehrmann-De~Ridder {\em et~al.\/},
  \href{http://doi.org/10.1103/PhysRevLett.99.132002}{Phys. Rev. Lett.}
  \href{http://doi.org/10.1103/PhysRevLett.99.132002}{{\bf 99}, 132002} (2007),
  \href{https://arxiv.org/abs/0707.1285}{[arXiv:0707.1285]}\relax
\mciteBstWouldAddEndPuncttrue
\mciteSetBstMidEndSepPunct{\mcitedefaultmidpunct}
{\mcitedefaultendpunct}{\mcitedefaultseppunct}\relax
\EndOfBibitem
\bibitem{GehrmannDeRidder:2007hr}
A.~Gehrmann-De~Ridder {\em et~al.\/},
  \href{http://doi.org/10.1088/1126-6708/2007/12/094}{JHEP}
  \href{http://doi.org/10.1088/1126-6708/2007/12/094}{{\bf 12}, 094} (2007),
  \href{https://arxiv.org/abs/0711.4711}{[arXiv:0711.4711]}\relax
\mciteBstWouldAddEndPuncttrue
\mciteSetBstMidEndSepPunct{\mcitedefaultmidpunct}
{\mcitedefaultendpunct}{\mcitedefaultseppunct}\relax
\EndOfBibitem
\bibitem{GehrmannDeRidder:2008ug}
A.~Gehrmann-De~Ridder {\em et~al.\/},
  \href{http://doi.org/10.1103/PhysRevLett.100.172001}{Phys. Rev. Lett.}
  \href{http://doi.org/10.1103/PhysRevLett.100.172001}{{\bf 100}, 172001}
  (2008), \href{https://arxiv.org/abs/0802.0813}{[arXiv:0802.0813]}\relax
\mciteBstWouldAddEndPuncttrue
\mciteSetBstMidEndSepPunct{\mcitedefaultmidpunct}
{\mcitedefaultendpunct}{\mcitedefaultseppunct}\relax
\EndOfBibitem
\bibitem{Gehrmann-DeRidder:2014hxk}
A.~Gehrmann-De~Ridder {\em et~al.\/},
  \href{http://doi.org/10.1016/j.cpc.2014.07.024}{Comput. Phys. Commun.}
  \href{http://doi.org/10.1016/j.cpc.2014.07.024}{{\bf 185}, 3331} (2014),
  \href{https://arxiv.org/abs/1402.4140}{[arXiv:1402.4140]}\relax
\mciteBstWouldAddEndPuncttrue
\mciteSetBstMidEndSepPunct{\mcitedefaultmidpunct}
{\mcitedefaultendpunct}{\mcitedefaultseppunct}\relax
\EndOfBibitem
\bibitem{qcd:Gehrmann-DeRidder:2014hxk-url}
\url{https://eerad3.hepforge.org/}\relax
\mciteBstWouldAddEndPuncttrue
\mciteSetBstMidEndSepPunct{\mcitedefaultmidpunct}
{\mcitedefaultendpunct}{\mcitedefaultseppunct}\relax
\EndOfBibitem
\bibitem{Weinzierl:2008iv}
S.~Weinzierl, \href{http://doi.org/10.1103/PhysRevLett.101.162001}{Phys. Rev.
  Lett.} \href{http://doi.org/10.1103/PhysRevLett.101.162001}{{\bf 101},
  162001} (2008),
  \href{https://arxiv.org/abs/0807.3241}{[arXiv:0807.3241]}\relax
\mciteBstWouldAddEndPuncttrue
\mciteSetBstMidEndSepPunct{\mcitedefaultmidpunct}
{\mcitedefaultendpunct}{\mcitedefaultseppunct}\relax
\EndOfBibitem
\bibitem{Weinzierl:2009ms}
S.~Weinzierl, \href{http://doi.org/10.1088/1126-6708/2009/06/041}{JHEP}
  \href{http://doi.org/10.1088/1126-6708/2009/06/041}{{\bf 06}, 041} (2009),
  \href{https://arxiv.org/abs/0904.1077}{[arXiv:0904.1077]}\relax
\mciteBstWouldAddEndPuncttrue
\mciteSetBstMidEndSepPunct{\mcitedefaultmidpunct}
{\mcitedefaultendpunct}{\mcitedefaultseppunct}\relax
\EndOfBibitem
\bibitem{DelDuca:2016ily}
V.~Del~Duca {\em et~al.\/},
  \href{http://doi.org/10.1103/PhysRevD.94.074019}{Phys. Rev. D}
  \href{http://doi.org/10.1103/PhysRevD.94.074019}{{\bf 94}, 7, 074019} (2016),
  \href{https://arxiv.org/abs/1606.03453}{[arXiv:1606.03453]}\relax
\mciteBstWouldAddEndPuncttrue
\mciteSetBstMidEndSepPunct{\mcitedefaultmidpunct}
{\mcitedefaultendpunct}{\mcitedefaultseppunct}\relax
\EndOfBibitem
\bibitem{DelDuca:2016csb}
V.~Del~Duca {\em et~al.\/},
  \href{http://doi.org/10.1103/PhysRevLett.117.152004}{Phys. Rev. Lett.}
  \href{http://doi.org/10.1103/PhysRevLett.117.152004}{{\bf 117}, 15, 152004}
  (2016), \href{https://arxiv.org/abs/1603.08927}{[arXiv:1603.08927]}\relax
\mciteBstWouldAddEndPuncttrue
\mciteSetBstMidEndSepPunct{\mcitedefaultmidpunct}
{\mcitedefaultendpunct}{\mcitedefaultseppunct}\relax
\EndOfBibitem
\bibitem{Currie:2016ytq}
J.~Currie, T.~Gehrmann and J.~Niehues,
  \href{http://doi.org/10.1103/PhysRevLett.117.042001}{Phys. Rev. Lett.}
  \href{http://doi.org/10.1103/PhysRevLett.117.042001}{{\bf 117}, 4, 042001}
  (2016), \href{https://arxiv.org/abs/1606.03991}{[arXiv:1606.03991]}\relax
\mciteBstWouldAddEndPuncttrue
\mciteSetBstMidEndSepPunct{\mcitedefaultmidpunct}
{\mcitedefaultendpunct}{\mcitedefaultseppunct}\relax
\EndOfBibitem
\bibitem{Currie:2017tpe}
J.~Currie {\em et~al.\/}, \href{http://doi.org/10.1007/JHEP07(2017)018}{JHEP}
  \href{http://doi.org/10.1007/JHEP07(2017)018}{{\bf 07}, 018} (2017),
  [Erratum: JHEP 12, 042 (2020)],
  \href{https://arxiv.org/abs/1703.05977}{[arXiv:1703.05977]}\relax
\mciteBstWouldAddEndPuncttrue
\mciteSetBstMidEndSepPunct{\mcitedefaultmidpunct}
{\mcitedefaultendpunct}{\mcitedefaultseppunct}\relax
\EndOfBibitem
\bibitem{Currie:2018fgr}
J.~Currie {\em et~al.\/}, \href{http://doi.org/10.1007/JHEP05(2018)209}{JHEP}
  \href{http://doi.org/10.1007/JHEP05(2018)209}{{\bf 05}, 209} (2018),
  \href{https://arxiv.org/abs/1803.09973}{[arXiv:1803.09973]}\relax
\mciteBstWouldAddEndPuncttrue
\mciteSetBstMidEndSepPunct{\mcitedefaultmidpunct}
{\mcitedefaultendpunct}{\mcitedefaultseppunct}\relax
\EndOfBibitem
\bibitem{Gehrmann:2018odt}
T.~Gehrmann {\em et~al.\/},
  \href{http://doi.org/10.1016/j.physletb.2019.03.003}{Phys. Lett.}
  \href{http://doi.org/10.1016/j.physletb.2019.03.003}{{\bf B792}, 182} (2019),
  \href{https://arxiv.org/abs/1812.06104}{[arXiv:1812.06104]}\relax
\mciteBstWouldAddEndPuncttrue
\mciteSetBstMidEndSepPunct{\mcitedefaultmidpunct}
{\mcitedefaultendpunct}{\mcitedefaultseppunct}\relax
\EndOfBibitem
\bibitem{Melnikov:2006kv}
K.~Melnikov and F.~Petriello,
  \href{http://doi.org/10.1103/PhysRevD.74.114017}{Phys. Rev.}
  \href{http://doi.org/10.1103/PhysRevD.74.114017}{{\bf D74}, 114017} (2006),
  \href{https://arxiv.org/abs/hep-ph/0609070}{[hep-ph/0609070]}\relax
\mciteBstWouldAddEndPuncttrue
\mciteSetBstMidEndSepPunct{\mcitedefaultmidpunct}
{\mcitedefaultendpunct}{\mcitedefaultseppunct}\relax
\EndOfBibitem
\bibitem{qcd:Melnikov:2016kv-url}
\url{http://gate.hep.anl.gov/fpetriello/FEWZ.html}\relax
\mciteBstWouldAddEndPuncttrue
\mciteSetBstMidEndSepPunct{\mcitedefaultmidpunct}
{\mcitedefaultendpunct}{\mcitedefaultseppunct}\relax
\EndOfBibitem
\bibitem{Catani:2009sm}
S.~Catani {\em et~al.\/},
  \href{http://doi.org/10.1103/PhysRevLett.103.082001}{Phys. Rev. Lett.}
  \href{http://doi.org/10.1103/PhysRevLett.103.082001}{{\bf 103}, 082001}
  (2009), \href{https://arxiv.org/abs/0903.2120}{[arXiv:0903.2120]}\relax
\mciteBstWouldAddEndPuncttrue
\mciteSetBstMidEndSepPunct{\mcitedefaultmidpunct}
{\mcitedefaultendpunct}{\mcitedefaultseppunct}\relax
\EndOfBibitem
\bibitem{qcd:Catani:2009sm-url}
\url{http://theory.fi.infn.it/gazzani/dy.html}\relax
\mciteBstWouldAddEndPuncttrue
\mciteSetBstMidEndSepPunct{\mcitedefaultmidpunct}
{\mcitedefaultendpunct}{\mcitedefaultseppunct}\relax
\EndOfBibitem
\bibitem{Anastasiou:2005qj}
C.~Anastasiou, K.~Melnikov and F.~Petriello,
  \href{http://doi.org/10.1016/j.nuclphysb.2005.06.036}{Nucl. Phys.}
  \href{http://doi.org/10.1016/j.nuclphysb.2005.06.036}{{\bf B724}, 197}
  (2005), \href{https://arxiv.org/abs/hep-ph/0501130}{[hep-ph/0501130]}\relax
\mciteBstWouldAddEndPuncttrue
\mciteSetBstMidEndSepPunct{\mcitedefaultmidpunct}
{\mcitedefaultendpunct}{\mcitedefaultseppunct}\relax
\EndOfBibitem
\bibitem{qcd:Anastasiou:2005qj-url}
\url{http://www.phys.ethz.ch/~pheno/fehipro/}\relax
\mciteBstWouldAddEndPuncttrue
\mciteSetBstMidEndSepPunct{\mcitedefaultmidpunct}
{\mcitedefaultendpunct}{\mcitedefaultseppunct}\relax
\EndOfBibitem
\bibitem{Czakon:2021yub}
M.~Czakon {\em et~al.\/}  (2021),
  \href{https://arxiv.org/abs/2105.04436}{[arXiv:2105.04436]}\relax
\mciteBstWouldAddEndPuncttrue
\mciteSetBstMidEndSepPunct{\mcitedefaultmidpunct}
{\mcitedefaultendpunct}{\mcitedefaultseppunct}\relax
\EndOfBibitem
\bibitem{Grazzini:2017ckn}
M.~Grazzini {\em et~al.\/}, \href{http://doi.org/10.1007/JHEP05(2017)139}{JHEP}
  \href{http://doi.org/10.1007/JHEP05(2017)139}{{\bf 05}, 139} (2017),
  \href{https://arxiv.org/abs/1703.09065}{[arXiv:1703.09065]}\relax
\mciteBstWouldAddEndPuncttrue
\mciteSetBstMidEndSepPunct{\mcitedefaultmidpunct}
{\mcitedefaultendpunct}{\mcitedefaultseppunct}\relax
\EndOfBibitem
\bibitem{Catani:2011qz}
S.~Catani {\em et~al.\/},
  \href{http://doi.org/10.1103/PhysRevLett.108.072001}{Phys. Rev. Lett.}
  \href{http://doi.org/10.1103/PhysRevLett.108.072001}{{\bf 108}, 072001}
  (2012), [Erratum: Phys. Rev. Lett.117,no.8,089901(2016)],
  \href{https://arxiv.org/abs/1110.2375}{[arXiv:1110.2375]}\relax
\mciteBstWouldAddEndPuncttrue
\mciteSetBstMidEndSepPunct{\mcitedefaultmidpunct}
{\mcitedefaultendpunct}{\mcitedefaultseppunct}\relax
\EndOfBibitem
\bibitem{Campbell:2016yrh}
J.~M. Campbell {\em et~al.\/},
  \href{http://doi.org/10.1007/JHEP07(2016)148}{JHEP}
  \href{http://doi.org/10.1007/JHEP07(2016)148}{{\bf 07}, 148} (2016),
  \href{https://arxiv.org/abs/1603.02663}{[arXiv:1603.02663]}\relax
\mciteBstWouldAddEndPuncttrue
\mciteSetBstMidEndSepPunct{\mcitedefaultmidpunct}
{\mcitedefaultendpunct}{\mcitedefaultseppunct}\relax
\EndOfBibitem
\bibitem{Grazzini:2013bna}
M.~Grazzini {\em et~al.\/},
  \href{http://doi.org/10.1016/j.physletb.2014.02.037}{Phys. Lett.}
  \href{http://doi.org/10.1016/j.physletb.2014.02.037}{{\bf B731}, 204} (2014),
  \href{https://arxiv.org/abs/1309.7000}{[arXiv:1309.7000]}\relax
\mciteBstWouldAddEndPuncttrue
\mciteSetBstMidEndSepPunct{\mcitedefaultmidpunct}
{\mcitedefaultendpunct}{\mcitedefaultseppunct}\relax
\EndOfBibitem
\bibitem{Grazzini:2015nwa}
M.~Grazzini, S.~Kallweit and D.~Rathlev,
  \href{http://doi.org/10.1007/JHEP07(2015)085}{JHEP}
  \href{http://doi.org/10.1007/JHEP07(2015)085}{{\bf 07}, 085} (2015),
  \href{https://arxiv.org/abs/1504.01330}{[arXiv:1504.01330]}\relax
\mciteBstWouldAddEndPuncttrue
\mciteSetBstMidEndSepPunct{\mcitedefaultmidpunct}
{\mcitedefaultendpunct}{\mcitedefaultseppunct}\relax
\EndOfBibitem
\bibitem{Campbell:1999ah}
J.~M. Campbell and R.~K. Ellis,
  \href{http://doi.org/10.1103/PhysRevD.60.113006}{Phys. Rev. D}
  \href{http://doi.org/10.1103/PhysRevD.60.113006}{{\bf 60}, 113006} (1999),
  \href{https://arxiv.org/abs/hep-ph/9905386}{[hep-ph/9905386]}\relax
\mciteBstWouldAddEndPuncttrue
\mciteSetBstMidEndSepPunct{\mcitedefaultmidpunct}
{\mcitedefaultendpunct}{\mcitedefaultseppunct}\relax
\EndOfBibitem
\bibitem{Boughezal:2016wmq}
R.~Boughezal {\em et~al.\/},
  \href{http://doi.org/10.1140/epjc/s10052-016-4558-y}{Eur. Phys. J.}
  \href{http://doi.org/10.1140/epjc/s10052-016-4558-y}{{\bf C77}, 1, 7} (2017),
  \href{https://arxiv.org/abs/1605.08011}{[arXiv:1605.08011]}\relax
\mciteBstWouldAddEndPuncttrue
\mciteSetBstMidEndSepPunct{\mcitedefaultmidpunct}
{\mcitedefaultendpunct}{\mcitedefaultseppunct}\relax
\EndOfBibitem
\bibitem{Campbell:2016lzl}
J.~M. Campbell, R.~K. Ellis and C.~Williams,
  \href{http://doi.org/10.1103/PhysRevLett.118.222001}{Phys. Rev. Lett.}
  \href{http://doi.org/10.1103/PhysRevLett.118.222001}{{\bf 118}, 22, 222001}
  (2017), \href{https://arxiv.org/abs/1612.04333}{[arXiv:1612.04333]}\relax
\mciteBstWouldAddEndPuncttrue
\mciteSetBstMidEndSepPunct{\mcitedefaultmidpunct}
{\mcitedefaultendpunct}{\mcitedefaultseppunct}\relax
\EndOfBibitem
\bibitem{Chen:2019zmr}
X.~Chen {\em et~al.\/}, Submitted to: J. High Energy Phys.  (2019),
  \href{https://arxiv.org/abs/1904.01044}{[arXiv:1904.01044]}\relax
\mciteBstWouldAddEndPuncttrue
\mciteSetBstMidEndSepPunct{\mcitedefaultmidpunct}
{\mcitedefaultendpunct}{\mcitedefaultseppunct}\relax
\EndOfBibitem
\bibitem{Campbell:2017dqk}
J.~M. Campbell, R.~K. Ellis and C.~Williams,
  \href{http://doi.org/10.1103/PhysRevD.96.014037}{Phys. Rev.}
  \href{http://doi.org/10.1103/PhysRevD.96.014037}{{\bf D96}, 1, 014037}
  (2017), \href{https://arxiv.org/abs/1703.10109}{[arXiv:1703.10109]}\relax
\mciteBstWouldAddEndPuncttrue
\mciteSetBstMidEndSepPunct{\mcitedefaultmidpunct}
{\mcitedefaultendpunct}{\mcitedefaultseppunct}\relax
\EndOfBibitem
\bibitem{Gehrmann-DeRidder:2017mvr}
A.~Gehrmann-De~Ridder {\em et~al.\/},
  \href{http://doi.org/10.1103/PhysRevLett.120.122001}{Phys. Rev. Lett.}
  \href{http://doi.org/10.1103/PhysRevLett.120.122001}{{\bf 120}, 12, 122001}
  (2018), \href{https://arxiv.org/abs/1712.07543}{[arXiv:1712.07543]}\relax
\mciteBstWouldAddEndPuncttrue
\mciteSetBstMidEndSepPunct{\mcitedefaultmidpunct}
{\mcitedefaultendpunct}{\mcitedefaultseppunct}\relax
\EndOfBibitem
\bibitem{Ridder:2015dxa}
A.~Gehrmann-De~Ridder {\em et~al.\/},
  \href{http://doi.org/10.1103/PhysRevLett.117.022001}{Phys. Rev. Lett.}
  \href{http://doi.org/10.1103/PhysRevLett.117.022001}{{\bf 117}, 2, 022001}
  (2016), \href{https://arxiv.org/abs/1507.02850}{[arXiv:1507.02850]}\relax
\mciteBstWouldAddEndPuncttrue
\mciteSetBstMidEndSepPunct{\mcitedefaultmidpunct}
{\mcitedefaultendpunct}{\mcitedefaultseppunct}\relax
\EndOfBibitem
\bibitem{Boughezal:2015ded}
R.~Boughezal {\em et~al.\/},
  \href{http://doi.org/10.1103/PhysRevLett.116.152001}{Phys. Rev. Lett.}
  \href{http://doi.org/10.1103/PhysRevLett.116.152001}{{\bf 116}, 15, 152001}
  (2016), \href{https://arxiv.org/abs/1512.01291}{[arXiv:1512.01291]}\relax
\mciteBstWouldAddEndPuncttrue
\mciteSetBstMidEndSepPunct{\mcitedefaultmidpunct}
{\mcitedefaultendpunct}{\mcitedefaultseppunct}\relax
\EndOfBibitem
\bibitem{Boughezal:2015dra}
R.~Boughezal {\em et~al.\/},
  \href{http://doi.org/10.1103/PhysRevLett.115.082003}{Phys. Rev. Lett.}
  \href{http://doi.org/10.1103/PhysRevLett.115.082003}{{\bf 115}, 8, 082003}
  (2015), \href{https://arxiv.org/abs/1504.07922}{[arXiv:1504.07922]}\relax
\mciteBstWouldAddEndPuncttrue
\mciteSetBstMidEndSepPunct{\mcitedefaultmidpunct}
{\mcitedefaultendpunct}{\mcitedefaultseppunct}\relax
\EndOfBibitem
\bibitem{Boughezal:2015aha}
R.~Boughezal {\em et~al.\/},
  \href{http://doi.org/10.1016/j.physletb.2015.06.055}{Phys. Lett.}
  \href{http://doi.org/10.1016/j.physletb.2015.06.055}{{\bf B748}, 5} (2015),
  \href{https://arxiv.org/abs/1505.03893}{[arXiv:1505.03893]}\relax
\mciteBstWouldAddEndPuncttrue
\mciteSetBstMidEndSepPunct{\mcitedefaultmidpunct}
{\mcitedefaultendpunct}{\mcitedefaultseppunct}\relax
\EndOfBibitem
\bibitem{Caola:2015wna}
F.~Caola, K.~Melnikov and M.~Schulze,
  \href{http://doi.org/10.1103/PhysRevD.92.074032}{Phys. Rev.}
  \href{http://doi.org/10.1103/PhysRevD.92.074032}{{\bf D92}, 7, 074032}
  (2015), \href{https://arxiv.org/abs/1508.02684}{[arXiv:1508.02684]}\relax
\mciteBstWouldAddEndPuncttrue
\mciteSetBstMidEndSepPunct{\mcitedefaultmidpunct}
{\mcitedefaultendpunct}{\mcitedefaultseppunct}\relax
\EndOfBibitem
\bibitem{Chen:2016zka}
X.~Chen {\em et~al.\/}, \href{http://doi.org/10.1007/JHEP10(2016)066}{JHEP}
  \href{http://doi.org/10.1007/JHEP10(2016)066}{{\bf 10}, 066} (2016),
  \href{https://arxiv.org/abs/1607.08817}{[arXiv:1607.08817]}\relax
\mciteBstWouldAddEndPuncttrue
\mciteSetBstMidEndSepPunct{\mcitedefaultmidpunct}
{\mcitedefaultendpunct}{\mcitedefaultseppunct}\relax
\EndOfBibitem
\bibitem{Ferrera:2011bk}
G.~Ferrera, M.~Grazzini and F.~Tramontano,
  \href{http://doi.org/10.1103/PhysRevLett.107.152003}{Phys. Rev. Lett.}
  \href{http://doi.org/10.1103/PhysRevLett.107.152003}{{\bf 107}, 152003}
  (2011), \href{https://arxiv.org/abs/1107.1164}{[arXiv:1107.1164]}\relax
\mciteBstWouldAddEndPuncttrue
\mciteSetBstMidEndSepPunct{\mcitedefaultmidpunct}
{\mcitedefaultendpunct}{\mcitedefaultseppunct}\relax
\EndOfBibitem
\bibitem{Ferrera:2014lca}
G.~Ferrera, M.~Grazzini and F.~Tramontano,
  \href{http://doi.org/10.1016/j.physletb.2014.11.040}{Phys. Lett.}
  \href{http://doi.org/10.1016/j.physletb.2014.11.040}{{\bf B740}, 51} (2015),
  \href{https://arxiv.org/abs/1407.4747}{[arXiv:1407.4747]}\relax
\mciteBstWouldAddEndPuncttrue
\mciteSetBstMidEndSepPunct{\mcitedefaultmidpunct}
{\mcitedefaultendpunct}{\mcitedefaultseppunct}\relax
\EndOfBibitem
\bibitem{Liu:2018gxa}
Z.~L. Liu and J.~Gao, \href{http://doi.org/10.1103/PhysRevD.98.071501}{Phys.
  Rev. D} \href{http://doi.org/10.1103/PhysRevD.98.071501}{{\bf 98}, 7, 071501}
  (2018), \href{https://arxiv.org/abs/1807.03835}{[arXiv:1807.03835]}\relax
\mciteBstWouldAddEndPuncttrue
\mciteSetBstMidEndSepPunct{\mcitedefaultmidpunct}
{\mcitedefaultendpunct}{\mcitedefaultseppunct}\relax
\EndOfBibitem
\bibitem{Brucherseifer:2014ama}
M.~Brucherseifer, F.~Caola and K.~Melnikov,
  \href{http://doi.org/10.1016/j.physletb.2014.06.075}{Phys. Lett.}
  \href{http://doi.org/10.1016/j.physletb.2014.06.075}{{\bf B736}, 58} (2014),
  \href{https://arxiv.org/abs/1404.7116}{[arXiv:1404.7116]}\relax
\mciteBstWouldAddEndPuncttrue
\mciteSetBstMidEndSepPunct{\mcitedefaultmidpunct}
{\mcitedefaultendpunct}{\mcitedefaultseppunct}\relax
\EndOfBibitem
\bibitem{Berger:2016oht}
E.~L. Berger {\em et~al.\/},
  \href{http://doi.org/10.1103/PhysRevD.94.071501}{Phys. Rev.}
  \href{http://doi.org/10.1103/PhysRevD.94.071501}{{\bf D94}, 7, 071501}
  (2016), \href{https://arxiv.org/abs/1606.08463}{[arXiv:1606.08463]}\relax
\mciteBstWouldAddEndPuncttrue
\mciteSetBstMidEndSepPunct{\mcitedefaultmidpunct}
{\mcitedefaultendpunct}{\mcitedefaultseppunct}\relax
\EndOfBibitem
\bibitem{Berger:2017zof}
E.~L. Berger, J.~Gao and H.~X. Zhu,
  \href{http://doi.org/10.1007/JHEP11(2017)158}{JHEP}
  \href{http://doi.org/10.1007/JHEP11(2017)158}{{\bf 11}, 158} (2017),
  \href{https://arxiv.org/abs/1708.09405}{[arXiv:1708.09405]}\relax
\mciteBstWouldAddEndPuncttrue
\mciteSetBstMidEndSepPunct{\mcitedefaultmidpunct}
{\mcitedefaultendpunct}{\mcitedefaultseppunct}\relax
\EndOfBibitem
\bibitem{Campbell:2020fhf}
J.~Campbell, T.~Neumann and Z.~Sullivan,
  \href{http://doi.org/10.1007/JHEP02(2021)040}{JHEP}
  \href{http://doi.org/10.1007/JHEP02(2021)040}{{\bf 02}, 040} (2021),
  \href{https://arxiv.org/abs/2012.01574}{[arXiv:2012.01574]}\relax
\mciteBstWouldAddEndPuncttrue
\mciteSetBstMidEndSepPunct{\mcitedefaultmidpunct}
{\mcitedefaultendpunct}{\mcitedefaultseppunct}\relax
\EndOfBibitem
\bibitem{Czakon:2014xsa}
M.~Czakon, P.~Fiedler and A.~Mitov,
  \href{http://doi.org/10.1103/PhysRevLett.115.052001}{Phys. Rev. Lett.}
  \href{http://doi.org/10.1103/PhysRevLett.115.052001}{{\bf 115}, 5, 052001}
  (2015), \href{https://arxiv.org/abs/1411.3007}{[arXiv:1411.3007]}\relax
\mciteBstWouldAddEndPuncttrue
\mciteSetBstMidEndSepPunct{\mcitedefaultmidpunct}
{\mcitedefaultendpunct}{\mcitedefaultseppunct}\relax
\EndOfBibitem
\bibitem{Czakon:2016ckf}
M.~Czakon {\em et~al.\/}, \href{http://doi.org/10.1007/JHEP05(2016)034}{JHEP}
  \href{http://doi.org/10.1007/JHEP05(2016)034}{{\bf 05}, 034} (2016),
  \href{https://arxiv.org/abs/1601.05375}{[arXiv:1601.05375]}\relax
\mciteBstWouldAddEndPuncttrue
\mciteSetBstMidEndSepPunct{\mcitedefaultmidpunct}
{\mcitedefaultendpunct}{\mcitedefaultseppunct}\relax
\EndOfBibitem
\bibitem{Currie:2016bfm}
J.~Currie, E.~W.~N. Glover and J.~Pires,
  \href{http://doi.org/10.1103/PhysRevLett.118.072002}{Phys. Rev. Lett.}
  \href{http://doi.org/10.1103/PhysRevLett.118.072002}{{\bf 118}, 7, 072002}
  (2017), \href{https://arxiv.org/abs/1611.01460}{[arXiv:1611.01460]}\relax
\mciteBstWouldAddEndPuncttrue
\mciteSetBstMidEndSepPunct{\mcitedefaultmidpunct}
{\mcitedefaultendpunct}{\mcitedefaultseppunct}\relax
\EndOfBibitem
\bibitem{Chen:2022tpk}
X.~Chen {\em et~al.\/}, \href{http://doi.org/10.1007/JHEP09(2022)025}{JHEP}
  \href{http://doi.org/10.1007/JHEP09(2022)025}{{\bf 09}, 025} (2022),
  \href{https://arxiv.org/abs/2204.10173}{[arXiv:2204.10173]}\relax
\mciteBstWouldAddEndPuncttrue
\mciteSetBstMidEndSepPunct{\mcitedefaultmidpunct}
{\mcitedefaultendpunct}{\mcitedefaultseppunct}\relax
\EndOfBibitem
\bibitem{Czakon:2020coa}
M.~Czakon {\em et~al.\/}  (2020),
  \href{https://arxiv.org/abs/2011.01011}{[arXiv:2011.01011]}\relax
\mciteBstWouldAddEndPuncttrue
\mciteSetBstMidEndSepPunct{\mcitedefaultmidpunct}
{\mcitedefaultendpunct}{\mcitedefaultseppunct}\relax
\EndOfBibitem
\bibitem{Czakon:2022khx}
M.~Czakon {\em et~al.\/}, \href{http://doi.org/10.1007/JHEP02(2023)241}{JHEP}
  \href{http://doi.org/10.1007/JHEP02(2023)241}{{\bf 02}, 241} (2023),
  \href{https://arxiv.org/abs/2212.00467}{[arXiv:2212.00467]}\relax
\mciteBstWouldAddEndPuncttrue
\mciteSetBstMidEndSepPunct{\mcitedefaultmidpunct}
{\mcitedefaultendpunct}{\mcitedefaultseppunct}\relax
\EndOfBibitem
\bibitem{Czakon:2021ohs}
M.~L. Czakon {\em et~al.\/}  (2021),
  \href{https://arxiv.org/abs/2102.08267}{[arXiv:2102.08267]}\relax
\mciteBstWouldAddEndPuncttrue
\mciteSetBstMidEndSepPunct{\mcitedefaultmidpunct}
{\mcitedefaultendpunct}{\mcitedefaultseppunct}\relax
\EndOfBibitem
\bibitem{Cruz-Martinez:2018rod}
J.~Cruz-Martinez {\em et~al.\/},
  \href{http://doi.org/10.1016/j.physletb.2018.04.046}{Phys. Lett.}
  \href{http://doi.org/10.1016/j.physletb.2018.04.046}{{\bf B781}, 672} (2018),
  \href{https://arxiv.org/abs/1802.02445}{[arXiv:1802.02445]}\relax
\mciteBstWouldAddEndPuncttrue
\mciteSetBstMidEndSepPunct{\mcitedefaultmidpunct}
{\mcitedefaultendpunct}{\mcitedefaultseppunct}\relax
\EndOfBibitem
\bibitem{Liu:2019tuy}
T.~Liu, K.~Melnikov and A.~A. Penin  (2019),
  \href{https://arxiv.org/abs/1906.10899}{[arXiv:1906.10899]}\relax
\mciteBstWouldAddEndPuncttrue
\mciteSetBstMidEndSepPunct{\mcitedefaultmidpunct}
{\mcitedefaultendpunct}{\mcitedefaultseppunct}\relax
\EndOfBibitem
\bibitem{Chawdhry:2019bji}
H.~A. Chawdhry {\em et~al.\/},
  \href{http://doi.org/10.1007/JHEP02(2020)057}{JHEP}
  \href{http://doi.org/10.1007/JHEP02(2020)057}{{\bf 02}, 057} (2020),
  \href{https://arxiv.org/abs/1911.00479}{[arXiv:1911.00479]}\relax
\mciteBstWouldAddEndPuncttrue
\mciteSetBstMidEndSepPunct{\mcitedefaultmidpunct}
{\mcitedefaultendpunct}{\mcitedefaultseppunct}\relax
\EndOfBibitem
\bibitem{Kallweit:2020gcp}
S.~Kallweit, V.~Sotnikov and M.~Wiesemann,
  \href{http://doi.org/10.1016/j.physletb.2020.136013}{Phys. Lett. B}
  \href{http://doi.org/10.1016/j.physletb.2020.136013}{{\bf 812}, 136013}
  (2021), \href{https://arxiv.org/abs/2010.04681}{[arXiv:2010.04681]}\relax
\mciteBstWouldAddEndPuncttrue
\mciteSetBstMidEndSepPunct{\mcitedefaultmidpunct}
{\mcitedefaultendpunct}{\mcitedefaultseppunct}\relax
\EndOfBibitem
\bibitem{Chawdhry:2021hkp}
H.~A. Chawdhry {\em et~al.\/},
  \href{http://doi.org/10.1007/JHEP09(2021)093}{JHEP}
  \href{http://doi.org/10.1007/JHEP09(2021)093}{{\bf 09}, 093} (2021),
  \href{https://arxiv.org/abs/2105.06940}{[arXiv:2105.06940]}\relax
\mciteBstWouldAddEndPuncttrue
\mciteSetBstMidEndSepPunct{\mcitedefaultmidpunct}
{\mcitedefaultendpunct}{\mcitedefaultseppunct}\relax
\EndOfBibitem
\bibitem{Badger:2023mgf}
S.~Badger {\em et~al.\/}  (2023),
  \href{https://arxiv.org/abs/2304.06682}{[arXiv:2304.06682]}\relax
\mciteBstWouldAddEndPuncttrue
\mciteSetBstMidEndSepPunct{\mcitedefaultmidpunct}
{\mcitedefaultendpunct}{\mcitedefaultseppunct}\relax
\EndOfBibitem
\bibitem{Czakon:2021mjy}
M.~Czakon, A.~Mitov and R.~Poncelet,
  \href{http://doi.org/10.1103/PhysRevLett.127.152001}{Phys. Rev. Lett.}
  \href{http://doi.org/10.1103/PhysRevLett.127.152001}{{\bf 127}, 15, 152001}
  (2021), [Erratum: Phys.Rev.Lett. 129, 119901 (2022)],
  \href{https://arxiv.org/abs/2106.05331}{[arXiv:2106.05331]}\relax
\mciteBstWouldAddEndPuncttrue
\mciteSetBstMidEndSepPunct{\mcitedefaultmidpunct}
{\mcitedefaultendpunct}{\mcitedefaultseppunct}\relax
\EndOfBibitem
\bibitem{Hartanto:2022qhh}
H.~B. Hartanto {\em et~al.\/},
  \href{http://doi.org/10.1103/PhysRevD.106.074016}{Phys. Rev. D}
  \href{http://doi.org/10.1103/PhysRevD.106.074016}{{\bf 106}, 7, 074016}
  (2022), \href{https://arxiv.org/abs/2205.01687}{[arXiv:2205.01687]}\relax
\mciteBstWouldAddEndPuncttrue
\mciteSetBstMidEndSepPunct{\mcitedefaultmidpunct}
{\mcitedefaultendpunct}{\mcitedefaultseppunct}\relax
\EndOfBibitem
\bibitem{Alvarez:2023fhi}
M.~Alvarez {\em et~al.\/}, \href{http://doi.org/10.1007/JHEP03(2023)129}{JHEP}
  \href{http://doi.org/10.1007/JHEP03(2023)129}{{\bf 03}, 129} (2023),
  \href{https://arxiv.org/abs/2301.01086}{[arXiv:2301.01086]}\relax
\mciteBstWouldAddEndPuncttrue
\mciteSetBstMidEndSepPunct{\mcitedefaultmidpunct}
{\mcitedefaultendpunct}{\mcitedefaultseppunct}\relax
\EndOfBibitem
\bibitem{Salam:2021tbm}
G.~P. Salam and E.~Slade, \href{http://doi.org/10.1007/JHEP11(2021)220}{JHEP}
  \href{http://doi.org/10.1007/JHEP11(2021)220}{{\bf 11}, 220} (2021),
  \href{https://arxiv.org/abs/2106.08329}{[arXiv:2106.08329]}\relax
\mciteBstWouldAddEndPuncttrue
\mciteSetBstMidEndSepPunct{\mcitedefaultmidpunct}
{\mcitedefaultendpunct}{\mcitedefaultseppunct}\relax
\EndOfBibitem
\bibitem{Ebert:2020dfc}
M.~A. Ebert {\em et~al.\/}, \href{http://doi.org/10.1007/JHEP04(2021)102}{JHEP}
  \href{http://doi.org/10.1007/JHEP04(2021)102}{{\bf 04}, 102} (2021),
  \href{https://arxiv.org/abs/2006.11382}{[arXiv:2006.11382]}\relax
\mciteBstWouldAddEndPuncttrue
\mciteSetBstMidEndSepPunct{\mcitedefaultmidpunct}
{\mcitedefaultendpunct}{\mcitedefaultseppunct}\relax
\EndOfBibitem
\bibitem{Amoroso:2020lgh}
S.~Amoroso {\em et~al.\/}, in \enquote{{11th Les Houches Workshop on Physics at
  TeV Colliders}: {PhysTeV Les Houches},}  (2020),
  \href{https://arxiv.org/abs/2003.01700}{[arXiv:2003.01700]}\relax
\mciteBstWouldAddEndPuncttrue
\mciteSetBstMidEndSepPunct{\mcitedefaultmidpunct}
{\mcitedefaultendpunct}{\mcitedefaultseppunct}\relax
\EndOfBibitem
\bibitem{Catani:2022mfv}
S.~Catani {\em et~al.\/},
  \href{http://doi.org/10.1103/PhysRevLett.130.111902}{Phys. Rev. Lett.}
  \href{http://doi.org/10.1103/PhysRevLett.130.111902}{{\bf 130}, 11, 111902}
  (2023), \href{https://arxiv.org/abs/2210.07846}{[arXiv:2210.07846]}\relax
\mciteBstWouldAddEndPuncttrue
\mciteSetBstMidEndSepPunct{\mcitedefaultmidpunct}
{\mcitedefaultendpunct}{\mcitedefaultseppunct}\relax
\EndOfBibitem
\bibitem{Camarda:2019zyx}
S.~Camarda {\em et~al.\/},
  \href{http://doi.org/10.1140/epjc/s10052-020-7757-5}{Eur. Phys. J. C}
  \href{http://doi.org/10.1140/epjc/s10052-020-7757-5}{{\bf 80}, 3, 251}
  (2020), [Erratum: Eur.Phys.J.C 80, 440 (2020)],
  \href{https://arxiv.org/abs/1910.07049}{[arXiv:1910.07049]}\relax
\mciteBstWouldAddEndPuncttrue
\mciteSetBstMidEndSepPunct{\mcitedefaultmidpunct}
{\mcitedefaultendpunct}{\mcitedefaultseppunct}\relax
\EndOfBibitem
\bibitem{Gavin:2010az}
R.~Gavin {\em et~al.\/},
  \href{http://doi.org/10.1016/j.cpc.2011.06.008}{Comput. Phys. Commun.}
  \href{http://doi.org/10.1016/j.cpc.2011.06.008}{{\bf 182}, 2388} (2011),
  \href{https://arxiv.org/abs/1011.3540}{[arXiv:1011.3540]}\relax
\mciteBstWouldAddEndPuncttrue
\mciteSetBstMidEndSepPunct{\mcitedefaultmidpunct}
{\mcitedefaultendpunct}{\mcitedefaultseppunct}\relax
\EndOfBibitem
\bibitem{Camarda:2022wti}
S.~Camarda {\em et~al.\/},
  \href{http://doi.org/10.1140/epjc/s10052-022-10436-4}{Eur. Phys. J. C}
  \href{http://doi.org/10.1140/epjc/s10052-022-10436-4}{{\bf 82}, 5, 492}
  (2022), \href{https://arxiv.org/abs/2202.10343}{[arXiv:2202.10343]}\relax
\mciteBstWouldAddEndPuncttrue
\mciteSetBstMidEndSepPunct{\mcitedefaultmidpunct}
{\mcitedefaultendpunct}{\mcitedefaultseppunct}\relax
\EndOfBibitem
\bibitem{Harlander:2012pb}
R.~V. Harlander, S.~Liebler and H.~Mantler,
  \href{http://doi.org/10.1016/j.cpc.2013.02.006}{Comput. Phys. Commun.}
  \href{http://doi.org/10.1016/j.cpc.2013.02.006}{{\bf 184}, 1605} (2013),
  \href{https://arxiv.org/abs/1212.3249}{[arXiv:1212.3249]}\relax
\mciteBstWouldAddEndPuncttrue
\mciteSetBstMidEndSepPunct{\mcitedefaultmidpunct}
{\mcitedefaultendpunct}{\mcitedefaultseppunct}\relax
\EndOfBibitem
\bibitem{Carli:2010rw}
T.~Carli {\em et~al.\/},
  \href{http://doi.org/10.1140/epjc/s10052-010-1255-0}{Eur. Phys. J. C}
  \href{http://doi.org/10.1140/epjc/s10052-010-1255-0}{{\bf 66}, 503} (2010),
  \href{https://arxiv.org/abs/0911.2985}{[arXiv:0911.2985]}\relax
\mciteBstWouldAddEndPuncttrue
\mciteSetBstMidEndSepPunct{\mcitedefaultmidpunct}
{\mcitedefaultendpunct}{\mcitedefaultseppunct}\relax
\EndOfBibitem
\bibitem{Kluge:2006xs}
T.~Kluge, K.~Rabbertz and M.~Wobisch, in \enquote{14th International Workshop
  on Deep Inelastic Scattering ({DIS} 2006),} 483, Tsukuba, Japan, April 20-24
  (2006), \href{https://arxiv.org/abs/hep-ph/0609285}{[hep-ph/0609285]}\relax
\mciteBstWouldAddEndPuncttrue
\mciteSetBstMidEndSepPunct{\mcitedefaultmidpunct}
{\mcitedefaultendpunct}{\mcitedefaultseppunct}\relax
\EndOfBibitem
\bibitem{Britzger:2012bs}
D.~Britzger {\em et~al.\/}, in \enquote{Proceedings, XX.\ International
  Workshop on Deep-Inelastic Scattering and Related Subjects ({DIS} 2012),}
  217, Bonn, Germany, March 26-30 (2012),
  \href{https://arxiv.org/abs/1208.3641}{[arXiv:1208.3641]}\relax
\mciteBstWouldAddEndPuncttrue
\mciteSetBstMidEndSepPunct{\mcitedefaultmidpunct}
{\mcitedefaultendpunct}{\mcitedefaultseppunct}\relax
\EndOfBibitem
\bibitem{Carrazza:2020gss}
S.~Carrazza {\em et~al.\/}, \href{http://doi.org/10.1007/JHEP12(2020)108}{JHEP}
  \href{http://doi.org/10.1007/JHEP12(2020)108}{{\bf 12}, 108} (2020),
  \href{https://arxiv.org/abs/2008.12789}{[arXiv:2008.12789]}\relax
\mciteBstWouldAddEndPuncttrue
\mciteSetBstMidEndSepPunct{\mcitedefaultmidpunct}
{\mcitedefaultendpunct}{\mcitedefaultseppunct}\relax
\EndOfBibitem
\bibitem{Czakon:2017dip}
M.~Czakon, D.~Heymes and A.~Mitov  (2017),
  \href{https://arxiv.org/abs/1704.08551}{[arXiv:1704.08551]}\relax
\mciteBstWouldAddEndPuncttrue
\mciteSetBstMidEndSepPunct{\mcitedefaultmidpunct}
{\mcitedefaultendpunct}{\mcitedefaultseppunct}\relax
\EndOfBibitem
\bibitem{Britzger:2019kkb}
D.~Britzger {\em et~al.\/},
  \href{http://doi.org/10.1140/epjc/s10052-021-09688-3}{Eur. Phys. J. C}
  \href{http://doi.org/10.1140/epjc/s10052-021-09688-3}{{\bf 79}, 10, 845}
  (2019), [Erratum: Eur.Phys.J.C 81, 957 (2021)],
  \href{https://arxiv.org/abs/1906.05303}{[arXiv:1906.05303]}\relax
\mciteBstWouldAddEndPuncttrue
\mciteSetBstMidEndSepPunct{\mcitedefaultmidpunct}
{\mcitedefaultendpunct}{\mcitedefaultseppunct}\relax
\EndOfBibitem
\bibitem{Maitre:2020blv}
D.~Maître, \href{http://doi.org/10.1088/1742-6596/1525/1/012014}{J. Phys.
  Conf. Ser.} \href{http://doi.org/10.1088/1742-6596/1525/1/012014}{{\bf 1525},
  1, 012014} (2020)\relax
\mciteBstWouldAddEndPuncttrue
\mciteSetBstMidEndSepPunct{\mcitedefaultmidpunct}
{\mcitedefaultendpunct}{\mcitedefaultseppunct}\relax
\EndOfBibitem
\bibitem{Maitre:2016sov}
D.~Maître, G.~Heinrich and M.~Johnson,
  \href{http://doi.org/10.22323/1.260.0016}{PoS}
  \href{http://doi.org/10.22323/1.260.0016}{{\bf LL2016}, 016} (2016),
  \href{https://arxiv.org/abs/1607.06259}{[arXiv:1607.06259]}\relax
\mciteBstWouldAddEndPuncttrue
\mciteSetBstMidEndSepPunct{\mcitedefaultmidpunct}
{\mcitedefaultendpunct}{\mcitedefaultseppunct}\relax
\EndOfBibitem
\bibitem{Dokshitzer:1978hw}
Y.~L. Dokshitzer, D.~Diakonov and S.~I. Troian,
  \href{http://doi.org/10.1016/0370-1573(80)90043-5}{Phys. Rept.}
  \href{http://doi.org/10.1016/0370-1573(80)90043-5}{{\bf 58}, 269}
  (1980)\relax
\mciteBstWouldAddEndPuncttrue
\mciteSetBstMidEndSepPunct{\mcitedefaultmidpunct}
{\mcitedefaultendpunct}{\mcitedefaultseppunct}\relax
\EndOfBibitem
\bibitem{Parisi:1979se}
G.~Parisi and R.~Petronzio,
  \href{http://doi.org/10.1016/0550-3213(79)90040-3}{Nucl. Phys.}
  \href{http://doi.org/10.1016/0550-3213(79)90040-3}{{\bf B154}, 427}
  (1979)\relax
\mciteBstWouldAddEndPuncttrue
\mciteSetBstMidEndSepPunct{\mcitedefaultmidpunct}
{\mcitedefaultendpunct}{\mcitedefaultseppunct}\relax
\EndOfBibitem
\bibitem{Curci:1979bg}
G.~Curci, M.~Greco and Y.~Srivastava,
  \href{http://doi.org/10.1016/0550-3213(79)90345-6}{Nucl. Phys.}
  \href{http://doi.org/10.1016/0550-3213(79)90345-6}{{\bf B159}, 451}
  (1979)\relax
\mciteBstWouldAddEndPuncttrue
\mciteSetBstMidEndSepPunct{\mcitedefaultmidpunct}
{\mcitedefaultendpunct}{\mcitedefaultseppunct}\relax
\EndOfBibitem
\bibitem{Bassetto:1979nt}
A.~Bassetto, M.~Ciafaloni and G.~Marchesini,
  \href{http://doi.org/10.1016/0550-3213(80)90413-7}{Nucl. Phys.}
  \href{http://doi.org/10.1016/0550-3213(80)90413-7}{{\bf B163}, 477}
  (1980)\relax
\mciteBstWouldAddEndPuncttrue
\mciteSetBstMidEndSepPunct{\mcitedefaultmidpunct}
{\mcitedefaultendpunct}{\mcitedefaultseppunct}\relax
\EndOfBibitem
\bibitem{Collins:1981uk}
J.~C. Collins and D.~E. Soper,
  \href{http://doi.org/10.1016/0550-3213(81)90339-4}{Nucl. Phys.}
  \href{http://doi.org/10.1016/0550-3213(81)90339-4}{{\bf B193}, 381} (1981),
  [Erratum: Nucl. Phys.B213,545(1983)]\relax
\mciteBstWouldAddEndPuncttrue
\mciteSetBstMidEndSepPunct{\mcitedefaultmidpunct}
{\mcitedefaultendpunct}{\mcitedefaultseppunct}\relax
\EndOfBibitem
\bibitem{Collins:1981va}
J.~C. Collins and D.~E. Soper,
  \href{http://doi.org/10.1016/0550-3213(82)90453-9}{Nucl. Phys.}
  \href{http://doi.org/10.1016/0550-3213(82)90453-9}{{\bf B197}, 446}
  (1982)\relax
\mciteBstWouldAddEndPuncttrue
\mciteSetBstMidEndSepPunct{\mcitedefaultmidpunct}
{\mcitedefaultendpunct}{\mcitedefaultseppunct}\relax
\EndOfBibitem
\bibitem{Kodaira:1981nh}
J.~Kodaira and L.~Trentadue,
  \href{http://doi.org/10.1016/0370-2693(82)90907-8}{Phys. Lett.}
  \href{http://doi.org/10.1016/0370-2693(82)90907-8}{{\bf 112B}, 66}
  (1982)\relax
\mciteBstWouldAddEndPuncttrue
\mciteSetBstMidEndSepPunct{\mcitedefaultmidpunct}
{\mcitedefaultendpunct}{\mcitedefaultseppunct}\relax
\EndOfBibitem
\bibitem{Kodaira:1982az}
J.~Kodaira and L.~Trentadue,
  \href{http://doi.org/10.1016/0370-2693(83)91213-3}{Phys. Lett.}
  \href{http://doi.org/10.1016/0370-2693(83)91213-3}{{\bf 123B}, 335}
  (1983)\relax
\mciteBstWouldAddEndPuncttrue
\mciteSetBstMidEndSepPunct{\mcitedefaultmidpunct}
{\mcitedefaultendpunct}{\mcitedefaultseppunct}\relax
\EndOfBibitem
\bibitem{Collins:1984kg}
J.~C. Collins, D.~E. Soper and G.~F. Sterman,
  \href{http://doi.org/10.1016/0550-3213(85)90479-1}{Nucl. Phys.}
  \href{http://doi.org/10.1016/0550-3213(85)90479-1}{{\bf B250}, 199}
  (1985)\relax
\mciteBstWouldAddEndPuncttrue
\mciteSetBstMidEndSepPunct{\mcitedefaultmidpunct}
{\mcitedefaultendpunct}{\mcitedefaultseppunct}\relax
\EndOfBibitem
\bibitem{Catani:1992ua}
S.~Catani {\em et~al.\/},
  \href{http://doi.org/10.1016/0550-3213(93)90271-P}{Nucl. Phys.}
  \href{http://doi.org/10.1016/0550-3213(93)90271-P}{{\bf B407}, 3}
  (1993)\relax
\mciteBstWouldAddEndPuncttrue
\mciteSetBstMidEndSepPunct{\mcitedefaultmidpunct}
{\mcitedefaultendpunct}{\mcitedefaultseppunct}\relax
\EndOfBibitem
\bibitem{Bauer:2000yr}
C.~W. Bauer {\em et~al.\/},
  \href{http://doi.org/10.1103/PhysRevD.63.114020}{Phys. Rev.}
  \href{http://doi.org/10.1103/PhysRevD.63.114020}{{\bf D63}, 114020} (2001),
  \href{https://arxiv.org/abs/hep-ph/0011336}{[hep-ph/0011336]}\relax
\mciteBstWouldAddEndPuncttrue
\mciteSetBstMidEndSepPunct{\mcitedefaultmidpunct}
{\mcitedefaultendpunct}{\mcitedefaultseppunct}\relax
\EndOfBibitem
\bibitem{Bauer:2001yt}
C.~W. Bauer, D.~Pirjol and I.~W. Stewart,
  \href{http://doi.org/10.1103/PhysRevD.65.054022}{Phys. Rev.}
  \href{http://doi.org/10.1103/PhysRevD.65.054022}{{\bf D65}, 054022} (2002),
  \href{https://arxiv.org/abs/hep-ph/0109045}{[hep-ph/0109045]}\relax
\mciteBstWouldAddEndPuncttrue
\mciteSetBstMidEndSepPunct{\mcitedefaultmidpunct}
{\mcitedefaultendpunct}{\mcitedefaultseppunct}\relax
\EndOfBibitem
\bibitem{Becher:2014oda}
T.~Becher, A.~Broggio and A.~Ferroglia,
  \href{http://doi.org/10.1007/978-3-319-14848-9}{Lect. Notes Phys.}
  \href{http://doi.org/10.1007/978-3-319-14848-9}{{\bf 896}, pp.1} (2015),
  \href{https://arxiv.org/abs/1410.1892}{[arXiv:1410.1892]}\relax
\mciteBstWouldAddEndPuncttrue
\mciteSetBstMidEndSepPunct{\mcitedefaultmidpunct}
{\mcitedefaultendpunct}{\mcitedefaultseppunct}\relax
\EndOfBibitem
\bibitem{Catani:1991hj}
S.~Catani {\em et~al.\/},
  \href{http://doi.org/10.1016/0370-2693(91)90196-W}{Phys. Lett.}
  \href{http://doi.org/10.1016/0370-2693(91)90196-W}{{\bf B269}, 432}
  (1991)\relax
\mciteBstWouldAddEndPuncttrue
\mciteSetBstMidEndSepPunct{\mcitedefaultmidpunct}
{\mcitedefaultendpunct}{\mcitedefaultseppunct}\relax
\EndOfBibitem
\bibitem{Brown:1990nm}
N.~Brown and W.~J. Stirling,
  \href{http://doi.org/10.1016/0370-2693(90)90502-W}{Phys. Lett.}
  \href{http://doi.org/10.1016/0370-2693(90)90502-W}{{\bf B252}, 657}
  (1990)\relax
\mciteBstWouldAddEndPuncttrue
\mciteSetBstMidEndSepPunct{\mcitedefaultmidpunct}
{\mcitedefaultendpunct}{\mcitedefaultseppunct}\relax
\EndOfBibitem
\bibitem{Bartel:1986ua}
W.~Bartel {\em et~al.\/} (JADE), \href{http://doi.org/10.1007/BF01410449}{Z.
  Phys.} \href{http://doi.org/10.1007/BF01410449}{{\bf C33}, 23} (1986),
  [,53(1986)]\relax
\mciteBstWouldAddEndPuncttrue
\mciteSetBstMidEndSepPunct{\mcitedefaultmidpunct}
{\mcitedefaultendpunct}{\mcitedefaultseppunct}\relax
\EndOfBibitem
\bibitem{Kidonakis:1998nf}
N.~Kidonakis, G.~Oderda and G.~F. Sterman,
  \href{http://doi.org/10.1016/S0550-3213(98)00441-6}{Nucl. Phys.}
  \href{http://doi.org/10.1016/S0550-3213(98)00441-6}{{\bf B531}, 365} (1998),
  \href{https://arxiv.org/abs/hep-ph/9803241}{[hep-ph/9803241]}\relax
\mciteBstWouldAddEndPuncttrue
\mciteSetBstMidEndSepPunct{\mcitedefaultmidpunct}
{\mcitedefaultendpunct}{\mcitedefaultseppunct}\relax
\EndOfBibitem
\bibitem{Bonciani:2003nt}
R.~Bonciani {\em et~al.\/},
  \href{http://doi.org/10.1016/j.physletb.2003.09.068}{Phys. Lett.}
  \href{http://doi.org/10.1016/j.physletb.2003.09.068}{{\bf B575}, 268} (2003),
  \href{https://arxiv.org/abs/hep-ph/0307035}{[hep-ph/0307035]}\relax
\mciteBstWouldAddEndPuncttrue
\mciteSetBstMidEndSepPunct{\mcitedefaultmidpunct}
{\mcitedefaultendpunct}{\mcitedefaultseppunct}\relax
\EndOfBibitem
\bibitem{Banfi:2004yd}
A.~Banfi, G.~P. Salam and G.~Zanderighi,
  \href{http://doi.org/10.1088/1126-6708/2005/03/073}{JHEP}
  \href{http://doi.org/10.1088/1126-6708/2005/03/073}{{\bf 03}, 073} (2005),
  \href{https://arxiv.org/abs/hep-ph/0407286}{[hep-ph/0407286]}\relax
\mciteBstWouldAddEndPuncttrue
\mciteSetBstMidEndSepPunct{\mcitedefaultmidpunct}
{\mcitedefaultendpunct}{\mcitedefaultseppunct}\relax
\EndOfBibitem
\bibitem{deFlorian:2000pr}
D.~de~Florian and M.~Grazzini,
  \href{http://doi.org/10.1103/PhysRevLett.85.4678}{Phys. Rev. Lett.}
  \href{http://doi.org/10.1103/PhysRevLett.85.4678}{{\bf 85}, 4678} (2000),
  \href{https://arxiv.org/abs/hep-ph/0008152}{[hep-ph/0008152]}\relax
\mciteBstWouldAddEndPuncttrue
\mciteSetBstMidEndSepPunct{\mcitedefaultmidpunct}
{\mcitedefaultendpunct}{\mcitedefaultseppunct}\relax
\EndOfBibitem
\bibitem{Bozzi:2005wk}
G.~Bozzi {\em et~al.\/},
  \href{http://doi.org/10.1016/j.nuclphysb.2005.12.022}{Nucl. Phys.}
  \href{http://doi.org/10.1016/j.nuclphysb.2005.12.022}{{\bf B737}, 73} (2006),
  \href{https://arxiv.org/abs/hep-ph/0508068}{[hep-ph/0508068]}\relax
\mciteBstWouldAddEndPuncttrue
\mciteSetBstMidEndSepPunct{\mcitedefaultmidpunct}
{\mcitedefaultendpunct}{\mcitedefaultseppunct}\relax
\EndOfBibitem
\bibitem{qcd:Bozzi:2005wk-url}
\url{http://theory.fi.infn.it/grazzini/codes.html}\relax
\mciteBstWouldAddEndPuncttrue
\mciteSetBstMidEndSepPunct{\mcitedefaultmidpunct}
{\mcitedefaultendpunct}{\mcitedefaultseppunct}\relax
\EndOfBibitem
\bibitem{Bozzi:2010xn}
G.~Bozzi {\em et~al.\/},
  \href{http://doi.org/10.1016/j.physletb.2010.12.024}{Phys. Lett.}
  \href{http://doi.org/10.1016/j.physletb.2010.12.024}{{\bf B696}, 207} (2011),
  \href{https://arxiv.org/abs/1007.2351}{[arXiv:1007.2351]}\relax
\mciteBstWouldAddEndPuncttrue
\mciteSetBstMidEndSepPunct{\mcitedefaultmidpunct}
{\mcitedefaultendpunct}{\mcitedefaultseppunct}\relax
\EndOfBibitem
\bibitem{Becher:2010tm}
T.~Becher and M.~Neubert,
  \href{http://doi.org/10.1140/epjc/s10052-011-1665-7}{Eur. Phys. J.}
  \href{http://doi.org/10.1140/epjc/s10052-011-1665-7}{{\bf C71}, 1665} (2011),
  \href{https://arxiv.org/abs/1007.4005}{[arXiv:1007.4005]}\relax
\mciteBstWouldAddEndPuncttrue
\mciteSetBstMidEndSepPunct{\mcitedefaultmidpunct}
{\mcitedefaultendpunct}{\mcitedefaultseppunct}\relax
\EndOfBibitem
\bibitem{qcd:CuTe}
\url{http://cute.hepforge.org/}\relax
\mciteBstWouldAddEndPuncttrue
\mciteSetBstMidEndSepPunct{\mcitedefaultmidpunct}
{\mcitedefaultendpunct}{\mcitedefaultseppunct}\relax
\EndOfBibitem
\bibitem{deFlorian:2012mx}
D.~de~Florian {\em et~al.\/},
  \href{http://doi.org/10.1007/JHEP06(2012)132}{JHEP}
  \href{http://doi.org/10.1007/JHEP06(2012)132}{{\bf 06}, 132} (2012),
  \href{https://arxiv.org/abs/1203.6321}{[arXiv:1203.6321]}\relax
\mciteBstWouldAddEndPuncttrue
\mciteSetBstMidEndSepPunct{\mcitedefaultmidpunct}
{\mcitedefaultendpunct}{\mcitedefaultseppunct}\relax
\EndOfBibitem
\bibitem{qcd:deFlorian:2012mx-url}
\url{http://theory.fi.infn.it/grazzini/codes.html}\relax
\mciteBstWouldAddEndPuncttrue
\mciteSetBstMidEndSepPunct{\mcitedefaultmidpunct}
{\mcitedefaultendpunct}{\mcitedefaultseppunct}\relax
\EndOfBibitem
\bibitem{Balazs:1997xd}
C.~Balazs and C.~P. Yuan, \href{http://doi.org/10.1103/PhysRevD.56.5558}{Phys.
  Rev.} \href{http://doi.org/10.1103/PhysRevD.56.5558}{{\bf D56}, 5558} (1997),
  \href{https://arxiv.org/abs/hep-ph/9704258}{[hep-ph/9704258]}\relax
\mciteBstWouldAddEndPuncttrue
\mciteSetBstMidEndSepPunct{\mcitedefaultmidpunct}
{\mcitedefaultendpunct}{\mcitedefaultseppunct}\relax
\EndOfBibitem
\bibitem{Catani:2015vma}
S.~Catani {\em et~al.\/}, \href{http://doi.org/10.1007/JHEP12(2015)047}{JHEP}
  \href{http://doi.org/10.1007/JHEP12(2015)047}{{\bf 12}, 047} (2015),
  \href{https://arxiv.org/abs/1507.06937}{[arXiv:1507.06937]}\relax
\mciteBstWouldAddEndPuncttrue
\mciteSetBstMidEndSepPunct{\mcitedefaultmidpunct}
{\mcitedefaultendpunct}{\mcitedefaultseppunct}\relax
\EndOfBibitem
\bibitem{Banfi:2012du}
A.~Banfi {\em et~al.\/},
  \href{http://doi.org/10.1016/j.physletb.2012.07.035}{Phys. Lett.}
  \href{http://doi.org/10.1016/j.physletb.2012.07.035}{{\bf B715}, 152} (2012),
  \href{https://arxiv.org/abs/1205.4760}{[arXiv:1205.4760]}\relax
\mciteBstWouldAddEndPuncttrue
\mciteSetBstMidEndSepPunct{\mcitedefaultmidpunct}
{\mcitedefaultendpunct}{\mcitedefaultseppunct}\relax
\EndOfBibitem
\bibitem{Grazzini:2015wpa}
M.~Grazzini {\em et~al.\/}, \href{http://doi.org/10.1007/JHEP08(2015)154}{JHEP}
  \href{http://doi.org/10.1007/JHEP08(2015)154}{{\bf 08}, 154} (2015),
  \href{https://arxiv.org/abs/1507.02565}{[arXiv:1507.02565]}\relax
\mciteBstWouldAddEndPuncttrue
\mciteSetBstMidEndSepPunct{\mcitedefaultmidpunct}
{\mcitedefaultendpunct}{\mcitedefaultseppunct}\relax
\EndOfBibitem
\bibitem{Campbell:2022uzw}
J.~M. Campbell {\em et~al.\/},
  \href{http://doi.org/10.1007/JHEP03(2023)080}{JHEP}
  \href{http://doi.org/10.1007/JHEP03(2023)080}{{\bf 03}, 080} (2023),
  \href{https://arxiv.org/abs/2210.10724}{[arXiv:2210.10724]}\relax
\mciteBstWouldAddEndPuncttrue
\mciteSetBstMidEndSepPunct{\mcitedefaultmidpunct}
{\mcitedefaultendpunct}{\mcitedefaultseppunct}\relax
\EndOfBibitem
\bibitem{deFlorian:2004mp}
D.~de~Florian and M.~Grazzini,
  \href{http://doi.org/10.1016/j.nuclphysb.2004.10.051}{Nucl. Phys.}
  \href{http://doi.org/10.1016/j.nuclphysb.2004.10.051}{{\bf B704}, 387}
  (2005), \href{https://arxiv.org/abs/hep-ph/0407241}{[hep-ph/0407241]}\relax
\mciteBstWouldAddEndPuncttrue
\mciteSetBstMidEndSepPunct{\mcitedefaultmidpunct}
{\mcitedefaultendpunct}{\mcitedefaultseppunct}\relax
\EndOfBibitem
\bibitem{Becher:2012qc}
T.~Becher and G.~Bell, \href{http://doi.org/10.1007/JHEP11(2012)126}{JHEP}
  \href{http://doi.org/10.1007/JHEP11(2012)126}{{\bf 11}, 126} (2012),
  \href{https://arxiv.org/abs/1210.0580}{[arXiv:1210.0580]}\relax
\mciteBstWouldAddEndPuncttrue
\mciteSetBstMidEndSepPunct{\mcitedefaultmidpunct}
{\mcitedefaultendpunct}{\mcitedefaultseppunct}\relax
\EndOfBibitem
\bibitem{Banfi:2012jm}
A.~Banfi {\em et~al.\/},
  \href{http://doi.org/10.1103/PhysRevLett.109.202001}{Phys. Rev. Lett.}
  \href{http://doi.org/10.1103/PhysRevLett.109.202001}{{\bf 109}, 202001}
  (2012), \href{https://arxiv.org/abs/1206.4998}{[arXiv:1206.4998]}\relax
\mciteBstWouldAddEndPuncttrue
\mciteSetBstMidEndSepPunct{\mcitedefaultmidpunct}
{\mcitedefaultendpunct}{\mcitedefaultseppunct}\relax
\EndOfBibitem
\bibitem{Becher:2013xia}
T.~Becher, M.~Neubert and L.~Rothen,
  \href{http://doi.org/10.1007/JHEP10(2013)125}{JHEP}
  \href{http://doi.org/10.1007/JHEP10(2013)125}{{\bf 10}, 125} (2013),
  \href{https://arxiv.org/abs/1307.0025}{[arXiv:1307.0025]}\relax
\mciteBstWouldAddEndPuncttrue
\mciteSetBstMidEndSepPunct{\mcitedefaultmidpunct}
{\mcitedefaultendpunct}{\mcitedefaultseppunct}\relax
\EndOfBibitem
\bibitem{Stewart:2013faa}
I.~W. Stewart {\em et~al.\/},
  \href{http://doi.org/10.1103/PhysRevD.89.054001}{Phys. Rev. D}
  \href{http://doi.org/10.1103/PhysRevD.89.054001}{{\bf 89}, 5, 054001} (2014),
  \href{https://arxiv.org/abs/1307.1808}{[arXiv:1307.1808]}\relax
\mciteBstWouldAddEndPuncttrue
\mciteSetBstMidEndSepPunct{\mcitedefaultmidpunct}
{\mcitedefaultendpunct}{\mcitedefaultseppunct}\relax
\EndOfBibitem
\bibitem{Campbell:2023cha}
J.~M. Campbell {\em et~al.\/},
  \href{http://doi.org/10.1007/JHEP04(2023)106}{JHEP}
  \href{http://doi.org/10.1007/JHEP04(2023)106}{{\bf 04}, 106} (2023),
  \href{https://arxiv.org/abs/2301.11768}{[arXiv:2301.11768]}\relax
\mciteBstWouldAddEndPuncttrue
\mciteSetBstMidEndSepPunct{\mcitedefaultmidpunct}
{\mcitedefaultendpunct}{\mcitedefaultseppunct}\relax
\EndOfBibitem
\bibitem{Stewart:2010pd}
I.~W. Stewart, F.~J. Tackmann and W.~J. Waalewijn,
  \href{http://doi.org/10.1103/PhysRevLett.106.032001}{Phys. Rev. Lett.}
  \href{http://doi.org/10.1103/PhysRevLett.106.032001}{{\bf 106}, 032001}
  (2011), \href{https://arxiv.org/abs/1005.4060}{[arXiv:1005.4060]}\relax
\mciteBstWouldAddEndPuncttrue
\mciteSetBstMidEndSepPunct{\mcitedefaultmidpunct}
{\mcitedefaultendpunct}{\mcitedefaultseppunct}\relax
\EndOfBibitem
\bibitem{Chien:2012ur}
Y.-T. Chien {\em et~al.\/},
  \href{http://doi.org/10.1103/PhysRevD.87.014010}{Phys. Rev.}
  \href{http://doi.org/10.1103/PhysRevD.87.014010}{{\bf D87}, 1, 014010}
  (2013), \href{https://arxiv.org/abs/1208.0010}{[arXiv:1208.0010]}\relax
\mciteBstWouldAddEndPuncttrue
\mciteSetBstMidEndSepPunct{\mcitedefaultmidpunct}
{\mcitedefaultendpunct}{\mcitedefaultseppunct}\relax
\EndOfBibitem
\bibitem{Jouttenus:2013hs}
T.~T. Jouttenus {\em et~al.\/},
  \href{http://doi.org/10.1103/PhysRevD.88.054031}{Phys. Rev.}
  \href{http://doi.org/10.1103/PhysRevD.88.054031}{{\bf D88}, 5, 054031}
  (2013), \href{https://arxiv.org/abs/1302.0846}{[arXiv:1302.0846]}\relax
\mciteBstWouldAddEndPuncttrue
\mciteSetBstMidEndSepPunct{\mcitedefaultmidpunct}
{\mcitedefaultendpunct}{\mcitedefaultseppunct}\relax
\EndOfBibitem
\bibitem{Dasgupta:2012hg}
M.~Dasgupta {\em et~al.\/}, \href{http://doi.org/10.1007/JHEP10(2012)126}{JHEP}
  \href{http://doi.org/10.1007/JHEP10(2012)126}{{\bf 10}, 126} (2012),
  \href{https://arxiv.org/abs/1207.1640}{[arXiv:1207.1640]}\relax
\mciteBstWouldAddEndPuncttrue
\mciteSetBstMidEndSepPunct{\mcitedefaultmidpunct}
{\mcitedefaultendpunct}{\mcitedefaultseppunct}\relax
\EndOfBibitem
\bibitem{Ahrens:2010zv}
V.~Ahrens {\em et~al.\/}, \href{http://doi.org/10.1007/JHEP09(2010)097}{JHEP}
  \href{http://doi.org/10.1007/JHEP09(2010)097}{{\bf 09}, 097} (2010),
  \href{https://arxiv.org/abs/1003.5827}{[arXiv:1003.5827]}\relax
\mciteBstWouldAddEndPuncttrue
\mciteSetBstMidEndSepPunct{\mcitedefaultmidpunct}
{\mcitedefaultendpunct}{\mcitedefaultseppunct}\relax
\EndOfBibitem
\bibitem{Aliev:2010zk}
M.~Aliev {\em et~al.\/},
  \href{http://doi.org/10.1016/j.cpc.2010.12.040}{Comput. Phys. Commun.}
  \href{http://doi.org/10.1016/j.cpc.2010.12.040}{{\bf 182}, 1034} (2011),
  \href{https://arxiv.org/abs/1007.1327}{[arXiv:1007.1327]}\relax
\mciteBstWouldAddEndPuncttrue
\mciteSetBstMidEndSepPunct{\mcitedefaultmidpunct}
{\mcitedefaultendpunct}{\mcitedefaultseppunct}\relax
\EndOfBibitem
\bibitem{Kidonakis:2010dk}
N.~Kidonakis, \href{http://doi.org/10.1103/PhysRevD.82.114030}{Phys. Rev.}
  \href{http://doi.org/10.1103/PhysRevD.82.114030}{{\bf D82}, 114030} (2010),
  \href{https://arxiv.org/abs/1009.4935}{[arXiv:1009.4935]}\relax
\mciteBstWouldAddEndPuncttrue
\mciteSetBstMidEndSepPunct{\mcitedefaultmidpunct}
{\mcitedefaultendpunct}{\mcitedefaultseppunct}\relax
\EndOfBibitem
\bibitem{Becher:2011fc}
T.~Becher, C.~Lorentzen and M.~D. Schwartz,
  \href{http://doi.org/10.1103/PhysRevLett.108.012001}{Phys. Rev. Lett.}
  \href{http://doi.org/10.1103/PhysRevLett.108.012001}{{\bf 108}, 012001}
  (2012), \href{https://arxiv.org/abs/1106.4310}{[arXiv:1106.4310]}\relax
\mciteBstWouldAddEndPuncttrue
\mciteSetBstMidEndSepPunct{\mcitedefaultmidpunct}
{\mcitedefaultendpunct}{\mcitedefaultseppunct}\relax
\EndOfBibitem
\bibitem{Becher:2014aya}
T.~Becher {\em et~al.\/},
  \href{http://doi.org/10.1140/epjc/s10052-015-3368-y}{Eur. Phys. J.}
  \href{http://doi.org/10.1140/epjc/s10052-015-3368-y}{{\bf C75}, 4, 154}
  (2015), \href{https://arxiv.org/abs/1412.8408}{[arXiv:1412.8408]}\relax
\mciteBstWouldAddEndPuncttrue
\mciteSetBstMidEndSepPunct{\mcitedefaultmidpunct}
{\mcitedefaultendpunct}{\mcitedefaultseppunct}\relax
\EndOfBibitem
\bibitem{Gerwick:2014gya}
E.~Gerwick {\em et~al.\/}, \href{http://doi.org/10.1007/JHEP02(2015)106}{JHEP}
  \href{http://doi.org/10.1007/JHEP02(2015)106}{{\bf 02}, 106} (2015),
  \href{https://arxiv.org/abs/1411.7325}{[arXiv:1411.7325]}\relax
\mciteBstWouldAddEndPuncttrue
\mciteSetBstMidEndSepPunct{\mcitedefaultmidpunct}
{\mcitedefaultendpunct}{\mcitedefaultseppunct}\relax
\EndOfBibitem
\bibitem{Banfi:2014sua}
A.~Banfi {\em et~al.\/}, \href{http://doi.org/10.1007/JHEP05(2015)102}{JHEP}
  \href{http://doi.org/10.1007/JHEP05(2015)102}{{\bf 05}, 102} (2015),
  \href{https://arxiv.org/abs/1412.2126}{[arXiv:1412.2126]}\relax
\mciteBstWouldAddEndPuncttrue
\mciteSetBstMidEndSepPunct{\mcitedefaultmidpunct}
{\mcitedefaultendpunct}{\mcitedefaultseppunct}\relax
\EndOfBibitem
\bibitem{Becher:2008cf}
T.~Becher and M.~D. Schwartz,
  \href{http://doi.org/10.1088/1126-6708/2008/07/034}{JHEP}
  \href{http://doi.org/10.1088/1126-6708/2008/07/034}{{\bf 07}, 034} (2008),
  \href{https://arxiv.org/abs/0803.0342}{[arXiv:0803.0342]}\relax
\mciteBstWouldAddEndPuncttrue
\mciteSetBstMidEndSepPunct{\mcitedefaultmidpunct}
{\mcitedefaultendpunct}{\mcitedefaultseppunct}\relax
\EndOfBibitem
\bibitem{Hoang:2014wka}
A.~H. Hoang {\em et~al.\/},
  \href{http://doi.org/10.1103/PhysRevD.91.094017}{Phys. Rev.}
  \href{http://doi.org/10.1103/PhysRevD.91.094017}{{\bf D91}, 9, 094017}
  (2015), \href{https://arxiv.org/abs/1411.6633}{[arXiv:1411.6633]}\relax
\mciteBstWouldAddEndPuncttrue
\mciteSetBstMidEndSepPunct{\mcitedefaultmidpunct}
{\mcitedefaultendpunct}{\mcitedefaultseppunct}\relax
\EndOfBibitem
\bibitem{Chien:2010kc}
Y.-T. Chien and M.~D. Schwartz,
  \href{http://doi.org/10.1007/JHEP08(2010)058}{JHEP}
  \href{http://doi.org/10.1007/JHEP08(2010)058}{{\bf 08}, 058} (2010),
  \href{https://arxiv.org/abs/1005.1644}{[arXiv:1005.1644]}\relax
\mciteBstWouldAddEndPuncttrue
\mciteSetBstMidEndSepPunct{\mcitedefaultmidpunct}
{\mcitedefaultendpunct}{\mcitedefaultseppunct}\relax
\EndOfBibitem
\bibitem{Bizon:2017rah}
W.~Bizon {\em et~al.\/}, \href{http://doi.org/10.1007/JHEP02(2018)108}{JHEP}
  \href{http://doi.org/10.1007/JHEP02(2018)108}{{\bf 02}, 108} (2018),
  \href{https://arxiv.org/abs/1705.09127}{[arXiv:1705.09127]}\relax
\mciteBstWouldAddEndPuncttrue
\mciteSetBstMidEndSepPunct{\mcitedefaultmidpunct}
{\mcitedefaultendpunct}{\mcitedefaultseppunct}\relax
\EndOfBibitem
\bibitem{Chen:2018pzu}
X.~Chen {\em et~al.\/},
  \href{http://doi.org/10.1016/j.physletb.2018.11.037}{Phys. Lett. B}
  \href{http://doi.org/10.1016/j.physletb.2018.11.037}{{\bf 788}, 425} (2019),
  \href{https://arxiv.org/abs/1805.00736}{[arXiv:1805.00736]}\relax
\mciteBstWouldAddEndPuncttrue
\mciteSetBstMidEndSepPunct{\mcitedefaultmidpunct}
{\mcitedefaultendpunct}{\mcitedefaultseppunct}\relax
\EndOfBibitem
\bibitem{Bizon:2019zgf}
W.~Bizon {\em et~al.\/},
  \href{http://doi.org/10.1140/epjc/s10052-019-7324-0}{Eur. Phys. J. C}
  \href{http://doi.org/10.1140/epjc/s10052-019-7324-0}{{\bf 79}, 10, 868}
  (2019), \href{https://arxiv.org/abs/1905.05171}{[arXiv:1905.05171]}\relax
\mciteBstWouldAddEndPuncttrue
\mciteSetBstMidEndSepPunct{\mcitedefaultmidpunct}
{\mcitedefaultendpunct}{\mcitedefaultseppunct}\relax
\EndOfBibitem
\bibitem{Neumann:2022lft}
T.~Neumann and J.~Campbell,
  \href{http://doi.org/10.1103/PhysRevD.107.L011506}{Phys. Rev. D}
  \href{http://doi.org/10.1103/PhysRevD.107.L011506}{{\bf 107}, 1, L011506}
  (2023), \href{https://arxiv.org/abs/2207.07056}{[arXiv:2207.07056]}\relax
\mciteBstWouldAddEndPuncttrue
\mciteSetBstMidEndSepPunct{\mcitedefaultmidpunct}
{\mcitedefaultendpunct}{\mcitedefaultseppunct}\relax
\EndOfBibitem
\bibitem{Re:2021con}
E.~Re, L.~Rottoli and P.~Torrielli  (2021),
  \href{https://arxiv.org/abs/2104.07509}{[arXiv:2104.07509]}\relax
\mciteBstWouldAddEndPuncttrue
\mciteSetBstMidEndSepPunct{\mcitedefaultmidpunct}
{\mcitedefaultendpunct}{\mcitedefaultseppunct}\relax
\EndOfBibitem
\bibitem{Catani:2014uta}
S.~Catani {\em et~al.\/},
  \href{http://doi.org/10.1016/j.nuclphysb.2014.09.012}{Nucl. Phys.}
  \href{http://doi.org/10.1016/j.nuclphysb.2014.09.012}{{\bf B888}, 75} (2014),
  \href{https://arxiv.org/abs/1405.4827}{[arXiv:1405.4827]}\relax
\mciteBstWouldAddEndPuncttrue
\mciteSetBstMidEndSepPunct{\mcitedefaultmidpunct}
{\mcitedefaultendpunct}{\mcitedefaultseppunct}\relax
\EndOfBibitem
\bibitem{Banfi:2015pju}
A.~Banfi {\em et~al.\/}, \href{http://doi.org/10.1007/JHEP04(2016)049}{JHEP}
  \href{http://doi.org/10.1007/JHEP04(2016)049}{{\bf 04}, 049} (2016),
  \href{https://arxiv.org/abs/1511.02886}{[arXiv:1511.02886]}\relax
\mciteBstWouldAddEndPuncttrue
\mciteSetBstMidEndSepPunct{\mcitedefaultmidpunct}
{\mcitedefaultendpunct}{\mcitedefaultseppunct}\relax
\EndOfBibitem
\bibitem{Larkoski:2015uaa}
A.~J. Larkoski and I.~Moult,
  \href{http://doi.org/10.1103/PhysRevD.93.014017}{Phys. Rev.}
  \href{http://doi.org/10.1103/PhysRevD.93.014017}{{\bf D93}, 014017} (2016),
  \href{https://arxiv.org/abs/1510.08459}{[arXiv:1510.08459]}\relax
\mciteBstWouldAddEndPuncttrue
\mciteSetBstMidEndSepPunct{\mcitedefaultmidpunct}
{\mcitedefaultendpunct}{\mcitedefaultseppunct}\relax
\EndOfBibitem
\bibitem{Lustermans:2016nvk}
G.~Lustermans, W.~J. Waalewijn and L.~Zeune,
  \href{http://doi.org/10.1016/j.physletb.2016.09.060}{Phys. Lett.}
  \href{http://doi.org/10.1016/j.physletb.2016.09.060}{{\bf B762}, 447} (2016),
  \href{https://arxiv.org/abs/1605.02740}{[arXiv:1605.02740]}\relax
\mciteBstWouldAddEndPuncttrue
\mciteSetBstMidEndSepPunct{\mcitedefaultmidpunct}
{\mcitedefaultendpunct}{\mcitedefaultseppunct}\relax
\EndOfBibitem
\bibitem{Muselli:2017bad}
C.~Muselli, S.~Forte and G.~Ridolfi,
  \href{http://doi.org/10.1007/JHEP03(2017)106}{JHEP}
  \href{http://doi.org/10.1007/JHEP03(2017)106}{{\bf 03}, 106} (2017),
  \href{https://arxiv.org/abs/1701.01464}{[arXiv:1701.01464]}\relax
\mciteBstWouldAddEndPuncttrue
\mciteSetBstMidEndSepPunct{\mcitedefaultmidpunct}
{\mcitedefaultendpunct}{\mcitedefaultseppunct}\relax
\EndOfBibitem
\bibitem{Bizon:2018foh}
W.~Bizo\'n {\em et~al.\/}, \href{http://doi.org/10.1007/JHEP12(2018)132}{JHEP}
  \href{http://doi.org/10.1007/JHEP12(2018)132}{{\bf 12}, 132} (2018),
  \href{https://arxiv.org/abs/1805.05916}{[arXiv:1805.05916]}\relax
\mciteBstWouldAddEndPuncttrue
\mciteSetBstMidEndSepPunct{\mcitedefaultmidpunct}
{\mcitedefaultendpunct}{\mcitedefaultseppunct}\relax
\EndOfBibitem
\bibitem{Bonvini:2018ixe}
M.~Bonvini and S.~Marzani,
  \href{http://doi.org/10.1103/PhysRevLett.120.202003}{Phys. Rev. Lett.}
  \href{http://doi.org/10.1103/PhysRevLett.120.202003}{{\bf 120}, 20, 202003}
  (2018), \href{https://arxiv.org/abs/1802.07758}{[arXiv:1802.07758]}\relax
\mciteBstWouldAddEndPuncttrue
\mciteSetBstMidEndSepPunct{\mcitedefaultmidpunct}
{\mcitedefaultendpunct}{\mcitedefaultseppunct}\relax
\EndOfBibitem
\bibitem{Procura:2018zpn}
M.~Procura, W.~J. Waalewijn and L.~Zeune,
  \href{http://doi.org/10.1007/JHEP10(2018)098}{JHEP}
  \href{http://doi.org/10.1007/JHEP10(2018)098}{{\bf 10}, 098} (2018),
  \href{https://arxiv.org/abs/1806.10622}{[arXiv:1806.10622]}\relax
\mciteBstWouldAddEndPuncttrue
\mciteSetBstMidEndSepPunct{\mcitedefaultmidpunct}
{\mcitedefaultendpunct}{\mcitedefaultseppunct}\relax
\EndOfBibitem
\bibitem{Lustermans:2019plv}
G.~Lustermans {\em et~al.\/},
  \href{http://doi.org/10.1007/JHEP03(2019)124}{JHEP}
  \href{http://doi.org/10.1007/JHEP03(2019)124}{{\bf 03}, 124} (2019),
  \href{https://arxiv.org/abs/1901.03331}{[arXiv:1901.03331]}\relax
\mciteBstWouldAddEndPuncttrue
\mciteSetBstMidEndSepPunct{\mcitedefaultmidpunct}
{\mcitedefaultendpunct}{\mcitedefaultseppunct}\relax
\EndOfBibitem
\bibitem{Monni:2019yyr}
P.~F. Monni, L.~Rottoli and P.~Torrielli,
  \href{http://doi.org/10.1103/PhysRevLett.124.252001}{Phys. Rev. Lett.}
  \href{http://doi.org/10.1103/PhysRevLett.124.252001}{{\bf 124}, 25, 252001}
  (2020), \href{https://arxiv.org/abs/1909.04704}{[arXiv:1909.04704]}\relax
\mciteBstWouldAddEndPuncttrue
\mciteSetBstMidEndSepPunct{\mcitedefaultmidpunct}
{\mcitedefaultendpunct}{\mcitedefaultseppunct}\relax
\EndOfBibitem
\bibitem{Becher:2020ugp}
T.~Becher and T.~Neumann, \href{http://doi.org/10.1007/JHEP03(2021)199}{JHEP}
  \href{http://doi.org/10.1007/JHEP03(2021)199}{{\bf 03}, 199} (2021),
  \href{https://arxiv.org/abs/2009.11437}{[arXiv:2009.11437]}\relax
\mciteBstWouldAddEndPuncttrue
\mciteSetBstMidEndSepPunct{\mcitedefaultmidpunct}
{\mcitedefaultendpunct}{\mcitedefaultseppunct}\relax
\EndOfBibitem
\bibitem{Dasgupta:2013ihk}
M.~Dasgupta {\em et~al.\/}, \href{http://doi.org/10.1007/JHEP09(2013)029}{JHEP}
  \href{http://doi.org/10.1007/JHEP09(2013)029}{{\bf 09}, 029} (2013),
  \href{https://arxiv.org/abs/1307.0007}{[arXiv:1307.0007]}\relax
\mciteBstWouldAddEndPuncttrue
\mciteSetBstMidEndSepPunct{\mcitedefaultmidpunct}
{\mcitedefaultendpunct}{\mcitedefaultseppunct}\relax
\EndOfBibitem
\bibitem{Larkoski:2014wba}
A.~J. Larkoski {\em et~al.\/},
  \href{http://doi.org/10.1007/JHEP05(2014)146}{JHEP}
  \href{http://doi.org/10.1007/JHEP05(2014)146}{{\bf 05}, 146} (2014),
  \href{https://arxiv.org/abs/1402.2657}{[arXiv:1402.2657]}\relax
\mciteBstWouldAddEndPuncttrue
\mciteSetBstMidEndSepPunct{\mcitedefaultmidpunct}
{\mcitedefaultendpunct}{\mcitedefaultseppunct}\relax
\EndOfBibitem
\bibitem{Larkoski:2017jix}
A.~J. Larkoski, I.~Moult and B.~Nachman,
  \href{http://doi.org/10.1016/j.physrep.2019.11.001}{Phys. Rept.}
  \href{http://doi.org/10.1016/j.physrep.2019.11.001}{{\bf 841}, 1} (2020),
  \href{https://arxiv.org/abs/1709.04464}{[arXiv:1709.04464]}\relax
\mciteBstWouldAddEndPuncttrue
\mciteSetBstMidEndSepPunct{\mcitedefaultmidpunct}
{\mcitedefaultendpunct}{\mcitedefaultseppunct}\relax
\EndOfBibitem
\bibitem{Marzani:2019hun}
S.~Marzani, G.~Soyez and M.~Spannowsky  (2019), [Lect. Notes
  Phys.958,pp.(2019)],
  \href{https://arxiv.org/abs/1901.10342}{[arXiv:1901.10342]}\relax
\mciteBstWouldAddEndPuncttrue
\mciteSetBstMidEndSepPunct{\mcitedefaultmidpunct}
{\mcitedefaultendpunct}{\mcitedefaultseppunct}\relax
\EndOfBibitem
\bibitem{Kogler:2018hem}
R.~Kogler {\em et~al.\/},
  \href{http://doi.org/10.1103/RevModPhys.91.045003}{Rev. Mod. Phys.}
  \href{http://doi.org/10.1103/RevModPhys.91.045003}{{\bf 91}, 4, 045003}
  (2019), \href{https://arxiv.org/abs/1803.06991}{[arXiv:1803.06991]}\relax
\mciteBstWouldAddEndPuncttrue
\mciteSetBstMidEndSepPunct{\mcitedefaultmidpunct}
{\mcitedefaultendpunct}{\mcitedefaultseppunct}\relax
\EndOfBibitem
\bibitem{qcd:Basics}
Yu.L.~Dokshitzer \etal, {\it ``Basics of perturbative QCD,''}\rm \
  Gif-sur-Yvette, France: \'Editions fronti\`eres\ (1991), see also
  {\url{http://www.lpthe.jussieu.fr/\~{}yuri/} {BPQCD/cover.html}}\relax
\mciteBstWouldAddEndPuncttrue
\mciteSetBstMidEndSepPunct{\mcitedefaultmidpunct}
{\mcitedefaultendpunct}{\mcitedefaultseppunct}\relax
\EndOfBibitem
\bibitem{Sjostrand:2000wi}
T.~Sj\:ostrand {\em et~al.\/},
  \href{http://doi.org/10.1016/S0010-4655(00)00236-8}{Comput. Phys. Commun.}
  \href{http://doi.org/10.1016/S0010-4655(00)00236-8}{{\bf 135}, 238} (2001),
  \href{https://arxiv.org/abs/hep-ph/0010017}{[hep-ph/0010017]}\relax
\mciteBstWouldAddEndPuncttrue
\mciteSetBstMidEndSepPunct{\mcitedefaultmidpunct}
{\mcitedefaultendpunct}{\mcitedefaultseppunct}\relax
\EndOfBibitem
\bibitem{Sjostrand:2006za}
T.~Sj\:ostrand, S.~Mrenna and P.~Z. Skands,
  \href{http://doi.org/10.1088/1126-6708/2006/05/026}{JHEP}
  \href{http://doi.org/10.1088/1126-6708/2006/05/026}{{\bf 05}, 026} (2006),
  \href{https://arxiv.org/abs/hep-ph/0603175}{[hep-ph/0603175]}\relax
\mciteBstWouldAddEndPuncttrue
\mciteSetBstMidEndSepPunct{\mcitedefaultmidpunct}
{\mcitedefaultendpunct}{\mcitedefaultseppunct}\relax
\EndOfBibitem
\bibitem{qcd:Sjostrand:2006za-url}
\url{http://projects.hepforge.org/pythia6/}\relax
\mciteBstWouldAddEndPuncttrue
\mciteSetBstMidEndSepPunct{\mcitedefaultmidpunct}
{\mcitedefaultendpunct}{\mcitedefaultseppunct}\relax
\EndOfBibitem
\bibitem{Sjostrand:2014zea}
T.~Sjostrand {\em et~al.\/},
  \href{http://doi.org/10.1016/j.cpc.2015.01.024}{Comput. Phys. Commun.}
  \href{http://doi.org/10.1016/j.cpc.2015.01.024}{{\bf 191}, 159} (2015),
  \href{https://arxiv.org/abs/1410.3012}{[arXiv:1410.3012]}\relax
\mciteBstWouldAddEndPuncttrue
\mciteSetBstMidEndSepPunct{\mcitedefaultmidpunct}
{\mcitedefaultendpunct}{\mcitedefaultseppunct}\relax
\EndOfBibitem
\bibitem{qcd:Sjostrand:2014zea-url}
\url{http://home.thep.lu.se/~torbjorn/Pythia.html}\relax
\mciteBstWouldAddEndPuncttrue
\mciteSetBstMidEndSepPunct{\mcitedefaultmidpunct}
{\mcitedefaultendpunct}{\mcitedefaultseppunct}\relax
\EndOfBibitem
\bibitem{Webber:1983if}
B.~R. Webber, \href{http://doi.org/10.1016/0550-3213(84)90333-X}{Nucl. Phys.}
  \href{http://doi.org/10.1016/0550-3213(84)90333-X}{{\bf B238}, 492}
  (1984)\relax
\mciteBstWouldAddEndPuncttrue
\mciteSetBstMidEndSepPunct{\mcitedefaultmidpunct}
{\mcitedefaultendpunct}{\mcitedefaultseppunct}\relax
\EndOfBibitem
\bibitem{Corcella:2000bw}
G.~Corcella {\em et~al.\/},
  \href{http://doi.org/10.1088/1126-6708/2001/01/010}{JHEP}
  \href{http://doi.org/10.1088/1126-6708/2001/01/010}{{\bf 01}, 010} (2001),
  \href{https://arxiv.org/abs/hep-ph/0011363}{[hep-ph/0011363]}\relax
\mciteBstWouldAddEndPuncttrue
\mciteSetBstMidEndSepPunct{\mcitedefaultmidpunct}
{\mcitedefaultendpunct}{\mcitedefaultseppunct}\relax
\EndOfBibitem
\bibitem{qcd:Corcella:2000bw-url}
\url{http://www.hep.phy.cam.ac.uk/theory/webber/Herwig/}\relax
\mciteBstWouldAddEndPuncttrue
\mciteSetBstMidEndSepPunct{\mcitedefaultmidpunct}
{\mcitedefaultendpunct}{\mcitedefaultseppunct}\relax
\EndOfBibitem
\bibitem{Bahr:2008pv}
M.~Bahr {\em et~al.\/},
  \href{http://doi.org/10.1140/epjc/s10052-008-0798-9}{Eur. Phys. J.}
  \href{http://doi.org/10.1140/epjc/s10052-008-0798-9}{{\bf C58}, 639} (2008),
  \href{https://arxiv.org/abs/0803.0883}{[arXiv:0803.0883]}\relax
\mciteBstWouldAddEndPuncttrue
\mciteSetBstMidEndSepPunct{\mcitedefaultmidpunct}
{\mcitedefaultendpunct}{\mcitedefaultseppunct}\relax
\EndOfBibitem
\bibitem{qcd:Bahr:2008pv-url}
\url{http://projects.hepforge.org/herwig/}\relax
\mciteBstWouldAddEndPuncttrue
\mciteSetBstMidEndSepPunct{\mcitedefaultmidpunct}
{\mcitedefaultendpunct}{\mcitedefaultseppunct}\relax
\EndOfBibitem
\bibitem{Gleisberg:2008ta}
T.~Gleisberg {\em et~al.\/},
  \href{http://doi.org/10.1088/1126-6708/2009/02/007}{JHEP}
  \href{http://doi.org/10.1088/1126-6708/2009/02/007}{{\bf 02}, 007} (2009),
  \href{https://arxiv.org/abs/0811.4622}{[arXiv:0811.4622]}\relax
\mciteBstWouldAddEndPuncttrue
\mciteSetBstMidEndSepPunct{\mcitedefaultmidpunct}
{\mcitedefaultendpunct}{\mcitedefaultseppunct}\relax
\EndOfBibitem
\bibitem{Lonnblad:1992tz}
L.~Lonnblad, \href{http://doi.org/10.1016/0010-4655(92)90068-A}{Comput. Phys.
  Commun.} \href{http://doi.org/10.1016/0010-4655(92)90068-A}{{\bf 71}, 15}
  (1992)\relax
\mciteBstWouldAddEndPuncttrue
\mciteSetBstMidEndSepPunct{\mcitedefaultmidpunct}
{\mcitedefaultendpunct}{\mcitedefaultseppunct}\relax
\EndOfBibitem
\bibitem{Buckley:2011ms}
A.~Buckley {\em et~al.\/},
  \href{http://doi.org/10.1016/j.physrep.2011.03.005}{Phys. Rept.}
  \href{http://doi.org/10.1016/j.physrep.2011.03.005}{{\bf 504}, 145} (2011),
  \href{https://arxiv.org/abs/1101.2599}{[arXiv:1101.2599]}\relax
\mciteBstWouldAddEndPuncttrue
\mciteSetBstMidEndSepPunct{\mcitedefaultmidpunct}
{\mcitedefaultendpunct}{\mcitedefaultseppunct}\relax
\EndOfBibitem
\bibitem{Bellm:2019yyh}
J.~Bellm {\em et~al.\/},
  \href{http://doi.org/10.1140/epjc/s10052-019-7574-x}{Eur. Phys. J. C}
  \href{http://doi.org/10.1140/epjc/s10052-019-7574-x}{{\bf 80}, 2, 93} (2020),
  \href{https://arxiv.org/abs/1903.12563}{[arXiv:1903.12563]}\relax
\mciteBstWouldAddEndPuncttrue
\mciteSetBstMidEndSepPunct{\mcitedefaultmidpunct}
{\mcitedefaultendpunct}{\mcitedefaultseppunct}\relax
\EndOfBibitem
\bibitem{Andersson:1983ia}
B.~Andersson {\em et~al.\/},
  \href{http://doi.org/10.1016/0370-1573(83)90080-7}{Phys. Rept.}
  \href{http://doi.org/10.1016/0370-1573(83)90080-7}{{\bf 97}, 31} (1983)\relax
\mciteBstWouldAddEndPuncttrue
\mciteSetBstMidEndSepPunct{\mcitedefaultmidpunct}
{\mcitedefaultendpunct}{\mcitedefaultseppunct}\relax
\EndOfBibitem
\bibitem{Sjostrand:1984ic}
T.~Sjostrand, \href{http://doi.org/10.1016/0550-3213(84)90607-2}{Nucl. Phys.}
  \href{http://doi.org/10.1016/0550-3213(84)90607-2}{{\bf B248}, 469}
  (1984)\relax
\mciteBstWouldAddEndPuncttrue
\mciteSetBstMidEndSepPunct{\mcitedefaultmidpunct}
{\mcitedefaultendpunct}{\mcitedefaultseppunct}\relax
\EndOfBibitem
\bibitem{Sjostrand:1987su}
T.~Sjostrand and M.~van Zijl,
  \href{http://doi.org/10.1103/PhysRevD.36.2019}{Phys. Rev.}
  \href{http://doi.org/10.1103/PhysRevD.36.2019}{{\bf D36}, 2019} (1987)\relax
\mciteBstWouldAddEndPuncttrue
\mciteSetBstMidEndSepPunct{\mcitedefaultmidpunct}
{\mcitedefaultendpunct}{\mcitedefaultseppunct}\relax
\EndOfBibitem
\bibitem{Catani:2001cc}
S.~Catani {\em et~al.\/},
  \href{http://doi.org/10.1088/1126-6708/2001/11/063}{JHEP}
  \href{http://doi.org/10.1088/1126-6708/2001/11/063}{{\bf 11}, 063} (2001),
  \href{https://arxiv.org/abs/hep-ph/0109231}{[hep-ph/0109231]}\relax
\mciteBstWouldAddEndPuncttrue
\mciteSetBstMidEndSepPunct{\mcitedefaultmidpunct}
{\mcitedefaultendpunct}{\mcitedefaultseppunct}\relax
\EndOfBibitem
\bibitem{Alwall:2007fs}
J.~Alwall {\em et~al.\/},
  \href{http://doi.org/10.1140/epjc/s10052-007-0490-5}{Eur. Phys. J.}
  \href{http://doi.org/10.1140/epjc/s10052-007-0490-5}{{\bf C53}, 473} (2008),
  \href{https://arxiv.org/abs/0706.2569}{[arXiv:0706.2569]}\relax
\mciteBstWouldAddEndPuncttrue
\mciteSetBstMidEndSepPunct{\mcitedefaultmidpunct}
{\mcitedefaultendpunct}{\mcitedefaultseppunct}\relax
\EndOfBibitem
\bibitem{Frixione:2002ik}
S.~Frixione and B.~R. Webber,
  \href{http://doi.org/10.1088/1126-6708/2002/06/029}{JHEP}
  \href{http://doi.org/10.1088/1126-6708/2002/06/029}{{\bf 06}, 029} (2002),
  \href{https://arxiv.org/abs/hep-ph/0204244}{[hep-ph/0204244]}\relax
\mciteBstWouldAddEndPuncttrue
\mciteSetBstMidEndSepPunct{\mcitedefaultmidpunct}
{\mcitedefaultendpunct}{\mcitedefaultseppunct}\relax
\EndOfBibitem
\bibitem{Nason:2004rx}
P.~Nason, \href{http://doi.org/10.1088/1126-6708/2004/11/040}{JHEP}
  \href{http://doi.org/10.1088/1126-6708/2004/11/040}{{\bf 11}, 040} (2004),
  \href{https://arxiv.org/abs/hep-ph/0409146}{[hep-ph/0409146]}\relax
\mciteBstWouldAddEndPuncttrue
\mciteSetBstMidEndSepPunct{\mcitedefaultmidpunct}
{\mcitedefaultendpunct}{\mcitedefaultseppunct}\relax
\EndOfBibitem
\bibitem{Alioli:2010xd}
S.~Alioli {\em et~al.\/}, \href{http://doi.org/10.1007/JHEP06(2010)043}{JHEP}
  \href{http://doi.org/10.1007/JHEP06(2010)043}{{\bf 06}, 043} (2010),
  \href{https://arxiv.org/abs/1002.2581}{[arXiv:1002.2581]}\relax
\mciteBstWouldAddEndPuncttrue
\mciteSetBstMidEndSepPunct{\mcitedefaultmidpunct}
{\mcitedefaultendpunct}{\mcitedefaultseppunct}\relax
\EndOfBibitem
\bibitem{qcd:Alioli:2010xd-url}
\url{http://powhegbox.mib.infn.it/}\relax
\mciteBstWouldAddEndPuncttrue
\mciteSetBstMidEndSepPunct{\mcitedefaultmidpunct}
{\mcitedefaultendpunct}{\mcitedefaultseppunct}\relax
\EndOfBibitem
\bibitem{Platzer:2012bs}
S.~Plätzer, \href{http://doi.org/10.1007/JHEP08(2013)114}{JHEP}
  \href{http://doi.org/10.1007/JHEP08(2013)114}{{\bf 08}, 114} (2013),
  \href{https://arxiv.org/abs/1211.5467}{[arXiv:1211.5467]}\relax
\mciteBstWouldAddEndPuncttrue
\mciteSetBstMidEndSepPunct{\mcitedefaultmidpunct}
{\mcitedefaultendpunct}{\mcitedefaultseppunct}\relax
\EndOfBibitem
\bibitem{Frederix:2012ps}
R.~Frederix and S.~Frixione,
  \href{http://doi.org/10.1007/JHEP12(2012)061}{JHEP}
  \href{http://doi.org/10.1007/JHEP12(2012)061}{{\bf 12}, 061} (2012),
  \href{https://arxiv.org/abs/1209.6215}{[arXiv:1209.6215]}\relax
\mciteBstWouldAddEndPuncttrue
\mciteSetBstMidEndSepPunct{\mcitedefaultmidpunct}
{\mcitedefaultendpunct}{\mcitedefaultseppunct}\relax
\EndOfBibitem
\bibitem{Hamilton:2012rf}
K.~Hamilton {\em et~al.\/}, \href{http://doi.org/10.1007/JHEP05(2013)082}{JHEP}
  \href{http://doi.org/10.1007/JHEP05(2013)082}{{\bf 05}, 082} (2013),
  \href{https://arxiv.org/abs/1212.4504}{[arXiv:1212.4504]}\relax
\mciteBstWouldAddEndPuncttrue
\mciteSetBstMidEndSepPunct{\mcitedefaultmidpunct}
{\mcitedefaultendpunct}{\mcitedefaultseppunct}\relax
\EndOfBibitem
\bibitem{Monni:2019whf}
P.~F. Monni {\em et~al.\/}, \href{http://doi.org/10.1007/JHEP05(2020)143}{JHEP}
  \href{http://doi.org/10.1007/JHEP05(2020)143}{{\bf 05}, 143} (2020),
  \href{https://arxiv.org/abs/1908.06987}{[arXiv:1908.06987]}\relax
\mciteBstWouldAddEndPuncttrue
\mciteSetBstMidEndSepPunct{\mcitedefaultmidpunct}
{\mcitedefaultendpunct}{\mcitedefaultseppunct}\relax
\EndOfBibitem
\bibitem{Alioli:2013hqa}
S.~Alioli {\em et~al.\/}, \href{http://doi.org/10.1007/JHEP06(2014)089}{JHEP}
  \href{http://doi.org/10.1007/JHEP06(2014)089}{{\bf 06}, 089} (2014),
  \href{https://arxiv.org/abs/1311.0286}{[arXiv:1311.0286]}\relax
\mciteBstWouldAddEndPuncttrue
\mciteSetBstMidEndSepPunct{\mcitedefaultmidpunct}
{\mcitedefaultendpunct}{\mcitedefaultseppunct}\relax
\EndOfBibitem
\bibitem{Hamilton:2013fea}
K.~Hamilton {\em et~al.\/}, \href{http://doi.org/10.1007/JHEP10(2013)222}{JHEP}
  \href{http://doi.org/10.1007/JHEP10(2013)222}{{\bf 10}, 222} (2013),
  \href{https://arxiv.org/abs/1309.0017}{[arXiv:1309.0017]}\relax
\mciteBstWouldAddEndPuncttrue
\mciteSetBstMidEndSepPunct{\mcitedefaultmidpunct}
{\mcitedefaultendpunct}{\mcitedefaultseppunct}\relax
\EndOfBibitem
\bibitem{Karlberg:2014qua}
A.~Karlberg, E.~Re and G.~Zanderighi,
  \href{http://doi.org/10.1007/JHEP09(2014)134}{JHEP}
  \href{http://doi.org/10.1007/JHEP09(2014)134}{{\bf 09}, 134} (2014),
  \href{https://arxiv.org/abs/1407.2940}{[arXiv:1407.2940]}\relax
\mciteBstWouldAddEndPuncttrue
\mciteSetBstMidEndSepPunct{\mcitedefaultmidpunct}
{\mcitedefaultendpunct}{\mcitedefaultseppunct}\relax
\EndOfBibitem
\bibitem{Hoeche:2014aia}
S.~Höche, Y.~Li and S.~Prestel,
  \href{http://doi.org/10.1103/PhysRevD.91.074015}{Phys. Rev.}
  \href{http://doi.org/10.1103/PhysRevD.91.074015}{{\bf D91}, 7, 074015}
  (2015), \href{https://arxiv.org/abs/1405.3607}{[arXiv:1405.3607]}\relax
\mciteBstWouldAddEndPuncttrue
\mciteSetBstMidEndSepPunct{\mcitedefaultmidpunct}
{\mcitedefaultendpunct}{\mcitedefaultseppunct}\relax
\EndOfBibitem
\bibitem{Hoche:2014dla}
S.~Höche, Y.~Li and S.~Prestel,
  \href{http://doi.org/10.1103/PhysRevD.90.054011}{Phys. Rev.}
  \href{http://doi.org/10.1103/PhysRevD.90.054011}{{\bf D90}, 5, 054011}
  (2014), \href{https://arxiv.org/abs/1407.3773}{[arXiv:1407.3773]}\relax
\mciteBstWouldAddEndPuncttrue
\mciteSetBstMidEndSepPunct{\mcitedefaultmidpunct}
{\mcitedefaultendpunct}{\mcitedefaultseppunct}\relax
\EndOfBibitem
\bibitem{Alioli:2015toa}
S.~Alioli {\em et~al.\/},
  \href{http://doi.org/10.1103/PhysRevD.92.094020}{Phys. Rev.}
  \href{http://doi.org/10.1103/PhysRevD.92.094020}{{\bf D92}, 9, 094020}
  (2015), \href{https://arxiv.org/abs/1508.01475}{[arXiv:1508.01475]}\relax
\mciteBstWouldAddEndPuncttrue
\mciteSetBstMidEndSepPunct{\mcitedefaultmidpunct}
{\mcitedefaultendpunct}{\mcitedefaultseppunct}\relax
\EndOfBibitem
\bibitem{Alioli:2021qbf}
S.~Alioli {\em et~al.\/}  (2021),
  \href{https://arxiv.org/abs/2102.08390}{[arXiv:2102.08390]}\relax
\mciteBstWouldAddEndPuncttrue
\mciteSetBstMidEndSepPunct{\mcitedefaultmidpunct}
{\mcitedefaultendpunct}{\mcitedefaultseppunct}\relax
\EndOfBibitem
\bibitem{Alioli:2023har}
S.~Alioli {\em et~al.\/}, \href{http://doi.org/10.1007/JHEP05(2023)128}{JHEP}
  \href{http://doi.org/10.1007/JHEP05(2023)128}{{\bf 05}, 128} (2023),
  \href{https://arxiv.org/abs/2301.11875}{[arXiv:2301.11875]}\relax
\mciteBstWouldAddEndPuncttrue
\mciteSetBstMidEndSepPunct{\mcitedefaultmidpunct}
{\mcitedefaultendpunct}{\mcitedefaultseppunct}\relax
\EndOfBibitem
\bibitem{Astill:2016hpa}
W.~Astill {\em et~al.\/}, \href{http://doi.org/10.1007/JHEP06(2016)154}{JHEP}
  \href{http://doi.org/10.1007/JHEP06(2016)154}{{\bf 06}, 154} (2016),
  \href{https://arxiv.org/abs/1603.01620}{[arXiv:1603.01620]}\relax
\mciteBstWouldAddEndPuncttrue
\mciteSetBstMidEndSepPunct{\mcitedefaultmidpunct}
{\mcitedefaultendpunct}{\mcitedefaultseppunct}\relax
\EndOfBibitem
\bibitem{Astill:2018ivh}
W.~Astill {\em et~al.\/}, \href{http://doi.org/10.1007/JHEP11(2018)157}{JHEP}
  \href{http://doi.org/10.1007/JHEP11(2018)157}{{\bf 11}, 157} (2018),
  \href{https://arxiv.org/abs/1804.08141}{[arXiv:1804.08141]}\relax
\mciteBstWouldAddEndPuncttrue
\mciteSetBstMidEndSepPunct{\mcitedefaultmidpunct}
{\mcitedefaultendpunct}{\mcitedefaultseppunct}\relax
\EndOfBibitem
\bibitem{Re:2018vac}
E.~Re, M.~Wiesemann and G.~Zanderighi,
  \href{http://doi.org/10.1007/JHEP12(2018)121}{JHEP}
  \href{http://doi.org/10.1007/JHEP12(2018)121}{{\bf 12}, 121} (2018),
  \href{https://arxiv.org/abs/1805.09857}{[arXiv:1805.09857]}\relax
\mciteBstWouldAddEndPuncttrue
\mciteSetBstMidEndSepPunct{\mcitedefaultmidpunct}
{\mcitedefaultendpunct}{\mcitedefaultseppunct}\relax
\EndOfBibitem
\bibitem{Alioli:2019qzz}
S.~Alioli {\em et~al.\/},
  \href{http://doi.org/10.1103/PhysRevD.100.096016}{Phys. Rev. D}
  \href{http://doi.org/10.1103/PhysRevD.100.096016}{{\bf 100}, 9, 096016}
  (2019), \href{https://arxiv.org/abs/1909.02026}{[arXiv:1909.02026]}\relax
\mciteBstWouldAddEndPuncttrue
\mciteSetBstMidEndSepPunct{\mcitedefaultmidpunct}
{\mcitedefaultendpunct}{\mcitedefaultseppunct}\relax
\EndOfBibitem
\bibitem{Lombardi:2020wju}
D.~Lombardi, M.~Wiesemann and G.~Zanderighi  (2020),
  \href{https://arxiv.org/abs/2010.10478}{[arXiv:2010.10478]}\relax
\mciteBstWouldAddEndPuncttrue
\mciteSetBstMidEndSepPunct{\mcitedefaultmidpunct}
{\mcitedefaultendpunct}{\mcitedefaultseppunct}\relax
\EndOfBibitem
\bibitem{Lombardi:2021wug}
D.~Lombardi, M.~Wiesemann and G.~Zanderighi,
  \href{http://doi.org/10.1016/j.physletb.2021.136846}{Phys. Lett. B}
  \href{http://doi.org/10.1016/j.physletb.2021.136846}{{\bf 824}, 136846}
  (2022), \href{https://arxiv.org/abs/2108.11315}{[arXiv:2108.11315]}\relax
\mciteBstWouldAddEndPuncttrue
\mciteSetBstMidEndSepPunct{\mcitedefaultmidpunct}
{\mcitedefaultendpunct}{\mcitedefaultseppunct}\relax
\EndOfBibitem
\bibitem{Alioli:2020qrd}
S.~Alioli {\em et~al.\/}, \href{http://doi.org/10.1007/JHEP04(2021)041}{JHEP}
  \href{http://doi.org/10.1007/JHEP04(2021)041}{{\bf 04}, 041} (2021),
  \href{https://arxiv.org/abs/2010.10498}{[arXiv:2010.10498]}\relax
\mciteBstWouldAddEndPuncttrue
\mciteSetBstMidEndSepPunct{\mcitedefaultmidpunct}
{\mcitedefaultendpunct}{\mcitedefaultseppunct}\relax
\EndOfBibitem
\bibitem{Lombardi:2021rvg}
D.~Lombardi, M.~Wiesemann and G.~Zanderighi  (2021),
  \href{https://arxiv.org/abs/2103.12077}{[arXiv:2103.12077]}\relax
\mciteBstWouldAddEndPuncttrue
\mciteSetBstMidEndSepPunct{\mcitedefaultmidpunct}
{\mcitedefaultendpunct}{\mcitedefaultseppunct}\relax
\EndOfBibitem
\bibitem{Alioli:2021egp}
S.~Alioli {\em et~al.\/},
  \href{http://doi.org/10.1016/j.physletb.2021.136380}{Phys. Lett. B}
  \href{http://doi.org/10.1016/j.physletb.2021.136380}{{\bf 818}, 136380}
  (2021), \href{https://arxiv.org/abs/2103.01214}{[arXiv:2103.01214]}\relax
\mciteBstWouldAddEndPuncttrue
\mciteSetBstMidEndSepPunct{\mcitedefaultmidpunct}
{\mcitedefaultendpunct}{\mcitedefaultseppunct}\relax
\EndOfBibitem
\bibitem{Alioli:2022dkj}
S.~Alioli {\em et~al.\/}  (2022),
  \href{https://arxiv.org/abs/2212.10489}{[arXiv:2212.10489]}\relax
\mciteBstWouldAddEndPuncttrue
\mciteSetBstMidEndSepPunct{\mcitedefaultmidpunct}
{\mcitedefaultendpunct}{\mcitedefaultseppunct}\relax
\EndOfBibitem
\bibitem{Lindert:2022qdd}
J.~M. Lindert {\em et~al.\/},
  \href{http://doi.org/10.1007/JHEP11(2022)036}{JHEP}
  \href{http://doi.org/10.1007/JHEP11(2022)036}{{\bf 11}, 036} (2022),
  \href{https://arxiv.org/abs/2208.12660}{[arXiv:2208.12660]}\relax
\mciteBstWouldAddEndPuncttrue
\mciteSetBstMidEndSepPunct{\mcitedefaultmidpunct}
{\mcitedefaultendpunct}{\mcitedefaultseppunct}\relax
\EndOfBibitem
\bibitem{Gavardi:2022ixt}
A.~Gavardi, C.~Oleari and E.~Re,
  \href{http://doi.org/10.1007/JHEP09(2022)061}{JHEP}
  \href{http://doi.org/10.1007/JHEP09(2022)061}{{\bf 09}, 061} (2022),
  \href{https://arxiv.org/abs/2204.12602}{[arXiv:2204.12602]}\relax
\mciteBstWouldAddEndPuncttrue
\mciteSetBstMidEndSepPunct{\mcitedefaultmidpunct}
{\mcitedefaultendpunct}{\mcitedefaultseppunct}\relax
\EndOfBibitem
\bibitem{Buonocore:2021fnj}
L.~Buonocore {\em et~al.\/},
  \href{http://doi.org/10.1007/JHEP01(2022)072}{JHEP}
  \href{http://doi.org/10.1007/JHEP01(2022)072}{{\bf 01}, 072} (2022),
  \href{https://arxiv.org/abs/2108.05337}{[arXiv:2108.05337]}\relax
\mciteBstWouldAddEndPuncttrue
\mciteSetBstMidEndSepPunct{\mcitedefaultmidpunct}
{\mcitedefaultendpunct}{\mcitedefaultseppunct}\relax
\EndOfBibitem
\bibitem{Haisch:2022nwz}
U.~Haisch {\em et~al.\/}, \href{http://doi.org/10.1007/JHEP07(2022)054}{JHEP}
  \href{http://doi.org/10.1007/JHEP07(2022)054}{{\bf 07}, 054} (2022),
  \href{https://arxiv.org/abs/2204.00663}{[arXiv:2204.00663]}\relax
\mciteBstWouldAddEndPuncttrue
\mciteSetBstMidEndSepPunct{\mcitedefaultmidpunct}
{\mcitedefaultendpunct}{\mcitedefaultseppunct}\relax
\EndOfBibitem
\bibitem{Zanoli:2021iyp}
S.~Zanoli {\em et~al.\/}, \href{http://doi.org/10.1007/JHEP07(2022)008}{JHEP}
  \href{http://doi.org/10.1007/JHEP07(2022)008}{{\bf 07}, 008} (2022),
  \href{https://arxiv.org/abs/2112.04168}{[arXiv:2112.04168]}\relax
\mciteBstWouldAddEndPuncttrue
\mciteSetBstMidEndSepPunct{\mcitedefaultmidpunct}
{\mcitedefaultendpunct}{\mcitedefaultseppunct}\relax
\EndOfBibitem
\bibitem{Mazzitelli:2020jio}
J.~Mazzitelli {\em et~al.\/}  (2020),
  \href{https://arxiv.org/abs/2012.14267}{[arXiv:2012.14267]}\relax
\mciteBstWouldAddEndPuncttrue
\mciteSetBstMidEndSepPunct{\mcitedefaultmidpunct}
{\mcitedefaultendpunct}{\mcitedefaultseppunct}\relax
\EndOfBibitem
\bibitem{Mazzitelli:2021mmm}
J.~Mazzitelli {\em et~al.\/},
  \href{http://doi.org/10.1007/JHEP04(2022)079}{JHEP}
  \href{http://doi.org/10.1007/JHEP04(2022)079}{{\bf 04}, 079} (2022),
  \href{https://arxiv.org/abs/2112.12135}{[arXiv:2112.12135]}\relax
\mciteBstWouldAddEndPuncttrue
\mciteSetBstMidEndSepPunct{\mcitedefaultmidpunct}
{\mcitedefaultendpunct}{\mcitedefaultseppunct}\relax
\EndOfBibitem
\bibitem{Mazzitelli:2023znt}
J.~Mazzitelli {\em et~al.\/},
  \href{http://doi.org/10.1016/j.physletb.2023.137991}{Phys. Lett. B}
  \href{http://doi.org/10.1016/j.physletb.2023.137991}{{\bf 843}, 137991}
  (2023), \href{https://arxiv.org/abs/2302.01645}{[arXiv:2302.01645]}\relax
\mciteBstWouldAddEndPuncttrue
\mciteSetBstMidEndSepPunct{\mcitedefaultmidpunct}
{\mcitedefaultendpunct}{\mcitedefaultseppunct}\relax
\EndOfBibitem
\bibitem{Dasgupta:2018nvj}
M.~Dasgupta {\em et~al.\/}, \href{http://doi.org/10.1007/JHEP09(2018)033}{JHEP}
  \href{http://doi.org/10.1007/JHEP09(2018)033}{{\bf 09}, 033} (2018),
  [Erratum: JHEP 03, 083 (2020)],
  \href{https://arxiv.org/abs/1805.09327}{[arXiv:1805.09327]}\relax
\mciteBstWouldAddEndPuncttrue
\mciteSetBstMidEndSepPunct{\mcitedefaultmidpunct}
{\mcitedefaultendpunct}{\mcitedefaultseppunct}\relax
\EndOfBibitem
\bibitem{Hartgring:2013jma}
L.~Hartgring, E.~Laenen and P.~Skands,
  \href{http://doi.org/10.1007/JHEP10(2013)127}{JHEP}
  \href{http://doi.org/10.1007/JHEP10(2013)127}{{\bf 10}, 127} (2013),
  \href{https://arxiv.org/abs/1303.4974}{[arXiv:1303.4974]}\relax
\mciteBstWouldAddEndPuncttrue
\mciteSetBstMidEndSepPunct{\mcitedefaultmidpunct}
{\mcitedefaultendpunct}{\mcitedefaultseppunct}\relax
\EndOfBibitem
\bibitem{Li:2016yez}
H.~T. Li and P.~Skands,
  \href{http://doi.org/10.1016/j.physletb.2017.05.011}{Phys. Lett. B}
  \href{http://doi.org/10.1016/j.physletb.2017.05.011}{{\bf 771}, 59} (2017),
  \href{https://arxiv.org/abs/1611.00013}{[arXiv:1611.00013]}\relax
\mciteBstWouldAddEndPuncttrue
\mciteSetBstMidEndSepPunct{\mcitedefaultmidpunct}
{\mcitedefaultendpunct}{\mcitedefaultseppunct}\relax
\EndOfBibitem
\bibitem{Hoche:2017iem}
S.~H\"oche and S.~Prestel,
  \href{http://doi.org/10.1103/PhysRevD.96.074017}{Phys. Rev. D}
  \href{http://doi.org/10.1103/PhysRevD.96.074017}{{\bf 96}, 7, 074017} (2017),
  \href{https://arxiv.org/abs/1705.00742}{[arXiv:1705.00742]}\relax
\mciteBstWouldAddEndPuncttrue
\mciteSetBstMidEndSepPunct{\mcitedefaultmidpunct}
{\mcitedefaultendpunct}{\mcitedefaultseppunct}\relax
\EndOfBibitem
\bibitem{Dulat:2018vuy}
F.~Dulat, S.~H\"oche and S.~Prestel,
  \href{http://doi.org/10.1103/PhysRevD.98.074013}{Phys. Rev. D}
  \href{http://doi.org/10.1103/PhysRevD.98.074013}{{\bf 98}, 7, 074013} (2018),
  \href{https://arxiv.org/abs/1805.03757}{[arXiv:1805.03757]}\relax
\mciteBstWouldAddEndPuncttrue
\mciteSetBstMidEndSepPunct{\mcitedefaultmidpunct}
{\mcitedefaultendpunct}{\mcitedefaultseppunct}\relax
\EndOfBibitem
\bibitem{Banfi:2018mcq}
A.~Banfi, B.~K. El-Menoufi and P.~F. Monni,
  \href{http://doi.org/10.1007/JHEP01(2019)083}{JHEP}
  \href{http://doi.org/10.1007/JHEP01(2019)083}{{\bf 01}, 083} (2019),
  \href{https://arxiv.org/abs/1807.11487}{[arXiv:1807.11487]}\relax
\mciteBstWouldAddEndPuncttrue
\mciteSetBstMidEndSepPunct{\mcitedefaultmidpunct}
{\mcitedefaultendpunct}{\mcitedefaultseppunct}\relax
\EndOfBibitem
\bibitem{Bewick:2019rbu}
G.~Bewick {\em et~al.\/}, \href{http://doi.org/10.1007/JHEP04(2020)019}{JHEP}
  \href{http://doi.org/10.1007/JHEP04(2020)019}{{\bf 04}, 019} (2020),
  \href{https://arxiv.org/abs/1904.11866}{[arXiv:1904.11866]}\relax
\mciteBstWouldAddEndPuncttrue
\mciteSetBstMidEndSepPunct{\mcitedefaultmidpunct}
{\mcitedefaultendpunct}{\mcitedefaultseppunct}\relax
\EndOfBibitem
\bibitem{Dasgupta:2020fwr}
M.~Dasgupta {\em et~al.\/},
  \href{http://doi.org/10.1103/PhysRevLett.125.052002}{Phys. Rev. Lett.}
  \href{http://doi.org/10.1103/PhysRevLett.125.052002}{{\bf 125}, 5, 052002}
  (2020), \href{https://arxiv.org/abs/2002.11114}{[arXiv:2002.11114]}\relax
\mciteBstWouldAddEndPuncttrue
\mciteSetBstMidEndSepPunct{\mcitedefaultmidpunct}
{\mcitedefaultendpunct}{\mcitedefaultseppunct}\relax
\EndOfBibitem
\bibitem{Hamilton:2020rcu}
K.~Hamilton {\em et~al.\/}, \href{http://doi.org/10.1007/JHEP03(2021)041}{JHEP}
  \href{http://doi.org/10.1007/JHEP03(2021)041}{{\bf 03}, 041, 041} (2021),
  \href{https://arxiv.org/abs/2011.10054}{[arXiv:2011.10054]}\relax
\mciteBstWouldAddEndPuncttrue
\mciteSetBstMidEndSepPunct{\mcitedefaultmidpunct}
{\mcitedefaultendpunct}{\mcitedefaultseppunct}\relax
\EndOfBibitem
\bibitem{Forshaw:2020wrq}
J.~R. Forshaw, J.~Holguin and S.~Pl\"atzer,
  \href{http://doi.org/10.1007/JHEP09(2020)014}{JHEP}
  \href{http://doi.org/10.1007/JHEP09(2020)014}{{\bf 09}, 014} (2020),
  \href{https://arxiv.org/abs/2003.06400}{[arXiv:2003.06400]}\relax
\mciteBstWouldAddEndPuncttrue
\mciteSetBstMidEndSepPunct{\mcitedefaultmidpunct}
{\mcitedefaultendpunct}{\mcitedefaultseppunct}\relax
\EndOfBibitem
\bibitem{Nagy:2020rmk}
Z.~Nagy and D.~E. Soper,
  \href{http://doi.org/10.1103/PhysRevD.104.054049}{Phys. Rev. D}
  \href{http://doi.org/10.1103/PhysRevD.104.054049}{{\bf 104}, 5, 054049}
  (2021), \href{https://arxiv.org/abs/2011.04773}{[arXiv:2011.04773]}\relax
\mciteBstWouldAddEndPuncttrue
\mciteSetBstMidEndSepPunct{\mcitedefaultmidpunct}
{\mcitedefaultendpunct}{\mcitedefaultseppunct}\relax
\EndOfBibitem
\bibitem{Nagy:2020dvz}
Z.~Nagy and D.~E. Soper  (2020),
  \href{https://arxiv.org/abs/2011.04777}{[arXiv:2011.04777]}\relax
\mciteBstWouldAddEndPuncttrue
\mciteSetBstMidEndSepPunct{\mcitedefaultmidpunct}
{\mcitedefaultendpunct}{\mcitedefaultseppunct}\relax
\EndOfBibitem
\bibitem{Karlberg:2021kwr}
A.~Karlberg {\em et~al.\/},
  \href{http://doi.org/10.1140/epjc/s10052-021-09378-0}{Eur. Phys. J. C}
  \href{http://doi.org/10.1140/epjc/s10052-021-09378-0}{{\bf 81}, 8, 681}
  (2021), \href{https://arxiv.org/abs/2103.16526}{[arXiv:2103.16526]}\relax
\mciteBstWouldAddEndPuncttrue
\mciteSetBstMidEndSepPunct{\mcitedefaultmidpunct}
{\mcitedefaultendpunct}{\mcitedefaultseppunct}\relax
\EndOfBibitem
\bibitem{Hamilton:2021dyz}
K.~Hamilton {\em et~al.\/}, \href{http://doi.org/10.1007/JHEP03(2022)193}{JHEP}
  \href{http://doi.org/10.1007/JHEP03(2022)193}{{\bf 03}, 193} (2022),
  \href{https://arxiv.org/abs/2111.01161}{[arXiv:2111.01161]}\relax
\mciteBstWouldAddEndPuncttrue
\mciteSetBstMidEndSepPunct{\mcitedefaultmidpunct}
{\mcitedefaultendpunct}{\mcitedefaultseppunct}\relax
\EndOfBibitem
\bibitem{Herren:2022jej}
F.~Herren {\em et~al.\/}  (2022),
  \href{https://arxiv.org/abs/2208.06057}{[arXiv:2208.06057]}\relax
\mciteBstWouldAddEndPuncttrue
\mciteSetBstMidEndSepPunct{\mcitedefaultmidpunct}
{\mcitedefaultendpunct}{\mcitedefaultseppunct}\relax
\EndOfBibitem
\bibitem{Nagy:2022bph}
Z.~Nagy and D.~E. Soper,
  \href{http://doi.org/10.1103/PhysRevD.106.014024}{Phys. Rev. D}
  \href{http://doi.org/10.1103/PhysRevD.106.014024}{{\bf 106}, 1, 014024}
  (2022), \href{https://arxiv.org/abs/2204.05631}{[arXiv:2204.05631]}\relax
\mciteBstWouldAddEndPuncttrue
\mciteSetBstMidEndSepPunct{\mcitedefaultmidpunct}
{\mcitedefaultendpunct}{\mcitedefaultseppunct}\relax
\EndOfBibitem
\bibitem{vanBeekveld:2022zhl}
M.~van Beekveld {\em et~al.\/},
  \href{http://doi.org/10.1007/JHEP11(2022)019}{JHEP}
  \href{http://doi.org/10.1007/JHEP11(2022)019}{{\bf 11}, 019} (2022),
  \href{https://arxiv.org/abs/2205.02237}{[arXiv:2205.02237]}\relax
\mciteBstWouldAddEndPuncttrue
\mciteSetBstMidEndSepPunct{\mcitedefaultmidpunct}
{\mcitedefaultendpunct}{\mcitedefaultseppunct}\relax
\EndOfBibitem
\bibitem{vanBeekveld:2022ukn}
M.~van Beekveld {\em et~al.\/},
  \href{http://doi.org/10.1007/JHEP11(2022)020}{JHEP}
  \href{http://doi.org/10.1007/JHEP11(2022)020}{{\bf 11}, 020} (2022),
  \href{https://arxiv.org/abs/2207.09467}{[arXiv:2207.09467]}\relax
\mciteBstWouldAddEndPuncttrue
\mciteSetBstMidEndSepPunct{\mcitedefaultmidpunct}
{\mcitedefaultendpunct}{\mcitedefaultseppunct}\relax
\EndOfBibitem
\bibitem{vanBeekveld:2023lfu}
M.~van Beekveld and S.~Ferrario~Ravasio  (2023),
  \href{https://arxiv.org/abs/2305.08645}{[arXiv:2305.08645]}\relax
\mciteBstWouldAddEndPuncttrue
\mciteSetBstMidEndSepPunct{\mcitedefaultmidpunct}
{\mcitedefaultendpunct}{\mcitedefaultseppunct}\relax
\EndOfBibitem
\bibitem{Cacciari:2003fi}
M.~Cacciari {\em et~al.\/},
  \href{http://doi.org/10.1088/1126-6708/2004/04/068}{JHEP}
  \href{http://doi.org/10.1088/1126-6708/2004/04/068}{{\bf 04}, 068} (2004),
  \href{https://arxiv.org/abs/hep-ph/0303085}{[hep-ph/0303085]}\relax
\mciteBstWouldAddEndPuncttrue
\mciteSetBstMidEndSepPunct{\mcitedefaultmidpunct}
{\mcitedefaultendpunct}{\mcitedefaultseppunct}\relax
\EndOfBibitem
\bibitem{Cacciari:2011ze}
M.~Cacciari and N.~Houdeau, \href{http://doi.org/10.1007/JHEP09(2011)039}{JHEP}
  \href{http://doi.org/10.1007/JHEP09(2011)039}{{\bf 09}, 039} (2011),
  \href{https://arxiv.org/abs/1105.5152}{[arXiv:1105.5152]}\relax
\mciteBstWouldAddEndPuncttrue
\mciteSetBstMidEndSepPunct{\mcitedefaultmidpunct}
{\mcitedefaultendpunct}{\mcitedefaultseppunct}\relax
\EndOfBibitem
\bibitem{David:2013gaa}
A.~David and G.~Passarino,
  \href{http://doi.org/10.1016/j.physletb.2013.08.025}{Phys. Lett.}
  \href{http://doi.org/10.1016/j.physletb.2013.08.025}{{\bf B726}, 266} (2013),
  \href{https://arxiv.org/abs/1307.1843}{[arXiv:1307.1843]}\relax
\mciteBstWouldAddEndPuncttrue
\mciteSetBstMidEndSepPunct{\mcitedefaultmidpunct}
{\mcitedefaultendpunct}{\mcitedefaultseppunct}\relax
\EndOfBibitem
\bibitem{Bagnaschi:2014wea}
E.~Bagnaschi {\em et~al.\/},
  \href{http://doi.org/10.1007/JHEP02(2015)133}{JHEP}
  \href{http://doi.org/10.1007/JHEP02(2015)133}{{\bf 02}, 133} (2015),
  \href{https://arxiv.org/abs/1409.5036}{[arXiv:1409.5036]}\relax
\mciteBstWouldAddEndPuncttrue
\mciteSetBstMidEndSepPunct{\mcitedefaultmidpunct}
{\mcitedefaultendpunct}{\mcitedefaultseppunct}\relax
\EndOfBibitem
\bibitem{Duhr:2021mfd}
C.~Duhr {\em et~al.\/}, \href{http://doi.org/10.1007/JHEP09(2021)122}{JHEP}
  \href{http://doi.org/10.1007/JHEP09(2021)122}{{\bf 09}, 122} (2021),
  \href{https://arxiv.org/abs/2106.04585}{[arXiv:2106.04585]}\relax
\mciteBstWouldAddEndPuncttrue
\mciteSetBstMidEndSepPunct{\mcitedefaultmidpunct}
{\mcitedefaultendpunct}{\mcitedefaultseppunct}\relax
\EndOfBibitem
\bibitem{Ghosh:2022lrf}
A.~Ghosh {\em et~al.\/}  (2022),
  \href{https://arxiv.org/abs/2210.15167}{[arXiv:2210.15167]}\relax
\mciteBstWouldAddEndPuncttrue
\mciteSetBstMidEndSepPunct{\mcitedefaultmidpunct}
{\mcitedefaultendpunct}{\mcitedefaultseppunct}\relax
\EndOfBibitem
\bibitem{Soyez:2012hv}
G.~Soyez {\em et~al.\/},
  \href{http://doi.org/10.1103/PhysRevLett.110.162001}{Phys. Rev. Lett.}
  \href{http://doi.org/10.1103/PhysRevLett.110.162001}{{\bf 110}, 16, 162001}
  (2013), \href{https://arxiv.org/abs/1211.2811}{[arXiv:1211.2811]}\relax
\mciteBstWouldAddEndPuncttrue
\mciteSetBstMidEndSepPunct{\mcitedefaultmidpunct}
{\mcitedefaultendpunct}{\mcitedefaultseppunct}\relax
\EndOfBibitem
\bibitem{Dasgupta:2003iq}
M.~Dasgupta and G.~P. Salam,
  \href{http://doi.org/10.1088/0954-3899/30/5/R01}{J. Phys.}
  \href{http://doi.org/10.1088/0954-3899/30/5/R01}{{\bf G30}, R143} (2004),
  \href{https://arxiv.org/abs/hep-ph/0312283}{[hep-ph/0312283]}\relax
\mciteBstWouldAddEndPuncttrue
\mciteSetBstMidEndSepPunct{\mcitedefaultmidpunct}
{\mcitedefaultendpunct}{\mcitedefaultseppunct}\relax
\EndOfBibitem
\bibitem{Bierlich:2019rhm}
C.~Bierlich {\em et~al.\/},
  \href{http://doi.org/10.21468/SciPostPhys.8.2.026}{SciPost Phys.}
  \href{http://doi.org/10.21468/SciPostPhys.8.2.026}{{\bf 8}, 026} (2020),
  \href{https://arxiv.org/abs/1912.05451}{[arXiv:1912.05451]}\relax
\mciteBstWouldAddEndPuncttrue
\mciteSetBstMidEndSepPunct{\mcitedefaultmidpunct}
{\mcitedefaultendpunct}{\mcitedefaultseppunct}\relax
\EndOfBibitem
\bibitem{Moretti:1998qx}
S.~Moretti, L.~Lonnblad and T.~Sjostrand,
  \href{http://doi.org/10.1088/1126-6708/1998/08/001}{JHEP}
  \href{http://doi.org/10.1088/1126-6708/1998/08/001}{{\bf 08}, 001} (1998),
  \href{https://arxiv.org/abs/hep-ph/9804296}{[hep-ph/9804296]}\relax
\mciteBstWouldAddEndPuncttrue
\mciteSetBstMidEndSepPunct{\mcitedefaultmidpunct}
{\mcitedefaultendpunct}{\mcitedefaultseppunct}\relax
\EndOfBibitem
\bibitem{Salam:2009jx}
G.~P. Salam, \href{http://doi.org/10.1140/epjc/s10052-010-1314-6}{Eur. Phys.
  J.} \href{http://doi.org/10.1140/epjc/s10052-010-1314-6}{{\bf C67}, 637}
  (2010), \href{https://arxiv.org/abs/0906.1833}{[arXiv:0906.1833]}\relax
\mciteBstWouldAddEndPuncttrue
\mciteSetBstMidEndSepPunct{\mcitedefaultmidpunct}
{\mcitedefaultendpunct}{\mcitedefaultseppunct}\relax
\EndOfBibitem
\bibitem{Ellis:2007ib}
S.~D. Ellis {\em et~al.\/},
  \href{http://doi.org/10.1016/j.ppnp.2007.12.002}{Prog. Part. Nucl. Phys.}
  \href{http://doi.org/10.1016/j.ppnp.2007.12.002}{{\bf 60}, 484} (2008),
  \href{https://arxiv.org/abs/0712.2447}{[arXiv:0712.2447]}\relax
\mciteBstWouldAddEndPuncttrue
\mciteSetBstMidEndSepPunct{\mcitedefaultmidpunct}
{\mcitedefaultendpunct}{\mcitedefaultseppunct}\relax
\EndOfBibitem
\bibitem{Cacciari:2015jwa}
M.~Cacciari, \href{http://doi.org/10.1142/S0217751X1546001X}{Int. J. Mod.
  Phys.} \href{http://doi.org/10.1142/S0217751X1546001X}{{\bf A30}, 31,
  1546001} (2015),
  \href{https://arxiv.org/abs/1509.02272}{[arXiv:1509.02272]}\relax
\mciteBstWouldAddEndPuncttrue
\mciteSetBstMidEndSepPunct{\mcitedefaultmidpunct}
{\mcitedefaultendpunct}{\mcitedefaultseppunct}\relax
\EndOfBibitem
\bibitem{CDF:2005vhb}
A.~Abulencia {\em et~al.\/} (CDF),
  \href{http://doi.org/10.1103/PhysRevD.74.071103}{Phys. Rev. D}
  \href{http://doi.org/10.1103/PhysRevD.74.071103}{{\bf 74}, 071103} (2006),
  \href{https://arxiv.org/abs/hep-ex/0512020}{[hep-ex/0512020]}\relax
\mciteBstWouldAddEndPuncttrue
\mciteSetBstMidEndSepPunct{\mcitedefaultmidpunct}
{\mcitedefaultendpunct}{\mcitedefaultseppunct}\relax
\EndOfBibitem
\bibitem{Salam:2007xv}
G.~P. Salam and G.~Soyez,
  \href{http://doi.org/10.1088/1126-6708/2007/05/086}{JHEP}
  \href{http://doi.org/10.1088/1126-6708/2007/05/086}{{\bf 05}, 086} (2007),
  \href{https://arxiv.org/abs/0704.0292}{[arXiv:0704.0292]}\relax
\mciteBstWouldAddEndPuncttrue
\mciteSetBstMidEndSepPunct{\mcitedefaultmidpunct}
{\mcitedefaultendpunct}{\mcitedefaultseppunct}\relax
\EndOfBibitem
\bibitem{Catani:1993hr}
S.~Catani {\em et~al.\/},
  \href{http://doi.org/10.1016/0550-3213(93)90166-M}{Nucl. Phys.}
  \href{http://doi.org/10.1016/0550-3213(93)90166-M}{{\bf B406}, 187}
  (1993)\relax
\mciteBstWouldAddEndPuncttrue
\mciteSetBstMidEndSepPunct{\mcitedefaultmidpunct}
{\mcitedefaultendpunct}{\mcitedefaultseppunct}\relax
\EndOfBibitem
\bibitem{Ellis:1993tq}
S.~D. Ellis and D.~E. Soper,
  \href{http://doi.org/10.1103/PhysRevD.48.3160}{Phys. Rev.}
  \href{http://doi.org/10.1103/PhysRevD.48.3160}{{\bf D48}, 3160} (1993),
  \href{https://arxiv.org/abs/hep-ph/9305266}{[hep-ph/9305266]}\relax
\mciteBstWouldAddEndPuncttrue
\mciteSetBstMidEndSepPunct{\mcitedefaultmidpunct}
{\mcitedefaultendpunct}{\mcitedefaultseppunct}\relax
\EndOfBibitem
\bibitem{Dokshitzer:1997in}
Y.~L. Dokshitzer {\em et~al.\/},
  \href{http://doi.org/10.1088/1126-6708/1997/08/001}{JHEP}
  \href{http://doi.org/10.1088/1126-6708/1997/08/001}{{\bf 08}, 001} (1997),
  \href{https://arxiv.org/abs/hep-ph/9707323}{[hep-ph/9707323]}\relax
\mciteBstWouldAddEndPuncttrue
\mciteSetBstMidEndSepPunct{\mcitedefaultmidpunct}
{\mcitedefaultendpunct}{\mcitedefaultseppunct}\relax
\EndOfBibitem
\bibitem{Wobisch:1998wt}
M.~Wobisch and T.~Wengler, in \enquote{{Monte Carlo generators for HERA
  physics. Proceedings, Workshop, Hamburg, Germany, 1998-1999},} 270--279
  (1998), \href{https://arxiv.org/abs/hep-ph/9907280}{[hep-ph/9907280]}\relax
\mciteBstWouldAddEndPuncttrue
\mciteSetBstMidEndSepPunct{\mcitedefaultmidpunct}
{\mcitedefaultendpunct}{\mcitedefaultseppunct}\relax
\EndOfBibitem
\bibitem{Cacciari:2008gp}
M.~Cacciari, G.~P. Salam and G.~Soyez,
  \href{http://doi.org/10.1088/1126-6708/2008/04/063}{JHEP}
  \href{http://doi.org/10.1088/1126-6708/2008/04/063}{{\bf 04}, 063} (2008),
  \href{https://arxiv.org/abs/0802.1189}{[arXiv:0802.1189]}\relax
\mciteBstWouldAddEndPuncttrue
\mciteSetBstMidEndSepPunct{\mcitedefaultmidpunct}
{\mcitedefaultendpunct}{\mcitedefaultseppunct}\relax
\EndOfBibitem
\bibitem{Bethke:1991wk}
S.~Bethke {\em et~al.\/},
  \href{http://doi.org/10.1016/S0550-3213(98)00219-3}{Nucl. Phys.}
  \href{http://doi.org/10.1016/S0550-3213(98)00219-3}{{\bf B370}, 310} (1992),
  [Erratum: Nucl. Phys.B523,681(1998)]\relax
\mciteBstWouldAddEndPuncttrue
\mciteSetBstMidEndSepPunct{\mcitedefaultmidpunct}
{\mcitedefaultendpunct}{\mcitedefaultseppunct}\relax
\EndOfBibitem
\bibitem{Cacciari:2005hq}
M.~Cacciari and G.~P. Salam,
  \href{http://doi.org/10.1016/j.physletb.2006.08.037}{Phys. Lett.}
  \href{http://doi.org/10.1016/j.physletb.2006.08.037}{{\bf B641}, 57} (2006),
  \href{https://arxiv.org/abs/hep-ph/0512210}{[hep-ph/0512210]}\relax
\mciteBstWouldAddEndPuncttrue
\mciteSetBstMidEndSepPunct{\mcitedefaultmidpunct}
{\mcitedefaultendpunct}{\mcitedefaultseppunct}\relax
\EndOfBibitem
\bibitem{Cacciari:2011ma}
M.~Cacciari, G.~P. Salam and G.~Soyez,
  \href{http://doi.org/10.1140/epjc/s10052-012-1896-2}{Eur. Phys. J.}
  \href{http://doi.org/10.1140/epjc/s10052-012-1896-2}{{\bf C72}, 1896} (2012),
  \href{https://arxiv.org/abs/1111.6097}{[arXiv:1111.6097]}\relax
\mciteBstWouldAddEndPuncttrue
\mciteSetBstMidEndSepPunct{\mcitedefaultmidpunct}
{\mcitedefaultendpunct}{\mcitedefaultseppunct}\relax
\EndOfBibitem
\bibitem{Banfi:2006hf}
A.~Banfi, G.~P. Salam and G.~Zanderighi,
  \href{http://doi.org/10.1140/epjc/s2006-02552-4}{Eur. Phys. J. C}
  \href{http://doi.org/10.1140/epjc/s2006-02552-4}{{\bf 47}, 113} (2006),
  \href{https://arxiv.org/abs/hep-ph/0601139}{[hep-ph/0601139]}\relax
\mciteBstWouldAddEndPuncttrue
\mciteSetBstMidEndSepPunct{\mcitedefaultmidpunct}
{\mcitedefaultendpunct}{\mcitedefaultseppunct}\relax
\EndOfBibitem
\bibitem{Banfi:2007gu}
A.~Banfi, G.~P. Salam and G.~Zanderighi,
  \href{http://doi.org/10.1088/1126-6708/2007/07/026}{JHEP}
  \href{http://doi.org/10.1088/1126-6708/2007/07/026}{{\bf 07}, 026} (2007),
  \href{https://arxiv.org/abs/0704.2999}{[arXiv:0704.2999]}\relax
\mciteBstWouldAddEndPuncttrue
\mciteSetBstMidEndSepPunct{\mcitedefaultmidpunct}
{\mcitedefaultendpunct}{\mcitedefaultseppunct}\relax
\EndOfBibitem
\bibitem{Czakon:2022wam}
M.~Czakon, A.~Mitov and R.~Poncelet,
  \href{http://doi.org/10.1007/JHEP04(2023)138}{JHEP}
  \href{http://doi.org/10.1007/JHEP04(2023)138}{{\bf 04}, 138} (2023),
  \href{https://arxiv.org/abs/2205.11879}{[arXiv:2205.11879]}\relax
\mciteBstWouldAddEndPuncttrue
\mciteSetBstMidEndSepPunct{\mcitedefaultmidpunct}
{\mcitedefaultendpunct}{\mcitedefaultseppunct}\relax
\EndOfBibitem
\bibitem{Gauld:2022lem}
R.~Gauld, A.~Huss and G.~Stagnitto,
  \href{http://doi.org/10.1103/PhysRevLett.130.161901}{Phys. Rev. Lett.}
  \href{http://doi.org/10.1103/PhysRevLett.130.161901}{{\bf 130}, 16, 161901}
  (2023), \href{https://arxiv.org/abs/2208.11138}{[arXiv:2208.11138]}\relax
\mciteBstWouldAddEndPuncttrue
\mciteSetBstMidEndSepPunct{\mcitedefaultmidpunct}
{\mcitedefaultendpunct}{\mcitedefaultseppunct}\relax
\EndOfBibitem
\bibitem{Caletti:2022glq}
S.~Caletti {\em et~al.\/}, \href{http://doi.org/10.1007/JHEP10(2022)158}{JHEP}
  \href{http://doi.org/10.1007/JHEP10(2022)158}{{\bf 10}, 158} (2022),
  \href{https://arxiv.org/abs/2205.01117}{[arXiv:2205.01117]}\relax
\mciteBstWouldAddEndPuncttrue
\mciteSetBstMidEndSepPunct{\mcitedefaultmidpunct}
{\mcitedefaultendpunct}{\mcitedefaultseppunct}\relax
\EndOfBibitem
\bibitem{Caola:2023wpj}
F.~Caola {\em et~al.\/}  (2023),
  \href{https://arxiv.org/abs/2306.07314}{[arXiv:2306.07314]}\relax
\mciteBstWouldAddEndPuncttrue
\mciteSetBstMidEndSepPunct{\mcitedefaultmidpunct}
{\mcitedefaultendpunct}{\mcitedefaultseppunct}\relax
\EndOfBibitem
\bibitem{Brandt:1964sa}
S.~Brandt {\em et~al.\/},
  \href{http://doi.org/10.1016/0031-9163(64)91176-X}{Phys. Lett.}
  \href{http://doi.org/10.1016/0031-9163(64)91176-X}{{\bf 12}, 57} (1964)\relax
\mciteBstWouldAddEndPuncttrue
\mciteSetBstMidEndSepPunct{\mcitedefaultmidpunct}
{\mcitedefaultendpunct}{\mcitedefaultseppunct}\relax
\EndOfBibitem
\bibitem{Farhi:1977sg}
E.~Farhi, \href{http://doi.org/10.1103/PhysRevLett.39.1587}{Phys. Rev. Lett.}
  \href{http://doi.org/10.1103/PhysRevLett.39.1587}{{\bf 39}, 1587}
  (1977)\relax
\mciteBstWouldAddEndPuncttrue
\mciteSetBstMidEndSepPunct{\mcitedefaultmidpunct}
{\mcitedefaultendpunct}{\mcitedefaultseppunct}\relax
\EndOfBibitem
\bibitem{Biebel:2001dm}
O.~Biebel, \href{http://doi.org/10.1016/S0370-1573(00)00072-7}{Phys. Rept.}
  \href{http://doi.org/10.1016/S0370-1573(00)00072-7}{{\bf 340}, 165}
  (2001)\relax
\mciteBstWouldAddEndPuncttrue
\mciteSetBstMidEndSepPunct{\mcitedefaultmidpunct}
{\mcitedefaultendpunct}{\mcitedefaultseppunct}\relax
\EndOfBibitem
\bibitem{Kluth:2006bw}
S.~Kluth, \href{http://doi.org/10.1088/0034-4885/69/6/R04}{Rept. Prog. Phys.}
  \href{http://doi.org/10.1088/0034-4885/69/6/R04}{{\bf 69}, 1771} (2006),
  \href{https://arxiv.org/abs/hep-ex/0603011}{[hep-ex/0603011]}\relax
\mciteBstWouldAddEndPuncttrue
\mciteSetBstMidEndSepPunct{\mcitedefaultmidpunct}
{\mcitedefaultendpunct}{\mcitedefaultseppunct}\relax
\EndOfBibitem
\bibitem{Basham:1978bw}
C.~L. Basham {\em et~al.\/},
  \href{http://doi.org/10.1103/PhysRevLett.41.1585}{Phys. Rev. Lett.}
  \href{http://doi.org/10.1103/PhysRevLett.41.1585}{{\bf 41}, 1585}
  (1978)\relax
\mciteBstWouldAddEndPuncttrue
\mciteSetBstMidEndSepPunct{\mcitedefaultmidpunct}
{\mcitedefaultendpunct}{\mcitedefaultseppunct}\relax
\EndOfBibitem
\bibitem{Ali:1984yp}
A.~Ali, E.~Pietarinen and W.~J. Stirling,
  \href{http://doi.org/10.1016/0370-2693(84)90283-1}{Phys. Lett.}
  \href{http://doi.org/10.1016/0370-2693(84)90283-1}{{\bf 141B}, 447}
  (1984)\relax
\mciteBstWouldAddEndPuncttrue
\mciteSetBstMidEndSepPunct{\mcitedefaultmidpunct}
{\mcitedefaultendpunct}{\mcitedefaultseppunct}\relax
\EndOfBibitem
\bibitem{Stewart:2010tn}
I.~W. Stewart, F.~J. Tackmann and W.~J. Waalewijn,
  \href{http://doi.org/10.1103/PhysRevLett.105.092002}{Phys. Rev. Lett.}
  \href{http://doi.org/10.1103/PhysRevLett.105.092002}{{\bf 105}, 092002}
  (2010), \href{https://arxiv.org/abs/1004.2489}{[arXiv:1004.2489]}\relax
\mciteBstWouldAddEndPuncttrue
\mciteSetBstMidEndSepPunct{\mcitedefaultmidpunct}
{\mcitedefaultendpunct}{\mcitedefaultseppunct}\relax
\EndOfBibitem
\bibitem{Banfi:2004nk}
A.~Banfi, G.~P. Salam and G.~Zanderighi,
  \href{http://doi.org/10.1088/1126-6708/2004/08/062}{JHEP}
  \href{http://doi.org/10.1088/1126-6708/2004/08/062}{{\bf 08}, 062} (2004),
  \href{https://arxiv.org/abs/hep-ph/0407287}{[hep-ph/0407287]}\relax
\mciteBstWouldAddEndPuncttrue
\mciteSetBstMidEndSepPunct{\mcitedefaultmidpunct}
{\mcitedefaultendpunct}{\mcitedefaultseppunct}\relax
\EndOfBibitem
\bibitem{Banfi:2010xy}
A.~Banfi, G.~P. Salam and G.~Zanderighi,
  \href{http://doi.org/10.1007/JHEP06(2010)038}{JHEP}
  \href{http://doi.org/10.1007/JHEP06(2010)038}{{\bf 06}, 038} (2010),
  \href{https://arxiv.org/abs/1001.4082}{[arXiv:1001.4082]}\relax
\mciteBstWouldAddEndPuncttrue
\mciteSetBstMidEndSepPunct{\mcitedefaultmidpunct}
{\mcitedefaultendpunct}{\mcitedefaultseppunct}\relax
\EndOfBibitem
\bibitem{Becher:2015lmy}
T.~Becher, X.~Garcia~i Tormo and J.~Piclum,
  \href{http://doi.org/10.1103/PhysRevD.93.054038}{Phys. Rev.}
  \href{http://doi.org/10.1103/PhysRevD.93.054038}{{\bf D93}, 5, 054038}
  (2016), [Erratum: Phys. Rev.D93,no.7,079905(2016)],
  \href{https://arxiv.org/abs/1512.00022}{[arXiv:1512.00022]}\relax
\mciteBstWouldAddEndPuncttrue
\mciteSetBstMidEndSepPunct{\mcitedefaultmidpunct}
{\mcitedefaultendpunct}{\mcitedefaultseppunct}\relax
\EndOfBibitem
\bibitem{Gao:2019ojf}
A.~Gao {\em et~al.\/},
  \href{http://doi.org/10.1103/PhysRevLett.123.062001}{Phys. Rev. Lett.}
  \href{http://doi.org/10.1103/PhysRevLett.123.062001}{{\bf 123}, 6, 062001}
  (2019), \href{https://arxiv.org/abs/1901.04497}{[arXiv:1901.04497]}\relax
\mciteBstWouldAddEndPuncttrue
\mciteSetBstMidEndSepPunct{\mcitedefaultmidpunct}
{\mcitedefaultendpunct}{\mcitedefaultseppunct}\relax
\EndOfBibitem
\bibitem{Aaltonen:2011et}
T.~Aaltonen {\em et~al.\/} (CDF),
  \href{http://doi.org/10.1103/PhysRevD.83.112007}{Phys. Rev.}
  \href{http://doi.org/10.1103/PhysRevD.83.112007}{{\bf D83}, 112007} (2011),
  \href{https://arxiv.org/abs/1103.5143}{[arXiv:1103.5143]}\relax
\mciteBstWouldAddEndPuncttrue
\mciteSetBstMidEndSepPunct{\mcitedefaultmidpunct}
{\mcitedefaultendpunct}{\mcitedefaultseppunct}\relax
\EndOfBibitem
\bibitem{Aad:2012np}
G.~Aad {\em et~al.\/} (ATLAS),
  \href{http://doi.org/10.1140/epjc/s10052-012-2211-y}{Eur. Phys. J.}
  \href{http://doi.org/10.1140/epjc/s10052-012-2211-y}{{\bf C72}, 2211} (2012),
  \href{https://arxiv.org/abs/1206.2135}{[arXiv:1206.2135]}\relax
\mciteBstWouldAddEndPuncttrue
\mciteSetBstMidEndSepPunct{\mcitedefaultmidpunct}
{\mcitedefaultendpunct}{\mcitedefaultseppunct}\relax
\EndOfBibitem
\bibitem{Aad:2012fza}
G.~Aad {\em et~al.\/} (ATLAS),
  \href{http://doi.org/10.1103/PhysRevD.88.032004}{Phys. Rev.}
  \href{http://doi.org/10.1103/PhysRevD.88.032004}{{\bf D88}, 3, 032004}
  (2013), \href{https://arxiv.org/abs/1207.6915}{[arXiv:1207.6915]}\relax
\mciteBstWouldAddEndPuncttrue
\mciteSetBstMidEndSepPunct{\mcitedefaultmidpunct}
{\mcitedefaultendpunct}{\mcitedefaultseppunct}\relax
\EndOfBibitem
\bibitem{ATLAS:2015yaa}
G.~Aad {\em et~al.\/} (ATLAS),
  \href{http://doi.org/10.1016/j.physletb.2015.09.050}{Phys. Lett.}
  \href{http://doi.org/10.1016/j.physletb.2015.09.050}{{\bf B750}, 427} (2015),
  \href{https://arxiv.org/abs/1508.01579}{[arXiv:1508.01579]}\relax
\mciteBstWouldAddEndPuncttrue
\mciteSetBstMidEndSepPunct{\mcitedefaultmidpunct}
{\mcitedefaultendpunct}{\mcitedefaultseppunct}\relax
\EndOfBibitem
\bibitem{Aad:2016ria}
G.~Aad {\em et~al.\/} (ATLAS),
  \href{http://doi.org/10.1140/epjc/s10052-016-4176-8}{Eur. Phys. J.}
  \href{http://doi.org/10.1140/epjc/s10052-016-4176-8}{{\bf C76}, 7, 375}
  (2016), \href{https://arxiv.org/abs/1602.08980}{[arXiv:1602.08980]}\relax
\mciteBstWouldAddEndPuncttrue
\mciteSetBstMidEndSepPunct{\mcitedefaultmidpunct}
{\mcitedefaultendpunct}{\mcitedefaultseppunct}\relax
\EndOfBibitem
\bibitem{Aaboud:2017fml}
M.~Aaboud {\em et~al.\/} (ATLAS),
  \href{http://doi.org/10.1140/epjc/s10052-017-5442-0}{Eur. Phys. J.}
  \href{http://doi.org/10.1140/epjc/s10052-017-5442-0}{{\bf C77}, 12, 872}
  (2017), \href{https://arxiv.org/abs/1707.02562}{[arXiv:1707.02562]}\relax
\mciteBstWouldAddEndPuncttrue
\mciteSetBstMidEndSepPunct{\mcitedefaultmidpunct}
{\mcitedefaultendpunct}{\mcitedefaultseppunct}\relax
\EndOfBibitem
\bibitem{Aaboud:2018hie}
M.~Aaboud {\em et~al.\/} (ATLAS),
  \href{http://doi.org/10.1103/PhysRevD.98.092004}{Phys. Rev.}
  \href{http://doi.org/10.1103/PhysRevD.98.092004}{{\bf D98}, 9, 092004}
  (2018), \href{https://arxiv.org/abs/1805.04691}{[arXiv:1805.04691]}\relax
\mciteBstWouldAddEndPuncttrue
\mciteSetBstMidEndSepPunct{\mcitedefaultmidpunct}
{\mcitedefaultendpunct}{\mcitedefaultseppunct}\relax
\EndOfBibitem
\bibitem{ATLAS:2023tgo}
G.~Aad {\em et~al.\/} (ATLAS),
  \href{http://doi.org/10.1007/JHEP07(2023)085}{JHEP}
  \href{http://doi.org/10.1007/JHEP07(2023)085}{{\bf 07}, 085} (2023),
  \href{https://arxiv.org/abs/2301.09351}{[arXiv:2301.09351]}\relax
\mciteBstWouldAddEndPuncttrue
\mciteSetBstMidEndSepPunct{\mcitedefaultmidpunct}
{\mcitedefaultendpunct}{\mcitedefaultseppunct}\relax
\EndOfBibitem
\bibitem{Khachatryan:2011dx}
V.~Khachatryan {\em et~al.\/} (CMS),
  \href{http://doi.org/10.1016/j.physletb.2011.03.060}{Phys. Lett.}
  \href{http://doi.org/10.1016/j.physletb.2011.03.060}{{\bf B699}, 48} (2011),
  \href{https://arxiv.org/abs/1102.0068}{[arXiv:1102.0068]}\relax
\mciteBstWouldAddEndPuncttrue
\mciteSetBstMidEndSepPunct{\mcitedefaultmidpunct}
{\mcitedefaultendpunct}{\mcitedefaultseppunct}\relax
\EndOfBibitem
\bibitem{Chatrchyan:2013tna}
S.~Chatrchyan {\em et~al.\/} (CMS),
  \href{http://doi.org/10.1016/j.physletb.2013.04.025}{Phys. Lett.}
  \href{http://doi.org/10.1016/j.physletb.2013.04.025}{{\bf B722}, 238} (2013),
  \href{https://arxiv.org/abs/1301.1646}{[arXiv:1301.1646]}\relax
\mciteBstWouldAddEndPuncttrue
\mciteSetBstMidEndSepPunct{\mcitedefaultmidpunct}
{\mcitedefaultendpunct}{\mcitedefaultseppunct}\relax
\EndOfBibitem
\bibitem{Khachatryan:2014ika}
V.~Khachatryan {\em et~al.\/} (CMS),
  \href{http://doi.org/10.1007/JHEP10(2014)087}{JHEP}
  \href{http://doi.org/10.1007/JHEP10(2014)087}{{\bf 10}, 87} (2014),
  \href{https://arxiv.org/abs/1407.2856}{[arXiv:1407.2856]}\relax
\mciteBstWouldAddEndPuncttrue
\mciteSetBstMidEndSepPunct{\mcitedefaultmidpunct}
{\mcitedefaultendpunct}{\mcitedefaultseppunct}\relax
\EndOfBibitem
\bibitem{Sirunyan:2018adt}
A.~M. Sirunyan {\em et~al.\/} (CMS),
  \href{http://doi.org/10.1007/JHEP12(2018)117}{JHEP}
  \href{http://doi.org/10.1007/JHEP12(2018)117}{{\bf 12}, 117} (2018),
  \href{https://arxiv.org/abs/1811.00588}{[arXiv:1811.00588]}\relax
\mciteBstWouldAddEndPuncttrue
\mciteSetBstMidEndSepPunct{\mcitedefaultmidpunct}
{\mcitedefaultendpunct}{\mcitedefaultseppunct}\relax
\EndOfBibitem
\bibitem{Ali:2010tw}
A.~Ali and G.~Kramer, \href{http://doi.org/10.1140/epjh/e2011-10047-1}{Eur.
  Phys. J. H} \href{http://doi.org/10.1140/epjh/e2011-10047-1}{{\bf 36}, 245}
  (2011), \href{https://arxiv.org/abs/1012.2288}{[arXiv:1012.2288]}\relax
\mciteBstWouldAddEndPuncttrue
\mciteSetBstMidEndSepPunct{\mcitedefaultmidpunct}
{\mcitedefaultendpunct}{\mcitedefaultseppunct}\relax
\EndOfBibitem
\bibitem{Chatrchyan:2013kwa}
S.~Chatrchyan {\em et~al.\/} (CMS),
  \href{http://doi.org/10.1016/j.physletb.2014.01.042}{Phys. Lett.}
  \href{http://doi.org/10.1016/j.physletb.2014.01.042}{{\bf B730}, 243} (2014),
  \href{https://arxiv.org/abs/1310.0878}{[arXiv:1310.0878]}\relax
\mciteBstWouldAddEndPuncttrue
\mciteSetBstMidEndSepPunct{\mcitedefaultmidpunct}
{\mcitedefaultendpunct}{\mcitedefaultseppunct}\relax
\EndOfBibitem
\bibitem{Aad:2011kq}
G.~Aad {\em et~al.\/} (ATLAS),
  \href{http://doi.org/10.1103/PhysRevD.83.052003}{Phys. Rev.}
  \href{http://doi.org/10.1103/PhysRevD.83.052003}{{\bf D83}, 052003} (2011),
  \href{https://arxiv.org/abs/1101.0070}{[arXiv:1101.0070]}\relax
\mciteBstWouldAddEndPuncttrue
\mciteSetBstMidEndSepPunct{\mcitedefaultmidpunct}
{\mcitedefaultendpunct}{\mcitedefaultseppunct}\relax
\EndOfBibitem
\bibitem{Chatrchyan:2012mec}
S.~Chatrchyan {\em et~al.\/} (CMS),
  \href{http://doi.org/10.1007/JHEP06(2012)160}{JHEP}
  \href{http://doi.org/10.1007/JHEP06(2012)160}{{\bf 06}, 160} (2012),
  \href{https://arxiv.org/abs/1204.3170}{[arXiv:1204.3170]}\relax
\mciteBstWouldAddEndPuncttrue
\mciteSetBstMidEndSepPunct{\mcitedefaultmidpunct}
{\mcitedefaultendpunct}{\mcitedefaultseppunct}\relax
\EndOfBibitem
\bibitem{ALICE:2014dla}
B.~B. Abelev {\em et~al.\/} (ALICE),
  \href{http://doi.org/10.1103/PhysRevD.91.112012}{Phys. Rev.}
  \href{http://doi.org/10.1103/PhysRevD.91.112012}{{\bf D91}, 11, 112012}
  (2015), \href{https://arxiv.org/abs/1411.4969}{[arXiv:1411.4969]}\relax
\mciteBstWouldAddEndPuncttrue
\mciteSetBstMidEndSepPunct{\mcitedefaultmidpunct}
{\mcitedefaultendpunct}{\mcitedefaultseppunct}\relax
\EndOfBibitem
\bibitem{Aad:2013fba}
G.~Aad {\em et~al.\/} (ATLAS),
  \href{http://doi.org/10.1140/epjc/s10052-013-2676-3}{Eur. Phys. J.}
  \href{http://doi.org/10.1140/epjc/s10052-013-2676-3}{{\bf C73}, 12, 2676}
  (2013), \href{https://arxiv.org/abs/1307.5749}{[arXiv:1307.5749]}\relax
\mciteBstWouldAddEndPuncttrue
\mciteSetBstMidEndSepPunct{\mcitedefaultmidpunct}
{\mcitedefaultendpunct}{\mcitedefaultseppunct}\relax
\EndOfBibitem
\bibitem{Glasman:2008yf}
C.~Glasman (H1, ZEUS),
  \href{http://doi.org/10.1016/j.nuclphysbps.2009.03.119}{Nucl. Phys. Proc.
  Suppl.} \href{http://doi.org/10.1016/j.nuclphysbps.2009.03.119}{{\bf 191},
  121} (2009), \href{https://arxiv.org/abs/0812.0757}{[arXiv:0812.0757]}\relax
\mciteBstWouldAddEndPuncttrue
\mciteSetBstMidEndSepPunct{\mcitedefaultmidpunct}
{\mcitedefaultendpunct}{\mcitedefaultseppunct}\relax
\EndOfBibitem
\bibitem{Carli:2015qta}
T.~Carli, K.~Rabbertz and S.~Schumann, in T.~Schörner-Sadenius, editor,
  \enquote{The Large Hadron Collider: Harvest of Run 1,} 139--194 (2015),
  \href{https://arxiv.org/abs/1506.03239}{[arXiv:1506.03239]}\relax
\mciteBstWouldAddEndPuncttrue
\mciteSetBstMidEndSepPunct{\mcitedefaultmidpunct}
{\mcitedefaultendpunct}{\mcitedefaultseppunct}\relax
\EndOfBibitem
\bibitem{Gras:2017jty}
P.~Gras {\em et~al.\/}, \href{http://doi.org/10.1007/JHEP07(2017)091}{JHEP}
  \href{http://doi.org/10.1007/JHEP07(2017)091}{{\bf 07}, 091} (2017),
  \href{https://arxiv.org/abs/1704.03878}{[arXiv:1704.03878]}\relax
\mciteBstWouldAddEndPuncttrue
\mciteSetBstMidEndSepPunct{\mcitedefaultmidpunct}
{\mcitedefaultendpunct}{\mcitedefaultseppunct}\relax
\EndOfBibitem
\bibitem{Abdesselam:2010pt}
A.~Abdesselam {\em et~al.\/},
  \href{http://doi.org/10.1140/epjc/s10052-011-1661-y}{Eur. Phys. J.}
  \href{http://doi.org/10.1140/epjc/s10052-011-1661-y}{{\bf C71}, 1661} (2011),
  \href{https://arxiv.org/abs/1012.5412}{[arXiv:1012.5412]}\relax
\mciteBstWouldAddEndPuncttrue
\mciteSetBstMidEndSepPunct{\mcitedefaultmidpunct}
{\mcitedefaultendpunct}{\mcitedefaultseppunct}\relax
\EndOfBibitem
\bibitem{Krohn:2009th}
D.~Krohn, J.~Thaler and L.-T. Wang,
  \href{http://doi.org/10.1007/JHEP02(2010)084}{JHEP}
  \href{http://doi.org/10.1007/JHEP02(2010)084}{{\bf 02}, 084} (2010),
  \href{https://arxiv.org/abs/0912.1342}{[arXiv:0912.1342]}\relax
\mciteBstWouldAddEndPuncttrue
\mciteSetBstMidEndSepPunct{\mcitedefaultmidpunct}
{\mcitedefaultendpunct}{\mcitedefaultseppunct}\relax
\EndOfBibitem
\bibitem{Altheimer:2012mn}
A.~Altheimer {\em et~al.\/},
  \href{http://doi.org/10.1088/0954-3899/39/6/063001}{J. Phys.}
  \href{http://doi.org/10.1088/0954-3899/39/6/063001}{{\bf G39}, 063001}
  (2012), \href{https://arxiv.org/abs/1201.0008}{[arXiv:1201.0008]}\relax
\mciteBstWouldAddEndPuncttrue
\mciteSetBstMidEndSepPunct{\mcitedefaultmidpunct}
{\mcitedefaultendpunct}{\mcitedefaultseppunct}\relax
\EndOfBibitem
\bibitem{Altheimer:2013yza}
A.~Altheimer {\em et~al.\/},
  \href{http://doi.org/10.1140/epjc/s10052-014-2792-8}{Eur. Phys. J.}
  \href{http://doi.org/10.1140/epjc/s10052-014-2792-8}{{\bf C74}, 3, 2792}
  (2014), \href{https://arxiv.org/abs/1311.2708}{[arXiv:1311.2708]}\relax
\mciteBstWouldAddEndPuncttrue
\mciteSetBstMidEndSepPunct{\mcitedefaultmidpunct}
{\mcitedefaultendpunct}{\mcitedefaultseppunct}\relax
\EndOfBibitem
\bibitem{Stichel:2014oka}
P.~C. Stichel and W.~J. Zakrzewski,
  \href{http://doi.org/10.1140/epjc/s10052-014-3218-3}{Eur. Phys. J.}
  \href{http://doi.org/10.1140/epjc/s10052-014-3218-3}{{\bf C75}, 1, 9} (2015),
  \href{https://arxiv.org/abs/1409.1363}{[arXiv:1409.1363]}\relax
\mciteBstWouldAddEndPuncttrue
\mciteSetBstMidEndSepPunct{\mcitedefaultmidpunct}
{\mcitedefaultendpunct}{\mcitedefaultseppunct}\relax
\EndOfBibitem
\bibitem{Adams:2015hiv}
D.~Adams {\em et~al.\/},
  \href{http://doi.org/10.1140/epjc/s10052-015-3587-2}{Eur. Phys. J.}
  \href{http://doi.org/10.1140/epjc/s10052-015-3587-2}{{\bf C75}, 9, 409}
  (2015), \href{https://arxiv.org/abs/1504.00679}{[arXiv:1504.00679]}\relax
\mciteBstWouldAddEndPuncttrue
\mciteSetBstMidEndSepPunct{\mcitedefaultmidpunct}
{\mcitedefaultendpunct}{\mcitedefaultseppunct}\relax
\EndOfBibitem
\bibitem{deOliveira:2015xxd}
L.~de~Oliveira {\em et~al.\/},
  \href{http://doi.org/10.1007/JHEP07(2016)069}{JHEP}
  \href{http://doi.org/10.1007/JHEP07(2016)069}{{\bf 07}, 069} (2016),
  \href{https://arxiv.org/abs/1511.05190}{[arXiv:1511.05190]}\relax
\mciteBstWouldAddEndPuncttrue
\mciteSetBstMidEndSepPunct{\mcitedefaultmidpunct}
{\mcitedefaultendpunct}{\mcitedefaultseppunct}\relax
\EndOfBibitem
\bibitem{Dreyer:2018nbf}
F.~A. Dreyer, G.~P. Salam and G.~Soyez,
  \href{http://doi.org/10.1007/JHEP12(2018)064}{JHEP}
  \href{http://doi.org/10.1007/JHEP12(2018)064}{{\bf 12}, 064} (2018),
  \href{https://arxiv.org/abs/1807.04758}{[arXiv:1807.04758]}\relax
\mciteBstWouldAddEndPuncttrue
\mciteSetBstMidEndSepPunct{\mcitedefaultmidpunct}
{\mcitedefaultendpunct}{\mcitedefaultseppunct}\relax
\EndOfBibitem
\bibitem{ATLAS:2020bbn}
G.~Aad {\em et~al.\/} (ATLAS),
  \href{http://doi.org/10.1103/PhysRevLett.124.222002}{Phys. Rev. Lett.}
  \href{http://doi.org/10.1103/PhysRevLett.124.222002}{{\bf 124}, 22, 222002}
  (2020), \href{https://arxiv.org/abs/2004.03540}{[arXiv:2004.03540]}\relax
\mciteBstWouldAddEndPuncttrue
\mciteSetBstMidEndSepPunct{\mcitedefaultmidpunct}
{\mcitedefaultendpunct}{\mcitedefaultseppunct}\relax
\EndOfBibitem
\bibitem{Louppe:2017ipp}
G.~Louppe {\em et~al.\/}, \href{http://doi.org/10.1007/JHEP01(2019)057}{JHEP}
  \href{http://doi.org/10.1007/JHEP01(2019)057}{{\bf 01}, 057} (2019),
  \href{https://arxiv.org/abs/1702.00748}{[arXiv:1702.00748]}\relax
\mciteBstWouldAddEndPuncttrue
\mciteSetBstMidEndSepPunct{\mcitedefaultmidpunct}
{\mcitedefaultendpunct}{\mcitedefaultseppunct}\relax
\EndOfBibitem
\bibitem{Guest:2018yhq}
D.~Guest, K.~Cranmer and D.~Whiteson,
  \href{http://doi.org/10.1146/annurev-nucl-101917-021019}{Ann. Rev. Nucl.
  Part. Sci.} \href{http://doi.org/10.1146/annurev-nucl-101917-021019}{{\bf
  68}, 161} (2018),
  \href{https://arxiv.org/abs/1806.11484}{[arXiv:1806.11484]}\relax
\mciteBstWouldAddEndPuncttrue
\mciteSetBstMidEndSepPunct{\mcitedefaultmidpunct}
{\mcitedefaultendpunct}{\mcitedefaultseppunct}\relax
\EndOfBibitem
\bibitem{Kasieczka:2019dbj}
A.~Butter {\em et~al.\/},
  \href{http://doi.org/10.21468/SciPostPhys.7.1.014}{SciPost Phys.}
  \href{http://doi.org/10.21468/SciPostPhys.7.1.014}{{\bf 7}, 014} (2019),
  \href{https://arxiv.org/abs/1902.09914}{[arXiv:1902.09914]}\relax
\mciteBstWouldAddEndPuncttrue
\mciteSetBstMidEndSepPunct{\mcitedefaultmidpunct}
{\mcitedefaultendpunct}{\mcitedefaultseppunct}\relax
\EndOfBibitem
\bibitem{Komiske:2018cqr}
P.~T. Komiske, E.~M. Metodiev and J.~Thaler,
  \href{http://doi.org/10.1007/JHEP01(2019)121}{JHEP}
  \href{http://doi.org/10.1007/JHEP01(2019)121}{{\bf 01}, 121} (2019),
  \href{https://arxiv.org/abs/1810.05165}{[arXiv:1810.05165]}\relax
\mciteBstWouldAddEndPuncttrue
\mciteSetBstMidEndSepPunct{\mcitedefaultmidpunct}
{\mcitedefaultendpunct}{\mcitedefaultseppunct}\relax
\EndOfBibitem
\bibitem{SchornerSadenius:2012de}
T.~Schorner-Sadenius, \href{http://doi.org/10.1140/epjc/s10052-012-2060-8}{Eur.
  Phys. J.} \href{http://doi.org/10.1140/epjc/s10052-012-2060-8}{{\bf C72},
  2060} (2012), [Erratum: Eur. Phys. J.C72,2133(2012)]\relax
\mciteBstWouldAddEndPuncttrue
\mciteSetBstMidEndSepPunct{\mcitedefaultmidpunct}
{\mcitedefaultendpunct}{\mcitedefaultseppunct}\relax
\EndOfBibitem
\bibitem{Campbell:2006wx}
J.~M. Campbell, J.~W. Huston and W.~J. Stirling,
  \href{http://doi.org/10.1088/0034-4885/70/1/R02}{Rept. Prog. Phys.}
  \href{http://doi.org/10.1088/0034-4885/70/1/R02}{{\bf 70}, 89} (2007),
  \href{https://arxiv.org/abs/hep-ph/0611148}{[hep-ph/0611148]}\relax
\mciteBstWouldAddEndPuncttrue
\mciteSetBstMidEndSepPunct{\mcitedefaultmidpunct}
{\mcitedefaultendpunct}{\mcitedefaultseppunct}\relax
\EndOfBibitem
\bibitem{Mangano:2010zza}
M.~L. Mangano, \href{http://doi.org/10.3367/UFNe.0180.201002a.0113}{Phys. Usp.}
  \href{http://doi.org/10.3367/UFNe.0180.201002a.0113}{{\bf 53}, 109} (2010),
  [Usp. Fiz. Nauk180,113(2010)]\relax
\mciteBstWouldAddEndPuncttrue
\mciteSetBstMidEndSepPunct{\mcitedefaultmidpunct}
{\mcitedefaultendpunct}{\mcitedefaultseppunct}\relax
\EndOfBibitem
\bibitem{Butterworth:2012fj}
J.~M. Butterworth, G.~Dissertori and G.~P. Salam,
  \href{http://doi.org/10.1146/annurev-nucl-102711-094913}{Ann. Rev. Nucl.
  Part. Sci.} \href{http://doi.org/10.1146/annurev-nucl-102711-094913}{{\bf
  62}, 387} (2012),
  \href{https://arxiv.org/abs/1202.0583}{[arXiv:1202.0583]}\relax
\mciteBstWouldAddEndPuncttrue
\mciteSetBstMidEndSepPunct{\mcitedefaultmidpunct}
{\mcitedefaultendpunct}{\mcitedefaultseppunct}\relax
\EndOfBibitem
\bibitem{Carli:2010cg}
T.~Carli, T.~Gehrmann and S.~Hoeche,
  \href{http://doi.org/10.1140/epjc/s10052-010-1261-2}{Eur. Phys. J.}
  \href{http://doi.org/10.1140/epjc/s10052-010-1261-2}{{\bf C67}, 73} (2010),
  \href{https://arxiv.org/abs/0912.3715}{[arXiv:0912.3715]}\relax
\mciteBstWouldAddEndPuncttrue
\mciteSetBstMidEndSepPunct{\mcitedefaultmidpunct}
{\mcitedefaultendpunct}{\mcitedefaultseppunct}\relax
\EndOfBibitem
\bibitem{Chekanov:2007ab}
S.~Chekanov {\em et~al.\/} (ZEUS),
  \href{http://doi.org/10.1016/j.nuclphysb.2007.08.021}{Nucl. Phys.}
  \href{http://doi.org/10.1016/j.nuclphysb.2007.08.021}{{\bf B792}, 1} (2008),
  \href{https://arxiv.org/abs/0707.3749}{[arXiv:0707.3749]}\relax
\mciteBstWouldAddEndPuncttrue
\mciteSetBstMidEndSepPunct{\mcitedefaultmidpunct}
{\mcitedefaultendpunct}{\mcitedefaultseppunct}\relax
\EndOfBibitem
\bibitem{Chekanov:2007ac}
S.~Chekanov {\em et~al.\/} (ZEUS),
  \href{http://doi.org/10.1103/PhysRevD.76.072011}{Phys. Rev.}
  \href{http://doi.org/10.1103/PhysRevD.76.072011}{{\bf D76}, 072011} (2007),
  \href{https://arxiv.org/abs/0706.3809}{[arXiv:0706.3809]}\relax
\mciteBstWouldAddEndPuncttrue
\mciteSetBstMidEndSepPunct{\mcitedefaultmidpunct}
{\mcitedefaultendpunct}{\mcitedefaultseppunct}\relax
\EndOfBibitem
\bibitem{Aktas:2006qe}
A.~Aktas {\em et~al.\/} (H1),
  \href{http://doi.org/10.1016/j.physletb.2006.05.060}{Phys. Lett.}
  \href{http://doi.org/10.1016/j.physletb.2006.05.060}{{\bf B639}, 21} (2006),
  \href{https://arxiv.org/abs/hep-ex/0603014}{[hep-ex/0603014]}\relax
\mciteBstWouldAddEndPuncttrue
\mciteSetBstMidEndSepPunct{\mcitedefaultmidpunct}
{\mcitedefaultendpunct}{\mcitedefaultseppunct}\relax
\EndOfBibitem
\bibitem{ZEUS:2011aa}
H.~Abramowicz {\em et~al.\/} (ZEUS),
  \href{http://doi.org/10.1140/epjc/s10052-011-1659-5}{Eur. Phys. J.}
  \href{http://doi.org/10.1140/epjc/s10052-011-1659-5}{{\bf C71}, 1659} (2011),
  \href{https://arxiv.org/abs/1104.5444}{[arXiv:1104.5444]}\relax
\mciteBstWouldAddEndPuncttrue
\mciteSetBstMidEndSepPunct{\mcitedefaultmidpunct}
{\mcitedefaultendpunct}{\mcitedefaultseppunct}\relax
\EndOfBibitem
\bibitem{Abramowicz:2012jz}
H.~Abramowicz {\em et~al.\/} (ZEUS),
  \href{http://doi.org/10.1016/j.nuclphysb.2012.06.006}{Nucl. Phys.}
  \href{http://doi.org/10.1016/j.nuclphysb.2012.06.006}{{\bf B864}, 1} (2012),
  \href{https://arxiv.org/abs/1205.6153}{[arXiv:1205.6153]}\relax
\mciteBstWouldAddEndPuncttrue
\mciteSetBstMidEndSepPunct{\mcitedefaultmidpunct}
{\mcitedefaultendpunct}{\mcitedefaultseppunct}\relax
\EndOfBibitem
\bibitem{Aaron:2009vs}
F.~D. Aaron {\em et~al.\/} (H1),
  \href{http://doi.org/10.1140/epjc/s10052-009-1208-7}{Eur. Phys. J.}
  \href{http://doi.org/10.1140/epjc/s10052-009-1208-7}{{\bf C65}, 363} (2010),
  \href{https://arxiv.org/abs/0904.3870}{[arXiv:0904.3870]}\relax
\mciteBstWouldAddEndPuncttrue
\mciteSetBstMidEndSepPunct{\mcitedefaultmidpunct}
{\mcitedefaultendpunct}{\mcitedefaultseppunct}\relax
\EndOfBibitem
\bibitem{Aaron:2007xx}
F.~D. Aaron {\em et~al.\/} (H1),
  \href{http://doi.org/10.1140/epjc/s10052-008-0544-3}{Eur. Phys. J.}
  \href{http://doi.org/10.1140/epjc/s10052-008-0544-3}{{\bf C54}, 389} (2008),
  \href{https://arxiv.org/abs/0711.2606}{[arXiv:0711.2606]}\relax
\mciteBstWouldAddEndPuncttrue
\mciteSetBstMidEndSepPunct{\mcitedefaultmidpunct}
{\mcitedefaultendpunct}{\mcitedefaultseppunct}\relax
\EndOfBibitem
\bibitem{Chekanov:2007pa}
S.~Chekanov {\em et~al.\/} (ZEUS),
  \href{http://doi.org/10.1140/epjc/s10052-007-0418-0}{Eur. Phys. J.}
  \href{http://doi.org/10.1140/epjc/s10052-007-0418-0}{{\bf C52}, 515} (2007),
  \href{https://arxiv.org/abs/0707.3093}{[arXiv:0707.3093]}\relax
\mciteBstWouldAddEndPuncttrue
\mciteSetBstMidEndSepPunct{\mcitedefaultmidpunct}
{\mcitedefaultendpunct}{\mcitedefaultseppunct}\relax
\EndOfBibitem
\bibitem{Chekanov:2008af}
S.~Chekanov {\em et~al.\/} (ZEUS),
  \href{http://doi.org/10.1103/PhysRevD.78.032004}{Phys. Rev.}
  \href{http://doi.org/10.1103/PhysRevD.78.032004}{{\bf D78}, 032004} (2008),
  \href{https://arxiv.org/abs/0802.3955}{[arXiv:0802.3955]}\relax
\mciteBstWouldAddEndPuncttrue
\mciteSetBstMidEndSepPunct{\mcitedefaultmidpunct}
{\mcitedefaultendpunct}{\mcitedefaultseppunct}\relax
\EndOfBibitem
\bibitem{Abramowicz:2010cka}
H.~Abramowicz {\em et~al.\/} (ZEUS),
  \href{http://doi.org/10.1140/epjc/s10052-010-1504-2}{Eur. Phys. J.}
  \href{http://doi.org/10.1140/epjc/s10052-010-1504-2}{{\bf C70}, 965} (2010),
  \href{https://arxiv.org/abs/1010.6167}{[arXiv:1010.6167]}\relax
\mciteBstWouldAddEndPuncttrue
\mciteSetBstMidEndSepPunct{\mcitedefaultmidpunct}
{\mcitedefaultendpunct}{\mcitedefaultseppunct}\relax
\EndOfBibitem
\bibitem{Abramowicz:2010ke}
H.~Abramowicz {\em et~al.\/} (ZEUS),
  \href{http://doi.org/10.1016/j.physletb.2010.06.015}{Phys. Lett.}
  \href{http://doi.org/10.1016/j.physletb.2010.06.015}{{\bf B691}, 127} (2010),
  \href{https://arxiv.org/abs/1003.2923}{[arXiv:1003.2923]}\relax
\mciteBstWouldAddEndPuncttrue
\mciteSetBstMidEndSepPunct{\mcitedefaultmidpunct}
{\mcitedefaultendpunct}{\mcitedefaultseppunct}\relax
\EndOfBibitem
\bibitem{Chekanov:2008ab}
S.~Chekanov {\em et~al.\/} (ZEUS),
  \href{http://doi.org/10.1103/PhysRevD.85.052008}{Phys. Rev.}
  \href{http://doi.org/10.1103/PhysRevD.85.052008}{{\bf D85}, 052008} (2012),
  \href{https://arxiv.org/abs/0808.3783}{[arXiv:0808.3783]}\relax
\mciteBstWouldAddEndPuncttrue
\mciteSetBstMidEndSepPunct{\mcitedefaultmidpunct}
{\mcitedefaultendpunct}{\mcitedefaultseppunct}\relax
\EndOfBibitem
\bibitem{Aaron:2010ac}
F.~D. Aaron {\em et~al.\/} (H1),
  \href{http://doi.org/10.1140/epjc/s10052-010-1282-x}{Eur. Phys. J.}
  \href{http://doi.org/10.1140/epjc/s10052-010-1282-x}{{\bf C67}, 1} (2010),
  \href{https://arxiv.org/abs/0911.5678}{[arXiv:0911.5678]}\relax
\mciteBstWouldAddEndPuncttrue
\mciteSetBstMidEndSepPunct{\mcitedefaultmidpunct}
{\mcitedefaultendpunct}{\mcitedefaultseppunct}\relax
\EndOfBibitem
\bibitem{H1:2014cbm}
V.~Andreev {\em et~al.\/} (H1),
  \href{http://doi.org/10.1140/epjc/s10052-014-3223-6}{Eur. Phys. J.}
  \href{http://doi.org/10.1140/epjc/s10052-014-3223-6}{{\bf C75}, 2, 65}
  (2015), \href{https://arxiv.org/abs/1406.4709}{[arXiv:1406.4709]}\relax
\mciteBstWouldAddEndPuncttrue
\mciteSetBstMidEndSepPunct{\mcitedefaultmidpunct}
{\mcitedefaultendpunct}{\mcitedefaultseppunct}\relax
\EndOfBibitem
\bibitem{Andreev:2017vxu}
V.~Andreev {\em et~al.\/} (H1),
  \href{http://doi.org/10.1140/epjc/s10052-017-5314-7}{Eur. Phys. J.}
  \href{http://doi.org/10.1140/epjc/s10052-017-5314-7}{{\bf C77}, 11, 791}
  (2017), [Erratum: Eur. Phys. J. {\bf C81}, 8, 738 (2021)],
  \href{https://arxiv.org/abs/1709.07251}{[arXiv:1709.07251]}\relax
\mciteBstWouldAddEndPuncttrue
\mciteSetBstMidEndSepPunct{\mcitedefaultmidpunct}
{\mcitedefaultendpunct}{\mcitedefaultseppunct}\relax
\EndOfBibitem
\bibitem{qcd:ATLAS-xsec-overview}
\url{http://atlas.web.cern.ch/Atlas/GROUPS/PHYSICS/CombinedSummaryPlots/SM}\relax
\mciteBstWouldAddEndPuncttrue
\mciteSetBstMidEndSepPunct{\mcitedefaultmidpunct}
{\mcitedefaultendpunct}{\mcitedefaultseppunct}\relax
\EndOfBibitem
\bibitem{qcd:CMS-xsec-overview}
\url{http://twiki.cern.ch/twiki/bin/view/CMSPublic/PhysicsResultsCombined}\relax
\mciteBstWouldAddEndPuncttrue
\mciteSetBstMidEndSepPunct{\mcitedefaultmidpunct}
{\mcitedefaultendpunct}{\mcitedefaultseppunct}\relax
\EndOfBibitem
\bibitem{Abulencia:2007ez}
A.~Abulencia {\em et~al.\/} (CDF),
  \href{http://doi.org/10.1103/PhysRevD.75.119901}{Phys. Rev.}
  \href{http://doi.org/10.1103/PhysRevD.75.119901}{{\bf D75}, 092006} (2007),
  [Erratum: Phys. Rev.D75,119901(2007)],
  \href{https://arxiv.org/abs/hep-ex/0701051}{[hep-ex/0701051]}\relax
\mciteBstWouldAddEndPuncttrue
\mciteSetBstMidEndSepPunct{\mcitedefaultmidpunct}
{\mcitedefaultendpunct}{\mcitedefaultseppunct}\relax
\EndOfBibitem
\bibitem{Aaltonen:2008eq}
T.~Aaltonen {\em et~al.\/} (CDF),
  \href{http://doi.org/10.1103/PhysRevD.79.119902}{Phys. Rev.}
  \href{http://doi.org/10.1103/PhysRevD.79.119902}{{\bf D78}, 052006} (2008),
  [Erratum: Phys. Rev.D79,119902(2009)],
  \href{https://arxiv.org/abs/0807.2204}{[arXiv:0807.2204]}\relax
\mciteBstWouldAddEndPuncttrue
\mciteSetBstMidEndSepPunct{\mcitedefaultmidpunct}
{\mcitedefaultendpunct}{\mcitedefaultseppunct}\relax
\EndOfBibitem
\bibitem{Abazov:2008ae}
V.~M. Abazov {\em et~al.\/} (D0),
  \href{http://doi.org/10.1103/PhysRevLett.101.062001}{Phys. Rev. Lett.}
  \href{http://doi.org/10.1103/PhysRevLett.101.062001}{{\bf 101}, 062001}
  (2008), \href{https://arxiv.org/abs/0802.2400}{[arXiv:0802.2400]}\relax
\mciteBstWouldAddEndPuncttrue
\mciteSetBstMidEndSepPunct{\mcitedefaultmidpunct}
{\mcitedefaultendpunct}{\mcitedefaultseppunct}\relax
\EndOfBibitem
\bibitem{Abazov:2011vi}
V.~M. Abazov {\em et~al.\/} (D0),
  \href{http://doi.org/10.1103/PhysRevD.85.052006}{Phys. Rev.}
  \href{http://doi.org/10.1103/PhysRevD.85.052006}{{\bf D85}, 052006} (2012),
  \href{https://arxiv.org/abs/1110.3771}{[arXiv:1110.3771]}\relax
\mciteBstWouldAddEndPuncttrue
\mciteSetBstMidEndSepPunct{\mcitedefaultmidpunct}
{\mcitedefaultendpunct}{\mcitedefaultseppunct}\relax
\EndOfBibitem
\bibitem{Abelev:2013fn}
B.~Abelev {\em et~al.\/} (ALICE),
  \href{http://doi.org/10.1016/j.physletb.2013.04.026}{Phys. Lett.}
  \href{http://doi.org/10.1016/j.physletb.2013.04.026}{{\bf B722}, 262} (2013),
  \href{https://arxiv.org/abs/1301.3475}{[arXiv:1301.3475]}\relax
\mciteBstWouldAddEndPuncttrue
\mciteSetBstMidEndSepPunct{\mcitedefaultmidpunct}
{\mcitedefaultendpunct}{\mcitedefaultseppunct}\relax
\EndOfBibitem
\bibitem{Aad:2013lpa}
G.~Aad {\em et~al.\/} (ATLAS),
  \href{http://doi.org/10.1140/epjc/s10052-013-2509-4}{Eur. Phys. J.}
  \href{http://doi.org/10.1140/epjc/s10052-013-2509-4}{{\bf C73}, 8, 2509}
  (2013), \href{https://arxiv.org/abs/1304.4739}{[arXiv:1304.4739]}\relax
\mciteBstWouldAddEndPuncttrue
\mciteSetBstMidEndSepPunct{\mcitedefaultmidpunct}
{\mcitedefaultendpunct}{\mcitedefaultseppunct}\relax
\EndOfBibitem
\bibitem{Aad:2014vwa}
G.~Aad {\em et~al.\/} (ATLAS),
  \href{http://doi.org/10.1007/JHEP02(2015)153}{JHEP}
  \href{http://doi.org/10.1007/JHEP02(2015)153}{{\bf 02}, 153} (2015),
  [Erratum: JHEP09,141(2015)],
  \href{https://arxiv.org/abs/1410.8857}{[arXiv:1410.8857]}\relax
\mciteBstWouldAddEndPuncttrue
\mciteSetBstMidEndSepPunct{\mcitedefaultmidpunct}
{\mcitedefaultendpunct}{\mcitedefaultseppunct}\relax
\EndOfBibitem
\bibitem{Aaboud:2017dvo}
M.~Aaboud {\em et~al.\/} (ATLAS),
  \href{http://doi.org/10.1007/JHEP09(2017)020}{JHEP}
  \href{http://doi.org/10.1007/JHEP09(2017)020}{{\bf 09}, 020} (2017),
  \href{https://arxiv.org/abs/1706.03192}{[arXiv:1706.03192]}\relax
\mciteBstWouldAddEndPuncttrue
\mciteSetBstMidEndSepPunct{\mcitedefaultmidpunct}
{\mcitedefaultendpunct}{\mcitedefaultseppunct}\relax
\EndOfBibitem
\bibitem{Khachatryan:2015luy}
V.~Khachatryan {\em et~al.\/} (CMS),
  \href{http://doi.org/10.1140/epjc/s10052-016-4083-z}{Eur. Phys. J.}
  \href{http://doi.org/10.1140/epjc/s10052-016-4083-z}{{\bf C76}, 5, 265}
  (2016), \href{https://arxiv.org/abs/1512.06212}{[arXiv:1512.06212]}\relax
\mciteBstWouldAddEndPuncttrue
\mciteSetBstMidEndSepPunct{\mcitedefaultmidpunct}
{\mcitedefaultendpunct}{\mcitedefaultseppunct}\relax
\EndOfBibitem
\bibitem{Chatrchyan:2012bja}
S.~Chatrchyan {\em et~al.\/} (CMS),
  \href{http://doi.org/10.1103/PhysRevD.87.112002}{Phys. Rev.}
  \href{http://doi.org/10.1103/PhysRevD.87.112002}{{\bf D87}, 11, 112002}
  (2013), [Erratum: Phys. Rev.D87,no.11,119902(2013)],
  \href{https://arxiv.org/abs/1212.6660}{[arXiv:1212.6660]}\relax
\mciteBstWouldAddEndPuncttrue
\mciteSetBstMidEndSepPunct{\mcitedefaultmidpunct}
{\mcitedefaultendpunct}{\mcitedefaultseppunct}\relax
\EndOfBibitem
\bibitem{Khachatryan:2016mlc}
V.~Khachatryan {\em et~al.\/} (CMS),
  \href{http://doi.org/10.1007/JHEP03(2017)156}{JHEP}
  \href{http://doi.org/10.1007/JHEP03(2017)156}{{\bf 03}, 156} (2017),
  \href{https://arxiv.org/abs/1609.05331}{[arXiv:1609.05331]}\relax
\mciteBstWouldAddEndPuncttrue
\mciteSetBstMidEndSepPunct{\mcitedefaultmidpunct}
{\mcitedefaultendpunct}{\mcitedefaultseppunct}\relax
\EndOfBibitem
\bibitem{Khachatryan:2016wdh}
V.~Khachatryan {\em et~al.\/} (CMS),
  \href{http://doi.org/10.1140/epjc/s10052-016-4286-3}{Eur. Phys. J.}
  \href{http://doi.org/10.1140/epjc/s10052-016-4286-3}{{\bf C76}, 8, 451}
  (2016), \href{https://arxiv.org/abs/1605.04436}{[arXiv:1605.04436]}\relax
\mciteBstWouldAddEndPuncttrue
\mciteSetBstMidEndSepPunct{\mcitedefaultmidpunct}
{\mcitedefaultendpunct}{\mcitedefaultseppunct}\relax
\EndOfBibitem
\bibitem{ATLAS:2017ble}
M.~Aaboud {\em et~al.\/} (ATLAS),
  \href{http://doi.org/10.1007/JHEP05(2018)195}{JHEP}
  \href{http://doi.org/10.1007/JHEP05(2018)195}{{\bf 05}, 195} (2018),
  \href{https://arxiv.org/abs/1711.02692}{[arXiv:1711.02692]}\relax
\mciteBstWouldAddEndPuncttrue
\mciteSetBstMidEndSepPunct{\mcitedefaultmidpunct}
{\mcitedefaultendpunct}{\mcitedefaultseppunct}\relax
\EndOfBibitem
\bibitem{CMS:2020caw}
A.~M. Sirunyan {\em et~al.\/} (CMS),
  \href{http://doi.org/10.1007/JHEP12(2020)082}{JHEP}
  \href{http://doi.org/10.1007/JHEP12(2020)082}{{\bf 12}, 082} (2020),
  \href{https://arxiv.org/abs/2005.05159}{[arXiv:2005.05159]}\relax
\mciteBstWouldAddEndPuncttrue
\mciteSetBstMidEndSepPunct{\mcitedefaultmidpunct}
{\mcitedefaultendpunct}{\mcitedefaultseppunct}\relax
\EndOfBibitem
\bibitem{CMS:2021yzl}
A.~Tumasyan {\em et~al.\/} (CMS),
  \href{http://doi.org/10.1007/JHEP02(2022)142}{JHEP}
  \href{http://doi.org/10.1007/JHEP02(2022)142}{{\bf 02}, 142} (2022),
  [Addendum: JHEP 12, 035 (2022)],
  \href{https://arxiv.org/abs/2111.10431}{[arXiv:2111.10431]}\relax
\mciteBstWouldAddEndPuncttrue
\mciteSetBstMidEndSepPunct{\mcitedefaultmidpunct}
{\mcitedefaultendpunct}{\mcitedefaultseppunct}\relax
\EndOfBibitem
\bibitem{Currie:2018xkj}
J.~Currie {\em et~al.\/}, \href{http://doi.org/10.1007/JHEP10(2018)155}{JHEP}
  \href{http://doi.org/10.1007/JHEP10(2018)155}{{\bf 10}, 155} (2018),
  \href{https://arxiv.org/abs/1807.03692}{[arXiv:1807.03692]}\relax
\mciteBstWouldAddEndPuncttrue
\mciteSetBstMidEndSepPunct{\mcitedefaultmidpunct}
{\mcitedefaultendpunct}{\mcitedefaultseppunct}\relax
\EndOfBibitem
\bibitem{Schwartzman:2015ada}
A.~Schwartzman, \href{http://doi.org/10.1142/S0217751X15460021}{Int. J. Mod.
  Phys.} \href{http://doi.org/10.1142/S0217751X15460021}{{\bf A30}, 31,
  1546002} (2015),
  \href{https://arxiv.org/abs/1509.05459}{[arXiv:1509.05459]}\relax
\mciteBstWouldAddEndPuncttrue
\mciteSetBstMidEndSepPunct{\mcitedefaultmidpunct}
{\mcitedefaultendpunct}{\mcitedefaultseppunct}\relax
\EndOfBibitem
\bibitem{Hou:2019jgw}
T.-J. Hou {\em et~al.\/}  (2019),
  \href{https://arxiv.org/abs/1908.11238}{[arXiv:1908.11238]}\relax
\mciteBstWouldAddEndPuncttrue
\mciteSetBstMidEndSepPunct{\mcitedefaultmidpunct}
{\mcitedefaultendpunct}{\mcitedefaultseppunct}\relax
\EndOfBibitem
\bibitem{AbdulKhalek:2020jut}
R.~Abdul~Khalek {\em et~al.\/},
  \href{http://doi.org/10.1140/epjc/s10052-020-8328-5}{Eur. Phys. J. C}
  \href{http://doi.org/10.1140/epjc/s10052-020-8328-5}{{\bf 80}, 8, 797}
  (2020), \href{https://arxiv.org/abs/2005.11327}{[arXiv:2005.11327]}\relax
\mciteBstWouldAddEndPuncttrue
\mciteSetBstMidEndSepPunct{\mcitedefaultmidpunct}
{\mcitedefaultendpunct}{\mcitedefaultseppunct}\relax
\EndOfBibitem
\bibitem{Jing:2023isu}
X.~Jing {\em et~al.\/}  (2023),
  \href{https://arxiv.org/abs/2306.03918}{[arXiv:2306.03918]}\relax
\mciteBstWouldAddEndPuncttrue
\mciteSetBstMidEndSepPunct{\mcitedefaultmidpunct}
{\mcitedefaultendpunct}{\mcitedefaultseppunct}\relax
\EndOfBibitem
\bibitem{Aad:2013tea}
G.~Aad {\em et~al.\/} (ATLAS),
  \href{http://doi.org/10.1007/JHEP05(2014)059}{JHEP}
  \href{http://doi.org/10.1007/JHEP05(2014)059}{{\bf 05}, 059} (2014),
  \href{https://arxiv.org/abs/1312.3524}{[arXiv:1312.3524]}\relax
\mciteBstWouldAddEndPuncttrue
\mciteSetBstMidEndSepPunct{\mcitedefaultmidpunct}
{\mcitedefaultendpunct}{\mcitedefaultseppunct}\relax
\EndOfBibitem
\bibitem{Sirunyan:2017skj}
A.~M. Sirunyan {\em et~al.\/} (CMS),
  \href{http://doi.org/10.1140/epjc/s10052-017-5286-7}{Eur. Phys. J.}
  \href{http://doi.org/10.1140/epjc/s10052-017-5286-7}{{\bf C77}, 11, 746}
  (2017), \href{https://arxiv.org/abs/1705.02628}{[arXiv:1705.02628]}\relax
\mciteBstWouldAddEndPuncttrue
\mciteSetBstMidEndSepPunct{\mcitedefaultmidpunct}
{\mcitedefaultendpunct}{\mcitedefaultseppunct}\relax
\EndOfBibitem
\bibitem{Aaltonen:2008dn}
T.~Aaltonen {\em et~al.\/} (CDF),
  \href{http://doi.org/10.1103/PhysRevD.79.112002}{Phys. Rev.}
  \href{http://doi.org/10.1103/PhysRevD.79.112002}{{\bf D79}, 112002} (2009),
  \href{https://arxiv.org/abs/0812.4036}{[arXiv:0812.4036]}\relax
\mciteBstWouldAddEndPuncttrue
\mciteSetBstMidEndSepPunct{\mcitedefaultmidpunct}
{\mcitedefaultendpunct}{\mcitedefaultseppunct}\relax
\EndOfBibitem
\bibitem{Abazov:2009ac}
V.~M. Abazov {\em et~al.\/} (D0),
  \href{http://doi.org/10.1103/PhysRevLett.103.191803}{Phys. Rev. Lett.}
  \href{http://doi.org/10.1103/PhysRevLett.103.191803}{{\bf 103}, 191803}
  (2009), \href{https://arxiv.org/abs/0906.4819}{[arXiv:0906.4819]}\relax
\mciteBstWouldAddEndPuncttrue
\mciteSetBstMidEndSepPunct{\mcitedefaultmidpunct}
{\mcitedefaultendpunct}{\mcitedefaultseppunct}\relax
\EndOfBibitem
\bibitem{Chatrchyan:2012bf}
S.~Chatrchyan {\em et~al.\/} (CMS),
  \href{http://doi.org/10.1007/JHEP05(2012)055}{JHEP}
  \href{http://doi.org/10.1007/JHEP05(2012)055}{{\bf 05}, 055} (2012),
  \href{https://arxiv.org/abs/1202.5535}{[arXiv:1202.5535]}\relax
\mciteBstWouldAddEndPuncttrue
\mciteSetBstMidEndSepPunct{\mcitedefaultmidpunct}
{\mcitedefaultendpunct}{\mcitedefaultseppunct}\relax
\EndOfBibitem
\bibitem{Khachatryan:2014cja}
V.~Khachatryan {\em et~al.\/} (CMS),
  \href{http://doi.org/10.1016/j.physletb.2015.04.042}{Phys. Lett.}
  \href{http://doi.org/10.1016/j.physletb.2015.04.042}{{\bf B746}, 79} (2015),
  \href{https://arxiv.org/abs/1411.2646}{[arXiv:1411.2646]}\relax
\mciteBstWouldAddEndPuncttrue
\mciteSetBstMidEndSepPunct{\mcitedefaultmidpunct}
{\mcitedefaultendpunct}{\mcitedefaultseppunct}\relax
\EndOfBibitem
\bibitem{Sirunyan:2017ygf}
A.~M. Sirunyan {\em et~al.\/} (CMS),
  \href{http://doi.org/10.1007/JHEP07(2017)013}{JHEP}
  \href{http://doi.org/10.1007/JHEP07(2017)013}{{\bf 07}, 013} (2017),
  \href{https://arxiv.org/abs/1703.09986}{[arXiv:1703.09986]}\relax
\mciteBstWouldAddEndPuncttrue
\mciteSetBstMidEndSepPunct{\mcitedefaultmidpunct}
{\mcitedefaultendpunct}{\mcitedefaultseppunct}\relax
\EndOfBibitem
\bibitem{ATLAS:2012pu}
G.~Aad {\em et~al.\/} (ATLAS),
  \href{http://doi.org/10.1007/JHEP01(2013)029}{JHEP}
  \href{http://doi.org/10.1007/JHEP01(2013)029}{{\bf 01}, 029} (2013),
  \href{https://arxiv.org/abs/1210.1718}{[arXiv:1210.1718]}\relax
\mciteBstWouldAddEndPuncttrue
\mciteSetBstMidEndSepPunct{\mcitedefaultmidpunct}
{\mcitedefaultendpunct}{\mcitedefaultseppunct}\relax
\EndOfBibitem
\bibitem{Aaboud:2017yvp}
M.~Aaboud {\em et~al.\/} (ATLAS),
  \href{http://doi.org/10.1103/PhysRevD.96.052004}{Phys. Rev.}
  \href{http://doi.org/10.1103/PhysRevD.96.052004}{{\bf D96}, 5, 052004}
  (2017), \href{https://arxiv.org/abs/1703.09127}{[arXiv:1703.09127]}\relax
\mciteBstWouldAddEndPuncttrue
\mciteSetBstMidEndSepPunct{\mcitedefaultmidpunct}
{\mcitedefaultendpunct}{\mcitedefaultseppunct}\relax
\EndOfBibitem
\bibitem{Abazov:2004hm}
V.~M. Abazov {\em et~al.\/} (D0),
  \href{http://doi.org/10.1103/PhysRevLett.94.221801}{Phys. Rev. Lett.}
  \href{http://doi.org/10.1103/PhysRevLett.94.221801}{{\bf 94}, 221801} (2005),
  \href{https://arxiv.org/abs/hep-ex/0409040}{[hep-ex/0409040]}\relax
\mciteBstWouldAddEndPuncttrue
\mciteSetBstMidEndSepPunct{\mcitedefaultmidpunct}
{\mcitedefaultendpunct}{\mcitedefaultseppunct}\relax
\EndOfBibitem
\bibitem{Abazov:2012jhu}
V.~M. Abazov {\em et~al.\/} (D0),
  \href{http://doi.org/10.1016/j.physletb.2013.03.029}{Phys. Lett.}
  \href{http://doi.org/10.1016/j.physletb.2013.03.029}{{\bf B721}, 212} (2013),
  \href{https://arxiv.org/abs/1212.1842}{[arXiv:1212.1842]}\relax
\mciteBstWouldAddEndPuncttrue
\mciteSetBstMidEndSepPunct{\mcitedefaultmidpunct}
{\mcitedefaultendpunct}{\mcitedefaultseppunct}\relax
\EndOfBibitem
\bibitem{daCosta:2011ni}
G.~Aad {\em et~al.\/} (ATLAS),
  \href{http://doi.org/10.1103/PhysRevLett.106.172002}{Phys. Rev. Lett.}
  \href{http://doi.org/10.1103/PhysRevLett.106.172002}{{\bf 106}, 172002}
  (2011), \href{https://arxiv.org/abs/1102.2696}{[arXiv:1102.2696]}\relax
\mciteBstWouldAddEndPuncttrue
\mciteSetBstMidEndSepPunct{\mcitedefaultmidpunct}
{\mcitedefaultendpunct}{\mcitedefaultseppunct}\relax
\EndOfBibitem
\bibitem{Khachatryan:2011zj}
V.~Khachatryan {\em et~al.\/} (CMS),
  \href{http://doi.org/10.1103/PhysRevLett.106.122003}{Phys. Rev. Lett.}
  \href{http://doi.org/10.1103/PhysRevLett.106.122003}{{\bf 106}, 122003}
  (2011), \href{https://arxiv.org/abs/1101.5029}{[arXiv:1101.5029]}\relax
\mciteBstWouldAddEndPuncttrue
\mciteSetBstMidEndSepPunct{\mcitedefaultmidpunct}
{\mcitedefaultendpunct}{\mcitedefaultseppunct}\relax
\EndOfBibitem
\bibitem{Khachatryan:2016hkr}
V.~Khachatryan {\em et~al.\/} (CMS),
  \href{http://doi.org/10.1140/epjc/s10052-016-4346-8}{Eur. Phys. J.}
  \href{http://doi.org/10.1140/epjc/s10052-016-4346-8}{{\bf C76}, 10, 536}
  (2016), \href{https://arxiv.org/abs/1602.04384}{[arXiv:1602.04384]}\relax
\mciteBstWouldAddEndPuncttrue
\mciteSetBstMidEndSepPunct{\mcitedefaultmidpunct}
{\mcitedefaultendpunct}{\mcitedefaultseppunct}\relax
\EndOfBibitem
\bibitem{Sirunyan:2017jnl}
A.~M. Sirunyan {\em et~al.\/} (CMS),
  \href{http://doi.org/10.1140/epjc/s10052-018-6033-4}{Eur. Phys. J.}
  \href{http://doi.org/10.1140/epjc/s10052-018-6033-4}{{\bf C78}, 7, 566}
  (2018), \href{https://arxiv.org/abs/1712.05471}{[arXiv:1712.05471]}\relax
\mciteBstWouldAddEndPuncttrue
\mciteSetBstMidEndSepPunct{\mcitedefaultmidpunct}
{\mcitedefaultendpunct}{\mcitedefaultseppunct}\relax
\EndOfBibitem
\bibitem{Sirunyan:2019rpc}
A.~M. Sirunyan {\em et~al.\/} (CMS)  (2019),
  \href{https://arxiv.org/abs/1902.04374}{[arXiv:1902.04374]}\relax
\mciteBstWouldAddEndPuncttrue
\mciteSetBstMidEndSepPunct{\mcitedefaultmidpunct}
{\mcitedefaultendpunct}{\mcitedefaultseppunct}\relax
\EndOfBibitem
\bibitem{Kokkas:2015gfa}
P.~Kokkas, \href{http://doi.org/10.1142/S0217751X15460045}{Int. J. Mod. Phys.}
  \href{http://doi.org/10.1142/S0217751X15460045}{{\bf A30}, 31, 1546004}
  (2015), \href{https://arxiv.org/abs/1509.02144}{[arXiv:1509.02144]}\relax
\mciteBstWouldAddEndPuncttrue
\mciteSetBstMidEndSepPunct{\mcitedefaultmidpunct}
{\mcitedefaultendpunct}{\mcitedefaultseppunct}\relax
\EndOfBibitem
\bibitem{ATLAS:2020vup}
G.~Aad {\em et~al.\/} (ATLAS),
  \href{http://doi.org/10.1007/JHEP01(2021)188}{JHEP}
  \href{http://doi.org/10.1007/JHEP01(2021)188}{{\bf 01}, 188} (2021),
  [Erratum: JHEP 12, 053 (2021)],
  \href{https://arxiv.org/abs/2007.12600}{[arXiv:2007.12600]}\relax
\mciteBstWouldAddEndPuncttrue
\mciteSetBstMidEndSepPunct{\mcitedefaultmidpunct}
{\mcitedefaultendpunct}{\mcitedefaultseppunct}\relax
\EndOfBibitem
\bibitem{Aaboud:2016btc}
M.~Aaboud {\em et~al.\/} (ATLAS),
  \href{http://doi.org/10.1140/epjc/s10052-017-4911-9}{Eur. Phys. J.}
  \href{http://doi.org/10.1140/epjc/s10052-017-4911-9}{{\bf C77}, 6, 367}
  (2017), \href{https://arxiv.org/abs/1612.03016}{[arXiv:1612.03016]}\relax
\mciteBstWouldAddEndPuncttrue
\mciteSetBstMidEndSepPunct{\mcitedefaultmidpunct}
{\mcitedefaultendpunct}{\mcitedefaultseppunct}\relax
\EndOfBibitem
\bibitem{Aad:2016naf}
G.~Aad {\em et~al.\/} (ATLAS),
  \href{http://doi.org/10.1016/j.physletb.2016.06.023}{Phys. Lett.}
  \href{http://doi.org/10.1016/j.physletb.2016.06.023}{{\bf B759}, 601} (2016),
  \href{https://arxiv.org/abs/1603.09222}{[arXiv:1603.09222]}\relax
\mciteBstWouldAddEndPuncttrue
\mciteSetBstMidEndSepPunct{\mcitedefaultmidpunct}
{\mcitedefaultendpunct}{\mcitedefaultseppunct}\relax
\EndOfBibitem
\bibitem{Aaij:2015gna}
R.~Aaij {\em et~al.\/} (LHCb),
  \href{http://doi.org/10.1007/JHEP08(2015)039}{JHEP}
  \href{http://doi.org/10.1007/JHEP08(2015)039}{{\bf 08}, 039} (2015),
  \href{https://arxiv.org/abs/1505.07024}{[arXiv:1505.07024]}\relax
\mciteBstWouldAddEndPuncttrue
\mciteSetBstMidEndSepPunct{\mcitedefaultmidpunct}
{\mcitedefaultendpunct}{\mcitedefaultseppunct}\relax
\EndOfBibitem
\bibitem{CMS:2011aa}
S.~Chatrchyan {\em et~al.\/} (CMS),
  \href{http://doi.org/10.1007/JHEP10(2011)132}{JHEP}
  \href{http://doi.org/10.1007/JHEP10(2011)132}{{\bf 10}, 132} (2011),
  \href{https://arxiv.org/abs/1107.4789}{[arXiv:1107.4789]}\relax
\mciteBstWouldAddEndPuncttrue
\mciteSetBstMidEndSepPunct{\mcitedefaultmidpunct}
{\mcitedefaultendpunct}{\mcitedefaultseppunct}\relax
\EndOfBibitem
\bibitem{Chatrchyan:2014mua}
S.~Chatrchyan {\em et~al.\/} (CMS),
  \href{http://doi.org/10.1103/PhysRevLett.112.191802}{Phys. Rev. Lett.}
  \href{http://doi.org/10.1103/PhysRevLett.112.191802}{{\bf 112}, 191802}
  (2014), \href{https://arxiv.org/abs/1402.0923}{[arXiv:1402.0923]}\relax
\mciteBstWouldAddEndPuncttrue
\mciteSetBstMidEndSepPunct{\mcitedefaultmidpunct}
{\mcitedefaultendpunct}{\mcitedefaultseppunct}\relax
\EndOfBibitem
\bibitem{Aaij:2012vn}
R.~Aaij {\em et~al.\/} (LHCb),
  \href{http://doi.org/10.1007/JHEP06(2012)058}{JHEP}
  \href{http://doi.org/10.1007/JHEP06(2012)058}{{\bf 06}, 058} (2012),
  \href{https://arxiv.org/abs/1204.1620}{[arXiv:1204.1620]}\relax
\mciteBstWouldAddEndPuncttrue
\mciteSetBstMidEndSepPunct{\mcitedefaultmidpunct}
{\mcitedefaultendpunct}{\mcitedefaultseppunct}\relax
\EndOfBibitem
\bibitem{Aaij:2015zlq}
R.~Aaij {\em et~al.\/} (LHCb),
  \href{http://doi.org/10.1007/JHEP01(2016)155}{JHEP}
  \href{http://doi.org/10.1007/JHEP01(2016)155}{{\bf 01}, 155} (2016),
  \href{https://arxiv.org/abs/1511.08039}{[arXiv:1511.08039]}\relax
\mciteBstWouldAddEndPuncttrue
\mciteSetBstMidEndSepPunct{\mcitedefaultmidpunct}
{\mcitedefaultendpunct}{\mcitedefaultseppunct}\relax
\EndOfBibitem
\bibitem{Aaij:2016mgv}
R.~Aaij {\em et~al.\/} (LHCb),
  \href{http://doi.org/10.1007/JHEP09(2016)136}{JHEP}
  \href{http://doi.org/10.1007/JHEP09(2016)136}{{\bf 09}, 136} (2016),
  \href{https://arxiv.org/abs/1607.06495}{[arXiv:1607.06495]}\relax
\mciteBstWouldAddEndPuncttrue
\mciteSetBstMidEndSepPunct{\mcitedefaultmidpunct}
{\mcitedefaultendpunct}{\mcitedefaultseppunct}\relax
\EndOfBibitem
\bibitem{ATLAS:2019zci}
G.~Aad {\em et~al.\/} (ATLAS),
  \href{http://doi.org/10.1140/epjc/s10052-020-8001-z}{Eur. Phys. J. C}
  \href{http://doi.org/10.1140/epjc/s10052-020-8001-z}{{\bf 80}, 7, 616}
  (2020), \href{https://arxiv.org/abs/1912.02844}{[arXiv:1912.02844]}\relax
\mciteBstWouldAddEndPuncttrue
\mciteSetBstMidEndSepPunct{\mcitedefaultmidpunct}
{\mcitedefaultendpunct}{\mcitedefaultseppunct}\relax
\EndOfBibitem
\bibitem{CMS:2019raw}
A.~M. Sirunyan {\em et~al.\/} (CMS),
  \href{http://doi.org/10.1007/JHEP12(2019)061}{JHEP}
  \href{http://doi.org/10.1007/JHEP12(2019)061}{{\bf 12}, 061} (2019),
  \href{https://arxiv.org/abs/1909.04133}{[arXiv:1909.04133]}\relax
\mciteBstWouldAddEndPuncttrue
\mciteSetBstMidEndSepPunct{\mcitedefaultmidpunct}
{\mcitedefaultendpunct}{\mcitedefaultseppunct}\relax
\EndOfBibitem
\bibitem{Chatrchyan:2011cm}
S.~Chatrchyan {\em et~al.\/} (CMS),
  \href{http://doi.org/10.1007/JHEP10(2011)007}{JHEP}
  \href{http://doi.org/10.1007/JHEP10(2011)007}{{\bf 10}, 007} (2011),
  \href{https://arxiv.org/abs/1108.0566}{[arXiv:1108.0566]}\relax
\mciteBstWouldAddEndPuncttrue
\mciteSetBstMidEndSepPunct{\mcitedefaultmidpunct}
{\mcitedefaultendpunct}{\mcitedefaultseppunct}\relax
\EndOfBibitem
\bibitem{CMS:2014jea}
V.~Khachatryan {\em et~al.\/} (CMS),
  \href{http://doi.org/10.1140/epjc/s10052-015-3364-2}{Eur. Phys. J.}
  \href{http://doi.org/10.1140/epjc/s10052-015-3364-2}{{\bf C75}, 4, 147}
  (2015), \href{https://arxiv.org/abs/1412.1115}{[arXiv:1412.1115]}\relax
\mciteBstWouldAddEndPuncttrue
\mciteSetBstMidEndSepPunct{\mcitedefaultmidpunct}
{\mcitedefaultendpunct}{\mcitedefaultseppunct}\relax
\EndOfBibitem
\bibitem{Aad:2013iua}
G.~Aad {\em et~al.\/} (ATLAS),
  \href{http://doi.org/10.1016/j.physletb.2013.07.049}{Phys. Lett.}
  \href{http://doi.org/10.1016/j.physletb.2013.07.049}{{\bf B725}, 223} (2013),
  \href{https://arxiv.org/abs/1305.4192}{[arXiv:1305.4192]}\relax
\mciteBstWouldAddEndPuncttrue
\mciteSetBstMidEndSepPunct{\mcitedefaultmidpunct}
{\mcitedefaultendpunct}{\mcitedefaultseppunct}\relax
\EndOfBibitem
\bibitem{Aad:2014qja}
G.~Aad {\em et~al.\/} (ATLAS),
  \href{http://doi.org/10.1007/JHEP06(2014)112}{JHEP}
  \href{http://doi.org/10.1007/JHEP06(2014)112}{{\bf 06}, 112} (2014),
  \href{https://arxiv.org/abs/1404.1212}{[arXiv:1404.1212]}\relax
\mciteBstWouldAddEndPuncttrue
\mciteSetBstMidEndSepPunct{\mcitedefaultmidpunct}
{\mcitedefaultendpunct}{\mcitedefaultseppunct}\relax
\EndOfBibitem
\bibitem{Aad:2016zzw}
G.~Aad {\em et~al.\/} (ATLAS),
  \href{http://doi.org/10.1007/JHEP08(2016)009}{JHEP}
  \href{http://doi.org/10.1007/JHEP08(2016)009}{{\bf 08}, 009} (2016),
  \href{https://arxiv.org/abs/1606.01736}{[arXiv:1606.01736]}\relax
\mciteBstWouldAddEndPuncttrue
\mciteSetBstMidEndSepPunct{\mcitedefaultmidpunct}
{\mcitedefaultendpunct}{\mcitedefaultseppunct}\relax
\EndOfBibitem
\bibitem{Sirunyan:2018owv}
A.~M. Sirunyan {\em et~al.\/} (CMS), Submitted to: JHEP  (2018),
  \href{https://arxiv.org/abs/1812.10529}{[arXiv:1812.10529]}\relax
\mciteBstWouldAddEndPuncttrue
\mciteSetBstMidEndSepPunct{\mcitedefaultmidpunct}
{\mcitedefaultendpunct}{\mcitedefaultseppunct}\relax
\EndOfBibitem
\bibitem{Aaboud:2017ffb}
M.~Aaboud {\em et~al.\/} (ATLAS),
  \href{http://doi.org/10.1007/JHEP12(2017)059}{JHEP}
  \href{http://doi.org/10.1007/JHEP12(2017)059}{{\bf 12}, 059} (2017),
  \href{https://arxiv.org/abs/1710.05167}{[arXiv:1710.05167]}\relax
\mciteBstWouldAddEndPuncttrue
\mciteSetBstMidEndSepPunct{\mcitedefaultmidpunct}
{\mcitedefaultendpunct}{\mcitedefaultseppunct}\relax
\EndOfBibitem
\bibitem{Chatrchyan:2013mza}
S.~Chatrchyan {\em et~al.\/} (CMS),
  \href{http://doi.org/10.1103/PhysRevD.90.032004}{Phys. Rev.}
  \href{http://doi.org/10.1103/PhysRevD.90.032004}{{\bf D90}, 3, 032004}
  (2014), \href{https://arxiv.org/abs/1312.6283}{[arXiv:1312.6283]}\relax
\mciteBstWouldAddEndPuncttrue
\mciteSetBstMidEndSepPunct{\mcitedefaultmidpunct}
{\mcitedefaultendpunct}{\mcitedefaultseppunct}\relax
\EndOfBibitem
\bibitem{Khachatryan:2016pev}
V.~Khachatryan {\em et~al.\/} (CMS),
  \href{http://doi.org/10.1140/epjc/s10052-016-4293-4}{Eur. Phys. J.}
  \href{http://doi.org/10.1140/epjc/s10052-016-4293-4}{{\bf C76}, 8, 469}
  (2016), \href{https://arxiv.org/abs/1603.01803}{[arXiv:1603.01803]}\relax
\mciteBstWouldAddEndPuncttrue
\mciteSetBstMidEndSepPunct{\mcitedefaultmidpunct}
{\mcitedefaultendpunct}{\mcitedefaultseppunct}\relax
\EndOfBibitem
\bibitem{Aad:2019rou}
G.~Aad {\em et~al.\/} (ATLAS), Submitted to: Eur. Phys. J.  (2019),
  \href{https://arxiv.org/abs/1904.05631}{[arXiv:1904.05631]}\relax
\mciteBstWouldAddEndPuncttrue
\mciteSetBstMidEndSepPunct{\mcitedefaultmidpunct}
{\mcitedefaultendpunct}{\mcitedefaultseppunct}\relax
\EndOfBibitem
\bibitem{LHCb:2016zpq}
R.~Aaij {\em et~al.\/} (LHCb),
  \href{http://doi.org/10.1007/JHEP10(2016)030}{JHEP}
  \href{http://doi.org/10.1007/JHEP10(2016)030}{{\bf 10}, 030} (2016),
  \href{https://arxiv.org/abs/1608.01484}{[arXiv:1608.01484]}\relax
\mciteBstWouldAddEndPuncttrue
\mciteSetBstMidEndSepPunct{\mcitedefaultmidpunct}
{\mcitedefaultendpunct}{\mcitedefaultseppunct}\relax
\EndOfBibitem
\bibitem{Aad:2015auj}
G.~Aad {\em et~al.\/} (ATLAS),
  \href{http://doi.org/10.1140/epjc/s10052-016-4070-4}{Eur. Phys. J.}
  \href{http://doi.org/10.1140/epjc/s10052-016-4070-4}{{\bf C76}, 5, 291}
  (2016), \href{https://arxiv.org/abs/1512.02192}{[arXiv:1512.02192]}\relax
\mciteBstWouldAddEndPuncttrue
\mciteSetBstMidEndSepPunct{\mcitedefaultmidpunct}
{\mcitedefaultendpunct}{\mcitedefaultseppunct}\relax
\EndOfBibitem
\bibitem{Khachatryan:2016nbe}
V.~Khachatryan {\em et~al.\/} (CMS),
  \href{http://doi.org/10.1007/JHEP02(2017)096}{JHEP}
  \href{http://doi.org/10.1007/JHEP02(2017)096}{{\bf 02}, 096} (2017),
  \href{https://arxiv.org/abs/1606.05864}{[arXiv:1606.05864]}\relax
\mciteBstWouldAddEndPuncttrue
\mciteSetBstMidEndSepPunct{\mcitedefaultmidpunct}
{\mcitedefaultendpunct}{\mcitedefaultseppunct}\relax
\EndOfBibitem
\bibitem{Sirunyan:2017igm}
A.~M. Sirunyan {\em et~al.\/} (CMS),
  \href{http://doi.org/10.1007/JHEP03(2018)172}{JHEP}
  \href{http://doi.org/10.1007/JHEP03(2018)172}{{\bf 03}, 172} (2018),
  \href{https://arxiv.org/abs/1710.07955}{[arXiv:1710.07955]}\relax
\mciteBstWouldAddEndPuncttrue
\mciteSetBstMidEndSepPunct{\mcitedefaultmidpunct}
{\mcitedefaultendpunct}{\mcitedefaultseppunct}\relax
\EndOfBibitem
\bibitem{Aaboud:2017soa}
M.~Aaboud {\em et~al.\/} (ATLAS),
  \href{http://doi.org/10.1007/JHEP05(2018)077}{JHEP}
  \href{http://doi.org/10.1007/JHEP05(2018)077}{{\bf 05}, 077} (2018),
  \href{https://arxiv.org/abs/1711.03296}{[arXiv:1711.03296]}\relax
\mciteBstWouldAddEndPuncttrue
\mciteSetBstMidEndSepPunct{\mcitedefaultmidpunct}
{\mcitedefaultendpunct}{\mcitedefaultseppunct}\relax
\EndOfBibitem
\bibitem{Blumenschein:2015iqa}
U.~Blumenschein, \href{http://doi.org/10.1142/S0217751X15460070}{Int. J. Mod.
  Phys.} \href{http://doi.org/10.1142/S0217751X15460070}{{\bf A30}, 31,
  1546007} (2015),
  \href{https://arxiv.org/abs/1509.04885}{[arXiv:1509.04885]}\relax
\mciteBstWouldAddEndPuncttrue
\mciteSetBstMidEndSepPunct{\mcitedefaultmidpunct}
{\mcitedefaultendpunct}{\mcitedefaultseppunct}\relax
\EndOfBibitem
\bibitem{Aad:2019hga}
G.~Aad {\em et~al.\/} (ATLAS)  (2019),
  \href{https://arxiv.org/abs/1907.06728}{[arXiv:1907.06728]}\relax
\mciteBstWouldAddEndPuncttrue
\mciteSetBstMidEndSepPunct{\mcitedefaultmidpunct}
{\mcitedefaultendpunct}{\mcitedefaultseppunct}\relax
\EndOfBibitem
\bibitem{Khachatryan:2016fue}
V.~Khachatryan {\em et~al.\/} (CMS),
  \href{http://doi.org/10.1103/PhysRevD.95.052002}{Phys. Rev.}
  \href{http://doi.org/10.1103/PhysRevD.95.052002}{{\bf D95}, 052002} (2017),
  \href{https://arxiv.org/abs/1610.04222}{[arXiv:1610.04222]}\relax
\mciteBstWouldAddEndPuncttrue
\mciteSetBstMidEndSepPunct{\mcitedefaultmidpunct}
{\mcitedefaultendpunct}{\mcitedefaultseppunct}\relax
\EndOfBibitem
\bibitem{Sirunyan:2017wgx}
A.~M. Sirunyan {\em et~al.\/} (CMS),
  \href{http://doi.org/10.1103/PhysRevD.96.072005}{Phys. Rev.}
  \href{http://doi.org/10.1103/PhysRevD.96.072005}{{\bf D96}, 7, 072005}
  (2017), \href{https://arxiv.org/abs/1707.05979}{[arXiv:1707.05979]}\relax
\mciteBstWouldAddEndPuncttrue
\mciteSetBstMidEndSepPunct{\mcitedefaultmidpunct}
{\mcitedefaultendpunct}{\mcitedefaultseppunct}\relax
\EndOfBibitem
\bibitem{Sirunyan:2018cpw}
A.~M. Sirunyan {\em et~al.\/} (CMS),
  \href{http://doi.org/10.1140/epjc/s10052-018-6373-0}{Eur. Phys. J.}
  \href{http://doi.org/10.1140/epjc/s10052-018-6373-0}{{\bf C78}, 11, 965}
  (2018), \href{https://arxiv.org/abs/1804.05252}{[arXiv:1804.05252]}\relax
\mciteBstWouldAddEndPuncttrue
\mciteSetBstMidEndSepPunct{\mcitedefaultmidpunct}
{\mcitedefaultendpunct}{\mcitedefaultseppunct}\relax
\EndOfBibitem
\bibitem{ATLAS:2022nrp}
G.~Aad {\em et~al.\/} (ATLAS),
  \href{http://doi.org/10.1007/JHEP06(2023)080}{JHEP}
  \href{http://doi.org/10.1007/JHEP06(2023)080}{{\bf 06}, 080} (2023),
  \href{https://arxiv.org/abs/2205.02597}{[arXiv:2205.02597]}\relax
\mciteBstWouldAddEndPuncttrue
\mciteSetBstMidEndSepPunct{\mcitedefaultmidpunct}
{\mcitedefaultendpunct}{\mcitedefaultseppunct}\relax
\EndOfBibitem
\bibitem{Aad:2013ysa}
G.~Aad {\em et~al.\/} (ATLAS),
  \href{http://doi.org/10.1007/JHEP07(2013)032}{JHEP}
  \href{http://doi.org/10.1007/JHEP07(2013)032}{{\bf 07}, 032} (2013),
  \href{https://arxiv.org/abs/1304.7098}{[arXiv:1304.7098]}\relax
\mciteBstWouldAddEndPuncttrue
\mciteSetBstMidEndSepPunct{\mcitedefaultmidpunct}
{\mcitedefaultendpunct}{\mcitedefaultseppunct}\relax
\EndOfBibitem
\bibitem{Bern:2013gka}
Z.~Bern {\em et~al.\/}, \href{http://doi.org/10.1103/PhysRevD.88.014025}{Phys.
  Rev.} \href{http://doi.org/10.1103/PhysRevD.88.014025}{{\bf D88}, 1, 014025}
  (2013), \href{https://arxiv.org/abs/1304.1253}{[arXiv:1304.1253]}\relax
\mciteBstWouldAddEndPuncttrue
\mciteSetBstMidEndSepPunct{\mcitedefaultmidpunct}
{\mcitedefaultendpunct}{\mcitedefaultseppunct}\relax
\EndOfBibitem
\bibitem{ATLAS:2022uav}
  (2022), \href{https://arxiv.org/abs/2204.12355}{[arXiv:2204.12355]}\relax
\mciteBstWouldAddEndPuncttrue
\mciteSetBstMidEndSepPunct{\mcitedefaultmidpunct}
{\mcitedefaultendpunct}{\mcitedefaultseppunct}\relax
\EndOfBibitem
\bibitem{Voutilainen:2015lqa}
M.~Voutilainen, \href{http://doi.org/10.1142/S0217751X15460082}{Int. J. Mod.
  Phys.} \href{http://doi.org/10.1142/S0217751X15460082}{{\bf A30}, 31,
  1546008} (2015),
  \href{https://arxiv.org/abs/1509.05026}{[arXiv:1509.05026]}\relax
\mciteBstWouldAddEndPuncttrue
\mciteSetBstMidEndSepPunct{\mcitedefaultmidpunct}
{\mcitedefaultendpunct}{\mcitedefaultseppunct}\relax
\EndOfBibitem
\bibitem{Aad:2016xcr}
G.~Aad {\em et~al.\/} (ATLAS),
  \href{http://doi.org/10.1007/JHEP08(2016)005}{JHEP}
  \href{http://doi.org/10.1007/JHEP08(2016)005}{{\bf 08}, 005} (2016),
  \href{https://arxiv.org/abs/1605.03495}{[arXiv:1605.03495]}\relax
\mciteBstWouldAddEndPuncttrue
\mciteSetBstMidEndSepPunct{\mcitedefaultmidpunct}
{\mcitedefaultendpunct}{\mcitedefaultseppunct}\relax
\EndOfBibitem
\bibitem{ATLAS:2019buk}
G.~Aad {\em et~al.\/} (ATLAS),
  \href{http://doi.org/10.1007/JHEP10(2019)203}{JHEP}
  \href{http://doi.org/10.1007/JHEP10(2019)203}{{\bf 10}, 203} (2019),
  \href{https://arxiv.org/abs/1908.02746}{[arXiv:1908.02746]}\relax
\mciteBstWouldAddEndPuncttrue
\mciteSetBstMidEndSepPunct{\mcitedefaultmidpunct}
{\mcitedefaultendpunct}{\mcitedefaultseppunct}\relax
\EndOfBibitem
\bibitem{Chatrchyan:2011ue}
S.~Chatrchyan {\em et~al.\/} (CMS),
  \href{http://doi.org/10.1103/PhysRevD.84.052011}{Phys. Rev.}
  \href{http://doi.org/10.1103/PhysRevD.84.052011}{{\bf D84}, 052011} (2011),
  \href{https://arxiv.org/abs/1108.2044}{[arXiv:1108.2044]}\relax
\mciteBstWouldAddEndPuncttrue
\mciteSetBstMidEndSepPunct{\mcitedefaultmidpunct}
{\mcitedefaultendpunct}{\mcitedefaultseppunct}\relax
\EndOfBibitem
\bibitem{Chatrchyan:2013mwa}
S.~Chatrchyan {\em et~al.\/} (CMS),
  \href{http://doi.org/10.1007/JHEP06(2014)009}{JHEP}
  \href{http://doi.org/10.1007/JHEP06(2014)009}{{\bf 06}, 009} (2014),
  \href{https://arxiv.org/abs/1311.6141}{[arXiv:1311.6141]}\relax
\mciteBstWouldAddEndPuncttrue
\mciteSetBstMidEndSepPunct{\mcitedefaultmidpunct}
{\mcitedefaultendpunct}{\mcitedefaultseppunct}\relax
\EndOfBibitem
\bibitem{Aad:2013zba}
G.~Aad {\em et~al.\/} (ATLAS),
  \href{http://doi.org/10.1103/PhysRevD.89.052004}{Phys. Rev.}
  \href{http://doi.org/10.1103/PhysRevD.89.052004}{{\bf D89}, 5, 052004}
  (2014), \href{https://arxiv.org/abs/1311.1440}{[arXiv:1311.1440]}\relax
\mciteBstWouldAddEndPuncttrue
\mciteSetBstMidEndSepPunct{\mcitedefaultmidpunct}
{\mcitedefaultendpunct}{\mcitedefaultseppunct}\relax
\EndOfBibitem
\bibitem{Aaboud:2017kff}
M.~Aaboud {\em et~al.\/} (ATLAS),
  \href{http://doi.org/10.1016/j.physletb.2018.03.035}{Phys. Lett.}
  \href{http://doi.org/10.1016/j.physletb.2018.03.035}{{\bf B780}, 578} (2018),
  \href{https://arxiv.org/abs/1801.00112}{[arXiv:1801.00112]}\relax
\mciteBstWouldAddEndPuncttrue
\mciteSetBstMidEndSepPunct{\mcitedefaultmidpunct}
{\mcitedefaultendpunct}{\mcitedefaultseppunct}\relax
\EndOfBibitem
\bibitem{Aaboud:2016sdm}
M.~Aaboud {\em et~al.\/} (ATLAS),
  \href{http://doi.org/10.1016/j.nuclphysb.2017.03.006}{Nucl. Phys.}
  \href{http://doi.org/10.1016/j.nuclphysb.2017.03.006}{{\bf B918}, 257}
  (2017), \href{https://arxiv.org/abs/1611.06586}{[arXiv:1611.06586]}\relax
\mciteBstWouldAddEndPuncttrue
\mciteSetBstMidEndSepPunct{\mcitedefaultmidpunct}
{\mcitedefaultendpunct}{\mcitedefaultseppunct}\relax
\EndOfBibitem
\bibitem{Sirunyan:2018gro}
A.~M. Sirunyan {\em et~al.\/} (CMS),
  \href{http://doi.org/10.1140/epjc/s10052-018-6482-9}{Eur. Phys. J.}
  \href{http://doi.org/10.1140/epjc/s10052-018-6482-9}{{\bf C79}, 1, 20}
  (2019), \href{https://arxiv.org/abs/1807.00782}{[arXiv:1807.00782]}\relax
\mciteBstWouldAddEndPuncttrue
\mciteSetBstMidEndSepPunct{\mcitedefaultmidpunct}
{\mcitedefaultendpunct}{\mcitedefaultseppunct}\relax
\EndOfBibitem
\bibitem{Sirunyan:2019uya}
A.~M. Sirunyan {\em et~al.\/} (CMS)  (2019),
  \href{https://arxiv.org/abs/1907.08155}{[arXiv:1907.08155]}\relax
\mciteBstWouldAddEndPuncttrue
\mciteSetBstMidEndSepPunct{\mcitedefaultmidpunct}
{\mcitedefaultendpunct}{\mcitedefaultseppunct}\relax
\EndOfBibitem
\bibitem{ATLAS:2019iaa}
G.~Aad {\em et~al.\/} (ATLAS),
  \href{http://doi.org/10.1007/JHEP03(2020)179}{JHEP}
  \href{http://doi.org/10.1007/JHEP03(2020)179}{{\bf 03}, 179} (2020),
  \href{https://arxiv.org/abs/1912.09866}{[arXiv:1912.09866]}\relax
\mciteBstWouldAddEndPuncttrue
\mciteSetBstMidEndSepPunct{\mcitedefaultmidpunct}
{\mcitedefaultendpunct}{\mcitedefaultseppunct}\relax
\EndOfBibitem
\bibitem{Aaboud:2017skj}
M.~Aaboud {\em et~al.\/} (ATLAS),
  \href{http://doi.org/10.1016/j.physletb.2017.11.054}{Phys. Lett.}
  \href{http://doi.org/10.1016/j.physletb.2017.11.054}{{\bf B776}, 295} (2018),
  \href{https://arxiv.org/abs/1710.09560}{[arXiv:1710.09560]}\relax
\mciteBstWouldAddEndPuncttrue
\mciteSetBstMidEndSepPunct{\mcitedefaultmidpunct}
{\mcitedefaultendpunct}{\mcitedefaultseppunct}\relax
\EndOfBibitem
\bibitem{Kuhn:2005gv}
J.~H. Kuhn {\em et~al.\/},
  \href{http://doi.org/10.1088/1126-6708/2006/03/059}{JHEP}
  \href{http://doi.org/10.1088/1126-6708/2006/03/059}{{\bf 03}, 059} (2006),
  \href{https://arxiv.org/abs/hep-ph/0508253}{[hep-ph/0508253]}\relax
\mciteBstWouldAddEndPuncttrue
\mciteSetBstMidEndSepPunct{\mcitedefaultmidpunct}
{\mcitedefaultendpunct}{\mcitedefaultseppunct}\relax
\EndOfBibitem
\bibitem{Lindert:2017olm}
J.~M. Lindert {\em et~al.\/},
  \href{http://doi.org/10.1140/epjc/s10052-017-5389-1}{Eur. Phys. J. C}
  \href{http://doi.org/10.1140/epjc/s10052-017-5389-1}{{\bf 77}, 12, 829}
  (2017), \href{https://arxiv.org/abs/1705.04664}{[arXiv:1705.04664]}\relax
\mciteBstWouldAddEndPuncttrue
\mciteSetBstMidEndSepPunct{\mcitedefaultmidpunct}
{\mcitedefaultendpunct}{\mcitedefaultseppunct}\relax
\EndOfBibitem
\bibitem{Khachatryan:2015ira}
V.~Khachatryan {\em et~al.\/} (CMS),
  \href{http://doi.org/10.1007/JHEP04(2016)010}{JHEP}
  \href{http://doi.org/10.1007/JHEP04(2016)010}{{\bf 10}, 128} (2015),
  [Erratum: JHEP04,010(2016)],
  \href{https://arxiv.org/abs/1505.06520}{[arXiv:1505.06520]}\relax
\mciteBstWouldAddEndPuncttrue
\mciteSetBstMidEndSepPunct{\mcitedefaultmidpunct}
{\mcitedefaultendpunct}{\mcitedefaultseppunct}\relax
\EndOfBibitem
\bibitem{Aad:2016wpd}
G.~Aad {\em et~al.\/} (ATLAS),
  \href{http://doi.org/10.1007/JHEP09(2016)029}{JHEP}
  \href{http://doi.org/10.1007/JHEP09(2016)029}{{\bf 09}, 029} (2016),
  \href{https://arxiv.org/abs/1603.01702}{[arXiv:1603.01702]}\relax
\mciteBstWouldAddEndPuncttrue
\mciteSetBstMidEndSepPunct{\mcitedefaultmidpunct}
{\mcitedefaultendpunct}{\mcitedefaultseppunct}\relax
\EndOfBibitem
\bibitem{Aaboud:2017qkn}
M.~Aaboud {\em et~al.\/} (ATLAS),
  \href{http://doi.org/10.1016/j.physletb.2017.08.047}{Phys. Lett.}
  \href{http://doi.org/10.1016/j.physletb.2017.08.047}{{\bf B773}, 354} (2017),
  \href{https://arxiv.org/abs/1702.04519}{[arXiv:1702.04519]}\relax
\mciteBstWouldAddEndPuncttrue
\mciteSetBstMidEndSepPunct{\mcitedefaultmidpunct}
{\mcitedefaultendpunct}{\mcitedefaultseppunct}\relax
\EndOfBibitem
\bibitem{Aaboud:2019nkz}
M.~Aaboud {\em et~al.\/} (ATLAS)  (2019),
  \href{https://arxiv.org/abs/1905.04242}{[arXiv:1905.04242]}\relax
\mciteBstWouldAddEndPuncttrue
\mciteSetBstMidEndSepPunct{\mcitedefaultmidpunct}
{\mcitedefaultendpunct}{\mcitedefaultseppunct}\relax
\EndOfBibitem
\bibitem{Khachatryan:2015sga}
V.~Khachatryan {\em et~al.\/} (CMS),
  \href{http://doi.org/10.1140/epjc/s10052-016-4219-1}{Eur. Phys. J.}
  \href{http://doi.org/10.1140/epjc/s10052-016-4219-1}{{\bf C76}, 7, 401}
  (2016), \href{https://arxiv.org/abs/1507.03268}{[arXiv:1507.03268]}\relax
\mciteBstWouldAddEndPuncttrue
\mciteSetBstMidEndSepPunct{\mcitedefaultmidpunct}
{\mcitedefaultendpunct}{\mcitedefaultseppunct}\relax
\EndOfBibitem
\bibitem{ATLAS:2019dny}
G.~Aad {\em et~al.\/} (ATLAS),
  \href{http://doi.org/10.1016/j.physletb.2019.134913}{Phys. Lett. B}
  \href{http://doi.org/10.1016/j.physletb.2019.134913}{{\bf 798}, 134913}
  (2019), \href{https://arxiv.org/abs/1903.10415}{[arXiv:1903.10415]}\relax
\mciteBstWouldAddEndPuncttrue
\mciteSetBstMidEndSepPunct{\mcitedefaultmidpunct}
{\mcitedefaultendpunct}{\mcitedefaultseppunct}\relax
\EndOfBibitem
\bibitem{ATLAS:2022xnu}
G.~Aad {\em et~al.\/} (ATLAS),
  \href{http://doi.org/10.1103/PhysRevLett.129.061803}{Phys. Rev. Lett.}
  \href{http://doi.org/10.1103/PhysRevLett.129.061803}{{\bf 129}, 6, 061803}
  (2022), \href{https://arxiv.org/abs/2201.13045}{[arXiv:2201.13045]}\relax
\mciteBstWouldAddEndPuncttrue
\mciteSetBstMidEndSepPunct{\mcitedefaultmidpunct}
{\mcitedefaultendpunct}{\mcitedefaultseppunct}\relax
\EndOfBibitem
\bibitem{CMS:2020hjs}
A.~M. Sirunyan {\em et~al.\/} (CMS),
  \href{http://doi.org/10.1103/PhysRevLett.125.151802}{Phys. Rev. Lett.}
  \href{http://doi.org/10.1103/PhysRevLett.125.151802}{{\bf 125}, 15, 151802}
  (2020), \href{https://arxiv.org/abs/2006.11191}{[arXiv:2006.11191]}\relax
\mciteBstWouldAddEndPuncttrue
\mciteSetBstMidEndSepPunct{\mcitedefaultmidpunct}
{\mcitedefaultendpunct}{\mcitedefaultseppunct}\relax
\EndOfBibitem
\bibitem{Aad:2012tba}
G.~Aad {\em et~al.\/} (ATLAS),
  \href{http://doi.org/10.1007/JHEP01(2013)086}{JHEP}
  \href{http://doi.org/10.1007/JHEP01(2013)086}{{\bf 01}, 086} (2013),
  \href{https://arxiv.org/abs/1211.1913}{[arXiv:1211.1913]}\relax
\mciteBstWouldAddEndPuncttrue
\mciteSetBstMidEndSepPunct{\mcitedefaultmidpunct}
{\mcitedefaultendpunct}{\mcitedefaultseppunct}\relax
\EndOfBibitem
\bibitem{Aaboud:2017vol}
M.~Aaboud {\em et~al.\/} (ATLAS),
  \href{http://doi.org/10.1103/PhysRevD.95.112005}{Phys. Rev.}
  \href{http://doi.org/10.1103/PhysRevD.95.112005}{{\bf D95}, 11, 112005}
  (2017), \href{https://arxiv.org/abs/1704.03839}{[arXiv:1704.03839]}\relax
\mciteBstWouldAddEndPuncttrue
\mciteSetBstMidEndSepPunct{\mcitedefaultmidpunct}
{\mcitedefaultendpunct}{\mcitedefaultseppunct}\relax
\EndOfBibitem
\bibitem{ATLAS:2021mbt}
G.~Aad {\em et~al.\/} (ATLAS),
  \href{http://doi.org/10.1007/JHEP11(2021)169}{JHEP}
  \href{http://doi.org/10.1007/JHEP11(2021)169}{{\bf 11}, 169} (2021),
  \href{https://arxiv.org/abs/2107.09330}{[arXiv:2107.09330]}\relax
\mciteBstWouldAddEndPuncttrue
\mciteSetBstMidEndSepPunct{\mcitedefaultmidpunct}
{\mcitedefaultendpunct}{\mcitedefaultseppunct}\relax
\EndOfBibitem
\bibitem{Chatrchyan:2014fsa}
S.~Chatrchyan {\em et~al.\/} (CMS),
  \href{http://doi.org/10.1140/epjc/s10052-014-3129-3}{Eur. Phys. J.}
  \href{http://doi.org/10.1140/epjc/s10052-014-3129-3}{{\bf C74}, 11, 3129}
  (2014), \href{https://arxiv.org/abs/1405.7225}{[arXiv:1405.7225]}\relax
\mciteBstWouldAddEndPuncttrue
\mciteSetBstMidEndSepPunct{\mcitedefaultmidpunct}
{\mcitedefaultendpunct}{\mcitedefaultseppunct}\relax
\EndOfBibitem
\bibitem{Aaboud:2017lxm}
M.~Aaboud {\em et~al.\/} (ATLAS),
  \href{http://doi.org/10.1016/j.physletb.2018.03.057}{Phys. Lett.}
  \href{http://doi.org/10.1016/j.physletb.2018.03.057}{{\bf B781}, 55} (2018),
  \href{https://arxiv.org/abs/1712.07291}{[arXiv:1712.07291]}\relax
\mciteBstWouldAddEndPuncttrue
\mciteSetBstMidEndSepPunct{\mcitedefaultmidpunct}
{\mcitedefaultendpunct}{\mcitedefaultseppunct}\relax
\EndOfBibitem
\bibitem{Kroninger:2015oma}
K.~Kröninger, A.~B. Meyer and P.~Uwer, in T.~Schörner-Sadenius, editor,
  \enquote{The Large Hadron Collider: Harvest of Run 1,} 259--300 (2015),
  \href{https://arxiv.org/abs/1506.02800}{[arXiv:1506.02800]}\relax
\mciteBstWouldAddEndPuncttrue
\mciteSetBstMidEndSepPunct{\mcitedefaultmidpunct}
{\mcitedefaultendpunct}{\mcitedefaultseppunct}\relax
\EndOfBibitem
\bibitem{Aaboud:2016iot}
M.~Aaboud {\em et~al.\/} (ATLAS),
  \href{http://doi.org/10.1103/PhysRevD.94.092003}{Phys. Rev.}
  \href{http://doi.org/10.1103/PhysRevD.94.092003}{{\bf D94}, 9, 092003}
  (2016), \href{https://arxiv.org/abs/1607.07281}{[arXiv:1607.07281]}\relax
\mciteBstWouldAddEndPuncttrue
\mciteSetBstMidEndSepPunct{\mcitedefaultmidpunct}
{\mcitedefaultendpunct}{\mcitedefaultseppunct}\relax
\EndOfBibitem
\bibitem{Aaboud:2016syx}
M.~Aaboud {\em et~al.\/} (ATLAS),
  \href{http://doi.org/10.1140/epjc/s10052-017-4821-x}{Eur. Phys. J.}
  \href{http://doi.org/10.1140/epjc/s10052-017-4821-x}{{\bf C77}, 5, 292}
  (2017), \href{https://arxiv.org/abs/1612.05220}{[arXiv:1612.05220]}\relax
\mciteBstWouldAddEndPuncttrue
\mciteSetBstMidEndSepPunct{\mcitedefaultmidpunct}
{\mcitedefaultendpunct}{\mcitedefaultseppunct}\relax
\EndOfBibitem
\bibitem{Sirunyan:2018goh}
A.~M. Sirunyan {\em et~al.\/} (CMS),
  \href{http://doi.org/10.1140/epjc/s10052-019-6863-8}{Eur. Phys. J.}
  \href{http://doi.org/10.1140/epjc/s10052-019-6863-8}{{\bf C79}, 5, 368}
  (2019), \href{https://arxiv.org/abs/1812.10505}{[arXiv:1812.10505]}\relax
\mciteBstWouldAddEndPuncttrue
\mciteSetBstMidEndSepPunct{\mcitedefaultmidpunct}
{\mcitedefaultendpunct}{\mcitedefaultseppunct}\relax
\EndOfBibitem
\bibitem{CMS:2019esx}
A.~M. Sirunyan {\em et~al.\/} (CMS),
  \href{http://doi.org/10.1140/epjc/s10052-020-7917-7}{Eur. Phys. J. C}
  \href{http://doi.org/10.1140/epjc/s10052-020-7917-7}{{\bf 80}, 7, 658}
  (2020), \href{https://arxiv.org/abs/1904.05237}{[arXiv:1904.05237]}\relax
\mciteBstWouldAddEndPuncttrue
\mciteSetBstMidEndSepPunct{\mcitedefaultmidpunct}
{\mcitedefaultendpunct}{\mcitedefaultseppunct}\relax
\EndOfBibitem
\bibitem{ATLAS:2023gsl}
G.~Aad {\em et~al.\/} (ATLAS),
  \href{http://doi.org/10.1007/JHEP07(2023)141}{JHEP}
  \href{http://doi.org/10.1007/JHEP07(2023)141}{{\bf 07}, 141} (2023),
  \href{https://arxiv.org/abs/2303.15340}{[arXiv:2303.15340]}\relax
\mciteBstWouldAddEndPuncttrue
\mciteSetBstMidEndSepPunct{\mcitedefaultmidpunct}
{\mcitedefaultendpunct}{\mcitedefaultseppunct}\relax
\EndOfBibitem
\bibitem{Czakon:2015owf}
M.~Czakon, D.~Heymes and A.~Mitov,
  \href{http://doi.org/10.1103/PhysRevLett.116.082003}{Phys. Rev. Lett.}
  \href{http://doi.org/10.1103/PhysRevLett.116.082003}{{\bf 116}, 8, 082003}
  (2016), \href{https://arxiv.org/abs/1511.00549}{[arXiv:1511.00549]}\relax
\mciteBstWouldAddEndPuncttrue
\mciteSetBstMidEndSepPunct{\mcitedefaultmidpunct}
{\mcitedefaultendpunct}{\mcitedefaultseppunct}\relax
\EndOfBibitem
\bibitem{Catani:2019hip}
S.~Catani {\em et~al.\/}, \href{http://doi.org/10.1007/JHEP07(2019)100}{JHEP}
  \href{http://doi.org/10.1007/JHEP07(2019)100}{{\bf 07}, 100} (2019),
  \href{https://arxiv.org/abs/1906.06535}{[arXiv:1906.06535]}\relax
\mciteBstWouldAddEndPuncttrue
\mciteSetBstMidEndSepPunct{\mcitedefaultmidpunct}
{\mcitedefaultendpunct}{\mcitedefaultseppunct}\relax
\EndOfBibitem
\bibitem{Czakon:2018nun}
M.~Czakon {\em et~al.\/}, \href{http://doi.org/10.1007/JHEP05(2018)149}{JHEP}
  \href{http://doi.org/10.1007/JHEP05(2018)149}{{\bf 05}, 149} (2018),
  \href{https://arxiv.org/abs/1803.07623}{[arXiv:1803.07623]}\relax
\mciteBstWouldAddEndPuncttrue
\mciteSetBstMidEndSepPunct{\mcitedefaultmidpunct}
{\mcitedefaultendpunct}{\mcitedefaultseppunct}\relax
\EndOfBibitem
\bibitem{ATLAS:2023ajo}
G.~Aad {\em et~al.\/} (ATLAS),
  \href{http://doi.org/10.1140/epjc/s10052-023-11573-0}{Eur. Phys. J. C}
  \href{http://doi.org/10.1140/epjc/s10052-023-11573-0}{{\bf 83}, 6, 496}
  (2023), \href{https://arxiv.org/abs/2303.15061}{[arXiv:2303.15061]}\relax
\mciteBstWouldAddEndPuncttrue
\mciteSetBstMidEndSepPunct{\mcitedefaultmidpunct}
{\mcitedefaultendpunct}{\mcitedefaultseppunct}\relax
\EndOfBibitem
\bibitem{Bevilacqua:2012em}
G.~Bevilacqua and M.~Worek, \href{http://doi.org/10.1007/JHEP07(2012)111}{JHEP}
  \href{http://doi.org/10.1007/JHEP07(2012)111}{{\bf 07}, 111} (2012),
  \href{https://arxiv.org/abs/1206.3064}{[arXiv:1206.3064]}\relax
\mciteBstWouldAddEndPuncttrue
\mciteSetBstMidEndSepPunct{\mcitedefaultmidpunct}
{\mcitedefaultendpunct}{\mcitedefaultseppunct}\relax
\EndOfBibitem
\bibitem{Chatrchyan:2013haa}
S.~Chatrchyan {\em et~al.\/} (CMS),
  \href{http://doi.org/10.1016/j.physletb.2014.08.040}{Phys. Lett.}
  \href{http://doi.org/10.1016/j.physletb.2014.08.040}{{\bf B728}, 496} (2014),
  [Erratum: Phys. Lett.B738,526(2014)],
  \href{https://arxiv.org/abs/1307.1907}{[arXiv:1307.1907]}\relax
\mciteBstWouldAddEndPuncttrue
\mciteSetBstMidEndSepPunct{\mcitedefaultmidpunct}
{\mcitedefaultendpunct}{\mcitedefaultseppunct}\relax
\EndOfBibitem
\bibitem{Klijnsma:2017eqp}
T.~Klijnsma {\em et~al.\/},
  \href{http://doi.org/10.1140/epjc/s10052-017-5340-5}{Eur. Phys. J.}
  \href{http://doi.org/10.1140/epjc/s10052-017-5340-5}{{\bf C77}, 11, 778}
  (2017), \href{https://arxiv.org/abs/1708.07495}{[arXiv:1708.07495]}\relax
\mciteBstWouldAddEndPuncttrue
\mciteSetBstMidEndSepPunct{\mcitedefaultmidpunct}
{\mcitedefaultendpunct}{\mcitedefaultseppunct}\relax
\EndOfBibitem
\bibitem{ATLAS:2014nxi}
G.~Aad {\em et~al.\/} (ATLAS),
  \href{http://doi.org/10.1140/epjc/s10052-016-4501-2}{Eur. Phys. J. C}
  \href{http://doi.org/10.1140/epjc/s10052-016-4501-2}{{\bf 74}, 10, 3109}
  (2014), [Addendum: Eur.Phys.J.C 76, 642 (2016)],
  \href{https://arxiv.org/abs/1406.5375}{[arXiv:1406.5375]}\relax
\mciteBstWouldAddEndPuncttrue
\mciteSetBstMidEndSepPunct{\mcitedefaultmidpunct}
{\mcitedefaultendpunct}{\mcitedefaultseppunct}\relax
\EndOfBibitem
\bibitem{Sirunyan:2017uhy}
A.~M. Sirunyan {\em et~al.\/} (CMS),
  \href{http://doi.org/10.1007/JHEP09(2017)051}{JHEP}
  \href{http://doi.org/10.1007/JHEP09(2017)051}{{\bf 09}, 051} (2017),
  \href{https://arxiv.org/abs/1701.06228}{[arXiv:1701.06228]}\relax
\mciteBstWouldAddEndPuncttrue
\mciteSetBstMidEndSepPunct{\mcitedefaultmidpunct}
{\mcitedefaultendpunct}{\mcitedefaultseppunct}\relax
\EndOfBibitem
\bibitem{ATLAS:2023tnc}
G.~Aad {\em et~al.\/} (ATLAS)  (2023),
  \href{https://arxiv.org/abs/2306.11379}{[arXiv:2306.11379]}\relax
\mciteBstWouldAddEndPuncttrue
\mciteSetBstMidEndSepPunct{\mcitedefaultmidpunct}
{\mcitedefaultendpunct}{\mcitedefaultseppunct}\relax
\EndOfBibitem
\bibitem{Aaboud:2018ezd}
M.~Aaboud {\em et~al.\/} (ATLAS),
  \href{http://doi.org/10.1016/j.physletb.2018.09.019}{Phys. Lett.}
  \href{http://doi.org/10.1016/j.physletb.2018.09.019}{{\bf B786}, 114} (2018),
  \href{https://arxiv.org/abs/1805.10197}{[arXiv:1805.10197]}\relax
\mciteBstWouldAddEndPuncttrue
\mciteSetBstMidEndSepPunct{\mcitedefaultmidpunct}
{\mcitedefaultendpunct}{\mcitedefaultseppunct}\relax
\EndOfBibitem
\bibitem{Sirunyan:2018sgc}
A.~M. Sirunyan {\em et~al.\/} (CMS),
  \href{http://doi.org/10.1016/j.physletb.2019.03.059}{Phys. Lett.}
  \href{http://doi.org/10.1016/j.physletb.2019.03.059}{{\bf B792}, 369} (2019),
  \href{https://arxiv.org/abs/1812.06504}{[arXiv:1812.06504]}\relax
\mciteBstWouldAddEndPuncttrue
\mciteSetBstMidEndSepPunct{\mcitedefaultmidpunct}
{\mcitedefaultendpunct}{\mcitedefaultseppunct}\relax
\EndOfBibitem
\bibitem{deFlorian:2016spz}
D.~de~Florian {\em et~al.\/} (LHC Higgs Cross Section Working Group)  (2016),
  \href{https://arxiv.org/abs/1610.07922}{[arXiv:1610.07922]}\relax
\mciteBstWouldAddEndPuncttrue
\mciteSetBstMidEndSepPunct{\mcitedefaultmidpunct}
{\mcitedefaultendpunct}{\mcitedefaultseppunct}\relax
\EndOfBibitem
\bibitem{Aaboud:2018xdt}
M.~Aaboud {\em et~al.\/} (ATLAS),
  \href{http://doi.org/10.1103/PhysRevD.98.052005}{Phys. Rev.}
  \href{http://doi.org/10.1103/PhysRevD.98.052005}{{\bf D98}, 052005} (2018),
  \href{https://arxiv.org/abs/1802.04146}{[arXiv:1802.04146]}\relax
\mciteBstWouldAddEndPuncttrue
\mciteSetBstMidEndSepPunct{\mcitedefaultmidpunct}
{\mcitedefaultendpunct}{\mcitedefaultseppunct}\relax
\EndOfBibitem
\bibitem{Sirunyan:2018kta}
A.~M. Sirunyan {\em et~al.\/} (CMS),
  \href{http://doi.org/10.1007/JHEP01(2019)183}{JHEP}
  \href{http://doi.org/10.1007/JHEP01(2019)183}{{\bf 01}, 183} (2019),
  \href{https://arxiv.org/abs/1807.03825}{[arXiv:1807.03825]}\relax
\mciteBstWouldAddEndPuncttrue
\mciteSetBstMidEndSepPunct{\mcitedefaultmidpunct}
{\mcitedefaultendpunct}{\mcitedefaultseppunct}\relax
\EndOfBibitem
\bibitem{ATLAS:2022fnp}
G.~Aad {\em et~al.\/} (ATLAS),
  \href{http://doi.org/10.1007/JHEP08(2022)027}{JHEP}
  \href{http://doi.org/10.1007/JHEP08(2022)027}{{\bf 08}, 027} (2022),
  \href{https://arxiv.org/abs/2202.00487}{[arXiv:2202.00487]}\relax
\mciteBstWouldAddEndPuncttrue
\mciteSetBstMidEndSepPunct{\mcitedefaultmidpunct}
{\mcitedefaultendpunct}{\mcitedefaultseppunct}\relax
\EndOfBibitem
\bibitem{CMS:2022wpo}
A.~Tumasyan {\em et~al.\/} (CMS),
  \href{http://doi.org/10.1007/JHEP07(2023)091}{JHEP}
  \href{http://doi.org/10.1007/JHEP07(2023)091}{{\bf 07}, 091} (2023),
  \href{https://arxiv.org/abs/2208.12279}{[arXiv:2208.12279]}\relax
\mciteBstWouldAddEndPuncttrue
\mciteSetBstMidEndSepPunct{\mcitedefaultmidpunct}
{\mcitedefaultendpunct}{\mcitedefaultseppunct}\relax
\EndOfBibitem
\bibitem{ParticleDataGroup:2022pth}
R.~L. Workman {\em et~al.\/} (Particle Data Group),
  \href{http://doi.org/10.1093/ptep/ptac097}{PTEP}
  \href{http://doi.org/10.1093/ptep/ptac097}{{\bf 2022}, 083C01} (2022)\relax
\mciteBstWouldAddEndPuncttrue
\mciteSetBstMidEndSepPunct{\mcitedefaultmidpunct}
{\mcitedefaultendpunct}{\mcitedefaultseppunct}\relax
\EndOfBibitem
\bibitem{dEnterria:2019its}
D.~d'Enterria {\em et~al.\/}, in \enquote{{Workshop on precision measurements
  of the QCD coupling constant (alphas-2019) Trento, Trentino, Italy, February
  11-15, 2019},}  (2019),
  \href{https://arxiv.org/abs/1907.01435}{[arXiv:1907.01435]}\relax
\mciteBstWouldAddEndPuncttrue
\mciteSetBstMidEndSepPunct{\mcitedefaultmidpunct}
{\mcitedefaultendpunct}{\mcitedefaultseppunct}\relax
\EndOfBibitem
\bibitem{dEnterria:2022hzv}
D.~d'Enterria {\em et~al.\/}  (2022),
  \href{https://arxiv.org/abs/2203.08271}{[arXiv:2203.08271]}\relax
\mciteBstWouldAddEndPuncttrue
\mciteSetBstMidEndSepPunct{\mcitedefaultmidpunct}
{\mcitedefaultendpunct}{\mcitedefaultseppunct}\relax
\EndOfBibitem
\bibitem{Salam:2017qdl}
G.~P. Salam, in A.~Levy, S.~Forte and G.~Ridolfi, editors, \enquote{From My
  Vast Repertoire ...: Guido Altarelli's Legacy,} 101--121 (2019),
  \href{https://arxiv.org/abs/1712.05165}{[arXiv:1712.05165]}\relax
\mciteBstWouldAddEndPuncttrue
\mciteSetBstMidEndSepPunct{\mcitedefaultmidpunct}
{\mcitedefaultendpunct}{\mcitedefaultseppunct}\relax
\EndOfBibitem
\bibitem{Pich:2018lmu}
A.~Pich {\em et~al.\/}, in \enquote{{13th Conference on Quark Confinement and
  the Hadron Spectrum (Confinement XIII) Maynooth, Ireland, July 31-August 6,
  2018},}  (2018),
  \href{https://arxiv.org/abs/1811.11801}{[arXiv:1811.11801]}\relax
\mciteBstWouldAddEndPuncttrue
\mciteSetBstMidEndSepPunct{\mcitedefaultmidpunct}
{\mcitedefaultendpunct}{\mcitedefaultseppunct}\relax
\EndOfBibitem
\bibitem{Pich:2020gzz}
A.~Pich, \href{http://doi.org/10.1016/j.ppnp.2020.103846}{Prog. Part. Nucl.
  Phys.} \href{http://doi.org/10.1016/j.ppnp.2020.103846}{{\bf 117}, 103846}
  (2021), \href{https://arxiv.org/abs/2012.04716}{[arXiv:2012.04716]}\relax
\mciteBstWouldAddEndPuncttrue
\mciteSetBstMidEndSepPunct{\mcitedefaultmidpunct}
{\mcitedefaultendpunct}{\mcitedefaultseppunct}\relax
\EndOfBibitem
\bibitem{FlavourLatticeAveragingGroupFLAG:2021npn}
Y.~Aoki {\em et~al.\/} (Flavour Lattice Averaging Group (FLAG)),
  \href{http://doi.org/10.1140/epjc/s10052-022-10536-1}{Eur. Phys. J. C}
  \href{http://doi.org/10.1140/epjc/s10052-022-10536-1}{{\bf 82}, 10, 869}
  (2022), \href{https://arxiv.org/abs/2111.09849}{[arXiv:2111.09849]}\relax
\mciteBstWouldAddEndPuncttrue
\mciteSetBstMidEndSepPunct{\mcitedefaultmidpunct}
{\mcitedefaultendpunct}{\mcitedefaultseppunct}\relax
\EndOfBibitem
\bibitem{Beneke:2008ad}
M.~Beneke and M.~Jamin,
  \href{http://doi.org/10.1088/1126-6708/2008/09/044}{JHEP}
  \href{http://doi.org/10.1088/1126-6708/2008/09/044}{{\bf 09}, 044} (2008),
  \href{https://arxiv.org/abs/0806.3156}{[arXiv:0806.3156]}\relax
\mciteBstWouldAddEndPuncttrue
\mciteSetBstMidEndSepPunct{\mcitedefaultmidpunct}
{\mcitedefaultendpunct}{\mcitedefaultseppunct}\relax
\EndOfBibitem
\bibitem{Maltman:2008nf}
K.~Maltman and T.~Yavin, \href{http://doi.org/10.1103/PhysRevD.78.094020}{Phys.
  Rev.} \href{http://doi.org/10.1103/PhysRevD.78.094020}{{\bf D78}, 094020}
  (2008), \href{https://arxiv.org/abs/0807.0650}{[arXiv:0807.0650]}\relax
\mciteBstWouldAddEndPuncttrue
\mciteSetBstMidEndSepPunct{\mcitedefaultmidpunct}
{\mcitedefaultendpunct}{\mcitedefaultseppunct}\relax
\EndOfBibitem
\bibitem{Narison:2009vy}
S.~Narison, \href{http://doi.org/10.1016/j.physletb.2009.01.062}{Phys. Lett.}
  \href{http://doi.org/10.1016/j.physletb.2009.01.062}{{\bf B673}, 30} (2009),
  \href{https://arxiv.org/abs/0901.3823}{[arXiv:0901.3823]}\relax
\mciteBstWouldAddEndPuncttrue
\mciteSetBstMidEndSepPunct{\mcitedefaultmidpunct}
{\mcitedefaultendpunct}{\mcitedefaultseppunct}\relax
\EndOfBibitem
\bibitem{Caprini:2009vf}
I.~Caprini and J.~Fischer,
  \href{http://doi.org/10.1140/epjc/s10052-009-1142-8}{Eur. Phys. J.}
  \href{http://doi.org/10.1140/epjc/s10052-009-1142-8}{{\bf C64}, 35} (2009),
  \href{https://arxiv.org/abs/0906.5211}{[arXiv:0906.5211]}\relax
\mciteBstWouldAddEndPuncttrue
\mciteSetBstMidEndSepPunct{\mcitedefaultmidpunct}
{\mcitedefaultendpunct}{\mcitedefaultseppunct}\relax
\EndOfBibitem
\bibitem{Pich:2013lsa}
A.~Pich, \href{http://doi.org/10.1016/j.ppnp.2013.11.002}{Prog. Part. Nucl.
  Phys.} \href{http://doi.org/10.1016/j.ppnp.2013.11.002}{{\bf 75}, 41} (2014),
  \href{https://arxiv.org/abs/1310.7922}{[arXiv:1310.7922]}\relax
\mciteBstWouldAddEndPuncttrue
\mciteSetBstMidEndSepPunct{\mcitedefaultmidpunct}
{\mcitedefaultendpunct}{\mcitedefaultseppunct}\relax
\EndOfBibitem
\bibitem{Boito:2014sta}
D.~Boito {\em et~al.\/}, \href{http://doi.org/10.1103/PhysRevD.91.034003}{Phys.
  Rev.} \href{http://doi.org/10.1103/PhysRevD.91.034003}{{\bf D91}, 3, 034003}
  (2015), \href{https://arxiv.org/abs/1410.3528}{[arXiv:1410.3528]}\relax
\mciteBstWouldAddEndPuncttrue
\mciteSetBstMidEndSepPunct{\mcitedefaultmidpunct}
{\mcitedefaultendpunct}{\mcitedefaultseppunct}\relax
\EndOfBibitem
\bibitem{Altarelli:2013bpa}
G.~Altarelli, \href{http://doi.org/10.22323/1.177.0002}{PoS}
  \href{http://doi.org/10.22323/1.177.0002}{{\bf Corfu2012}, 002} (2013),
  \href{https://arxiv.org/abs/1303.6065}{[arXiv:1303.6065]}\relax
\mciteBstWouldAddEndPuncttrue
\mciteSetBstMidEndSepPunct{\mcitedefaultmidpunct}
{\mcitedefaultendpunct}{\mcitedefaultseppunct}\relax
\EndOfBibitem
\bibitem{Pich:2016bdg}
A.~Pich and A.~Rodríguez-Sánchez,
  \href{http://doi.org/10.1103/PhysRevD.94.034027}{Phys. Rev.}
  \href{http://doi.org/10.1103/PhysRevD.94.034027}{{\bf D94}, 3, 034027}
  (2016), \href{https://arxiv.org/abs/1605.06830}{[arXiv:1605.06830]}\relax
\mciteBstWouldAddEndPuncttrue
\mciteSetBstMidEndSepPunct{\mcitedefaultmidpunct}
{\mcitedefaultendpunct}{\mcitedefaultseppunct}\relax
\EndOfBibitem
\bibitem{Davier:2013sfa}
M.~Davier {\em et~al.\/},
  \href{http://doi.org/10.1140/epjc/s10052-014-2803-9}{Eur. Phys. J.}
  \href{http://doi.org/10.1140/epjc/s10052-014-2803-9}{{\bf C74}, 3, 2803}
  (2014), \href{https://arxiv.org/abs/1312.1501}{[arXiv:1312.1501]}\relax
\mciteBstWouldAddEndPuncttrue
\mciteSetBstMidEndSepPunct{\mcitedefaultmidpunct}
{\mcitedefaultendpunct}{\mcitedefaultseppunct}\relax
\EndOfBibitem
\bibitem{Boito:2016oam}
D.~Boito {\em et~al.\/}, \href{http://doi.org/10.1103/PhysRevD.95.034024}{Phys.
  Rev.} \href{http://doi.org/10.1103/PhysRevD.95.034024}{{\bf D95}, 3, 034024}
  (2017), \href{https://arxiv.org/abs/1611.03457}{[arXiv:1611.03457]}\relax
\mciteBstWouldAddEndPuncttrue
\mciteSetBstMidEndSepPunct{\mcitedefaultmidpunct}
{\mcitedefaultendpunct}{\mcitedefaultseppunct}\relax
\EndOfBibitem
\bibitem{Boito:2019iwh}
D.~Boito {\em et~al.\/},
  \href{http://doi.org/10.1103/PhysRevD.100.074009}{Phys. Rev.}
  \href{http://doi.org/10.1103/PhysRevD.100.074009}{{\bf D100}, 7, 074009}
  (2019), \href{https://arxiv.org/abs/1907.03360}{[arXiv:1907.03360]}\relax
\mciteBstWouldAddEndPuncttrue
\mciteSetBstMidEndSepPunct{\mcitedefaultmidpunct}
{\mcitedefaultendpunct}{\mcitedefaultseppunct}\relax
\EndOfBibitem
\bibitem{Hoang:2020mkw}
A.~H. Hoang and C.~Regner,
  \href{http://doi.org/10.1103/PhysRevD.105.096023}{Phys. Rev. D}
  \href{http://doi.org/10.1103/PhysRevD.105.096023}{{\bf 105}, 9, 096023}
  (2022), \href{https://arxiv.org/abs/2008.00578}{[arXiv:2008.00578]}\relax
\mciteBstWouldAddEndPuncttrue
\mciteSetBstMidEndSepPunct{\mcitedefaultmidpunct}
{\mcitedefaultendpunct}{\mcitedefaultseppunct}\relax
\EndOfBibitem
\bibitem{Hoang:2021nlz}
A.~H. Hoang and C.~Regner,
  \href{http://doi.org/10.1140/epjs/s11734-021-00257-z}{Eur. Phys. J. ST}
  \href{http://doi.org/10.1140/epjs/s11734-021-00257-z}{{\bf 230}, 12-13, 2625}
  (2021), \href{https://arxiv.org/abs/2105.11222}{[arXiv:2105.11222]}\relax
\mciteBstWouldAddEndPuncttrue
\mciteSetBstMidEndSepPunct{\mcitedefaultmidpunct}
{\mcitedefaultendpunct}{\mcitedefaultseppunct}\relax
\EndOfBibitem
\bibitem{Golterman:2023oml}
M.~Golterman, K.~Maltman and S.~Peris,
  \href{http://doi.org/10.1103/PhysRevD.108.014007}{Phys. Rev. D}
  \href{http://doi.org/10.1103/PhysRevD.108.014007}{{\bf 108}, 1, 014007}
  (2023), \href{https://arxiv.org/abs/2305.10386}{[arXiv:2305.10386]}\relax
\mciteBstWouldAddEndPuncttrue
\mciteSetBstMidEndSepPunct{\mcitedefaultmidpunct}
{\mcitedefaultendpunct}{\mcitedefaultseppunct}\relax
\EndOfBibitem
\bibitem{Benitez-Rathgeb:2022hfj}
M.~A. Benitez-Rathgeb {\em et~al.\/},
  \href{http://doi.org/10.1007/JHEP09(2022)223}{JHEP}
  \href{http://doi.org/10.1007/JHEP09(2022)223}{{\bf 09}, 223} (2022),
  \href{https://arxiv.org/abs/2207.01116}{[arXiv:2207.01116]}\relax
\mciteBstWouldAddEndPuncttrue
\mciteSetBstMidEndSepPunct{\mcitedefaultmidpunct}
{\mcitedefaultendpunct}{\mcitedefaultseppunct}\relax
\EndOfBibitem
\bibitem{Benitez-Rathgeb:2022yqb}
M.~A. Benitez-Rathgeb {\em et~al.\/},
  \href{http://doi.org/10.1007/JHEP07(2022)016}{JHEP}
  \href{http://doi.org/10.1007/JHEP07(2022)016}{{\bf 07}, 016} (2022),
  \href{https://arxiv.org/abs/2202.10957}{[arXiv:2202.10957]}\relax
\mciteBstWouldAddEndPuncttrue
\mciteSetBstMidEndSepPunct{\mcitedefaultmidpunct}
{\mcitedefaultendpunct}{\mcitedefaultseppunct}\relax
\EndOfBibitem
\bibitem{Boito:2018yvl}
D.~Boito {\em et~al.\/}, \href{http://doi.org/10.1103/PhysRevD.98.074030}{Phys.
  Rev.} \href{http://doi.org/10.1103/PhysRevD.98.074030}{{\bf D98}, 7, 074030}
  (2018), \href{https://arxiv.org/abs/1805.08176}{[arXiv:1805.08176]}\relax
\mciteBstWouldAddEndPuncttrue
\mciteSetBstMidEndSepPunct{\mcitedefaultmidpunct}
{\mcitedefaultendpunct}{\mcitedefaultseppunct}\relax
\EndOfBibitem
\bibitem{Boito:2020xli}
D.~Boito {\em et~al.\/},
  \href{http://doi.org/10.1103/PhysRevD.103.034028}{Phys. Rev. D}
  \href{http://doi.org/10.1103/PhysRevD.103.034028}{{\bf 103}, 3, 034028}
  (2021), \href{https://arxiv.org/abs/2012.10440}{[arXiv:2012.10440]}\relax
\mciteBstWouldAddEndPuncttrue
\mciteSetBstMidEndSepPunct{\mcitedefaultmidpunct}
{\mcitedefaultendpunct}{\mcitedefaultseppunct}\relax
\EndOfBibitem
\bibitem{Ayala:2022cxo}
C.~Ayala, G.~Cvetic and D.~Teca,
  \href{http://doi.org/10.1088/1361-6471/acbd65}{J. Phys. G}
  \href{http://doi.org/10.1088/1361-6471/acbd65}{{\bf 50}, 4, 045004} (2023),
  \href{https://arxiv.org/abs/2206.05631}{[arXiv:2206.05631]}\relax
\mciteBstWouldAddEndPuncttrue
\mciteSetBstMidEndSepPunct{\mcitedefaultmidpunct}
{\mcitedefaultendpunct}{\mcitedefaultseppunct}\relax
\EndOfBibitem
\bibitem{Shen:2023qgz}
J.-M. Shen {\em et~al.\/}, \href{http://doi.org/10.1007/JHEP07(2023)109}{JHEP}
  \href{http://doi.org/10.1007/JHEP07(2023)109}{{\bf 07}, 109} (2023),
  \href{https://arxiv.org/abs/2303.11782}{[arXiv:2303.11782]}\relax
\mciteBstWouldAddEndPuncttrue
\mciteSetBstMidEndSepPunct{\mcitedefaultmidpunct}
{\mcitedefaultendpunct}{\mcitedefaultseppunct}\relax
\EndOfBibitem
\bibitem{Mateu:2017hlz}
V.~Mateu and P.~G. Ortega, \href{http://doi.org/10.1007/JHEP01(2018)122}{JHEP}
  \href{http://doi.org/10.1007/JHEP01(2018)122}{{\bf 01}, 122} (2018),
  \href{https://arxiv.org/abs/1711.05755}{[arXiv:1711.05755]}\relax
\mciteBstWouldAddEndPuncttrue
\mciteSetBstMidEndSepPunct{\mcitedefaultmidpunct}
{\mcitedefaultendpunct}{\mcitedefaultseppunct}\relax
\EndOfBibitem
\bibitem{Peset:2018ria}
C.~Peset, A.~Pineda and J.~Segovia,
  \href{http://doi.org/10.1007/JHEP09(2018)167}{JHEP}
  \href{http://doi.org/10.1007/JHEP09(2018)167}{{\bf 09}, 167} (2018),
  \href{https://arxiv.org/abs/1806.05197}{[arXiv:1806.05197]}\relax
\mciteBstWouldAddEndPuncttrue
\mciteSetBstMidEndSepPunct{\mcitedefaultmidpunct}
{\mcitedefaultendpunct}{\mcitedefaultseppunct}\relax
\EndOfBibitem
\bibitem{Narison:2018dcr}
S.~Narison, \href{http://doi.org/10.1142/S0217751X18500458}{Int. J. Mod. Phys.
  A} \href{http://doi.org/10.1142/S0217751X18500458}{{\bf 33}, 10, 1850045}
  (2018), [Addendum: Int. J. Mod. Phys. A 33, 1850045 (2018)],
  \href{https://arxiv.org/abs/1801.00592}{[arXiv:1801.00592]}\relax
\mciteBstWouldAddEndPuncttrue
\mciteSetBstMidEndSepPunct{\mcitedefaultmidpunct}
{\mcitedefaultendpunct}{\mcitedefaultseppunct}\relax
\EndOfBibitem
\bibitem{Narison:2018xbj}
S.~Narison, \href{http://doi.org/10.1142/S0217751X18920045}{Int. J. Mod. Phys.
  A} \href{http://doi.org/10.1142/S0217751X18920045}{{\bf 33}, 33, 1892004}
  (2018), \href{https://arxiv.org/abs/1812.09360}{[arXiv:1812.09360]}\relax
\mciteBstWouldAddEndPuncttrue
\mciteSetBstMidEndSepPunct{\mcitedefaultmidpunct}
{\mcitedefaultendpunct}{\mcitedefaultseppunct}\relax
\EndOfBibitem
\bibitem{Boito:2019pqp}
D.~Boito and V.~Mateu,
  \href{http://doi.org/10.1016/j.physletb.2020.135482}{Phys. Lett. B}
  \href{http://doi.org/10.1016/j.physletb.2020.135482}{{\bf 806}, 135482}
  (2020), \href{https://arxiv.org/abs/1912.06237}{[arXiv:1912.06237]}\relax
\mciteBstWouldAddEndPuncttrue
\mciteSetBstMidEndSepPunct{\mcitedefaultmidpunct}
{\mcitedefaultendpunct}{\mcitedefaultseppunct}\relax
\EndOfBibitem
\bibitem{Boito:2020lyp}
D.~Boito and V.~Mateu, \href{http://doi.org/10.1007/JHEP03(2020)094}{JHEP}
  \href{http://doi.org/10.1007/JHEP03(2020)094}{{\bf 03}, 094} (2020),
  \href{https://arxiv.org/abs/2001.11041}{[arXiv:2001.11041]}\relax
\mciteBstWouldAddEndPuncttrue
\mciteSetBstMidEndSepPunct{\mcitedefaultmidpunct}
{\mcitedefaultendpunct}{\mcitedefaultseppunct}\relax
\EndOfBibitem
\bibitem{Blumlein:2006be}
J.~Blumlein, H.~Bottcher and A.~Guffanti,
  \href{http://doi.org/10.1016/j.nuclphysb.2007.03.035}{Nucl. Phys.}
  \href{http://doi.org/10.1016/j.nuclphysb.2007.03.035}{{\bf B774}, 182}
  (2007), \href{https://arxiv.org/abs/hep-ph/0607200}{[hep-ph/0607200]}\relax
\mciteBstWouldAddEndPuncttrue
\mciteSetBstMidEndSepPunct{\mcitedefaultmidpunct}
{\mcitedefaultendpunct}{\mcitedefaultseppunct}\relax
\EndOfBibitem
\bibitem{Jimenez-Delgado:2014twa}
P.~Jimenez-Delgado and E.~Reya,
  \href{http://doi.org/10.1103/PhysRevD.89.074049}{Phys. Rev.}
  \href{http://doi.org/10.1103/PhysRevD.89.074049}{{\bf D89}, 7, 074049}
  (2014), \href{https://arxiv.org/abs/1403.1852}{[arXiv:1403.1852]}\relax
\mciteBstWouldAddEndPuncttrue
\mciteSetBstMidEndSepPunct{\mcitedefaultmidpunct}
{\mcitedefaultendpunct}{\mcitedefaultseppunct}\relax
\EndOfBibitem
\bibitem{Alekhin:2017kpj}
S.~Alekhin {\em et~al.\/},
  \href{http://doi.org/10.1103/PhysRevD.96.014011}{Phys. Rev.}
  \href{http://doi.org/10.1103/PhysRevD.96.014011}{{\bf D96}, 1, 014011}
  (2017), \href{https://arxiv.org/abs/1701.05838}{[arXiv:1701.05838]}\relax
\mciteBstWouldAddEndPuncttrue
\mciteSetBstMidEndSepPunct{\mcitedefaultmidpunct}
{\mcitedefaultendpunct}{\mcitedefaultseppunct}\relax
\EndOfBibitem
\bibitem{Alekhin:2018pai}
S.~Alekhin, J.~Blümlein and S.~Moch,
  \href{http://doi.org/10.1140/epjc/s10052-018-5947-1}{Eur. Phys. J.}
  \href{http://doi.org/10.1140/epjc/s10052-018-5947-1}{{\bf C78}, 6, 477}
  (2018), \href{https://arxiv.org/abs/1803.07537}{[arXiv:1803.07537]}\relax
\mciteBstWouldAddEndPuncttrue
\mciteSetBstMidEndSepPunct{\mcitedefaultmidpunct}
{\mcitedefaultendpunct}{\mcitedefaultseppunct}\relax
\EndOfBibitem
\bibitem{Cridge:2021qfd}
T.~Cridge {\em et~al.\/}  (2021),
  \href{https://arxiv.org/abs/2106.10289}{[arXiv:2106.10289]}\relax
\mciteBstWouldAddEndPuncttrue
\mciteSetBstMidEndSepPunct{\mcitedefaultmidpunct}
{\mcitedefaultendpunct}{\mcitedefaultseppunct}\relax
\EndOfBibitem
\bibitem{Ball:2018iqk}
R.~D. Ball {\em et~al.\/} (NNPDF),
  \href{http://doi.org/10.1140/epjc/s10052-018-5897-7}{Eur. Phys. J.}
  \href{http://doi.org/10.1140/epjc/s10052-018-5897-7}{{\bf C78}, 5, 408}
  (2018), \href{https://arxiv.org/abs/1802.03398}{[arXiv:1802.03398]}\relax
\mciteBstWouldAddEndPuncttrue
\mciteSetBstMidEndSepPunct{\mcitedefaultmidpunct}
{\mcitedefaultendpunct}{\mcitedefaultseppunct}\relax
\EndOfBibitem
\bibitem{H1:2021xxi}
I.~Abt {\em et~al.\/} (H1, ZEUS),
  \href{http://doi.org/10.1140/epjc/s10052-022-10083-9}{Eur. Phys. J. C}
  \href{http://doi.org/10.1140/epjc/s10052-022-10083-9}{{\bf 82}, 3, 243}
  (2022), \href{https://arxiv.org/abs/2112.01120}{[arXiv:2112.01120]}\relax
\mciteBstWouldAddEndPuncttrue
\mciteSetBstMidEndSepPunct{\mcitedefaultmidpunct}
{\mcitedefaultendpunct}{\mcitedefaultseppunct}\relax
\EndOfBibitem
\bibitem{Thorne:2011kq}
R.~S. Thorne and G.~Watt, \href{http://doi.org/10.1007/JHEP08(2011)100}{JHEP}
  \href{http://doi.org/10.1007/JHEP08(2011)100}{{\bf 08}, 100} (2011),
  \href{https://arxiv.org/abs/1106.5789}{[arXiv:1106.5789]}\relax
\mciteBstWouldAddEndPuncttrue
\mciteSetBstMidEndSepPunct{\mcitedefaultmidpunct}
{\mcitedefaultendpunct}{\mcitedefaultseppunct}\relax
\EndOfBibitem
\bibitem{Alekhin:2011ey}
S.~Alekhin, J.~Blumlein and S.~Moch,
  \href{http://doi.org/10.1140/epjc/s10052-011-1723-1}{Eur. Phys. J.}
  \href{http://doi.org/10.1140/epjc/s10052-011-1723-1}{{\bf C71}, 1723} (2011),
  \href{https://arxiv.org/abs/1101.5261}{[arXiv:1101.5261]}\relax
\mciteBstWouldAddEndPuncttrue
\mciteSetBstMidEndSepPunct{\mcitedefaultmidpunct}
{\mcitedefaultendpunct}{\mcitedefaultseppunct}\relax
\EndOfBibitem
\bibitem{NNPDF:2011aa}
R.~D. Ball {\em et~al.\/} (NNPDF),
  \href{http://doi.org/10.1016/j.physletb.2011.08.055}{Phys. Lett.}
  \href{http://doi.org/10.1016/j.physletb.2011.08.055}{{\bf B704}, 36} (2011),
  \href{https://arxiv.org/abs/1102.3182}{[arXiv:1102.3182]}\relax
\mciteBstWouldAddEndPuncttrue
\mciteSetBstMidEndSepPunct{\mcitedefaultmidpunct}
{\mcitedefaultendpunct}{\mcitedefaultseppunct}\relax
\EndOfBibitem
\bibitem{Ball:2013gsa}
R.~D. Ball {\em et~al.\/} (NNPDF),
  \href{http://doi.org/10.1016/j.physletb.2013.05.019}{Phys. Lett.}
  \href{http://doi.org/10.1016/j.physletb.2013.05.019}{{\bf B723}, 330} (2013),
  \href{https://arxiv.org/abs/1303.1189}{[arXiv:1303.1189]}\relax
\mciteBstWouldAddEndPuncttrue
\mciteSetBstMidEndSepPunct{\mcitedefaultmidpunct}
{\mcitedefaultendpunct}{\mcitedefaultseppunct}\relax
\EndOfBibitem
\bibitem{Thorne:2013hpa}
R.~S. Thorne, \href{http://doi.org/10.22323/1.191.0042}{PoS}
  \href{http://doi.org/10.22323/1.191.0042}{{\bf DIS2013}, 042} (2013),
  \href{https://arxiv.org/abs/1306.3907}{[arXiv:1306.3907]}\relax
\mciteBstWouldAddEndPuncttrue
\mciteSetBstMidEndSepPunct{\mcitedefaultmidpunct}
{\mcitedefaultendpunct}{\mcitedefaultseppunct}\relax
\EndOfBibitem
\bibitem{Dissertori:2009ik}
G.~Dissertori {\em et~al.\/},
  \href{http://doi.org/10.1088/1126-6708/2009/08/036}{JHEP}
  \href{http://doi.org/10.1088/1126-6708/2009/08/036}{{\bf 08}, 036} (2009),
  \href{https://arxiv.org/abs/0906.3436}{[arXiv:0906.3436]}\relax
\mciteBstWouldAddEndPuncttrue
\mciteSetBstMidEndSepPunct{\mcitedefaultmidpunct}
{\mcitedefaultendpunct}{\mcitedefaultseppunct}\relax
\EndOfBibitem
\bibitem{OPAL:2011aa}
G.~Abbiendi {\em et~al.\/} (OPAL),
  \href{http://doi.org/10.1140/epjc/s10052-011-1733-z}{Eur. Phys. J.}
  \href{http://doi.org/10.1140/epjc/s10052-011-1733-z}{{\bf C71}, 1733} (2011),
  \href{https://arxiv.org/abs/1101.1470}{[arXiv:1101.1470]}\relax
\mciteBstWouldAddEndPuncttrue
\mciteSetBstMidEndSepPunct{\mcitedefaultmidpunct}
{\mcitedefaultendpunct}{\mcitedefaultseppunct}\relax
\EndOfBibitem
\bibitem{Bethke:2008hf}
S.~Bethke {\em et~al.\/} (JADE),
  \href{http://doi.org/10.1140/epjc/s10052-009-1149-1}{Eur. Phys. J.}
  \href{http://doi.org/10.1140/epjc/s10052-009-1149-1}{{\bf C64}, 351} (2009),
  \href{https://arxiv.org/abs/0810.1389}{[arXiv:0810.1389]}\relax
\mciteBstWouldAddEndPuncttrue
\mciteSetBstMidEndSepPunct{\mcitedefaultmidpunct}
{\mcitedefaultendpunct}{\mcitedefaultseppunct}\relax
\EndOfBibitem
\bibitem{Dissertori:2009qa}
G.~Dissertori {\em et~al.\/},
  \href{http://doi.org/10.1103/PhysRevLett.104.072002}{Phys. Rev. Lett.}
  \href{http://doi.org/10.1103/PhysRevLett.104.072002}{{\bf 104}, 072002}
  (2010), \href{https://arxiv.org/abs/0910.4283}{[arXiv:0910.4283]}\relax
\mciteBstWouldAddEndPuncttrue
\mciteSetBstMidEndSepPunct{\mcitedefaultmidpunct}
{\mcitedefaultendpunct}{\mcitedefaultseppunct}\relax
\EndOfBibitem
\bibitem{Schieck:2012mp}
J.~Schieck {\em et~al.\/} (JADE),
  \href{http://doi.org/10.1140/epjc/s10052-013-2332-y}{Eur. Phys. J.}
  \href{http://doi.org/10.1140/epjc/s10052-013-2332-y}{{\bf C73}, 3, 2332}
  (2013), \href{https://arxiv.org/abs/1205.3714}{[arXiv:1205.3714]}\relax
\mciteBstWouldAddEndPuncttrue
\mciteSetBstMidEndSepPunct{\mcitedefaultmidpunct}
{\mcitedefaultendpunct}{\mcitedefaultseppunct}\relax
\EndOfBibitem
\bibitem{Verbytskyi:2019zhh}
A.~Verbytskyi {\em et~al.\/},
  \href{http://doi.org/10.1007/JHEP08(2019)129}{JHEP}
  \href{http://doi.org/10.1007/JHEP08(2019)129}{{\bf 08}, 129} (2019),
  \href{https://arxiv.org/abs/1902.08158}{[arXiv:1902.08158]}\relax
\mciteBstWouldAddEndPuncttrue
\mciteSetBstMidEndSepPunct{\mcitedefaultmidpunct}
{\mcitedefaultendpunct}{\mcitedefaultseppunct}\relax
\EndOfBibitem
\bibitem{Kardos:2018kqj}
A.~Kardos {\em et~al.\/},
  \href{http://doi.org/10.1140/epjc/s10052-018-5963-1}{Eur. Phys. J.}
  \href{http://doi.org/10.1140/epjc/s10052-018-5963-1}{{\bf C78}, 6, 498}
  (2018), \href{https://arxiv.org/abs/1804.09146}{[arXiv:1804.09146]}\relax
\mciteBstWouldAddEndPuncttrue
\mciteSetBstMidEndSepPunct{\mcitedefaultmidpunct}
{\mcitedefaultendpunct}{\mcitedefaultseppunct}\relax
\EndOfBibitem
\bibitem{Davison:2008vx}
R.~A. Davison and B.~R. Webber,
  \href{http://doi.org/10.1140/epjc/s10052-008-0836-7}{Eur. Phys. J.}
  \href{http://doi.org/10.1140/epjc/s10052-008-0836-7}{{\bf C59}, 13} (2009),
  \href{https://arxiv.org/abs/0809.3326}{[arXiv:0809.3326]}\relax
\mciteBstWouldAddEndPuncttrue
\mciteSetBstMidEndSepPunct{\mcitedefaultmidpunct}
{\mcitedefaultendpunct}{\mcitedefaultseppunct}\relax
\EndOfBibitem
\bibitem{Abbate:2010xh}
R.~Abbate {\em et~al.\/},
  \href{http://doi.org/10.1103/PhysRevD.83.074021}{Phys. Rev.}
  \href{http://doi.org/10.1103/PhysRevD.83.074021}{{\bf D83}, 074021} (2011),
  \href{https://arxiv.org/abs/1006.3080}{[arXiv:1006.3080]}\relax
\mciteBstWouldAddEndPuncttrue
\mciteSetBstMidEndSepPunct{\mcitedefaultmidpunct}
{\mcitedefaultendpunct}{\mcitedefaultseppunct}\relax
\EndOfBibitem
\bibitem{Gehrmann:2012sc}
T.~Gehrmann, G.~Luisoni and P.~F. Monni,
  \href{http://doi.org/10.1140/epjc/s10052-012-2265-x}{Eur. Phys. J.}
  \href{http://doi.org/10.1140/epjc/s10052-012-2265-x}{{\bf C73}, 1, 2265}
  (2013), \href{https://arxiv.org/abs/1210.6945}{[arXiv:1210.6945]}\relax
\mciteBstWouldAddEndPuncttrue
\mciteSetBstMidEndSepPunct{\mcitedefaultmidpunct}
{\mcitedefaultendpunct}{\mcitedefaultseppunct}\relax
\EndOfBibitem
\bibitem{Hoang:2015hka}
A.~H. Hoang {\em et~al.\/},
  \href{http://doi.org/10.1103/PhysRevD.91.094018}{Phys. Rev.}
  \href{http://doi.org/10.1103/PhysRevD.91.094018}{{\bf D91}, 9, 094018}
  (2015), \href{https://arxiv.org/abs/1501.04111}{[arXiv:1501.04111]}\relax
\mciteBstWouldAddEndPuncttrue
\mciteSetBstMidEndSepPunct{\mcitedefaultmidpunct}
{\mcitedefaultendpunct}{\mcitedefaultseppunct}\relax
\EndOfBibitem
\bibitem{Luisoni:2020efy}
G.~Luisoni, P.~F. Monni and G.~P. Salam,
  \href{http://doi.org/10.1140/epjc/s10052-021-08941-z}{Eur. Phys. J. C}
  \href{http://doi.org/10.1140/epjc/s10052-021-08941-z}{{\bf 81}, 2, 158}
  (2021), \href{https://arxiv.org/abs/2012.00622}{[arXiv:2012.00622]}\relax
\mciteBstWouldAddEndPuncttrue
\mciteSetBstMidEndSepPunct{\mcitedefaultmidpunct}
{\mcitedefaultendpunct}{\mcitedefaultseppunct}\relax
\EndOfBibitem
\bibitem{Caola:2021kzt}
F.~Caola {\em et~al.\/}, \href{http://doi.org/10.1007/JHEP01(2022)093}{JHEP}
  \href{http://doi.org/10.1007/JHEP01(2022)093}{{\bf 01}, 093} (2022),
  \href{https://arxiv.org/abs/2108.08897}{[arXiv:2108.08897]}\relax
\mciteBstWouldAddEndPuncttrue
\mciteSetBstMidEndSepPunct{\mcitedefaultmidpunct}
{\mcitedefaultendpunct}{\mcitedefaultseppunct}\relax
\EndOfBibitem
\bibitem{Caola:2022vea}
F.~Caola {\em et~al.\/}, \href{http://doi.org/10.1007/JHEP12(2022)062}{JHEP}
  \href{http://doi.org/10.1007/JHEP12(2022)062}{{\bf 12}, 062} (2022),
  \href{https://arxiv.org/abs/2204.02247}{[arXiv:2204.02247]}\relax
\mciteBstWouldAddEndPuncttrue
\mciteSetBstMidEndSepPunct{\mcitedefaultmidpunct}
{\mcitedefaultendpunct}{\mcitedefaultseppunct}\relax
\EndOfBibitem
\bibitem{Nason:2023asn}
P.~Nason and G.~Zanderighi, \href{http://doi.org/10.1007/JHEP06(2023)058}{JHEP}
  \href{http://doi.org/10.1007/JHEP06(2023)058}{{\bf 06}, 058} (2023),
  \href{https://arxiv.org/abs/2301.03607}{[arXiv:2301.03607]}\relax
\mciteBstWouldAddEndPuncttrue
\mciteSetBstMidEndSepPunct{\mcitedefaultmidpunct}
{\mcitedefaultendpunct}{\mcitedefaultseppunct}\relax
\EndOfBibitem
\bibitem{Frederix:2010ne}
R.~Frederix {\em et~al.\/}, \href{http://doi.org/10.1007/JHEP11(2010)050}{JHEP}
  \href{http://doi.org/10.1007/JHEP11(2010)050}{{\bf 11}, 050} (2010),
  \href{https://arxiv.org/abs/1008.5313}{[arXiv:1008.5313]}\relax
\mciteBstWouldAddEndPuncttrue
\mciteSetBstMidEndSepPunct{\mcitedefaultmidpunct}
{\mcitedefaultendpunct}{\mcitedefaultseppunct}\relax
\EndOfBibitem
\bibitem{Bolzoni:2013rsa}
P.~Bolzoni, B.~A. Kniehl and A.~V. Kotikov,
  \href{http://doi.org/10.1016/j.nuclphysb.2013.06.025}{Nucl. Phys.}
  \href{http://doi.org/10.1016/j.nuclphysb.2013.06.025}{{\bf B875}, 18} (2013),
  \href{https://arxiv.org/abs/1305.6017}{[arXiv:1305.6017]}\relax
\mciteBstWouldAddEndPuncttrue
\mciteSetBstMidEndSepPunct{\mcitedefaultmidpunct}
{\mcitedefaultendpunct}{\mcitedefaultseppunct}\relax
\EndOfBibitem
\bibitem{Currie:2017eqf}
J.~Currie {\em et~al.\/},
  \href{http://doi.org/10.1103/PhysRevLett.119.152001}{Phys. Rev. Lett.}
  \href{http://doi.org/10.1103/PhysRevLett.119.152001}{{\bf 119}, 15, 152001}
  (2017), \href{https://arxiv.org/abs/1705.10271}{[arXiv:1705.10271]}\relax
\mciteBstWouldAddEndPuncttrue
\mciteSetBstMidEndSepPunct{\mcitedefaultmidpunct}
{\mcitedefaultendpunct}{\mcitedefaultseppunct}\relax
\EndOfBibitem
\bibitem{Czakon:2019tmo}
M.~Czakon {\em et~al.\/}  (2019),
  \href{https://arxiv.org/abs/1907.12911}{[arXiv:1907.12911]}\relax
\mciteBstWouldAddEndPuncttrue
\mciteSetBstMidEndSepPunct{\mcitedefaultmidpunct}
{\mcitedefaultendpunct}{\mcitedefaultseppunct}\relax
\EndOfBibitem
\bibitem{Dittmaier:2012kx}
S.~Dittmaier, A.~Huss and C.~Speckner,
  \href{http://doi.org/10.1007/JHEP11(2012)095}{JHEP}
  \href{http://doi.org/10.1007/JHEP11(2012)095}{{\bf 11}, 095} (2012),
  \href{https://arxiv.org/abs/1210.0438}{[arXiv:1210.0438]}\relax
\mciteBstWouldAddEndPuncttrue
\mciteSetBstMidEndSepPunct{\mcitedefaultmidpunct}
{\mcitedefaultendpunct}{\mcitedefaultseppunct}\relax
\EndOfBibitem
\bibitem{Frederix:2016ost}
R.~Frederix {\em et~al.\/}, \href{http://doi.org/10.1007/JHEP04(2017)076}{JHEP}
  \href{http://doi.org/10.1007/JHEP04(2017)076}{{\bf 04}, 076} (2017),
  \href{https://arxiv.org/abs/1612.06548}{[arXiv:1612.06548]}\relax
\mciteBstWouldAddEndPuncttrue
\mciteSetBstMidEndSepPunct{\mcitedefaultmidpunct}
{\mcitedefaultendpunct}{\mcitedefaultseppunct}\relax
\EndOfBibitem
\bibitem{Czakon:2017wor}
M.~Czakon {\em et~al.\/}, \href{http://doi.org/10.1007/JHEP10(2017)186}{JHEP}
  \href{http://doi.org/10.1007/JHEP10(2017)186}{{\bf 10}, 186} (2017),
  \href{https://arxiv.org/abs/1705.04105}{[arXiv:1705.04105]}\relax
\mciteBstWouldAddEndPuncttrue
\mciteSetBstMidEndSepPunct{\mcitedefaultmidpunct}
{\mcitedefaultendpunct}{\mcitedefaultseppunct}\relax
\EndOfBibitem
\bibitem{Reyer:2019obz}
M.~Reyer, M.~Sch\"onherr and S.~Schumann,
  \href{http://doi.org/10.1140/epjc/s10052-019-6815-3}{Eur. Phys. J. C}
  \href{http://doi.org/10.1140/epjc/s10052-019-6815-3}{{\bf 79}, 4, 321}
  (2019), \href{https://arxiv.org/abs/1902.01763}{[arXiv:1902.01763]}\relax
\mciteBstWouldAddEndPuncttrue
\mciteSetBstMidEndSepPunct{\mcitedefaultmidpunct}
{\mcitedefaultendpunct}{\mcitedefaultseppunct}\relax
\EndOfBibitem
\bibitem{Johnson:2017ttl}
M.~Johnson and D.~Maître,
  \href{http://doi.org/10.1103/PhysRevD.97.054013}{Phys. Rev.}
  \href{http://doi.org/10.1103/PhysRevD.97.054013}{{\bf D97}, 5, 054013}
  (2018), \href{https://arxiv.org/abs/1711.01408}{[arXiv:1711.01408]}\relax
\mciteBstWouldAddEndPuncttrue
\mciteSetBstMidEndSepPunct{\mcitedefaultmidpunct}
{\mcitedefaultendpunct}{\mcitedefaultseppunct}\relax
\EndOfBibitem
\bibitem{Ridder:2016nkl}
A.~Gehrmann-De~Ridder {\em et~al.\/},
  \href{http://doi.org/10.1007/JHEP07(2016)133}{JHEP}
  \href{http://doi.org/10.1007/JHEP07(2016)133}{{\bf 07}, 133} (2016),
  \href{https://arxiv.org/abs/1605.04295}{[arXiv:1605.04295]}\relax
\mciteBstWouldAddEndPuncttrue
\mciteSetBstMidEndSepPunct{\mcitedefaultmidpunct}
{\mcitedefaultendpunct}{\mcitedefaultseppunct}\relax
\EndOfBibitem
\bibitem{Forte:2020pyp}
S.~Forte and Z.~Kassabov,
  \href{http://doi.org/10.1140/epjc/s10052-020-7748-6}{Eur. Phys. J. C}
  \href{http://doi.org/10.1140/epjc/s10052-020-7748-6}{{\bf 80}, 3, 182}
  (2020), \href{https://arxiv.org/abs/2001.04986}{[arXiv:2001.04986]}\relax
\mciteBstWouldAddEndPuncttrue
\mciteSetBstMidEndSepPunct{\mcitedefaultmidpunct}
{\mcitedefaultendpunct}{\mcitedefaultseppunct}\relax
\EndOfBibitem
\bibitem{Alekhin:2014irh}
S.~Alekhin {\em et~al.\/},
  \href{http://doi.org/10.1140/epjc/s10052-015-3480-z}{Eur. Phys. J. C}
  \href{http://doi.org/10.1140/epjc/s10052-015-3480-z}{{\bf 75}, 304} (2015),
  \href{https://arxiv.org/abs/1410.4412}{[arXiv:1410.4412]}\relax
\mciteBstWouldAddEndPuncttrue
\mciteSetBstMidEndSepPunct{\mcitedefaultmidpunct}
{\mcitedefaultendpunct}{\mcitedefaultseppunct}\relax
\EndOfBibitem
\bibitem{Lorkowski:2023lnc}
F.~Lorkowski (ZEUS), \href{http://doi.org/10.5506/APhysPolBSupp.16.5-A35}{Acta
  Phys. Polon. Supp.} \href{http://doi.org/10.5506/APhysPolBSupp.16.5-A35}{{\bf
  16}, 5, 35} (2023)\relax
\mciteBstWouldAddEndPuncttrue
\mciteSetBstMidEndSepPunct{\mcitedefaultmidpunct}
{\mcitedefaultendpunct}{\mcitedefaultseppunct}\relax
\EndOfBibitem
\bibitem{CMS-PAS-SMP-22-015}
{CMS Collaboration} (CMS), CMS Physics Analysis Summary CMS-PAS-SMP-22-015
  (2023), \urlprefix\url{http://cds.cern.ch/record/2866560}\relax
\mciteBstWouldAddEndPuncttrue
\mciteSetBstMidEndSepPunct{\mcitedefaultmidpunct}
{\mcitedefaultendpunct}{\mcitedefaultseppunct}\relax
\EndOfBibitem
\bibitem{CMS-PAS-SMP-22-005}
{CMS Collaboration} (CMS), CMS Physics Analysis Summary CMS-PAS-SMP-22-005
  (2023), \urlprefix\url{http://cds.cern.ch/record/2868568}\relax
\mciteBstWouldAddEndPuncttrue
\mciteSetBstMidEndSepPunct{\mcitedefaultmidpunct}
{\mcitedefaultendpunct}{\mcitedefaultseppunct}\relax
\EndOfBibitem
\bibitem{dEnterria:2019aat}
D.~d'Enterria and A.~Poldaru,
  \href{http://doi.org/10.1007/JHEP06(2020)016}{JHEP}
  \href{http://doi.org/10.1007/JHEP06(2020)016}{{\bf 06}, 016} (2020),
  \href{https://arxiv.org/abs/1912.11733}{[arXiv:1912.11733]}\relax
\mciteBstWouldAddEndPuncttrue
\mciteSetBstMidEndSepPunct{\mcitedefaultmidpunct}
{\mcitedefaultendpunct}{\mcitedefaultseppunct}\relax
\EndOfBibitem
\bibitem{Camarda:2022qdg}
S.~Camarda, G.~Ferrera and M.~Schott  (2022),
  \href{https://arxiv.org/abs/2203.05394}{[arXiv:2203.05394]}\relax
\mciteBstWouldAddEndPuncttrue
\mciteSetBstMidEndSepPunct{\mcitedefaultmidpunct}
{\mcitedefaultendpunct}{\mcitedefaultseppunct}\relax
\EndOfBibitem
\bibitem{ATLAS:2023lhg}
G.~Aad {\em et~al.\/} (ATLAS)  (2023),
  \href{https://arxiv.org/abs/2309.12986}{[arXiv:2309.12986]}\relax
\mciteBstWouldAddEndPuncttrue
\mciteSetBstMidEndSepPunct{\mcitedefaultmidpunct}
{\mcitedefaultendpunct}{\mcitedefaultseppunct}\relax
\EndOfBibitem
\bibitem{Affolder:2001hn}
T.~Affolder {\em et~al.\/} (CDF),
  \href{http://doi.org/10.1103/PhysRevLett.88.042001}{Phys. Rev. Lett.}
  \href{http://doi.org/10.1103/PhysRevLett.88.042001}{{\bf 88}, 042001} (2002),
  \href{https://arxiv.org/abs/hep-ex/0108034}{[hep-ex/0108034]}\relax
\mciteBstWouldAddEndPuncttrue
\mciteSetBstMidEndSepPunct{\mcitedefaultmidpunct}
{\mcitedefaultendpunct}{\mcitedefaultseppunct}\relax
\EndOfBibitem
\bibitem{Chekanov:2006yc}
S.~Chekanov {\em et~al.\/} (ZEUS),
  \href{http://doi.org/10.1016/j.physletb.2007.03.039}{Phys. Lett. B}
  \href{http://doi.org/10.1016/j.physletb.2007.03.039}{{\bf 649}, 12} (2007),
  \href{https://arxiv.org/abs/hep-ex/0701039}{[hep-ex/0701039]}\relax
\mciteBstWouldAddEndPuncttrue
\mciteSetBstMidEndSepPunct{\mcitedefaultmidpunct}
{\mcitedefaultendpunct}{\mcitedefaultseppunct}\relax
\EndOfBibitem
\bibitem{Abazov:2009nc}
V.~M. Abazov {\em et~al.\/} (D0),
  \href{http://doi.org/10.1103/PhysRevD.80.111107}{Phys. Rev.}
  \href{http://doi.org/10.1103/PhysRevD.80.111107}{{\bf D80}, 111107} (2009),
  \href{https://arxiv.org/abs/0911.2710}{[arXiv:0911.2710]}\relax
\mciteBstWouldAddEndPuncttrue
\mciteSetBstMidEndSepPunct{\mcitedefaultmidpunct}
{\mcitedefaultendpunct}{\mcitedefaultseppunct}\relax
\EndOfBibitem
\bibitem{Malaescu:2012ts}
B.~Malaescu and P.~Starovoitov,
  \href{http://doi.org/10.1140/epjc/s10052-012-2041-y}{Eur. Phys. J.}
  \href{http://doi.org/10.1140/epjc/s10052-012-2041-y}{{\bf C72}, 2041} (2012),
  \href{https://arxiv.org/abs/1203.5416}{[arXiv:1203.5416]}\relax
\mciteBstWouldAddEndPuncttrue
\mciteSetBstMidEndSepPunct{\mcitedefaultmidpunct}
{\mcitedefaultendpunct}{\mcitedefaultseppunct}\relax
\EndOfBibitem
\bibitem{Khachatryan:2014waa}
V.~Khachatryan {\em et~al.\/} (CMS),
  \href{http://doi.org/10.1140/epjc/s10052-015-3499-1}{Eur. Phys. J.}
  \href{http://doi.org/10.1140/epjc/s10052-015-3499-1}{{\bf C75}, 6, 288}
  (2015), \href{https://arxiv.org/abs/1410.6765}{[arXiv:1410.6765]}\relax
\mciteBstWouldAddEndPuncttrue
\mciteSetBstMidEndSepPunct{\mcitedefaultmidpunct}
{\mcitedefaultendpunct}{\mcitedefaultseppunct}\relax
\EndOfBibitem
\bibitem{Britzger:2017maj}
D.~Britzger {\em et~al.\/},
  \href{http://doi.org/10.1140/epjc/s10052-019-6551-8}{Eur. Phys. J.}
  \href{http://doi.org/10.1140/epjc/s10052-019-6551-8}{{\bf C79}, 1, 68}
  (2019), \href{https://arxiv.org/abs/1712.00480}{[arXiv:1712.00480]}\relax
\mciteBstWouldAddEndPuncttrue
\mciteSetBstMidEndSepPunct{\mcitedefaultmidpunct}
{\mcitedefaultendpunct}{\mcitedefaultseppunct}\relax
\EndOfBibitem
\bibitem{Chekanov:2005ve}
S.~Chekanov {\em et~al.\/} (ZEUS),
  \href{http://doi.org/10.1140/epjc/s2005-02347-1}{Eur. Phys. J.}
  \href{http://doi.org/10.1140/epjc/s2005-02347-1}{{\bf C44}, 183} (2005),
  \href{https://arxiv.org/abs/hep-ex/0502007}{[hep-ex/0502007]}\relax
\mciteBstWouldAddEndPuncttrue
\mciteSetBstMidEndSepPunct{\mcitedefaultmidpunct}
{\mcitedefaultendpunct}{\mcitedefaultseppunct}\relax
\EndOfBibitem
\bibitem{Chatrchyan:2013txa}
S.~Chatrchyan {\em et~al.\/} (CMS),
  \href{http://doi.org/10.1140/epjc/s10052-013-2604-6}{Eur. Phys. J.}
  \href{http://doi.org/10.1140/epjc/s10052-013-2604-6}{{\bf C73}, 10, 2604}
  (2013), \href{https://arxiv.org/abs/1304.7498}{[arXiv:1304.7498]}\relax
\mciteBstWouldAddEndPuncttrue
\mciteSetBstMidEndSepPunct{\mcitedefaultmidpunct}
{\mcitedefaultendpunct}{\mcitedefaultseppunct}\relax
\EndOfBibitem
\bibitem{Abazov:2012lua}
V.~M. Abazov {\em et~al.\/} (D0),
  \href{http://doi.org/10.1016/j.physletb.2012.10.003}{Phys. Lett.}
  \href{http://doi.org/10.1016/j.physletb.2012.10.003}{{\bf B718}, 56} (2012),
  \href{https://arxiv.org/abs/1207.4957}{[arXiv:1207.4957]}\relax
\mciteBstWouldAddEndPuncttrue
\mciteSetBstMidEndSepPunct{\mcitedefaultmidpunct}
{\mcitedefaultendpunct}{\mcitedefaultseppunct}\relax
\EndOfBibitem
\bibitem{CMS:2014mna}
V.~Khachatryan {\em et~al.\/} (CMS),
  \href{http://doi.org/10.1140/epjc/s10052-015-3376-y}{Eur. Phys. J.}
  \href{http://doi.org/10.1140/epjc/s10052-015-3376-y}{{\bf C75}, 5, 186}
  (2015), \href{https://arxiv.org/abs/1412.1633}{[arXiv:1412.1633]}\relax
\mciteBstWouldAddEndPuncttrue
\mciteSetBstMidEndSepPunct{\mcitedefaultmidpunct}
{\mcitedefaultendpunct}{\mcitedefaultseppunct}\relax
\EndOfBibitem
\bibitem{Andreev:2016tgi}
V.~Andreev {\em et~al.\/} (H1),
  \href{http://doi.org/10.1140/epjc/s10052-017-4717-9}{Eur. Phys. J. C}
  \href{http://doi.org/10.1140/epjc/s10052-017-4717-9}{{\bf 77}, 4, 215}
  (2017), \href{https://arxiv.org/abs/1611.03421}{[arXiv:1611.03421]}\relax
\mciteBstWouldAddEndPuncttrue
\mciteSetBstMidEndSepPunct{\mcitedefaultmidpunct}
{\mcitedefaultendpunct}{\mcitedefaultseppunct}\relax
\EndOfBibitem
\bibitem{Haller:2018nnx}
J.~Haller {\em et~al.\/},
  \href{http://doi.org/10.1140/epjc/s10052-018-6131-3}{Eur. Phys. J.}
  \href{http://doi.org/10.1140/epjc/s10052-018-6131-3}{{\bf C78}, 8, 675}
  (2018), \href{https://arxiv.org/abs/1803.01853}{[arXiv:1803.01853]}\relax
\mciteBstWouldAddEndPuncttrue
\mciteSetBstMidEndSepPunct{\mcitedefaultmidpunct}
{\mcitedefaultendpunct}{\mcitedefaultseppunct}\relax
\EndOfBibitem
\bibitem{deBlas:2022hdk}
J.~de~Blas {\em et~al.\/},
  \href{http://doi.org/10.1103/PhysRevLett.129.271801}{Phys. Rev. Lett.}
  \href{http://doi.org/10.1103/PhysRevLett.129.271801}{{\bf 129}, 27, 271801}
  (2022), [Supplemental Material: Phys.Rev.Lett. 129, 271801 (2022)],
  \href{https://arxiv.org/abs/2204.04204}{[arXiv:2204.04204]}\relax
\mciteBstWouldAddEndPuncttrue
\mciteSetBstMidEndSepPunct{\mcitedefaultmidpunct}
{\mcitedefaultendpunct}{\mcitedefaultseppunct}\relax
\EndOfBibitem
\bibitem{deBlas:2021wap}
J.~de~Blas {\em et~al.\/},
  \href{http://doi.org/10.1103/PhysRevD.106.033003}{Phys. Rev. D}
  \href{http://doi.org/10.1103/PhysRevD.106.033003}{{\bf 106}, 3, 033003}
  (2022), \href{https://arxiv.org/abs/2112.07274}{[arXiv:2112.07274]}\relax
\mciteBstWouldAddEndPuncttrue
\mciteSetBstMidEndSepPunct{\mcitedefaultmidpunct}
{\mcitedefaultendpunct}{\mcitedefaultseppunct}\relax
\EndOfBibitem
\bibitem{CDF:2022hxs}
T.~Aaltonen {\em et~al.\/} (CDF),
  \href{http://doi.org/10.1126/science.abk1781}{Science}
  \href{http://doi.org/10.1126/science.abk1781}{{\bf 376}, 6589, 170}
  (2022)\relax
\mciteBstWouldAddEndPuncttrue
\mciteSetBstMidEndSepPunct{\mcitedefaultmidpunct}
{\mcitedefaultendpunct}{\mcitedefaultseppunct}\relax
\EndOfBibitem
\bibitem{ALEPH:2005ab}
S.~Schael {\em et~al.\/} (ALEPH, DELPHI, L3, OPAL, SLD, LEP Electroweak Working
  Group, SLD Electroweak Group, SLD Heavy Flavour Group),
  \href{http://doi.org/10.1016/j.physrep.2005.12.006}{Phys. Rept.}
  \href{http://doi.org/10.1016/j.physrep.2005.12.006}{{\bf 427}, 257} (2006),
  \href{https://arxiv.org/abs/hep-ex/0509008}{[hep-ex/0509008]}\relax
\mciteBstWouldAddEndPuncttrue
\mciteSetBstMidEndSepPunct{\mcitedefaultmidpunct}
{\mcitedefaultendpunct}{\mcitedefaultseppunct}\relax
\EndOfBibitem
\bibitem{DelDebbio:2021ryq}
L.~Del~Debbio and A.~Ramos,
  \href{http://doi.org/10.1016/j.physrep.2021.03.005}{Physics Reports}
  \href{http://doi.org/10.1016/j.physrep.2021.03.005}{{\bf 920}, 1} (2021),
  ISSN 0370-1573,
  \href{https://arxiv.org/abs/2101.04762}{[arXiv:2101.04762]}\relax
\mciteBstWouldAddEndPuncttrue
\mciteSetBstMidEndSepPunct{\mcitedefaultmidpunct}
{\mcitedefaultendpunct}{\mcitedefaultseppunct}\relax
\EndOfBibitem
\bibitem{FlavourLatticeAveragingGroup:2019iem}
S.~Aoki {\em et~al.\/} (Flavour Lattice Averaging Group),
  \href{http://doi.org/10.1140/epjc/s10052-019-7354-7}{Eur. Phys. J. C}
  \href{http://doi.org/10.1140/epjc/s10052-019-7354-7}{{\bf 80}, 2, 113}
  (2020), \href{https://arxiv.org/abs/1902.08191}{[arXiv:1902.08191]}\relax
\mciteBstWouldAddEndPuncttrue
\mciteSetBstMidEndSepPunct{\mcitedefaultmidpunct}
{\mcitedefaultendpunct}{\mcitedefaultseppunct}\relax
\EndOfBibitem
\bibitem{Bruno:2017gxd}
M.~Bruno {\em et~al.\/} (ALPHA),
  \href{http://doi.org/10.1103/PhysRevLett.119.102001}{Phys. Rev. Lett.}
  \href{http://doi.org/10.1103/PhysRevLett.119.102001}{{\bf 119}, 10, 102001}
  (2017), \href{https://arxiv.org/abs/1706.03821}{[arXiv:1706.03821]}\relax
\mciteBstWouldAddEndPuncttrue
\mciteSetBstMidEndSepPunct{\mcitedefaultmidpunct}
{\mcitedefaultendpunct}{\mcitedefaultseppunct}\relax
\EndOfBibitem
\bibitem{Aoki:2009tf}
S.~Aoki {\em et~al.\/} (PACS-CS),
  \href{http://doi.org/10.1088/1126-6708/2009/10/053}{JHEP}
  \href{http://doi.org/10.1088/1126-6708/2009/10/053}{{\bf 10}, 053} (2009),
  \href{https://arxiv.org/abs/0906.3906}{[arXiv:0906.3906]}\relax
\mciteBstWouldAddEndPuncttrue
\mciteSetBstMidEndSepPunct{\mcitedefaultmidpunct}
{\mcitedefaultendpunct}{\mcitedefaultseppunct}\relax
\EndOfBibitem
\bibitem{McNeile:2010ji}
C.~McNeile {\em et~al.\/},
  \href{http://doi.org/10.1103/PhysRevD.82.034512}{Phys. Rev.}
  \href{http://doi.org/10.1103/PhysRevD.82.034512}{{\bf D82}, 034512} (2010),
  \href{https://arxiv.org/abs/1004.4285}{[arXiv:1004.4285]}\relax
\mciteBstWouldAddEndPuncttrue
\mciteSetBstMidEndSepPunct{\mcitedefaultmidpunct}
{\mcitedefaultendpunct}{\mcitedefaultseppunct}\relax
\EndOfBibitem
\bibitem{Maltman:2008bx}
K.~Maltman {\em et~al.\/},
  \href{http://doi.org/10.1103/PhysRevD.78.114504}{Phys. Rev.}
  \href{http://doi.org/10.1103/PhysRevD.78.114504}{{\bf D78}, 114504} (2008),
  \href{https://arxiv.org/abs/0807.2020}{[arXiv:0807.2020]}\relax
\mciteBstWouldAddEndPuncttrue
\mciteSetBstMidEndSepPunct{\mcitedefaultmidpunct}
{\mcitedefaultendpunct}{\mcitedefaultseppunct}\relax
\EndOfBibitem
\bibitem{Chakraborty:2014aca}
B.~Chakraborty {\em et~al.\/},
  \href{http://doi.org/10.1103/PhysRevD.91.054508}{Phys. Rev.}
  \href{http://doi.org/10.1103/PhysRevD.91.054508}{{\bf D91}, 5, 054508}
  (2015), \href{https://arxiv.org/abs/1408.4169}{[arXiv:1408.4169]}\relax
\mciteBstWouldAddEndPuncttrue
\mciteSetBstMidEndSepPunct{\mcitedefaultmidpunct}
{\mcitedefaultendpunct}{\mcitedefaultseppunct}\relax
\EndOfBibitem
\bibitem{Ayala:2020odx}
C.~Ayala, X.~Lobregat and A.~Pineda,
  \href{http://doi.org/10.1007/JHEP09(2020)016}{JHEP}
  \href{http://doi.org/10.1007/JHEP09(2020)016}{{\bf 09}, 016} (2020),
  \href{https://arxiv.org/abs/2005.12301}{[arXiv:2005.12301]}\relax
\mciteBstWouldAddEndPuncttrue
\mciteSetBstMidEndSepPunct{\mcitedefaultmidpunct}
{\mcitedefaultendpunct}{\mcitedefaultseppunct}\relax
\EndOfBibitem
\bibitem{Bazavov:2019qoo}
A.~Bazavov {\em et~al.\/} (TUMQCD),
  \href{http://doi.org/10.1103/PhysRevD.100.114511}{Phys. Rev. D}
  \href{http://doi.org/10.1103/PhysRevD.100.114511}{{\bf 100}, 11, 114511}
  (2019), \href{https://arxiv.org/abs/1907.11747}{[arXiv:1907.11747]}\relax
\mciteBstWouldAddEndPuncttrue
\mciteSetBstMidEndSepPunct{\mcitedefaultmidpunct}
{\mcitedefaultendpunct}{\mcitedefaultseppunct}\relax
\EndOfBibitem
\bibitem{Cali:2020hrj}
S.~Cali {\em et~al.\/},
  \href{http://doi.org/10.1103/PhysRevLett.125.242002}{Phys. Rev. Lett.}
  \href{http://doi.org/10.1103/PhysRevLett.125.242002}{{\bf 125}, 242002}
  (2020), \href{https://arxiv.org/abs/2003.05781}{[arXiv:2003.05781]}\relax
\mciteBstWouldAddEndPuncttrue
\mciteSetBstMidEndSepPunct{\mcitedefaultmidpunct}
{\mcitedefaultendpunct}{\mcitedefaultseppunct}\relax
\EndOfBibitem
\bibitem{DallaBrida:2019mqg}
M.~Dalla~Brida {\em et~al.\/} (ALPHA),
  \href{http://doi.org/10.1016/j.physletb.2020.135571}{Phys. Lett. B}
  \href{http://doi.org/10.1016/j.physletb.2020.135571}{{\bf 807}, 135571}
  (2020), \href{https://arxiv.org/abs/1912.06001}{[arXiv:1912.06001]}\relax
\mciteBstWouldAddEndPuncttrue
\mciteSetBstMidEndSepPunct{\mcitedefaultmidpunct}
{\mcitedefaultendpunct}{\mcitedefaultseppunct}\relax
\EndOfBibitem
\end{mcitethebibliography}

%%% Local Variables:
%%% TeX-master: "qcd"
%%% End:

\end{bibunit}
\fi

% Generate index for draft mode
\ifdefined\isdraft
\clearpage
\renewcommand{\twocolumn}[1][]{
     \twocolumngrid
     #1
}
\printindex
\fi

\end{document}